\documentclass[a4paper,12pt]{article}

\usepackage{amsmath,amsthm,amsfonts,amssymb,mathtools,bbm,mathdots,tensor}

\usepackage{pdfsync}
\usepackage[sans]{dsfont}
\usepackage{graphicx,import}
\usepackage[latin1]{inputenc}
\usepackage{hyperref}
\usepackage{empheq}
\usepackage[all]{xy}
\usepackage{stmaryrd}
\usepackage{rotating}
\usepackage{slashed}
\usepackage{array, makecell} %
\usepackage{comment}
\usepackage{hhline}
\usepackage{cancel}
\usepackage{dirtytalk}
\usepackage{cite}
\usepackage{amssymb,amsmath,amsthm,mathbbol,mathrsfs,mathtools,slashed,mathrsfs}
\usepackage{stmaryrd}
\usepackage{enumerate}

\usepackage[dvipsnames]{xcolor}
\usepackage{marginnote}
\usepackage{longtable}
\usepackage{euscript}
\makeatletter

\usepackage[left=.8in,right=.8in,top=.8in,bottom=.8in]{geometry}                  

\usepackage{sectsty}
\allsectionsfont{\bfseries\sffamily} 

\DeclareFontFamily{U}{matha}{\hyphenchar\font45}
\DeclareFontShape{U}{matha}{m}{n}{
      <5> <6> <7> <8> <9> <10> gen * matha
      <10.95> matha10 <12> <14.4> <17.28> <20.74> <24.88> matha12
      }{}
\DeclareSymbolFont{matha}{U}{matha}{m}{n}
\DeclareMathSymbol{\oright}       {2}{matha}{"69}

\def\be#1\ee{\begin{align}#1\end{align}} 
\newcommand{\bea}{\begin{eqnarray}}
\newcommand{\eea}{\end{eqnarray}}
\newcommand{\bsa}{\begin{subequations} \begin{align}}
\newcommand{\esa}{\end{align}  \end{subequations} }

\def\acc{a}
\def\heq{\cong}
\newcommand{\cC}{{\mathcal C}}

\newcommand{\cN}{{\mathcal N}}
\newcommand{\cF}{{\mathcal F}}

\newcommand{\bq}{\boldsymbol{q}}
\newcommand{\bj}{\boldsymbol{j}}
\newcommand{\bC}{\boldsymbol{C}}
\newcommand{\ba}{\boldsymbol{a}}
\newcommand{\bfA}{\boldsymbol{A}}

\newcommand{\cL}{{\mathcal L}}
\newcommand{\e}{{\mathrm{e}}}

\renewcommand{\d}{{\mathrm{d}}}
\newcommand{\D}{{\mathrm{D}}}

\newcommand{\s}{{\mathrm{s}}}

\newcommand{\pp}{{\partial}}

\renewcommand{\bar}{\overline}

\renewcommand{\hat}{\widehat}

\renewcommand{\ker}{{\mathrm{ker}}}

\newcommand{\hxi}{{\hat\xi}{}}

\def\W{w}

\newcommand{\Sc}{\nu }

\newcommand{\bb}{{\beta}}

\renewcommand{\a}{b}

\newcommand{\Y}{\Upsilon}
\renewcommand{\O}{\Omega}
\def\rom{\mathring{\omega}}
\newcommand{\LL}{\cL}

\newcommand{\hn}{\hat{n}}

\newcommand{\qqand}{\qquad\text{and}\qquad}

\newcommand{\rd}{{\mathrm{d}}}

\newcommand{\scri}{{\mathscr{I}}}
\newcommand{\p}{{\partial}}
\newcommand{\po}{{\partial_\Omega}}
\renewcommand{\L}{{\mathcal L}}

\renewcommand{\d}{\mathrm{d}}
\renewcommand{\O}{{\Omega}}

\newcommand{\N}{{\mathcal N}}
\newcommand{\hN}{\hat{\mathcal N}}
\newcommand{\rN}{\mathring{\mathcal N}}
\newcommand{\rtN}{\dot{\mathcal N}}
\newcommand{\SN}{\mathring{\mathcal N}}
\def\mN{\mathsf{N}}
\def\rmN{\mathring{\mathsf{N}}}
\def\rtmN{\dot{\mathsf{N}}}
\def\hmN{\hat{\mathsf{N}}}

\newcommand{\hV}{\hat{\mathcal V}}
\newcommand{\rV}{\mathring{\mathcal V}}

\def\rmV{\mathring{\mathsf{V}}}

\newcommand{\T}{{\mathcal T}}
\newcommand{\hT}{\hat{\mathcal T}}
\newcommand{\rT}{\mathring{\mathcal T}}
\newcommand{\rtT}{\dot{\mathcal T}}
\newcommand{\sT}{\mathscr{T}}
\def\mT{\mathsf{T}}
\def\rmT{\mathring{\mathsf{T}}}
\def\Tm{T^{\mathsf{mat}}}
\def\tm{\mathscr{T}^{\mathsf{mat}}}
\def\Tm{T^{\mathsf{mat}}}

\newcommand{\hK}{\hat{\mathcal K}}
\def\mK{\mathsf{K}}
\def\bmK{\bar{\mathsf{K}}}
\def\rmK{\mathring{\mathsf{K}}}

\def\sA{\mathsf{a}}
\def\rA{\mathring{A}}
\def\kA{A_K}

\def\bA{\bar{A}}

\def\sJ{\mathring\iota}
\def\rJ{\mathring{J}}

\def\rY{\mathring{Y}}
\def\hY{\hat{Y}}

\def\hmN{\hat{\mathsf{N}}}

\newcommand{\K}{{\mathcal K}}

\newcommand{\bZ}{\boldsymbol{Z}}
\newcommand{\btheta}{\bar{\theta}}
\newcommand{\bvartheta}{\boldsymbol{\vartheta}}
\newcommand{\bbeta}{\boldsymbol{\beta}{}}

\newcommand{\bftheta}{\boldsymbol{\theta}{}}

\newcommand{\bTheta}{\bar{\Theta}{}}

\newcommand{\bgamma}{\boldsymbol{\gamma}{}}
\newcommand{\bL}{\mathbf{L}{}}
\newcommand{\bE}{\mathbf{E}{}}

\newcommand{\bell}{\boldsymbol{\ell}{}}
\def\s{\sigma}
\def\pa{\partial}
\def\U{\Upsilon}
\def\k{k }

\newcommand{\Snabla}{\mathring{\nabla}}
\def\Sn{\mathring{\nabla}}
\def\hn{\hat{\nabla}}

\newcommand{\diff}{{\mathfrak{diff}}}

\newcommand{\tl}{\widetilde}

\newcommand{\SCRIeq}{\,\stackrel{\scri}{=}\,}

\newcommand{\scrieq}{\,\stackrel{\scri}{=}\,}

\newcommand{\E}{\mathcal{E}}

\newcommand{\lbr}{\llbracket}
\newcommand{\rbr}{\rrbracket}

\newcommand{\EOMeq}{\,\hat=\,}

\usepackage{authblk}

\title{\sffamily  Renormalization of conformal infinity  \\ as a stretched horizon}

\author[1]{\sffamily Laurent Freidel}
\author[1,2]{\sffamily Aldo Riello}

\affil[1]{\small \textit{Perimeter Institute for Theoretical Physics, Waterloo, ON, Canada}}
\affil[2]{\small \textit{Dpt. of Applied Mathematics, University of Waterloo, Waterloo, ON, Canada}}

\date{\today}

\begin{document}
\maketitle

\abstract{
In this paper, we provide a  comprehensive study of asymptotically flat spacetime in even dimensions $d\geq 4$.
We analyze the most general boundary condition and asymptotic  symmetry compatible with Penrose's definition of asymptotic null infinity $\scri$ through conformal compactification.
Following Penrose's prescription and using  a minimal version of the Bondi-Sachs gauge, we show that $\scri$ is naturally equipped with a Carrollian stress tensor whose radial derivative defines the asymptotic Weyl tensor. This analysis describes asymptotic infinity as a stretched horizon in the conformally compactified spacetime. We establish that charge aspects conservation can be written as Carrollian Bianchi identities for the asymptotic Weyl tensor. We then provide a covariant renormalization for the asymptotic symplectic potential, which results in a finite symplectic flux and asymptotic charges.  The renormalization scheme works even in the presence of logarithmic anomalies.
}

\begin{center}
	\rule{8cm}{0.4pt}
\end{center}

\newpage
\tableofcontents

\section{Introduction}

Some of the most exciting questions in studying classical and quantum gravity are: What is radiation? Can we define the gravitational S-matrix? What are the asymptotic symmetries of gravity? Is gravity holographic?
Answering these questions requires understanding the gravitational field's behaviour when approaching infinity, especially null infinity for asymptotically flat spacetimes. 

The celebrated works of Bondi,  van der Burg, Metzner  \cite{Bondi:1960jsa, Bondi62}  and Sachs \cite{Sachs:1961zz, Sachs:1962wk} 
first established that radiation is well defined and that there exists an infinite-dimensional symmetry group: the BMS group governing the asymptotic behaviour 
of gravitational observables. 

One of the key subsequent developments is the introduction by Penrose of the concept of conformal  compactification of spacetime, 
obtained  by a 
conformal transformation  of the physical spacetime with conformal factor $\O=1/r$
vanishing at  infinity
\cite{Penrose:1962ij,Penrose:1965am,PenroseRindler2}. 
This fundamental picture, which allows understanding null infinity, denoted $\scri$, as a  finite null boundary in the conformal spacetime, was further developed by Geroch and Friedrich \cite{Geroch:1977jn, friedrich2002conformal}. 
This idea, which is central to our work, allows us to comprehend asymptotic infinity in terms of the geometry of embedded null surfaces in the \emph{conformal} spacetime,  and hence in terms of \emph{Carrollian} geometry \cite{LevyLeblond1965, Duval:2014uva}.   

Since these foundational works, many essential studies on asymptotic infinity have taken place, 
too many for us to give an exhaustive account. We refer to 
\cite{Geroch:1977jn, Adamo:2009vu, Madler:2016xju} for enlightening reviews and references on the early history.

Three types of developments are especially relevant for us: 
First, the proper definition of asymptotic null infinity and the most general asymptotic behaviour has been revisited.
 In most of the literature, the conformal metric is assumed to be smooth near $\scri$ (or at least $C^3$, in $d=4$ \cite{Penrose:1965am}); this assumption leads to the celebrated ``Peeling theorems" \cite{Sachs:1961zz,Sachs:1962wk,NP62,Penrose:1962ij,Penrose:1965am}.
Friedrich\cite{friedrich1983cauchy} and  Winicour \cite{winicour1985logarithmic} noted that this hypothesis may be too strong. Chrusciel et al.  \cite{andersson1993hyperboloidal, chrusciel1993gravitational}  generalized the Bondi-Sachs formalism to 
a polyhomogeneous expansion  of the metric  that admits logarithmic terms $ \O^{n} \log^k \O$ (see \cite{Friedrich:2017cjg, Tafel:2021bxa,kehrberger2021case} for recent discussion of this issue). 

Second, the foundational work of Ashtekar and Streubel \cite{Ashtekar:1978zz, Ashtekar:1981sf, Ashtekar:1981bq, Ashtekar:1990gc} established that  $\scri$, defined in terms of the  Bondi-Sachs boundary conditions, is equipped with a canonical symplectic structure and that the BMS charges are canonical generators of symmetry for this symplectic structure \cite{Wald:1999wa}. It also emphasized that the radiation field can be understood as defining a connection on $\scri$, now understood to be a Carrollian connection in modern terminology \cite{Ciambelli:2018xat, Chandrasekaran:2021hxc, Freidel:2022vjq, Mars:2022gsa, Freidel:2023bnj}.

Third, there is the seminal work of Strominger and collaborators \cite{Strominger:2013jfa, Strominger:2014pwa} who established that the leading Weinberg soft theorems \cite{Weinberg:1965nx} can be understood as the expression of Ward identities for the BMS group. 
This led to the proposal of celestial holography and a wealth of new developments. In particular, it was shown that the validity of the subleading soft theorem requires an extension of the BMS group that includes superrotations \cite{Pasterski:2015tva, Pasterski:2016qvg, Kapec:2016jld, Donnay:2020guq} and that the subsubleading ones need an even bigger extension including higher spin symmetry charges \cite{Cachazo:2014fwa, Guevara:2021abz, Strominger:2021lvk, Freidel:2021ytz} (see \cite{Strominger:2017zoo, Raclariu:2021zjz, Pasterski:2021rjz, Donnay:2023mrd} for reviews).

These results revitalized the study of asymptotic infinity, its symmetries and asymptotic symplectic structure. In particular, they required to relax the usual Bondi-Sachs boundary conditions in order to include superrotation symmetries \cite{Barnich:2010eb, Barnich:2016lyg, Campiglia:2015yka, Campiglia:2020qvc}. Including these necessitate to extend the gravitational phase space and devise a symplectic renormalization procedure  e.g. to make the symplectic potential and associated charges finite \cite{Compere:2008us, Flanagan:2015pxa, Compere:2018ylh, Compere:2020lrt, Freidel:2021fxf, Chandrasekaran:2021vyu}.  

Another progress alongside the celestial holography project was the development of a notion of Carrollian holography 
and an understanding of the relationship between the two complementary descriptions 
\cite{Ciambelli:2018wre,  Ciambelli:2019lap,  Ciambelli:2018ojf, Donnay:2022aba, Donnay:2022wvx, Bagchi:2022emh}.

A common theme to all these advancements is the notion of Carrollian geometry. 
The Carrollian perspective on null physics was first revealed in \cite{LevyLeblond1965, Gupta1966, Henneaux1979a, Ashtekar:1981bq, Duval:2014uva}.
Besides its appearance in flat space holography,  its importance in studying the geometry of horizons and null surfaces has been revealed in 
\cite{Penna:2018gfx, Donnay:2019jiz}. Meanwhile, the development of the symplectic geometry of null horizons and the description of Einstein's dynamics as charge conservation laws were initiated in \cite{Parattu:2015gga, Donnay:2015abr,  Donnay:2016ejv, Parattu:2016trq, Lehner:2016vdi, Hopfmuller:2016scf} 
and then further developed in a wealth of recent works that have established a complete understanding of the canonical symmetries, the associated corner charges and the covariance property of the symplectic structure 
\cite{Wieland:2017cmf, Wieland:2017zkf, Hopfmuller:2018fni, Chandrasekaran:2018aop, Oliveri:2019gvm, Adami:2020ugu, Adami:2021nnf, Ashtekar:2021kqj, 
Chandrasekaran:2020wwn, Chandrasekaran:2021hxc, Freidel:2022vjq, Odak:2022ndm, Sheikh-Jabbari:2022mqi, Ciambelli:2023mir, Chandrasekaran:2023vzb, Odak:2023pga}. 
One of the main outcomes of these progresses has been the construction of a null Carrollian connection and an associated 
Carrollian energy momentum-tensor \cite{Chandrasekaran:2021hxc}, which generalizes the Brow-York construction \cite{Brown:1992br} to null surfaces.
The construction of a Carrollian connection and energy-momentum tensor and the realization that Einstein's equation are conservation laws for the Carrollian  stress tensor has been extended to stretched horizons in \cite{Jai-akson:2022gwg, Freidel:2022vjq} which introduced the notion of the stretched Carrollian geometry associated with near null surface.

This connects to one of the main results of our paper: The fact that null infinity can be understood in terms of the geometry of a stretched horizon.
In particular, we show that an asymptotic Carrollian stress tensor\footnote{We call it asymptotic stress tensor for short.} exists whose conservation equations are equivalent to the evolution equations of the mass, angular  momentum, and spin-$2$ charge aspects along a near-infinity hypersurface. This nicely merges the study of asymptotic infinity with the study of finite null surfaces, reinforcing the original picture introduced by Penrose.

In our work, following the four-dimensional seminal study of Geiller and Zwikel \cite{Geiller:2022vto},  we also analyze the most general set of asymptotic boundary conditions, and symmetries, compatible with the definition of asymptotic infinity devised by Penrose. 
In particular, we find an equivalence between the finiteness of the Weyl tensor at $\scri$ and the conformal metric being of polyhomogenous order $k=d-1$ in $\O$. Moreover,
in order to ensure that the asymptotic symplectic potential is well-defined, we use a general and covariant renormalization scheme that works in any dimension and even in the presence of logarithmic divergences.
This general scheme was first introduced by us and Hopfm\"ueller in \cite{Freidel:2019ohg, Hopfmueller:2020yqj}. For gravity, a similar procedure was developed by McNees and Zwikel in \cite{McNees:2023tus};  see also \cite{Geiller:2024amx}  where a renormalization of the variational charges was achieved for gravity in $d=4$  with the boundary conditions of \cite{Geiller:2022vto}. In $d=4$ our renormalization scheme works in the presence of the logarithmic anomaly first identified by Winicour \cite{winicour1985logarithmic}.
In $d=6$ and higher we identify two types of anomalies, that we call Coulombic and radiative, and show that the vanishing of the radiative anomaly, which is a necessary condition for the finiteness of the asymptotic Weyl tensor, allows our renormalization scheme to work.

The  renormalization of the symplectic structure is obtained in two steps:  First, using the Carrollian structure, we choose a boundary Lagrangian that eliminates the leading divergence and rotates the Einstein-Hilbert symplectic potential into a canonical form where the Carrollian geometry elements are the configuration variables and the asymptotic stress tensor components are the conjugated momenta variables. Then, the renormalized symplectic potential at $\scri$ is obtained simply as the finite element in the $\O$ expansion of the canonical symplectic potential.

The study of asymptotic infinity in  dimensions higher than $4$,  the construction of BMS  charges and memory effects have been discussed in 
\cite{Hollands:2003xp, Godazgar:2012zq, Hollands:2016oma, Kapec:2017gsg, Pate:2017fgt, Cameron:2021fhd, Capone:2021ouo, Fuentealba:2022yqt, Capone:2023roc, Colferai:2020rte}. 
It was also realized that odd and even dimensions behave differently and that in order to allow for radiation in odd dimensions, one needs to relax the
analyticity requirement in the $\O$ expansion \cite{Hollands:2003xp, Hollands:2004ac, Tanabe:2011es}. For this reason, we restrict our analysis to even dimensions.
In the original works, it was argued that the asymptotic symmetry group does not get extended beyond Poincar\'e and that there is no memory effect in $d>4$ \cite{Hollands:2003xp, Hollands:2004ac, Garfinkle:2017fre}, in seeming contradiction with the presence of soft theorems 
\cite{Kapec:2015vwa, Pate:2017fgt }. The resolution of the memory puzzle comes from the fact that the memory effects happen at Coulombic order $ \O^{(d-3)}$ and not at the radiative order  $\O^{\frac{(d-2)}{2}}$ \cite{Pate:2017fgt }. The resolution of the absence of the BMS symmetries puzzle relies on allowing boundary conditions which are not as restrictive \cite{Kapec:2015vwa} as in the original works.
Recently, Capone, Mitra, Poole and Tomova have performed in \cite{Capone:2023roc} a comprehensive analysis of symplectic renormalization in six dimensions and constructed a phase space carrying a non-trivial representation of BMS symmetries. Our analysis, which includes Penrose's conformal perspective, allows a more general boundary condition and includes logarithms, extends theirs.

The last main result we obtain is the connection between the radial derivative of the asymptotic  stress tensor and the pull-back of the rescaled Weyl tensor at $\scri$. We also show that this rescaled Weyl tensor satisfies Carrollian Bianchi identities that express the conservation laws of energy and angular momenta at $\scri$.
This result provides an asymptotic  Carrolllian perspective on the foundational results of Geld-Held-Penrose (GHP), who gave, in 4d, a description of asymptotic evolution in terms of the Bianchi identities. An extension of GHP formalism was achieved in \cite{Durkee:2010xq} for dimensions greater than $4$, and our results should also provide a Carrollian perspective on it. Let us mention that this part of our work connects naturally with the work of Prabhu et al., who have developed the conformal perspective of charge conservation in dimension $4$ and analyzed in great detail the charge aspects conservation at null and spacelike infinity in terms of the Bianchi identities \cite{Grant:2021sxk, Prabhu:2021cgk, Mohamed:2023jwv}. 

\subsection{Plan}
In section \ref{sec:Penrose}, we describe the Penrose compactification scheme in any dimension and the Penrose boundary conditions (Pbc) that define asymptotically flat spacetime \`a la Penrose. We show that a conformal  stress tensor exists in the conformal spacetime, which sources the  Einstein's equation in the conformally compactified spacetime. We show that the Pbc implies the existence of a  $(2,1)$ tensor $Y_{ab}{}^c$, which encodes the rescaling of the Weyl tensor contracted with the normal, and we present the conformal dynamics of Einstein's equations as Bianchi identities for a rescaled  Weyl tensor.

In section \ref{sec:Bondi}, we present the Bondi-Sachs gauge conditions and introduce the notion of stretched Carrollian geometry on each constant $\O$ slice.
We describe elements of the Carrollian connection tangent to the slices and describe its associated shear tensor and boost connection one form. We also describe the notion of asymptotic diffeomorphisms preserving the Bondi-Sachs gauge.

In section \ref{sec:radexp}, we give the construction of the conformal  stress tensor and present the Einstein's equations that determine the metric radial evolution along $\O$.

In section \ref{sec:radex}, we present the polyhomogeneous expansion of the metric coefficients in all dimensions. And we explicitly construct the first three orders in the series. We discuss the relation of our framework with previously studied asymptotic boundary conditions.

In section \ref{sec:Weyl}, we study the evolution equations of the charge aspect along the hypersurfaces of constant $\O$.
We first give the conservation for the asymptotic asymptotic   stress tensor $\hT_a{}^b$ obtained by projection along $\scri$ of the conformal stress tensor. This conservation law is extracted from a similar conservation law for the Carrollian stress tensor associated with stretched horizons and we give an independent derivation of this key result. We then relate the radial derivative of the asymptotic  stress tensor   to the projection along a null vector $K^a$ transverse to the hypersurface
of  $\hY_{ab}{}^c$.
Then, we derive the conservation laws for  $\hY_{ab}{}^c$ from the Bianchi identities.

In section \ref{sec:SP}, we present a covariant renormalization scheme that allows the construction of a finite symplectic potential in all dimensions and even in the presence of a polyhomogeneous expansion. Along the way, we show that the asymptotic symplectic potential admits a description as the symplectic potential of a stretched Carrollian fluid. We then perform the renormalization explicitly in dimension $4$ and compare it with previous results. One of the main result of this paper is the expression  \eqref{rsymp} for the renormalized symplectic potential. We finally discuss the renormalization of symmetry charges.

\subsection{Notations}

Although the details of the constructions will be introduced in due time, we collect here the most relevant notations, together with a short (and loose) description of their meaning. 

We consider a physical spacetime $(\tl M, \tl g_{ab})$ solution of the Einstein's equation with vanishing cosmological constant,
\be
\Lambda=0.
\ee
We assume that $(\tl M, \tl g_{ab})$ admits an asymptotically flat conformal compactification $(M,g)$ with
\be
g_{ab} = \O^2 \tl g_{ab} 
\ee
for a positive scalar function $\O$ on $M$ that vanishes only at 
\be
\scri := \pp M.
\ee
We also assume that (in a neighborhood of $\scri$) the scalar function $\O$ defines a \emph{timelike} foliation of $M$, denoted $\Sigma_\O$, which limits to the null surface $\scri := \Sigma_{\O \to 0}$. 
Note that although we will keep this restriction throughout the paper it is  only necessary for the fluid-gravity duality. The advantage of the formalism we develop here is that it applies to asymptotic foliations of any causal type. 

Alongside $\O$, we assume that (in a neighborhood of $\scri$) a second scalar function $u$ is given such that the codimension-2 level surfaces $\Sigma_{(u,\O)}$ are spacelike spheres. 
 This restricts the topology of $\scri$ to be given by (two copies of) $S^{d-2} \times \mathbb{R}$.

We then have codimension 1 and 2 projectors $h_a{}^b$ and $\gamma_a{}^b$ on the tangent and cotangent spaces of $\Sigma_\O$ and $\Sigma_{(u,\O)}$ respectively. We then denote the induce metrics
\be
h_{ab} := h_a{}^c h_b{}^d g_{cd}, \quad \gamma_{ab}:= \gamma_a{}^c\gamma_b{}^d g_{cd}.
\ee

We introduce the conormal to the $\Sigma_\O$ foliation,
\be
N_a := \pp_a \O,
\ee
and $K^a $ the null vector transverse to $\Sigma_\O$ and such that $K^a N_a = 1$. (We will choose a gauge in which $K = \pp_\O$ and $K_a \propto du$.) We also introduce $V^a = N^a h_a{}^b$ and assume that it is a \emph{causal} vector (i.e. timelike everywhere except that on $\scri$ where it is null and equal to $N^a$). Whence
\be
h_a{}^b = \delta_a{}^b - N_a K^b,\quad \gamma_a{}^b = h_a{}^b - K_a V^b.
\ee
We use $(a,b,\dots)$, $(i,j,\dots)$, $(A,B,\dots)$ as abstract indices to denote tensors (\emph{1}) over $(M,g_{ab})$, (\emph{2}) in the image of the projector $h_a{}^b$,  and (\emph{3}) in the image of $\gamma_a{}^b$, respectively.

At $\scri$, it is convenient to introduce rescaled version of $N^a$ (or $V^a$) and $K_a$, which at $\O$ are both null, denoted $\ell^a$ and $k_a$ respectively, such that
\be
\underline{k} = du , \quad \ell^a k_a = 1.
\ee
More explicitly, in $(d-2)+2$ coordinates where the metric takes the form
\be
\d s^2 = 2 e^\beta \d u (\d \O - \O \Phi \d u) + \gamma_{AB}( \d \sigma^A - \Upsilon^A \d u) (\d \sigma^B - \Upsilon^B \d u),
\ee
the normal vectors to the sphere $S_{(\O,u)}$ read:
\be
\begin{cases}
K = \pp_\O,\\
N = e^{-\beta} (\pp_u + \Upsilon^A \pp_A + 2\O \Phi \pp_\O),\\
L = e^{-\beta} (\pp_u + \Upsilon^A \pp_A + \O \Phi \pp_\O),\\
V = e^{-\beta} (\pp_u + \Upsilon^A \pp_A).
\end{cases}
\ee
The vector $V$ is tangent to the hypersurfaces $\Sigma_\O$; its norm-squared is $V^2 = -2\Omega \Sc$ with $\Sc := e^{-\beta} \Phi $.  Accordingly $V$ is null on scri and causal off $\scri$ when $\Sc  > 0$, which is a restriction we impose throughout.
The vectors $(L,K)$ are null everywhere and normalized w.r.t. $g_{ab}$ so that $L\cdot K = 1$.   
At $\scri$, $N^a$, $L^a$, and $N^a$ are all equal to each other. There, it is convenient to introduce a rescaled version of these vectors, denoted $\ell^a$ and given by 
\be
\ell := e^\beta N |_\scri = e^\beta V |_\scri = e^\beta L |_\scri = \pp_u + U^A \pp_A, \quad U^A := \Upsilon^A|_\scri .
\ee
The corresponding normal forms to the sphere $S_{(\O,u)}$ are
\be
N_a = \nabla_a \O,
\quad
K_a = e^\a \nabla_a u,
\quad
V_a = N_a - 2\O \nu K_a,\quad
L_a = N_a - \O \nu K_a.
\ee

We denote with a $\widehat\bullet$ codimension-1 projections of tensors via $h_a{}^b$, e.g. $\hN_i{}^j := h_i{}^a \cN_a{}^b h_b{}^j$, and with a $\mathring\bullet$ tensors codimension-2 projections via $\gamma_a{}^b$, e.g. $\mathring{\cN}{}_A{}^B := \gamma_A{}^a \cN_a{}^b \gamma_b{}^B$. Finally, we also denote $\mathring\cN := \mathring\cN_A{}^A$ 
its trace. $\dot\cN$ denote the traceless tensor with component 
$\dot\cN_A{}^B := \cN_{\langle A}{}^{B \rangle}$, where the bracket notation denotes  
 the symmetrized, traceless, part of a codimension-2 tensor:
\be
X_{\langle A}{}^{B \rangle} := \tfrac12 (X_A{}^B+ X^B{}_A) - \frac{1}{(d-2)} X_C{}^C \delta_A{}^B.
\ee
If $X_{AB}=X_{BA}$ is symmetric we have that $ {X}_{\langle A}{}^{B \rangle}  =X_{\langle A C \rangle} \gamma^{CB}= \gamma^{BC}X_{\langle  C A \rangle} $. 
If traces are specifically taken with respect to the (asymptotic) metric $q_{AB} := \gamma_{AB}|_{\O=0}$ (instead of $\gamma_{AB}$ itself), this will be denoted by a $q$ sub-index, as in $X_{\langle AB\rangle_q}$.

Associated to the four metrics $(\tl g_{ab}, g_{ab}, h_{ij}, \gamma_{AB})$ we have four different  covariant derivatives: $(\tl\nabla_a, \nabla_a , \hn_i, \Sn_A)$. 
$(\tl\nabla_a, \nabla_a)$ are Levi-Civita connection, while $\hn_i $ is the codimension $1$ rigged connection and $\Sn_A$ the codimension $2$ sphere connection.
They are related as:
\be
\hn_i X^j = h_i{}^a (\nabla_a X^b)h_b{}^j, \quad 
\Sn_A Y^B = \gamma_A{}^a (\nabla_a Y^b) \gamma_b{}^B,
\ee
where $X^b = X^a h_a{}^b$ and $Y^b = Y^a\gamma_a{}^b$.
One exception to this notational rule concerns the asymptotic Carrollian  stress tensor $\hT_i{}^j :=\hN\delta_i^j- \hN_i{}^j$ introduced in equation \eqref{eq:projectedCSET}.

We  use shortcut notations for contractions $\omega_V := \omega_a V^a$, $\xi_K := \xi^a K_a $ and $\xi_N := \xi^a N_a$. We also use the single and double-dot notation to denote contractions, e.g. $\bTheta\!:\!\bTheta = \bTheta_A{}^B \bTheta_B{}^A$ and
$(N \!\cdot\! \bTheta)_A{}^B = N_A{}^C \bTheta_C{}^A$. 

We denote $\heq$ an equality taken after imposing the Einstein's equation while $\SCRIeq$ denotes an equality taken as one pulls back to $\scri = \Sigma_{\O=0}$.

Note that we differentiate the notation for the leading order of each metric coefficient.
The Levi-Civita covariant derivative associated to $q_{AB}$ is denoted:
\be
D_A :\SCRIeq \Sn_A.
\ee

Finally, we note that we  define different stress tensor: The Conformal stress tensor denoted $\T_{ab}=\cN g_{ab}-\cN_{AB}$ sources the Einstein's tensor of the conformal frame metric; it is a purely geometrical quantity. Its codimension-1 counterpart, denoted $\hT_i{}^j :=\hN\delta_i^j- \hN_i{}^j$, is the asymptotic stress tensor. The asymptotic stress tensor can in turn be understood as the subleading component  of the Carrollian fluid stress tensor denoted 
$
\mT_i{}^j.
$
Finally, we denote the matter stress tensor as $\tl T^{\mathsf{mat}}_{ab} =\O^{d-3} \Tm_{ab}$. In most of the article the matter stress tensor will be set to zero.

\section{Conformal infinity \`a la Penrose}\label{sec:Penrose}

Consider an  even-dimensional (see below) smooth spacetime manifold $\tl M$ of dimension $d\geq 4$  with  Lorentzian metric $\tl g_{ab}$ satisfying the Einstein equations with vanishing cosmological constant, $\Lambda=0$.

The manifold $(\tl M, \tl g_{ab})$ is said \textit{asymptotically simple}  \cite{Penrose:1962ij, Penrose:1965am, PenroseRindler2, Geroch:1977jn, frauendiener2004conformal} if it admits a notion of conformal infinity, i.e. a conformal compactification of $(M,g_{ab})$.
This means that there exists a differentiable Lorentzian manifold $(M,g_{ab})$ with boundary,
\be
\scri :=\pp M,
\ee
and a positive (smooth) scalar field $\Omega$ on $M$ such that the following  three conditions are satisfied:
\begin{enumerate}[(i)]
\item[(0)]  $\scri$ has two connected components, each with topology $S^{d-2}\times \mathbb{R}$;
\item There exists a diffeomorphism from $\tl M$ to the interior of $M$, $\mathring{M} := M\setminus \scri$, that we will  use to  implicitly identify the two manifolds  from now on; 
\item The metrics on $\tl M$  and $\mathring M$ are related by
    \be
    g_{ab} = \Omega^2\tl g_{ab};
    \ee
\item At $\scri$, $\Omega = 0$ and the ingoing co-normal
    \be\label{eq:Na}
    N_a := \nabla_a \Omega
    \ee
    does not vanish.
    \item   The metric $g_{ab}$ is differentiable of class  $d-1$, $ g_{ab}\in C_{(d-1)}(M)$;
      
    \item  When matter is present, the matter energy-momentum tensor is at most of order $\O^{d-3}$ at $\scri$,  i.e. $\tl T^{\mathrm{mat}}_{ab} = \O^{d-3} T^{\mathrm{mat}}_{ab} $, for a  $T^{\mathrm{mat}}_{ab}$ that is finite at $\scri$.
\end{enumerate}

 The seemingly technical, assumption (iv)  means that  all geometrical quantities expressed in the ``conformal frame'' $(M,g_{ab})$ can be expanded off of $\scri$ in powers of $\Omega$ up to and including order $(d-1)$.
 It has important physical consequences: which we briefly discuss here to argue that it can be readily replaced by condition (iv$'$) below, which is more general.

Assumption (iv) was shown by Penrose and Geroch \cite{Penrose:1980yx, Geroch:1977jn} to imply, in $d=4$, the peeling theorem, which implies that the Weyl tensor vanishes at $\scri$ (see sec \ref{sec:Pbc}).  In dimension $d \geq  5 $, this assumption  implies the unwanted conclusion that there is no radiation.
This is why we relax condition (iv) into condition (iv$'$) below. 
 
In odd dimensions, the condition (iv$'$)  still implies that no radiation can be present \cite{Hollands:2004ac}, but to include radiation we would have to consider an expansion for metric in terms of $\sqrt{\O}$, which is beyond the scope of our paper. This is why we assume, from now on, that $d$ is \emph{even} and strictly larger than 2.\footnote{Similarly, in an asymptotically flat spacetime with a sufficiently differentiable electromagnetic field, the same condition of $d$ even and larger than $4$ must be required for the existence of electromagnetic radiation. See \cite{Freidel:2019ohg}  (especially eq. (41e) (45) and (49)) where the analysis is performed in $d >4$ because electromagnetism is conformal in $d=4$. Similarly, in odd dimensions, one should expand in powers of $\sqrt{\Omega}$ to have non-trivial radiation \cite{Tanabe:2011es}.
\label{fnt:evenD} }

Related to the peeling theorem \cite{PenroseRindler2, Sachs:1961zz, Sachs:1962wk, NP62, Penrose:1962ij, Penrose:1965am}, condition (iv) was first argued to be too restrictive by Winicour \cite{winicour1985logarithmic} and Chrusciel et al. \cite{chrusciel1993gravitational} (see \cite{Friedrich:2017cjg} for an interesting historical perspective on this issue). Indeed, in dimension $4$, it is physical to allow, in the expansion of the metric, terms involving logarithms starting at order $\O^3\ln\O$. In this case, $g_{ab}$ is said \emph{polyhomogeneous} of class $3$ and denote $g_{ab} \in C^{\mathrm{poly}}_3$ (see section \ref{sec:radex} for a precise definition). The corresponding spacetimes are said to be  \emph{logarithmically simple.}

In even dimensions higher than 4, two possible logarithmic behaviours are possible: either the first logarithm in the metric expansions is due to \emph{Coulombic} fields appearing at order $\O^{d-1}\ln\O$, in which case $g_{ab}\in C^{\mathrm{poly}}_{d-1}$ and we say that the spacetime is \emph{logarithmically simple}; or the first logarithm in the metric expansion is due to radiation modes appearing at order $\O^{\frac{d-2}{2}} \ln \O$. This is the most general case for even $d\geq 6$.\footnote{Due to a technicality explained in detail in section \ref{subsec:radexp}, such radiative logarithms are not present in $d=4$. hence in dimension $4$ there is only one type of possible logarithm at order $\O^{3} \ln \O$,  first studied by Winicour \cite{winicour1985logarithmic}.}  
We will show in section \ref{sec:polyexp} that the vanishing of the  radiative logarithm term is \emph{equivalent} to the condition of logarithmic simplicity (iv$'$).
We will also see, in section \ref{sec:Weyl}, that if such radiative logarithms show up, i.e. if the spacetime fails to be logarithmically simple, then the Weyl tensor does not admit a finite limit at $\scri$ when $d=6$. 
So, only logarithmically simple spacetimes are conformally finite in $d=6$. In the following, we will restrict our study of asymptotic renormalization  to logarithmically simple spacetimes in even dimensions.

Condition (v) prescribes the asymptotic behaviour of the matter stress tensor (see e.g \cite{Fernandez-Alvarez:2021zmp, Bieri:2020pee}). For increasing $d$, it requires a stronger decay of the matter energy-momentum tensor near infinity.
 Here we note that the stress tensor $\tilde T_{ab}^{\mathsf{mat}}$ of the Maxwell field in $d=4$, which is conformally invariant, is of order $\O^2$ at $\scri$ (this amounts to asking that the Maxwell field strength $F_{ab}$ has a finite limit at $\scri$ \cite{Geroch:1973am,BongaPrabhu2020}).\footnote{In $d=4$, conformal, and massless non-conformal, scalar fields yield the same asymptotic behaviour for $\tilde T_{ab}$ as discussed by Geroch in \cite{Geroch:1973am}. Note: Geroch uses a different numerical convention for the Schouten tensor, and $L_{ab}|_\text{Geroch} = 2\O^3\sT^\mathsf{mat}_{ab}|_\text{here}$ which in turn scales like $\O^2\tl{T}^\mathsf{mat}_{ab}$ (cf. equations \eqref{sTdef} and \eqref{eq:sT=sTmat}). Therefore, his result $L_{ab}=O( \O^4)$ means $\tl{T}^\mathsf{mat}_{ab} =O(\O^2)$.} Our requirement is one-order weaker in $d=4$, and it is taken to scale linearly with $d$ from there.

Finally, let us note that in dimensions $4$ and $6$ it is known that imposing stationarity of the initial data at spacelike infinity $\iota_0$ implies asymptotic simplicity
\cite{Blanchet:1986dk,friedrich2004smoothness,acena2011conformal,Capone:2023roc}, but it is not known whether the reverse statement holds or not.
Another relevant question is whether, in some appropriate sense, asymptotically simple spacetimes are dense in a set of physically relevant asymptotically flat solutions \cite{Friedrich:2017cjg}. In this work, we forgo these questions and simply propose a renormalization scheme that includes logarithmically simple spacetimes.

Henceforth, we relax assumption (iv) and replace it with:

\begin{enumerate}[(i)]
\item[(iv$'$)] $g_{ab} \in  C^{\mathrm{poly}}_{d-1}$.
\end{enumerate}

A spacetime satisfying assumptions (0 , i, ii, iii, iv$'$, v, vi) is said \emph{logarithmically simple}.  In Section \ref{subsec:Weyl} will prove that in a logarithmically simple spacetime the Weyl tensor is guaranteed to have a finite limit at $\scri$.

In the following, we refer to $(\tl M, \tl g_{ab})$ as the {\it physical frame}, and to $(M,g_{ab})$ as the {\it conformal frame}. 
The advantage of the conformal frame is that $\scri$ appears as a ``finite'' boundary located at $\Omega=0$, which is null  if the cosmological constant vanishes.

We adopt the convention that tilded tensor refer to the physical frame and non-tilded ones refer to the conformal frame. All contraction of non-tilded tensorial indices is performed in the conformal frame, i.e. by means of $g_{ab}$ and its inverse. Notice that $g^{ab} = \Omega^{-2}\tl g^{ab}$.

To conclude, we note that our results hold in the Bondi-Sachs gauge (section \ref{sec:Bondi}) where the conformal metric $g_{ab}$ takes the form
\be
\d s ^2 = 2 e^\bb \d u ( \d \Omega - \Omega \Phi \d u) + \gamma_{AB}(\d \sigma^A - \Y^A \d u) (\d \sigma^B - \Y^B \d u), 
\ee
Under the hypothesis of logarithmic simplicity, the Bondi-Sachs metric coefficients have the following $\O$-expansion (section \ref{sec:polyexp}): 
\begin{subequations}
\begin{align}
\gamma_{AB}&= q_{AB}+ \Omega\,  \gamma_{1AB} +  \cdots + \O^{d-1} (\gamma_{(d-1)AB}+  \gamma_{((d-1),1) AB}\ln\O)+\cdots
\\
\Upsilon^A &=U^A +\Omega \Y_1^A +  \cdots + \O^{d-1} ( \Y_{(d-1)}^A+ \Y^A_{(d-1,1)} \ln \O)+\dots \\
\Phi &= F  + \Omega \Phi_1 +  \cdots + \O^{d-2}( \Phi_{(d-2)} + \Phi_{(d-2,1)} \ln \O)+\dots  \\
\bb &= \a + \O^2 \bb_2 +  \cdots + \Omega^{d}( \bb_{d} + \bb_{(d,1)} \ln \O) + \dots
\end{align}
\label{eq:expansion}%
\end{subequations}%
In $d=4$, the first logarithm for $\Phi$ appears one order higher, i.e. at $\Phi_{(d-1,1)} = \Phi_{(3,1)}$, since $\Phi_{(2,1)}=0$.

\subsection{Penrose's boundary conditions and conformal geometry }

\paragraph{Schouten tensor} Useful references for this section are 
\cite{Geroch:1977jn, curry2018introduction,Fernandez-Alvarez:2021yog, Fernandez-Alvarez:2021zmp, Herfray:2021xyp}.
An ingredient that is essential for the understanding of conformal geometry is the Schouten tensor.
It is given in terms of the Ricci tensor as 
\be\label{Scdef}
S_{ab}:=  \frac1{(d-2)}\left( R_{ab} - \frac{g_{ab}}{2(d-1)} R\right).
\ee
This tensor is such that $ S= \frac{1}{2(d-1)} R$ and relates to the  Einstein and Riemann curvature tensor as follows
\bea \label{ScG}
 G_{ab} &=& (d-2)(S_{ab}- g_{ab} S),
\\\label{ScR}
R_{abc}{}^d &=& W_{abc}{}^d + \left( g_{ac} S_{b}{}^d - g_{bc} S_{a}{}^d + S_{ac} \delta_b{}^d - S_{ac}\delta_a{}^d  \right),\label{WeylR}
\eea
where $W_{abc}{}^d$ is the Weyl tensor.
They satisfy the Bianchi identities 
\be\label{ScB}
\nabla_d W_{abc}{}^d = {(3-d)} (\nabla_a S_{bc} - \nabla_{b} S_{ac}), 
\qquad
 \nabla_b S_{a}{}^b - \nabla_{a} S =0.
\ee
The Weyl tensor is invariant under conformal transformations $\tl W_{abc}{}^d =W_{abc}{}^d$ while the Schouten tensors  transforms under conformal rescaling as
\be\label{Sctrans}
\tl S_{ab} =  S_{ab} + \Omega^{-1} \nabla_a  \nabla_b \Omega - \frac12 g_{ab} \Omega^{-2} \nabla_c \Omega \nabla^c \Omega,
\ee
where contraction of indices of non-tilded tensors (i.e. tensors in the conformal frame) is done with the non-tilded metric (i.e. the conformal metric) or its inverse. Notice that $ g^{ab} = \Omega^{-2}\tl g^{ab}$.

\paragraph{Einstein's equations}
We now express the physical Einstein equations of motion for $(\tl M, \tl g_{ab})$ in the conformal frame \cite{PenroseRindler2}. 
In doing so, we will treat the conformal factor $\Omega$ as a background structure, i.e. a quantity that is not dynamical. E.g. when considering variations of the field in an action principle or a symplectic structure, we will always set
\be
\delta \Omega \equiv 0.
\label{eq:deltaO}
\ee
This allows us to use $\O$ to define, at least in a neighbourhood of $\scri$, a foliation of $(M,g_{ab})$ by its level surfaces $\Sigma_\O$. 
As long as the induced metric on $\Sigma_\O$ is nondegenerate, the spacelike vs. timelike nature of the foliation is irrelevant for most of our results. However, to leverage the fluid-gravity duality at $\Sigma_\O$, we will henceforth assume $\Sigma_\O$ to be timelike. This way, neglecting what happens around space-like infinity, we will think of $\Sigma_\O$ as the timelike boundary of the spacetime, eventually taking the limit $\Sigma_{\O\to0}=\scri$. (Note: infinitesimal variations of the metric, which is all we consider in this paper, cannot change the spacelike vs. timelike nature of the foliation.\footnote{ The nature of the foliation is encoded in the sign of the ``conformal norm'' introduced in equation \eqref{sTdef}. See also equation \eqref{delg} for how the action of a (Bondi) diffeomorphism on $g_{ab}$ mimics an infinitesimal change in the choice conformal factor, $\O \mapsto \O(1+\epsilon w_\xi+ O(\epsilon^2))$.})

Einstein's gravity in absence of matter and with vanishing cosmological constant is described by the Einstein--Hilbert Lagrangian\footnote{For brevity, $\sqrt{\tl g} := \sqrt{-\det(\tl g)}$ and similarly for $\sqrt{g}$.}
\be
\tl{\bL}_{\mathsf{EH}} = \tfrac12 \sqrt{\tl g}\,  \tl R .
\ee
The variation $\delta \tl \bL_{\mathsf{EH}}$ gives the equations of motion $\tl \bE := \tl \bE{}_{ab}\delta g^{ab}$ and  the Einstein-Hilbert symplectic potential\footnote{We label the symplectic potential by  $\mathsf{EH}$,  standing for Einstein-Hilbert, as we will encounter  different symplectic potentials through the course of renormalization.} current ${\tl\bftheta}{}_{\mathsf{EH}}^a= \tfrac12 \sqrt{\tl g} ( \tl g{}^{bc} \delta \tl\Gamma^a_{bc} - \tl g^{ab} \delta \tl\Gamma^c_{bc} )$:
\be
\delta \tl \bL_{\mathsf{EH}} :=  \tl \bE + \pp_a {\tl\bftheta}{}_{\mathsf{EH}}^a.
\label{eq:deltaL}
\ee
In this section, we  focus on the action and equations of motion; the study of the symplectic potential is the subject of section \ref{sec:SP}.
The vacuum equations of motion are
\be
\tl \bE  = \tfrac12 \sqrt{\tl g}\,\tl G_{ab}  \delta \tl g^{ab} =  \Omega^{2-d} \bE_{ab} \delta g^{ab}
\label{eq:eom}
\ee
where $\tl G_{ab} := \tl R_{ab} - \tfrac12 \tl R \,\tl g_{ab}$ is the Einstein tensor.
$\bE_{ab}$ represents the vacuum equations of motions expressed in terms of the conformal frame metric, they read
\be \label{eq:eom_conf1}
\bE_{ab} = \frac12\sqrt{g}\Big[
G_{ab} +  
\Omega^{-1}(d-2) \Big(\nabla_a N_b - g_{ab} \nabla_c N^c \Big) + \Omega^{-2} (d-1)(d-2) \tfrac12 N^2 \,g_{ab}
\Big],
\ee
where $G_{ab}$ is the Einstein tensor of $g_{ab}$, $\nabla_a$ is its Levi-Civita derivative, and all indices are raised with $g^{ab}$.

The Einstein tensor $G_{ab}$ is regular at $\scri$, the following two terms, inside the parenthesis, diverge at $\scri$.
Demanding that these divergent parts vanish gives rise to the first two {\it Penrose's boundary conditions} (Pbc) \cite{PenroseRindler2} and ensures that \eqref{eq:eom_conf1} is regular at $\scri$.
Owing to the prescribed decay of the matter stress energy tensor (v), the first $d-4$ orders in the expansion the Einstein's equations $\bE_{ab}$ \eqref{eq:eom_conf1} do not depend on the matter configuration and we introduce a third Penrose boundary condition to impose them. The rationale behind this choice is that the gravitational field in the presence and well-behaved matter field configuration must satisfy these conditions, which can therefore be implemented as a restriction on the accessible phase space of gravity in asymptotically flat spacetimes.

The Penrose's boundary conditions\footnote{In his book, Penrose refers to the first two boundary conditions as the ``Asymptotic Einstein conditions.'' We renamed them to emphasize their role in the following as a minimal set of boundary conditions, and thus contrast them with different choices named after Bondi-Sachs \cite{Bondi:1960jsa,Sachs:1962wk}, Barnich-Troessaert \cite{Barnich:2009se,Barnich:2011mi},  Comp\`ere et al. \cite{Compere:2018ylh}, Freidel et al. \cite{Freidel:2021fxf}. \label{foot:bc}} encodes the physical content of the abstract definition of asymptotic infinity.
We call Penrose Boundary conditions (Pbc) the sets of asymptotic 
Einstein's equation 
 which are satisfied even in the presence of matter fields. As their name suggests, they should be understood as \emph{boundary conditions} that make the dynamic well defined at $\scri$: their imposition restricts the available phase space to those metrics for which the conformal compactification is physically meaningful. 
Most of the literature imposes (see footnote \ref{foot:bc}) a much more restricted set of boundary conditions which does not follow from the demand of asymptotic infinity. The first exception and systematic study of the Penrose boundary conditions and the associated symmetries has been done by Geiller and Zwikel in \cite{Geiller:2022vto} (see also Capone \cite{Capone:2021ouo}). In this work, we extend the analysis,  from within the conformal frame, of the generalized boundary conditions they first studied.

The Penrose boundary condition are however not enough to ensure that the symplectic potential of the theory is asymptotically renormalizable. They leave us with equation of motions $\bE_{ab} = O( \O^{d-3})$. As we will see in more details in section \ref{subsec:renor2}, for the renormalization of the symplectic potential (here performed explicitly only in vacuum) to go through we need to impose the equations of motion one order further so that $ \tilde\bE_{ab}$ itself admits a finite limit at $\scri$, i.e. we need 
\be \label{Elim} 
\bE_{ab} = O( \O^{d-2}). 
\quad 
\ee This boundary condition is satisfied in the presence of matter for the more restricted fall-off of \cite{Geroch:1973am,BongaPrabhu2020}.
It is also obviously satisfied for pure gravity. If matter is present at this order, a more detailed investigation is needed to understand what restriction it implies on the matter content.

\paragraph{Penrose's boundary conditions}\label{sec:Pbc}
Taking the limit of $\bE_{ab}$ onto $\scri$ implies 
\be\label{Pbc1}
N^2  \SCRIeq 0 
\qquad [\text{Pbc1}].
\ee
Here and in the following, $\SCRIeq$ means that the equality holds up to terms of order  $\Omega $ or higher, which vanish at $\scri$.
As well known, this is the equation that fixes the null nature of $\scri$ in the absence of a cosmological constant.\footnote{In the presence of a non-vanishing $\Lambda$, equation \eqref{Pbc1} reads instead
$$
N^2 \SCRIeq - \frac{2\Lambda}{(d-1)(d-2)},
$$
which fixes the signature of $\scri$: null if $\Lambda=0$, timelike if $\Lambda < 0$, and spacelike if $\Lambda >0$. \label{fnt:lambda}}
This is the first of the Pbc.

The first Pbc  (together with assumption (iv')) allows us to define what we call the \textit{conformal norm}
\be
\Sc :=  \tfrac12 \Omega^{-1} N^2 ,
\label{eq:Delta}
\ee
which has a finite limit onto $\scri$. With this definition, the limit of $\O^{d-1} \tl \bE_{ab}$ implies that
\be\label{Pbc2}
\nabla_a N_b - \Sc g_{ab} \SCRIeq 0.
\qquad [\text{Pbc2}]
\ee
This means that  $\scri$ is shear-free, since  Pbc2 states that $\nabla_a N_b$ is at leading order of pure-trace type. 
It also states that at leading order $\Sc$ is proportional to the expansion of $\scri$ in the given conformal frame, i.e.  $\nabla_a N^a= d \Sc$.

Of the $d(d+1)/2$ components of  Pbc2 \eqref{Pbc2}, $d$ turn out to be already implied by the first Pbc \eqref{Pbc1}. 
These $d$ components are those obtained by contracting \eqref{Pbc2} with $N^b$: Using the definition of $N_a$ \eqref{eq:Na} and that of the conformal norm \eqref{eq:Delta}, one obtains
\be
N^b(\nabla_a N_b-\Sc g_{ab})= \Omega \nabla_a \Sc  \SCRIeq 0.
\label{eq:Pbc.2-redundant}
\ee
Therefore the Pbc1 and Pbc2 impose in total $1 + d(d-1)/2$ conditions on the conformal metric.

The Pbc2 also allows us to define what we call the \textit{generalized news tensor}: 
\be
\N_{ab} := \Omega^{-1}\Big( \nabla_a N_b - \Sc g_{ab} \Big).
\label{eq_gennews}
\ee
This tensor  has a finite limit at $\scri$ and is such that 
\be \label{NaN}
\N_{ab}=\N_{ba},\qquad \N_{aN}=\nabla_a\nu.
\ee 
Notice that $\N_{ab} = \Omega^{-1}(\frac12 \pounds_N g_{ab} - \Sc g_{ab})$  encodes information about the extrinsic curvature of $\scri$.
 Therefore, one expects that its projection along $\scri$  is related to the canonical momentum conjugate to the degrees of freedom on $\scri$.
This expectation will be confirmed in the following.

  Comparing with equation \eqref{Sctrans}, we see that the physical and conformal Schouten tensors are simply related by
\be \label{tlS}
\tl S_{ab} = S_{ab} +\N_{ab}.
\ee
Condition (v) -- valid even in the presence of matter fields (see \cite{Fernandez-Alvarez:2021zmp}) -- provides us with the third set of Penrose boundary conditions. In $d=4$ they are:
\footnote{In dimension $d\geq 4$  they are instead: 
\be\label{Pbcd}
\O^{4-d} (S_{ab} +\N_{ab}) \SCRIeq 0. \qquad [\text{Pbc$d$}]
\ee
}

\be\label{Pbc3}
( S_{ab} +\N_{ab} ) \SCRIeq 0. \qquad [\text{Pbc3}]
\ee 
The validity of  the three Pbc therefore implies that $\Omega^{-1}(S_{ab} +\N_{ab} )$ is regular at $\scri$ and we define
\be 
\label{sTdef}
 \sT_{ab}:= \O^{-1} (S_{ab} +\N_{ab} ).
\ee

The main consequence of the Pcb3 is the following identity (a full proof can be found in appendix \ref{app:Bianchi}, see below for a sketch and \cite{Geroch:1977jn, Fernandez-Alvarez:2021zmp} for a derivation in $d=4$): 
\be \label{WN}
 W_{abc N} = 2\O (  \nabla_{[a} \N_{b]c} + N_{[a} \sT_{b]c} -g_{c[a}\sT_{b] N} ),
\ee 
which implies, in particular, that 
\be\label{WN+}
W_{abcN} \SCRIeq 0
\ee
and that the rescaled Weyl tensor $Y_{abc}$ admits a finite limit at $\scri$,
\be\label{Ydef}
Y_{abc} := \O^{-1}  W_{abcN}.
\ee

Identity  \eqref{WN} is obtained by combining the Pbc3 \eqref{Pbc3}, the Bianchi identity \eqref{ScB}, and the conformal invariance of the Weyl tensor $W_{abc}{}^d = \widetilde{W}_{abc}{}^d$. In particular, the conformal invariance means that the left hand side of this equation can be computed in the two frame, giving (See also  appendix \ref{app:Bianchi}) :
\be 
\nabla_d (\Omega^{3-d}W_{abc}{}^d)
&= (3-d)\Omega^{3-d} \left(Y_{abc}  +2 \nabla_{[a} S_{b]c}\right),\cr
&= 2(3-d) \Omega^{3-d} \left(   \O \nabla_{[a} \sT_{b]c} +  2 N_{[a}   \sT_{b] c}  -  g_{c[a} \sT_{b] N}   
\right).
\label{Weyl-Bianchi}
\ee 
The result then follows from comparing the two right-hand sides.

In dimension $4$, this condition implies  that the only non-vanishing component at $\scri$ of the Weyl tensor is of Petrov type $N$ which means that it is encoded into a  symmetric tensor $D_{ab}$  tangential to $\scri$, i.e.  $D_{aN}=0$,  and such that the Weyl tensor reads
$W_{abc}{}^d \SCRIeq D_{a c } N_b N^d +N_{a} N_{c } D_b{}^d - D_{b c } N_a N^d -N_{a} N_{c } D_b{}^d $. 
If  moreover, we assume that $\scri$ is  asymptotically simple, i.e. sufficiently differentiable in the sense of (iv), we can then project the Weyl Bianchi identities and show that  the divergence of $D_{ab}$ vanishes at $\scri$, and therefore that $D_{ab}\scrieq 0$ if $\scri$ has topology of $S_2 \times \mathbb{R}$ \cite{Geroch:1977jn,Fernandez-Alvarez:2021zmp}:  So, in $d=4$, assumption (iv)  implies  that 
\be\label{Pbc4}
 W_{abc}{}^d \SCRIeq 0,
\ee
which is equivalent to the peeling theorem.
However, as we have discussed, the condition of asymptotic simplicity  can be too restrictive. We replaced it with the condition of logarithmic simplicity (iv$'$).
As we will see in section \ref{subsec:radexp}, in $d=4$,  the non-vanishing of the Weyl tensor at $\scri$ is  indeed related to the presence of a   \emph{logarithmic anomaly} of order $\O^{3} \ln\O$ in the $g_{ab}$ expansion, which is sourced by the divergence of $D_{ab}$  \cite{winicour1985logarithmic,chrusciel1993gravitational}.

Summarizing, in logarithmically simple spacetimes satisfying (iv$'$), $D_{ab}$ does not  necessarily vanish at $\scri$, and, in this case, neither does $W_{abc}{}^d$.  This violates the peeling theorem. In all cases, \eqref{WN+} remains valid.

\paragraph{Einstein's equations in Friedrich's conformal form} 
In the presence of matter fields we demand that $\tl T^{\mathsf{mat}}_{ab} =\O^{d-3} \Tm_{ab}$  for a $\Tm_{ab}$ that is finite at $\scri$  as per condition (v) in the definition of asymptotic infinity.  With the notation introduced so far, the full set of Einstein's equation reads 
\be\label{eq:sT=sTmat}
\sT_{ab}\heq \O^{d-4} \tm_{ab}, \qquad \tm_{ab}:= \frac{1}{(d-2)} \left( \Tm_{ab} - \frac{g_{ab}}{(d-1)} \Tm \right).
\ee
We have introduced the symbol $\heq$ to denote that the equality is valid on-shell.
We see that $\tm_{ab}$ is to $\Tm_{ab}$ what $S_{ab}$ is to $G_{ab}$.

 Putting everything together, we get Einstein's equation in Friedrich's conformal form \cite{friedrich2002conformal, Paetz:2013nga}
\begin{align}
\begin{dcases}
\nabla_a\O & =  N_a \\
\nabla_a N_b &\heq \nu g_{ab} - \O S_{ab} + \O^{d-3} \tm_{ab} \\
\nabla_a \nu & \heq - S_{a N} +\O^{d-3} \tm_{aN}
\end{dcases}
\end{align}
These equations, quite remarkably, imply that $ \nabla_a ( 2\O \nu - N_bN^b) \heq 0$.  The vanishing of the cosmological constant\footnote{More generally, when the cosmological constant is non vanishing we have that
$$ 
2\O \nu - N_bN^b = \frac{2\Lambda}{(d-1)(d-2)}.
$$} then sets for us the boundary condition Pcb1, which implies:
\be
2\O \nu - N_bN^b \heq 0.
\ee

These data determine the Weyl components  via  equation \eqref{Weyl-Bianchi}
\be
Y_{abc} + 2\nabla_{[a} S_{b]c} & \heq \tm_{ab c},  \label{Bianchi1}
\\
 \nabla_d (\O^{3-d} W_{abc}{}^d)  &\heq (3-d)\tm_{ab c},\label{Bianchi2}
\ee
where we introduced 
\be
\tm_{ab c} := 2\O \left(  (d-2) N_{[a}  \tm_{b]c} -g_{c[a} \tm_{b] N} + \O \nabla_{[a} \tm_{b]c}\right).
\ee
From now on, we  restrict to vacuum Einstein gravity and  therefore assume that 
\be 
\tm_{ab} \equiv 0 \equiv \tm_{abc}. \qquad \text{[vacuum]}
\ee

\paragraph{Conformal stress tensor} For our purposes, it is more convenient to write Einstein's equations in the conformal frame rather than in Friedrich's conformal form. This brings us to introduce a central object in our analysis, the ``conformal stress tensor". 

We start by employing the first two Pcb to write  the rescaled and manifestly finite Einstein's equations in the conformal frame  in terms of the conformal frame Schouten and generalized news tensors:
\be
\bE_{ab} := \Omega^{d} \tl \bE_{ab}  
= \frac{(d-2)}2 \sqrt{g}\Big(  (S_{ab}+\N_{ab}) - g_{ab}(S+ \N) \Big) \heq 0.
\label{eq:confEinsteqs}
\ee
 {Thus we see that, $\bE_{ab} \heq 0$ if and only if $S_{ab} + \N_{ab} \heq 0$ \cite{Geroch:1977jn,Fernandez-Alvarez:2021zmp}.
 Alternatively, one can introduce  the \emph{conformal stress tensor}
\be\label{CEST}
\T_{ab} :=  \Big(\N g_{ab}- \N_{a b} \Big),
\ee
which acts as a source of the Einstein equations when expressed in the conformal frame:
\be 
\bE_{ab} = \sqrt{g} \left( G_{ab} - (d-2) \T_{ab} \right) \heq 0.
\ee
 Notice that  the conformal stress tensor $\T_{ab}$ is a purely geometrical quantity: in the presence of matter one should add to it a matter contribution $\tl T^{\mathsf{mat}}_{ab}$, which we neglect here.

We conclude this section with an expression of the Einstein--Hilbert Lagrangian, when written in terms of the conformal frame metric and the generalized news:
\be
\tl \bL_{\mathsf{EH}}=  \Omega^{2-d}\sqrt{g}\Big( 
\tfrac12 R + (d-1)\N\Big)  = (d-1)\O^{2-d} \sqrt{g} \big( S + \N\big) = (d-1) \O^{3-d}\sqrt{g}\sT.
\label{eq:tildeL-EH}
\ee
Notice that we are assuming the validity of Pbc1\&2 for $\N$ to be everywhere finite, and in the last equality also the validity of the Pbc3. However, This does not suffice to make the off-shell action finite close to $\scri$,   since a priori it still diverges as $\O^{3-d}$.  Further assuming Pcb3 makes the equations of motion finite and the off-shell action logarithmically divergent, at most.  On the other hand, in the absence of a cosmological constant and matter, the on-shell action vanishes. 

\section{The Bondi-Sachs gauge }\label{sec:Bondi}

Given a choice of conformal factor $\Omega$, we consider the foliation $\Sigma_\Omega$ whose leaves are level surfaces of $\O$. The surfaces $\Sigma_\O$ approach\footnote{More precisely, since in the absence of a cosmological constant $\scri$ decomposes as the union  of future and past asymptotic infinity, $\scri= \scri^+\cup \scri^-$, we must here restrict our attention to either $\scri^+$ or $\scri^-$. The Bondi-Sachs coordinates $y^i=(u,\sigma^A)$ used here only cover $\scri^+$, while $\scri^-$ is obtained as the limit $ u\to -\infty$ of the set $(\O, u,\sigma^A)$. In the following we still keep the notation $\scri$ for simplicity. No confusion should arise since we are only studying the retarded Bondi-Sachs frame in this paper.} $\scri$ in the limit $\Omega\to 0$, i.e. $\Sigma_{\Omega=0}=\scri$. It will be practical to introduce foliation-adapted coordinates over a neighborhood of $\scri$:
\be
x^a = (\Omega, y^i),
\ee
where $\pp_i:=\frac{\pa}{\pa {y^i}}$ are vectors  tangent to $\Sigma_\Omega$.

To maintain full covariance in the choice of the foliation, we avoid restricting to a specific conformal frame, i.e. we do \textit{not} use the rescaling freedom $\Omega\mapsto e^\phi\Omega$ to fix a privileged conformal factor.\footnote{ With the following exception: When we talk about the fluid-gravity duality, we assume that the foliation is timelike. This is however an open condition: $\nu> 0$ when $\Omega>0$,  not a gauge fixing. Arbitrary small perturbations do not affect it. On the other hand under a finite conformal rescaling $\Omega \to e^\phi \Omega$ we have that  \[
 \nu  \mapsto    e^{-\phi}\left(\nu + \nabla_N \phi + \tfrac12 \Omega (\nabla \phi)^2 \right), \] so it implies a restriction on allowed finite conformal rescaling. As shown in \eqref{eq:conformalrescaling} radial diffeomorphism at $\scri$ have the same effect as conformal rescalings.
 }
As shown in sections \ref{subsec:diffeo} the conformal symmetry of $\scri$ naturally emerges as a residual {\it radial} diffeomorphism symmetry of  the conformal frame manifold $M$. 

Said differently, we consider all metrics $\tl g_{ab}$ which are compatible with a certain conformal compactification defined by the scalar $\Omega$. These metrics can still be acted upon by arbitrary diffeomorphisms, compatibly with the given compactification. Some of these diffeomorphisms, namely the radial diffeomorphisms such that $\xi^a\pp_a \O \neq 0$, act on the rescaled metric $g_{ab}= \O^2\tl g_{ab}$ in a way that \emph{mimics} a change in the choice of conformal factor.  In our framework, diffeomorphism covariance of $\tl g_{ab}$ subsumes the rescaling symmetry  of (the universal structure of) $\scri$, cf. sections \ref{subsec:diffeo} and \ref{sec:sym}.

Similarly, we leave the tangential coordinates $y^i$ at $\scri$ arbitrary, that is we do {\it not} use diffeomorphisms to implement restrictive conditions on components of the metric. The only restrictions we impose on our coordinate systems concern how a specific boundary reference frame is transported radially into the bulk.

\subsection{Bondi-Sachs coordinates}

 To transport the boundary reference frame into the bulk we use three conditions first introduced by Bondi and Sachs in the study of gravitational radiation \cite{Bondi:1960jsa,Bondi62,Sachs:1962wk}.

{First, we demand that the radial $\O$-evolution off of $\scri$ unfolds along null geodesics with tangent vector 
\be
K^a\pa_a := \pa_\O, \qquad K^2 = 0.
\ee
This choice allows us to define the asymptotic radiation field as a component in the radial expansion of the shear of $K$.

Second, among the $y^i$, we choose a retarded time coordinate $u$ whose level surfaces are null outgoing cones.
We will assume that the topology of $M$ is such that the intersection of the 2-foliation   parametrized by $(u,\O)$ defines $(d-2)$-spheres $S_{(u,\O)}$,  themselves coordinatized by $\sigma^A$. 

This leads us to the third and last choice: we demand that the area element of these spheres (in the conformal frame) is constant under radial evolution. This means that  $\O$ represents an areal distance.\footnote{Another customary choice is to demand that the radial coordinate is an affine parameter \cite{newman1962behavior}. It turns out that the Bondi-Sachs choice drastically simplifies the analysis of the radial Einstein's equations. This is why we use it here, although the symmetry algebra is potentially richer in the Newman-Unti case \cite{Barnich:2011ty, Geiller:2022vto}.} 

We  refer to the resulting coordinate system as the {\it non-rotating radiation frame} (first and second condition) expressed in the {\it radial Bondi-Sachs gauge} (third condition) -- or, for brevity, simply the {\it Bondi-Sachs gauge}  \cite{Madler:2016xju}.

To summarize,  we use (the conformal-frame analogue of) Bondi-Sachs coordinates for the conformal metric $x^a=(\Omega,u ,\sigma^A)$, where 
\begin{enumerate}[(i)]
    \item $u$ labels null outgoing geodesic congruences,
    \item $\Omega$ is an areal  distance parameter along these geodesics,
    \item $\s^A$ are $d-2$ coordinates on the $(d-2)$-spheres $S_{(u,\O)}$.
\end{enumerate} 
It is important to appreciate that while we use the Bondi-Sachs gauge conditions to transport the boundary reference frame at $\scri$ into the bulk, we \emph{do not} use the Bondi-Sachs asymptotic boundary conditions, which are too restrictive for our purpose.  Our terminology disentangles the gauge  from the boundary conditions.
}

In  the Bondi-Sachs coordinate system, the most general conformal-frame spacetime metric (which becomes null at $\scri$) reads:\footnote{In the physical frame, with $r = \Omega^{-1}$, this metric corresponds to:
$$
\d \tl s ^2 = -2 e^\bb \d u ( \d r + r \Phi \d u) + r^2\gamma_{AB}(\d \sigma^A - \Y^A \d u) (\d \sigma^B - \Y^B \d u) .
$$ 
Note: what we call $\bb$ here is what is called $2\beta$ in \cite{Barnich:2010eb,Madler:2016xju,Compere:2018ylh,Freidel:2021fxf,Geiller:2022vto,Capone:2023roc}.}
\be
\d s ^2 = 2 e^\bb \d u ( \d \Omega - \Omega \Phi \d u) + \gamma_{AB}(\d \sigma^A - \Y^A \d u) (\d \sigma^B - \Y^B \d u), 
\label{eq:ds2}
\ee
where the three choices above correspond respectively to the following conditions: 
\be
g^{uu} \equiv 0, \qquad g^{uA} \equiv 0, \qqand \pp_\O \sqrt{\gamma} \equiv 0. 
\label{eq:bondiradframe}
\ee
 Note that, as desired, the vector
$ K^a \pa_a  := \pp_\O$
associated to the 1-form $K_a = e^{\bb} \pp_a u$ is the tangent vector to a null radial geodesic.  The affine parameter $\lambda$ along this geodesic is such that  $\pa_\O \lambda = e^\bb$. This follows from the evaluation of the radial acceleration\footnote{Indeed,  $\nabla_K K_a = \nabla_K K_a - \tfrac12 \nabla_a K^2 = K^b( \nabla_b K_a - \nabla_a K_b) = K^b ( \nabla_b \bb K_a- \nabla_a \bb K_b)= K[\bb]  K_a.$ So the affine vector is $\pa_\lambda = e^{-\bb} \pa_\O$}
\be \label{rada}
\nabla_K K^a = \bar\kappa K^a, \qquad \bar\kappa = \pa_\O\bb.
\ee 

The normalization we have chosen for $K$ is such that  $K^a N_a=1$ and therefore, we can define a projector 
\be
h_a{}^b =\delta_a{}^b -  N_a K^b. 
\ee
This projector is such that $h_a{}^b N_b = 0 = K^a h_a{}^b$. 
 In other words, we have onto maps $h : TM \to T\Sigma_\O$ and $h^*: T^*M \to T^*\Sigma_\O$, defined respectively by $X^a \mapsto X^b h_b{}^a$ and $\alpha_a \mapsto h_a{}^b\alpha_b$. In particular, vectors in the image of $P$ are tangent to the hypersurfaces  $\Sigma_{\O}$  and satisfy $ X^a N_a=0$.

In our adapted coordinates $x^a=(\Omega, y^i)$, 
\be 
h_a{}^\Omega = 0, \qquad h_{\O}{}^a=0,\qquad h_{i}{}^j = \delta_i^j.
\ee 
Such a projector is called a \emph{rigging projector} by Mars and Senovilla \cite{Mars:1993mj}, and the null vector $K^a$ is the rigging vector. 

 Central to our analysis is the projection of the generalized news tensor given by 
\be\label{Nproj}
\hN_i{}^j := h_i{}^a\N_a{}^b h_b{}^j.
\ee
There are three natural vectors which can be used to complete the basis for the  2-dimensional normal  distribution $(T S)^\perp$ alongside $K^a$. 
The first choice is the normal $N^a = g^{ab}N_b$, i.e. 
\be\label{N-vector-BS}
N^a = g^{a \O} = e^{-\bb}( \pp_u + \Y^A \pp_A + 2\O\Phi \pp_{\O}), \qquad \Sc =  e^{-\bb}\Phi.
\ee
 The vector $N^a$ has norm $N^2 = 2\O \Sc$, and is therefore guaranteed to be null only at $\scri$. In order to ensure that the hypersurface $\Sigma_\Omega$ is timelike we restrict the sign of $\nu$ to be strictly positive when $\Omega>0$.
The second choice is that of an {\it everywhere null} vector 
\be
L = e^{-\bb}(\pp_u + \Y^A\pp_A  + \O\Phi\po),
\ee
which by convention has unit inner product with $K$, $L^a K_a = 1$.
Finally, the third choice is that of the rigging projection $ V^b := N^a h_a{}^b$ of the normal vector $N^a$.
This vector is tangent to the hypersurfaces  $\Sigma_\O$, orthogonal to the spheres $S_{(u,\Omega)}$, 
and it is normalized to have a unit inner product with $K_a$. In Bondi-Sachs coordinates, it is given by 
\be
  V^a \pa_a= e^{-\bb}(\pp_u + \Y^A\pp_A ).
\ee
The vector $V$ represents the natural time evolution along $\Sigma_\O$ and, by extension, along $\scri$. 
Since $V^2 =-2 \O \nu$, and $\nu$ is assumed to be positive, it is everywhere timelike except at $\scri$ where it becomes null. 
The relationship between these vectors can be summarised by
\be\label{NVK}
N^a&= L^a +\O \nu K^a,\cr
V^a&= L^a-\O\nu K^a.
\ee
At $\scri$, the three vectors degenerate into each other, yielding the null vector:
\be\label{asymptnull}
\ell : \SCRIeq   e^\bb L \SCRIeq e^\bb V \SCRIeq e^\bb N = \pp_u + U^A\pp_A , \quad U^A :\SCRIeq \Y^A. 
\ee
This null vector is normal to the cuts $S_{(u,\O=0)}$. It defines a time derivative over $\scri$.

The physical interpretation of the Bondi-Sachs metric components is as follows:
\begin{enumerate}[(a)]
\item $e^{\bb}$ is the size of the normal geometry to the $(d-2)$-spheres $S$, so that 
\be\label{determinants}
\sqrt{g} = e^\bb \sqrt{\gamma} .
\ee
Moreover, it also represents the redshift factor that measures the rate of change of the affine parameter along the null geodesics with $\O$, since 
$\pa_\O\lambda = e^{\bb} $.  For this reason, we call $\bb$ the {\it redshift coefficient}.
\item $\Phi$, multiplied by $e^{-\bb}$, gives the conformal norm $\Sc$. This equals, in particular the acceleration of the observers moving along $V$, since  $\nabla_V V^a \SCRIeq \nu V^a$.
Therefore, it is not surprising that its $\O$-expansion contains 
 information on the Newtonian component of the gravitational field, i.e. Bondi's mass aspect. We call $\Phi$ the {\it Newtonian coefficient}.
\item $\gamma_{AB}$ is the metric induced on the spheres $S_{(u,\O)}$. In the Bondi-Sachs gauge, it is such that $\pp_\O \sqrt{\gamma} \equiv 0$.  We call $\gamma_{AB}$ the {\it metric coefficient}. Its  $\O$-expansion contains information on the spin-2 gravitational degrees of freedom, or radiation aspect.
\item $\Y^A$, or more precisely its radial derivative, encodes the dragging of the (rotating) frame of the observers moving along $L$. We call $\Y^A$ the {\it shift coefficient}. Its  $\O$-expansion contains information on the momentum aspect.
\end{enumerate}
These quantities, together with their radial derivatives, define the dynamical arena of an asymptotically flat spacetime.

To conclude (for now) with the interpretation of the metric coefficients, we note here the following rewriting of the trace part of the Pbc 2, in terms of the leading metric coefficients  $\Phi = F + O(\O)$,  $\Upsilon^a = U^A + O(\O)$, $\gamma_{AB} = q_{AB} + O(\O)$, and of the leading order covariant derivative $\Sn_A = D_A + O(\O)$:\footnote{Recall, in the Bondi-Sachs gauge $\sqrt{q} = \sqrt{\gamma}$.} 
\be
q^{ij} \mathcal{L}_\ell h_{ij} \equiv \pp_u \ln \sqrt{q} + D_A U^A \ \stackrel{\text{[Pbc2]}}{=} \ (d-2) F .
\label{eq:theta=F0}
\ee
In other words, $(d-2)F$ is the expansion of $\ell$.

\subsection{Carrollian geometry \label{sec:Carroll}}

As we discussed in the previous section, the Bondi-Sachs gauge features a double foliation of the asymptotic spacetime by means of two coordinate functions $\O$ and $u$. What is also true, but maybe  under-appreciated, is the fact that the family of hypersufaces $\Sigma_\O$ carries, near $\scri$ a \emph{ruled stretched Carrollian structure} 
\be
\label{eq:stretchedCarroll}
\mathscr{C}:=( h, V, \underline{K}, \Sc).
\ee
A ruled stretched Carrollian structure requires on each $\Sigma_\O$ a metric $h$, a tangent vector $V$, a 1-form $\underline{K}$, and a scalar $\nu$ which satisfy certain compatibility conditions we now detail. Here, $h_{ij}$ is the induced metric on $\Sigma_\O$, i.e. $h_{ij} = h_i{}^a h_j^b g_{ab}$, while all other quantities are defined as in the previous section.   We assume that the integral curves of $V=V^i\pa_i$ are complete on $\Sigma_\O$, which is valid in a neighborhood of $\scri$.

To define what a \emph{ruled stretched Carrollian structure} is, a few ingredients are needed. First, each $\Sigma_\O $ is viewed as a line bundle $\pi: \Sigma_\O\to S$ over the $d-2$ sphere.
The fibers of this Carrollian bundle  are given by the integral curves of $V$  i.e. $\ker(\rd \pi) = \mathrm{Span}(V)$.
A metrized bundle $(\pi:\Sigma \to S, h)$ defines a Carroll structure if $V$ is null. Here, we have, for each $\Omega$, a \emph{stretched} Carroll structure, because we have a family of metrized bundles parametrized by $\O$ which give us a Carroll structure only  in the limit $\O \to 0$.
Indeed, as explained in the previous section, at $\scri$,  $V \SCRIeq \ell$ is null. This is formalized by requiring that the norm of the vector field $V$ is given by the scalar $\Sc$ \emph{multiplied by} $\O$, i.e. 
\be h_{ij}V^iV^j = - 2\O \Sc \to 0.
\ee

This means that the metric  $h_{ij}$ on $\Sigma_\O$ becomes null as $\O\to 0$. This is emphasized by the equation:
\be
h_{ij}V^j = - 2\O \Sc K_i \to 0.
\ee
In this limit the stretched Carroll structure becomes a Carroll structure.

We now turn to the \emph{ruling} of the (stretched) Carroll structure. This is a choice of Ehresmann connection $\underline{K}:=K_i \rd y^i$ on $\pi:\Sigma_\O \to S$ which satisfies $K_i V^i=1$.

Having an Ehresmann connection gives us a notion of \emph{horizontal vectors}, defined as vectors $X^i$ in the kernel of $\underline{K}$, $X^iK_i = 0$. Then, we can define $\gamma^{ij}$ by means of the following two defining properties: first, $K_i$ is in its kernel, i.e.
\be
\gamma^{ij}K_j = 0,
\ee
and, second, the mixed tensor $\gamma_i{}^j$ is a projector on the space of horizontal vectors, i.e.
\be
\gamma_i{}^j := h_{ik}\gamma^{kj} = \delta_i{}^j - K_i V^j.
\ee
In particular $V^i\gamma_i{}^j = 0 = \gamma_i{}^j K_j$.

\emph{Horizontal forms} on the Carrollian bundle $\pi : \Sigma_\O \to S$ are, on the other hand, canonically defined as pullback along $\pi$ of forms on the base space $S$. Since $V^i$ is by construction in their kernel, we see that $\gamma_i{}^j$ is also a projector on the space of horizontal forms.

Later, starting in section \ref{sec:2+2EEq}, we will use coordinate frame fields $e^A=\pi^*(\rd \sigma^A)$ which are obtained from the pull-back of the base sphere coordinates (see \cite{Freidel:2022bai} for more details).  

The tensor $\gamma_{ij} := h_{ik}h_{jl}\gamma^{kl}$ is related to the Carrollian metric $h_{ij}$ as\footnote{When $\O\neq 0$, we denote $h^{ij}$ the inverse of $h_{ij}$. It is a tensor that becomes singular at $\scri$ which for $\O\neq 0$ reads
$$
 h^{ij} =\gamma^{ij} - \frac{V^iV^j}{2\O \nu}.
$$ }
\be
h_{ij}= \gamma_{ij} -2\O \nu K_i K_j.
\ee 

In general the image of $\gamma_i{}^j: T\Sigma_\O \to T\Sigma_\O$ is a non integrable distribution.
In the following, however, we will restrict our attention to Ehresmann connections that are dual to a foliation, i.e. such that $\rd \underline{K} = \rd \bb \wedge \underline{K}$ for some function $\bb$. In the Bondi-Sachs case, $\bb$ is precisely the redshift factor since $\underline{K}= e^{\beta}\rd u$.
When this is the case, we have that the horizontal vectors are vectors tangent to the spheres $S_{(u,\O)}$. In other words the distribution $\mathrm{Im}(\gamma_i{}^j)$ is integrable in this case and $\gamma^{AB}=\gamma^{ij} e_i{}^A e_j{}^B$ denotes the  (inverse) induced metric  on $S_{(u,\O)}$.


The  geometrical data $ \mathscr{C}=(h_{ij}, V^i,K_j,\nu)$ and the study of its connection was independently introduced and studied by Jai-akson and one of the authors in \cite{Jai-akson:2022gwg, Freidel:2022vjq, Freidel:2022bai}, as the geometrical structure of \emph{stretched horizons},  and by Mars, S\'anchez-P\'erez and Manzano in \cite{Mars:2022gsa,Mars:2023hty,Manzano:2023oub} where it is called a geometrical hypersurface data with null rigging. 

These data can be understood as either a geometrical structure intrinsic to $\Sigma_\O$ \cite{Ciambelli:2023mir} or as coming from the embedding of a rigged hypersurface \cite{Mars:1993mj}.
In the last perspective, we have that the ruling form is the image of the rigging vector: $ K_a =  g_{bc} K^b$, while the Carrolian vector is the image of the normal $ V^a = N_c g^{cb} h_b{}^a $.

The Carrollian perspective on null physics was first revealed in \cite{LevyLeblond1965, Gupta1966, Henneaux1979a, Ashtekar:1981bq}.
its importance in  Flat space holography, Celestial holography and the study of horizons has been the emphasis of many recent works 
\cite{Duval:2014uva, Duval:2014uoa,  Penna:2018gfx,  Ciambelli:2018wre,  Ciambelli:2019lap, Donnay:2019jiz, Ciambelli:2018ojf, Adami:2020amw, Chandrasekaran:2020wwn, Adami:2021kvx, Freidel:2022bai, Petkou:2022bmz}

\paragraph{Carrollian Connection\label{sec:Carroll-connection}}
As shown in \cite{Mars:1993mj, Chandrasekaran:2021hxc,Freidel:2023bnj,Freidel:2022vjq}, 
the embedding perspective defines a connection on $\Sigma_\O$ denoted $\hn$.
This connection is called a \emph{rigging connection}  when defined from an embedding and a \emph{Carrollian connection} when viewed as an intrinsic structure on $\Sigma_\O$ (see  also \cite{Mars:2022gsa,Mars:2023hty, Ciambelli:2023mir}).
It is given from the embedding as 
\be
\hn_i X^j := h_i{}{}^a (\nabla_a X^b) h_b{}^j, \qquad \hn_i \alpha_j := h_i{}{}^a  h_j{}^b \nabla_a \alpha_b,
\ee
for vectors $X^a= X^ah_a{}^b$ and form $\alpha_a = h_a{}^b \alpha_b$ on $\Sigma_\O$.
$\hn$ is a torsionless connection.

From the Carrollian connection, the Carrollian (vertical) vector $V$, and the Ehresmann connection $\underline{K}$, one can define the \emph{boost connection}:
\be
\omega_i := K_j \hn_i V^j = - (\hn_V K_i + \mathring\pa_i \bb),
\ee
where we used the notation $\mathring\pa_i =\gamma_i{}^j \pa_j$ and the second identity follows from  $d \underline{K} = d\bb \wedge \underline{K}$.
With these data we can introduce the covariant tensor 
\be\label{eq:barK1}
\bmK_{ij} := (\hn_i + \omega_i) K_j.
\ee
This tensor is not symmetric. It can be decomposed into its horizontal and vertical components as 
\be
\bmK_{ij} =: \bTheta_{ij} - K_i \bar\eta_j 
\ee
where both $\bTheta_{ij}$ and $\bar\eta_j$ are horizontal, i.e. vanish when contracted with $V$ (from the definition it is easy to see that $\bmK_{ij}V^j = 0$). The horizontal component $\bTheta_{ij}$ is symmetric and is called the \emph{shear tensor} of the connection, while
\be
\bar\eta_i := \gamma_i{}^{j}\bar\eta_j = \gamma_i{}^{j}(\omega_j + \mathring\pa_j\beta) = \gamma_{i}{}^{j}\hn_V K_j
\ee
is not an independent piece of data.
Note that, alternatively, one could have introduced the mixed tensor $\bmK_i{}^j = \bmK_{il}\gamma^{lj}$ by means of the following identity:
\be
\hn_i \gamma^{jl} = -( \bmK_i{}^jV^l+ \bmK_i{}^l V^j).
\ee

As shown in \cite{Mars:2022gsa, Freidel:2022vjq, Ciambelli:2023mir}, 
one can reverse the logic presented here and define the Carrollian connection as the unique torsionless connection given a connection one form and shear tensor $\hn_i \leftrightarrow (\omega_i,\bTheta_{ij})$. In particular, these results show that the action of the connection is uniquely determined from $(\omega_i, \bmK_{ij})$ and $\cL_{V} h_{ab}$.\footnote{For instance in the the null case we have that  
$$ 
-2 \hn_i h_{jl} =  (\cL_V h_{ij}) K_l + (\cL_V h_{il}) K_j. 
$$ 
}

 At  $\scri = \Sigma_{\O=0} $ and in $d=4$, something special happens with the Carrollian connection $\hn_i \leftrightarrow (\omega_i, \bTheta_{ij})$. First, as a consequence of the first two Penrose boundary condition (valid in all $d\geq4$), one shows that\footnote{This is equation is obtained by combining the expression above for $\omega_i$ with \eqref{eq_gennews} and the definition \eqref{Nproj} of the projected generalized news tensor $\hN_i{}^j$.}
\be\label{omN}
\omega_i = \nu K_i + \O\, \hN_{i K},
\ee
which means that $\omega_i \SCRIeq \Sc K_i$ and thus $\bar\eta_i \SCRIeq \mathring\pa_i\bb$. Second, and more interestingly, if we specialize to $d=4$, one finds that the remaining piece of data encoding the Carrollian connection, i.e. the connection's shear $\bTheta_{ij}$, is precisely the \emph{gravitational radiation shear}.\footnote{When we impose the Bondi \emph{boundary conditions} we have that $\Sc \SCRIeq 0$ and that the Bondi shear $C_{AB} $ is twice the Carrollian shear evaluated at $\scri$: $C_{AB} \SCRIeq 2 \bTheta_{AB}$. Note that the vanishing of $\nu$ is \emph{not} a conformally invariant condition.} 
This result,  proved in section \ref{sec:News}, is in line with the visionary work of Ashtekar-Streubel, who defined radiation, in the restricted context Bondi frame, as a choice of asymptotic connection \cite{Ashtekar:1981bq}. This perspective was developed further by Herfray in \cite{Herfray:2020rvq,Herfray:2021xyp}, who presented radiation as a conformal tractor connection.

 We conclude by noting that, from the embedding perspective $\Sigma_\O \hookrightarrow M$, the tensor $\bar{\mathsf{K}}_{i}{}^{j}=\bar{\mathsf{K}}_{il}\gamma^{lj}$ can also be introduced as the projection of the covariant derivative of $K^a$:
\be\label{eq:barK2}
\bar{\mathsf{K}}_{i}{}^j = h_{i}{}^a(\nabla_a K^b) h_b{}^j.
\ee

\paragraph{Accelerations and generalized news}
Before concluding this section,  we relate the  tensor $\bmK_i{}^j$ to the projected news tensor  $\hN_i{}^j$: 
\be 
\hn_i V^j + 2 \O \nu \bmK_i{}^j = \nu \delta_i^j + \O \hN_i{}^j. 
\ee
Later, we will need to  compute the components of this relation. 
In particular, we have that the observer's acceleration is  (see appendices \ref{acc2}, \ref{subsec:covder}) 
 \be
\hn_V V^i & =  \nu V^i  + \O \left(  \kA V^i  + \rA^i \right).
\ee
where the acceleration components are 
\be\label{omegacomp}
\kA = (\pa_\O + 2\pa_\O \bb) \nu, \qquad 
\rA^i  =  \gamma^{ij} (\mathring\pa_j +2\mathring\pa_j \bb) \nu.
\ee
Knowledge of these structures is essential to organize the asymptotic gravitational symmetries at $\scri$.

For completeness, let us mention that we could consider another projection of the generalized news tensor given by defining
\be 
\hat{\mathcal{K}}_{ij} := h_i{}^a h_j{}^b \N_{ab},
\ee
This projection represents the extrinsic curvature tensor of $\Sigma_\O$. It is an interesting geometrical object but does not play a significant role in the symplectic analysis where the key player is the projected news \eqref{Nproj}. The relationship between the extrinsic curvature and the projected news is 
(see appendix \ref{app:news-news})
\be 
\hat{\mathcal{K}}_{ij} =\hN_{(i}{}^k h_{j) k} + \hn_i \nu K_j. 
\ee 
Finally we note that the relationship between the bulk  and Carrollian divergences  involves the  boost connection:  for a tangential vector $V^i = V^a h_a{}^i$,  we have
\be \label{div}
 \nabla_a V^a
&= (\hn_i -\omega_i) V^i,
\ee
which is easily proved noting that in light of \eqref{NVK}, $\omega_i = h_i{}^a K_b \nabla_a N^b$.

\subsection{Spacetime diffeomorphisms in the conformal frame}\label{subsec:diffeo}

This section analyzes the structure of diffeomorphisms compatible with the  Penrose boundary condition  and the Bondi-Sachs gauge.

\paragraph{Diffeomorphisms in the conformal frame}

We denote the space of metrics $\tl g_{ab}$ on $\tl M$ by $\tl{\mathscr{G}} = \{\tl g_{ab}\}$. Here, we understand $(a,b,x)$ as ``indices'' for the ``coordinate'' $\tl g_{ab}(x)$ on the infinite dimensional space $\tl{\mathscr{G}}$. Consider a spacetime vector field $\xi\in \mathfrak{X}\big( \tl M\big)$. This vector field defines the infinitesimal variation of the metric $\tl g_{ab}$, 
\be
\tl g_{ab} \mapsto \tl g_{ab} + \epsilon \cL_{\xi} \tl g_{ab} = \tl g_{ab} + \epsilon \left(\xi^v \pa_c \tl g_{ab} + 2\tl g_{c(a} \pp_{b)}\xi^c\right).
\ee
Reinterpreting this as a variation of the  ``coordinates'' $\tilde g_{ab}(x)$ on $\tl{\mathscr{G}}$, we have that each vector field $\xi$ on $\tl M$ induces a vector field $\delta_{\xi}$ over $\tl{\mathscr{G}}$: for an arbitrary functional $F$ on $\tl{\mathscr{G}}$, one has
\be
(\delta_\xi F)(\tl g_{ab}(x)) = \lim_{\epsilon\to0} \frac1\epsilon\Big(F\left(\tl g_{ab}(x) + \epsilon \cL_\xi \tl g_{ab}(x)\right) - F(\tl g_{ab}(x)\Big).
\ee
Or, for short:
\be
\delta_\xi \tl g_{ab}(x) = \cL_\xi \tl g_{ab}(x).
\ee

We now wish to pass to the conformal frame manifold $(M,g_{ab})$. First, we introduce the space of metrics $\mathscr{G} \subset \tl{\mathscr{G}}$ as the space of metrics $\tl{g}_{ab}$ which are logarithmically simple \emph{w.r.t. the choice of a conformal factor $\O$}. We assume this conformal factor to be fixed once and for all. It is in other words a ``background structure". On $\mathscr{G}$ we can pick coordinates $g_{ab}(x)$, related to $\tl g_{ab}(x)$ by the usual rescaling 
\be
g_{ab}(x) = \O^2 \tl g_{ab}(x).
\ee
A crucial point is the following: from the space-time viewpoint, $\O$ is a scalar, while from the point of view of field space it is a ``background structure'' i.e. it is fixed once and for all independently of $g_{ab}$. Therefore:
\be\label{eq:OmegaBackground}
0 = \delta_\xi \O \neq \cL_\xi \O = \xi^a\pa_a \O.
\ee
Whence, in the conformal frame coordinates over $\mathscr{G}$ we have 
\be
\delta_\xi g_{ab}(x) = \O^2 \delta_\xi \tl g_{ab}(x) = \O^2 \cL_\xi \tl g_{ab}(x) = \O^2 \cL_{\xi} (\O^{-2} g)_{ab}(x),
\ee
That is
\be\label{delg}
\delta_\xi g_{ab}(x) = \cL_\xi g_{ab}(x) - 2 \W_\xi g_{ab}(x), \qquad \W_\xi : = \xi[\ln \O] = \O^{-1}\xi^\O.
\ee
where $\cL_\xi g_{ab}(x)$ denotes the standard Lie derivative of a tensor and $\W_\xi$ is the rescaled radial component of $\xi$.

We stress here the following: \emph{when expressed in the conformal frame, diffeomorphisms of the physical metric do not necessarily act by Lie derivatives.} We will come back to this later.

Another lesson we can draw from the above formula for $\delta_\xi g_{ab}(x)$ is that not all vector fields $\xi\in\mathfrak{X}(\tl M )$ preserve the finiteness of $g_{ab}(x)$ as $\O\to 0$, but only those such that
\be
\xi^\O \SCRIeq 0.
\ee
By an active interpretation of diffeomorphism symmetry, these vector fields correspond to the infinitesimal diffeomorphisms that do \emph{not} displace $\scri = \Omega_{\O=0}$. 
Therefore, if we want to study the symmetries of logarithmically simple spacetimes, which by definition admit a conformal compactification, then we need to restrict to the study of vector fields in
\be
\mathfrak{X}_0(M) := \{\xi \ | \ \xi \in C_{d-1}^\text{poly} \text{and } \xi^\O \SCRIeq 0\},
\ee
which are of the form:
\be
\xi = \hxi{}^i \pp_i + \W_\xi   \O \pp_\O .
\ee
The space of these vector fields, equipped with their Lie bracket defines a Lie algebra:
\be
\mathfrak{diff}_0(M) = \left( \mathfrak{X}_0(M) ,  [\cdot,\cdot]_\mathrm{Lie} \right) \subset \mathfrak{diff}(M).
\ee

Here we focused on the Lie algebra of the vector fields over $M$. What about the Lie algebra of the corresponding vector fields over $\mathscr{G}$? 
Before turning to this, we want to mention a generalization that will play a crucial role in the context of diffeormophisms compatible with the Bondi-Sachs gauge conditions. 

At each metric $g \in \mathscr{G}$, a vector field $\xi \in \mathfrak{X}_0(M)$ defines a tangent vector $\delta_\xi|_g \in T_g \mathscr{G}$. If the same $\xi$ is chosen for all $g$, then we say that $\xi$ is field-\emph{in}dependent and we write $\delta\xi=0$. However, if at different $g$'s we choose different $\xi$'s in $\mathfrak{X}_0(M)$, then the right hand side of equation \eqref{delg} still defines a viable vector field $\delta_\xi \in \mathfrak{X}(\mathscr{G})$. We refer to such $\xi$'s as field-dependent vector fields -- we will need them later.\footnote{This is a standard construction, which from the action of a Lie algebra on a manifold constructs the associated \emph{action} (or, transformation) \emph{Lie algebroid}  with bracket $\lbr\cdot,\cdot\rbr$ \eqref{algebroid-def}. See e.g. the first few pages of \cite{LOJAFERNANDES2002119} and, in the context of field theories and gauge transformations, \cite{BarnichAlgebroid, GomesRielloScipost, RielloSchiavinaATMP}.}

Considering the larger class of field-dependent vector fields $\delta_\xi$, a straightforward computation yields (notice the minus sign, details in appendix \ref{app:brackets}):
\be
[\delta_\xi, \delta_\eta] = - \delta_{\lbr \xi,\eta\rbr} 
\qquad\text{where}\qquad
\lbr \xi,\eta\rbr = [\xi ,\eta]_\text{Lie} - \delta_\xi \eta  +\delta_\eta\xi.
\label{algebroid-def}
\ee
For field-{\it in}dependent diffeomorphisms, $\delta \xi = 0$, the ``modified bracket'' $\lbr \cdot,\cdot\rbr$ reduces to the usual Lie algebra defined by the Lie bracket between spacetime vector fields,  $[\cdot,\cdot]_\text{Lie}$. However, in the more general case where $\xi$ is field dependent, the bracket $\lbr \cdot,\cdot\rbr$ is still a Lie bracket, i.e. it is skew-symmetric and satisfies the Jacobi identity.
 In the context of asymptotic infinity, the modified bracket $\lbr\cdot,\cdot\rbr$ was introduced by \cite{Barnich:2010eb}; see \cite{bergmannkomar1972coordinate, Salisbury1983} for early uses in general relativity.

\paragraph{Bondi-Sachs diffeomorphisms\label{sec:BondiSachsDiffeo}}
 We now turn to the class of
diffeomorphisms compatible with the Bondi-Sachs gauge condition \eqref{eq:bondiradframe}, which we call  {\it Bondi-Sachs diffeomorphisms}.  As we will see, Bondi-Sachs diffeomorphisms are necessarily field-dependent.

 Vector fields $\xi\in\mathfrak X_0(M)$ that preserve the Bondi-Sachs gauge conditions \eqref{eq:bondiradframe} must satisfy:
\be
\delta_\xi(g^{uu}) = \delta_\xi(g^{uA}) = \delta_\xi(\pp_\O \sqrt{\gamma}) = 0.
\label{Bondigaugecond}
\ee
 We call them Bondi-Sachs vector fields and denote them 
\be
\mathfrak{X}_B(M) := \{\xi \in \mathfrak X_0(M) \ : \ \text{Equations \eqref{Bondigaugecond} hold} \}.
\ee
Below we will see how they form an algebra, the algebra of Bondi-Sachs diffeomorphisms.
To describe explicitly the Bondi-Sachs diffeomorphisms, it is convenient to introduce the quantity 
\be
R_\xi : =  \tfrac12 \gamma^{ab}{\LL}_\xi g_{ab}  
= \xi^u \pp_u \ln \sqrt\gamma - \U^A\pp_A \xi^u +  \Snabla_A \xi^A,
\ee
which gives  the Lie derivative of $\ln \sqrt{\gamma}$ along $\xi$.
 Importantly, $R_\xi = R_{\hxi}$ 
is independent of the radial component on $\xi$: it depends only on the projection $\hxi$ of $\xi$ along $\pp_i $.

The preservation of the Bondi-Sachs gauge conditions \eqref{Bondigaugecond} is then seen to be equivalent to the radial evolution conditions\footnote{The first two conditions can be covariantly written $ \pa_\O \hxi^i = -e^{\bb} \gamma^{ij} \pa_j \xi^u$. }
\cite{Barnich:2011mi}
\be\label{evol}
\xi \in \mathfrak{X}_B(M) \iff 
\begin{cases}
\pa_\O \xi^u=0,\\
\pa_\O\xi^A =-e^\bb \gamma^{AB}\pa_B \xi^u,\\
\pa_\O\big((d-2) \W_\xi - R_\xi\big)=0.
\end{cases}
\ee
The radial evolution conditions for the Bondi-Sachs diffeomorphisms \eqref{evol} then mean that $\xi \in \mathfrak X_B(M)$ is \emph{uniquely} specified by the   values $(X_\xi, W_\xi)$ of $(\hxi,w_\xi)$ at $\scri =\Sigma_{\Omega=0}$. The function $W_\xi$ is usually called the  Weyl parameter of $\xi$ \cite{Freidel:2021fxf}.
Thus, we have a bijective map
\bea
\mathfrak{X}_B(M) &\to& \mathfrak{X}(\scri) \times C^\infty(\scri)\cr
\xi &\mapsto& (X,W) = (X_\xi,W_\xi) :\SCRIeq (\hxi,w_\xi)
\label{initial}
\eea
with inverse
\bea
\mathfrak{X}(\scri) \times C^\infty(\scri) &\to&\mathfrak{X}_B(M)\cr
(X,W) &\mapsto& \xi = \xi_{(X,W)}.
\eea
This map can in fact be promoted to an isomorphism of Lie algebras. 

First, we must notice that the radial evolution equations \eqref{evol} mean that the vector field $ \xi$ explicitly depends on the metric $\tl g \in \mathscr{G}$.  They do not close under the Lie bracket $[\cdot,\cdot]_\text{Lie}$ of vector fields over $M$. They do, however, still form a Lie algebra when equipped with the bracket $\lbr\cdot,\cdot\rbr$ \eqref{algebroid-def} form a Lie algebra. When the field-dependence is solely due to the radial evolution equations due to the Bondi-Sachs gauge -- i.e. when the initial conditions are field-independent, $\delta X_\xi=\delta W_\xi=0$ -- we call the ensuing Lie algebra the Lie algebra of Bondi-Sachs diffeomorphism $\diff_B(M)$.

We now want to describe the Lie algebra $\diff_B(M)$ in terms of a Lie algebra structure $\mathfrak{B}$ over $\mathfrak{X}(\scri) \times C^\infty(\scri)$.
Since the $\lbr\cdot,\cdot\rbr$-bracket of two Bondi-Sachs diffeomorphisms gives another Bondi-Sachs diffeomorphism, and since these are in 1-to-1 correspondance with elements of $\mathfrak{X}(\scri)\times C^\infty(\scri)$, the following quantities fully determine $\lbr \xi_1,\xi_2\rbr$:\footnote{These can be computed by explicitly working out the commutator $[\delta_{\xi_1},\delta_{\xi_2}] g_{ab}$, for two Bondi-Sachs vector fields $\xi_1$ and $\xi_2$ acting on an arbitrary Bondi-Sachs metric $g_{ab}$. Of course, in view of the uniqueness of the radial evolution of Bondi-Sachs diffeomorphisms, one can read the result from the leading order terms.} 
\bea
&&X_{\lbr \xi_1, \xi_2\rbr} :\SCRIeq \widehat{\lbr \xi_1, \xi_2\rbr} \SCRIeq [X_{\xi_1}, X_{\xi_2}]_\text{Lie} , \cr
&&W_{\lbr \xi_1, \xi_2\rbr} :\SCRIeq \O^{-1} \lbr \xi_1, \xi_2\rbr^\O \SCRIeq X_{\xi_1}[W_{\xi_2}] - X_{\xi_2}[W_{\xi_1}],
\eea
where $[X_1, X_2]_\text{Lie}$ is the Lie bracket between elements of $\mathfrak{X}(\scri)$. Thus, defining the Lie algebra $\mathfrak{B}$ as the semidirect-sum Lie algebra\footnote{$\mathfrak{B}$ stands for Bondi or Boundary.}
\be\label{BondiSachsAlg}
\mathfrak{B} := \diff(\scri) \oright \mathbb{R}^\scri,
\quad
[(X_1,W_1),(X_2,W_2)]_{\mathfrak{B}} :=  ([X_1,X_2]_\text{Lie}, X_1[W_2]-X_2[W_1]),
\ee
where $\mathbb{R}^\scri $ denotes the Abelian Lie algebra over $C^\infty(\scri)$, one readily gets the isomorphism of Lie algebras
\be
\diff_B(M) \simeq \mathfrak{B},
\quad
\label{morphism}
\begin{dcases}
\lbr \xi_{(X_1,W_1)}, \xi_{(X_2,W_2)} \rbr = \xi_{([(X_1,W_1),(X_2,W_2)]_{\mathfrak{B}})} \\
[(X_{\xi_1},W_{\xi_1}), (X_{\xi_2},W_{\xi_2})]_\mathfrak{B} = (X_{\lbr \xi_1, \xi_2\rbr},W_{\lbr \xi_1, \xi_2\rbr}).
\end{dcases}
\ee
The algebra $\mathfrak{B}$ already appeared in the study of asymptotic infinity in \cite{Geiller:2022vto} and in the study of horizons and null hypersurfaces in \cite{Adami:2020ugu, Odak:2023pga, Chandrasekaran:2023vzb}.

}

 We conclude this section with an explicit expression of the map $(X,W) \mapsto \xi$. It turns out that we can simplify the ensuing expressions by using the Carrollian data on $\scri$, i.e.\footnote{These are related to the data introduced in the last section by means of a rescaling (i.e. boost) by $e^\a:\SCRIeq e^\bb$, that is: $\ell \SCRIeq e^\a V$ and $k \SCRIeq e^{-\a} \underline{K}$.}
\be
\ell = \pp_u + U^A \pp_A , \quad \underline{k} = du.
\ee
The Ehresmann connection $\underline{k}$ defines a unique vertical/horizontal decomposition of any $X \in \mathfrak{X}(\scri)$, characterized by a scalar $\tau$ and a horizontal vector field $Y^A$:
\be
X = \tau \ell + Y, \quad Y^i k_i \equiv Y[u] = 0.
\ee
We call $\tau$ the \emph{supertranslation}, and $Y$ the \emph{super-Lorentz}, parameter of the Bondi-Sachs diffeomorphism $X$.

Using this split, the components of the Bondi-Sachs vector field $\xi_{(X,W)}$ are then given by\footnote{Recall: $\gamma_{AB} = q_{AB} + O(\O) $, $\Upsilon^A = U^A + O(\O)$, $\Phi = F + O(\O)$, $\bb = \a+ O(\O)$.}
\be
\begin{dcases}
\hxi_{X}^u = \tau,\\
\hxi_{X}^A  = Y^A + \tau U^A - G^{AB}\pp_B  \tau ,\\
\W_{({X},W)} = W
+  \tfrac{1}{d-2} R_{\tau}.
\end{dcases}
\label{eq:xi.e.xpansion}
\ee
where $\tau$, $Y^A$, and $W$ are $\O$-{\it in}dependent quantities (and so is $U^A$),  $G^{AB}:= \int_0^\O  e^{\bb} \gamma^{AB}\rd \Omega'$ is the integrated inverse metric, and $R_\tau$ is  defined by 
 \be
R_\tau&  := (U^A-\U^A)\pp_A \tau 
  - D_A (G^{AB} \pa_B\tau).
  \ee 
 Consistently with the initial condition $w_{(X,W)} \SCRIeq W$ \eqref{initial}, 
it is clear that $R_\tau$ vanishes at $\scri$, and only depends on the  supertranslation  parameter $\tau$.
It is related to $R_\xi$ via  $ R_{\xi_{(X,W)}} =  (d-2) \tau F + D_A Y^A   +  R_\tau,$ where we use the second Penrose boundary condition  in the form \eqref{eq:theta=F0} as well as the Bondi-Sachs gauge condition $\pp_\O \sqrt{\gamma} = 0$. 
 The construction given here shows that the Bondi vector fields near $\O=0$ can be constructed order by order in $\O$ as an expansion that depends on the radial expansion of the metric coefficients near $\O=0$. We will come back to this point.

\subsection{Symmetry action on the Carrollian geometry and anomalies}\label{sec:sym}

We now analyze the action of the symmetry on the components of the stretched Carrollian geometry $ (h_{ij}, \gamma^{ij},K_i, V^i, \nu)$.
To perform this analysis, it is essential to introduce the concept of anomaly $\Delta_\xi$, which is the operator that measures how much the action of the symmetry {$\delta_\xi$} on each component differs from the spacetime geometrical action  \cite{Hopfmuller:2018fni, Chandrasekaran:2020wwn, Freidel:2021cbc}. 

For example, recall that the symmetry transformations leave  invariant the  Bondi-Sachs background structure  (cf. \eqref{eq:OmegaBackground}), i.e. they are such that 
 \be
\delta _\xi\O=0,\qquad \delta_\xi N_a =0, \qquad  \delta_\xi K^b =0, \qquad \delta_\xi h_a{}^b=0.  
 \ee
It is thus obvious that in general the $\delta_\xi$ does not act geometrically on these quantities.

To formalize   the anomaly operator, let us first recall that \emph{in the conformal frame} the geometric action of the vector field $\xi =\hxi + w \O\pa\O $ on an operator $O$ is characterized by the tensor type of the operator $O$, that is whether it is a scalar, vector, form etc. on $\Sigma_\O$ and the conformal weight $s$ of $O$, which measures how $O$ transforms at $\scri$ under a constant rescaling $\O\to  e^\phi\O$ then $O \to  e^{s\phi}O $. 
Given such a tensor  of weight $s$, we have that the transformation under field space diffeomorphism is 
 \be\label{eq:anomalyop}
 \delta_\xi O= \cL_\hxi O + \W_\xi (\O\pa_\O - s)O +   \Delta_\xi O 
 \ee
 where $ \Delta_\xi $ is the anomaly operator. For a background structure--on which by definition $\delta_\xi =0$--the anomaly operator  equals the opposite of the geometry action.
 
 In appendix \ref{app:symaction}, we evaluate the diffeomorphism action on the Carrollian geometry elements, and we find that 
  $ (h_{ij}, \gamma^{ij},K_i, V^i, \nu)$ are respectively of conformal weight $(+2,-2,+1,-1, -1)$. 
 We also find that the anomaly operator depends on a one form $H_\xi$, a horizontal vector $G_\xi$ and a scalar
 $B_\xi$ given by
 \be 
 H_{\xi i } := \hat\pa_i \W_\xi, \qquad  G_{\xi}^i=-\pa_\O\hxi^i  = e^\bb \gamma^{ij}\hat\pa_j\tau, \qquad B_{\xi} := \pa_\O \W_\xi, \qquad 
 \ee
and its  action on the Carrollian geometry components is
\be  \label{anomalies}
 \Delta_\xi h_{ij} &= \O( K_i H_{\xi j}+ K_j H_{\xi i}),  \cr
 \Delta_\xi \gamma^{ij} &=  V^i  G_\xi^j  + V^j G_\xi^i,\cr
  \Delta_\xi K_i&= + \O B_\xi K_i - G_{\xi i},\cr
\Delta_\xi V^i &= - \O( B_\xi V^i  + H_{\xi }^i - 2 \nu G_{\xi }^i),\cr
  \Delta_\xi \nu &=   -(   V^i H_{\xi i}  + 2 \O \nu B_\xi ) 
 \ee
 where $G_{\xi i}= h_{ij} G_\xi^j$ and $H_\xi^i =\gamma^{ij} H_{j\xi}$.
 The parameters $(B_\xi, G_\xi^i, H_{i \xi})$  can be viewed as parametrizing the bulk Lorentz transformations  as viewed from the boundary. 
 One can check (cf. appendix \ref{app:symaction}) that these transformations preserve the defining relations of the stretched Carrollian geometry
 \be\label{Carrolrel}
 V_i K^i=1, \qquad \gamma^{ij} K_j=0,\qquad h_{ij} V^j = -2\O\nu K_i.
 \ee 
 As an application of this  formalism, one can evaluate the transformation of the ruling form, which is given by 
  $\delta_\xi K_a = (\delta_\xi\bb) K_a$ where $e^{-\bb}$ transform as an anomalous scalar of weight $-1$
 \be 
 \delta_\xi[e^{-\bb}] = \hxi[e^{-\bb}] + \W_\xi (\O\pa_\O  + 1) e^{-\bb}  -   V[\xi^u] - e^{-\bb} \O\pa_\O  \W_\xi.
 \ee 
 
At $\scri$ these   transformations simplify: when $\O=0$, $V^i$ and $h_{ij}$ are non anomalous  and their transformations read
 \be
 \delta_{(X,W)}V_0^i& = [X,V_0]^i_{\text{Lie}} + W V_0^i,\label{dV}\\
 \delta_{(X,W)}q_{ij}& = \cL_X q_{ij} -2 W q_{ij}.
 \ee
 where $V_0^i= e^{-\a} \ell^i$ and $h_{0ij}=q_{ij}$ denote the  evaluation of $(V,h)$  at $\scri$. The scalars, on the other hand, transform anomalously: using that 
  $\nu_0= e^{-\a} F$ (this follows from equation \eqref{eq:Delta}, a consequence of the Pbc1) we have 
 \be
 \delta_{(X,W)}[e^{-\a}] &= X[e^{-\a}] + W e^{-\a}  -   e^{-\a} \ell[\tau], \label{db}\\
\delta_{(X,W)}[e^{-\a} F]&=  X[e^{-\a} F] + W e^{-\a} F  - e^{-\a} \ell[W],\label{dF}\\
\delta_{(X,W)}[\sqrt{q}]&=  ( D_A Y^A - (d-2) [W-\tau F])\sqrt{q}. \label{dq}
 \ee
 where we used that the  expression \eqref{eq:theta=F0} for the  asymptotic expansion.
 The asymptotic ruling form $k$ conjugate to $\ell$ and the dual metric also transform anomalously, with
 \be
  \Delta_{(X,W)} k_{i}=  -  \mathring\pa_i \tau, \qquad 
 \Delta_{(X,W)} q^{ij} =   \ell^i  \mathring\pa^j \tau  + \ell^j \mathring\pa^j \tau.
  \ee
 We recognize in this transformation  the usual shift symmetry of a null Carrollian structure \cite{Ciambelli:2023mir}.\footnote{It is called a class I transformation in \cite{Odak:2023pga} and a Carrollian boost in \cite{Hansen:2021fxi}.}
To conclude, let us note that the above results mean that the variation $\delta_{(X,W)}$ acts on the universal structure $(h_{ij}, N^i)$ of $\scri$ as well as on the conformal norm $\nu = \frac12 \O^{-1} N^\O$ as expected, namely with $X^i$ acting as a diffeomorphism, i.e. by Lie derivative, and $W$ by the appropriate conformal rescaling:
\be\label{eq:conformalrescaling}
\begin{dcases}
\delta_{(X,W)} h_{ij} \SCRIeq - 2 W h_{ij} + \mathcal{L}_X h_{ij},\\
\delta_{(X,W)} N^i  \SCRIeq W N^i + \mathcal{L}_X N^i,
\\
\delta_{(X,W)} \nu \SCRIeq W \nu - \nabla_{N}\nu + \nabla_X \nu.
\end{dcases}
\ee

\section{Radial evolution from Einstein's equations}\label{sec:radexp}

In this section, we analyze the radial Einsteins equation in the Bondi-Sachs gauge and in the conformal frame. We construct explicitly the  (purely geometric) conformal stress tensor, which serves as a source of the conformal Einstein tensor.
In the end, the analysis done in this section  is a generalization of those in \cite{Barnich:2010eb,Madler:2016xju}.
It is similar to  the analysis done by Geiller and Zwikel in \cite{Geiller:2022vto} 
who  studied the radial Einstein's equation for the physical metric in Bondi-Sachs and Newman-Unti gauge with Penrose boundary conditions.
The main difference is that we keep explicit the codimension $2$ geometrical meaning of the quantities involved in the radial expansion.

\subsection{Codimension $2$ decomposition of Einstein's equation}\label{sec:2+2EEq}

In this subsection we describe the codimension $2$ decomposition of Einstein's equation 
 associated with a collection of  embedded codimension $2$ surfaces $S\equiv S_{(u,\O)}$.
We  introduce a frame basis of the normal subspace $(\mathrm TS)^\perp$ in terms of a pair of orthonormal null vectors $(\ell,\k)$, by which we mean 
\be
\ell_a \ell^a = 0 =  \k_a \k^a 
\qquad\text{and}\qquad
\ell_a \k^a = 1.
\ee
This implies that if $\ell$ points toward the future, then $\k$ points toward the past. The notation is such that $(\ell,k)$ denotes an arbitrary null frame of $(TS)^\perp$ while $(L,K)$ denotes the particular Bondi frame described in section \ref{sec:Bondi}.
We  first write the Einstein tensor  in a general frame before evaluating  it and its source terms \eqref{CEST}  in the Bondi frame.

The null basis $(\ell, k)$ of $(\mathrm TS)^\perp$ can be used to decompose the  spacetime metric as
\be
g_{ab} = \ell_a \k_b + \k_a \ell_b + \gamma_{ab},
\label{eq_2+2metric_general}
\ee
where $\gamma_{ab}$ is the $(d-2)$-metric on $S$. 
In the following we will work with a horizontal frame $e_A= e_A{}^a\pa_a \in TS $ which we assume to be coordinate, i.e. such that $[e_A,e_B]_\text{Lie}=0$, and a  coframe $e^A= \rd x^a e_a{}^A$ normal to the null vectors $(\ell,k)$ and such that $\iota_{e_A} e^B=\delta_A{}^B$. 
The horizontal frame and coframes are such that $e_A{}^a\gamma_a{}^b= e_A{}^b$ and $\gamma_A{}^b e_b{}^A= e_a{}^A$. More generally,  
we have that  $Y\in \mathrm TS$ if and only if $Y^a = {\gamma^a}_b Y^b$, and for such a  $Y$ we define
\be
\Snabla_A Y^B := {e_A}^a  \nabla_a Y^b e_b{}^B.
\ee
Since $\Snabla$ is torsion-free and compatible with $\gamma_{ab}$, it is just the covariant Levi-Civita derivative on $S$.

In the following we use two notational conventions: first, we use the short-cut notation $\nabla_A :=e_A{}^a \nabla_a$ and $\pa_A:= e_A{}^a\pa_a$ to denote horizontal derivatives. Similarly we denote 
$Y^A := e^A{}_a Y^a$ horizontal vectors and $\eta_A =e_A{}^a\eta_a$ horizontal forms.
Horizontal indices $(A,B,\dots)$ of tensors living in the tangent space to $S$ are lowered and raised with $\gamma_{AB}=  e_A{}^a e_B{}^b\gamma_{ab}$ and its inverse $\gamma^{AB}= \gamma^{ab} e_a{}^A e_b{}^B$. 

Since $\ell$ and $\k$ are on the same footing,  we define the tensors associated with $\k$ as the ``barred'' version of the tensors associated with $\ell$  (cf. e.g. equation \eqref{eq:ThetaThetabar}). Any equation that holds true for $\ell$ will also hold in its barred version. The spacetime geometry, as perceived by the family of observers $\{S\}$, can be efficiently encoded in a series of quantities we generically denote as the $(2+2)$-coefficients. All of them encode a different aspect of the observers' motion.

The first coefficients we introduce are the \textit{deformation tensors}, $\Theta_{AB}$ and $\bar\Theta_{AB}$, and their traces, the tangential and transverse expansions $\theta$ and $\btheta$:
\be\label{eq:ThetaThetabar}
\bar\Theta_{AB} : = {e_A}^a {e_B}^b \nabla_a \k_b , &
\qquad \Theta_{AB} : = {e_A}^a {e_B}^b \nabla_a \ell_b ,
\cr
\bar\theta := \gamma^{AB}\bar \Theta_{AB},&\qquad
 \theta := \gamma^{AB} \Theta_{AB}.
\ee
The deformation tensors $\bar\Theta_{AB},\Theta_{AB}$ are symmetric thanks to the surface orthogonality conditions $\k_a [X,Y]^a=0=\ell_a [X,Y]^a$  when $X,Y \in TS$. The traceless part (with respect to $\gamma_{AB}$) of the deformation tensors are called the \textit{shear tensors} and are denoted
\be
\bar \Sigma_{AB} := \bar\Theta_{\langle AB\rangle} \equiv \bar\Theta_{AB} - \tfrac{1}{d-2}\gamma_{AB} \bar\theta,
\qquad
\Sigma_{AB} := \Theta_{\langle AB\rangle} \equiv \Theta_{AB} - \tfrac{1}{d-2}\gamma_{AB}\theta.
\ee
Next, we introduce the \textit{tangential accelerations}
\be
\bar \acc^A := (\nabla_\k \k^a) e_a{}^A,\qquad \acc^A :=  (\nabla_\ell \ell^a) e_a{}^A
\ee
and the \textit{twist vectors}
\be
\bar\eta^A := - (\nabla_{\ell} \k^a)e_a{}^A, \qquad \eta^A := - (\nabla_{\k} \ell^a)e_a{}^A.
\ee
Notably, $\acc^A$ and $\eta^A$ can be treated as vectors on $S$. 

 To shed light on the geometrical meaning of $\eta$, we need to study two questions. 
The first one is the integrability (in the sense of Frobenius theorem) of the normal distribution $(\mathrm T S)^\perp$ given by the planes spanned by $(\ell, k)$. 
The obstruction to the integrability is measured by the \textit{anholonomicity}  defined as the projection on $\mathrm T S$ of the Lie bracket between $\ell$ and $ k$:
\be
 \gamma_{Aa}[ \k, \ell]^a = \bar \eta_A-\eta_A .
\label{eq_f}
\ee
Therefore, the condition that $(\mathrm TS)^\perp$ is integrable, i.e. is the tangent space to a submanifold of $M$, is simply $\eta=\bar \eta$.
The second problem, eventually clarifying the meaning of the symmetric combination $\eta^A + \bar\eta^A$, regards the variation of the size of the normal geometry.
Define
\be
\e^{2\bb} := - \frac{\det g}{\det \gamma}. 
\label{eq_aa}
\ee
Then the Lie derivative of $\bb$ with respect to any $X\in \mathrm TS$, can be calculated using the decomposition of $g_{ab}$, and is found to be
\be
 \cL_X \bb = \tfrac12 g^{ab}\cL_X g_{ab} -  \tfrac12 \gamma^{ab} \cL_X \gamma_{ab} = \ell_a\nabla_{\bar \ell} X^a + \bar \ell_a \nabla_\ell X^a = X^a (\eta_a + \bar \eta_a). 
\ee
Therefore, using the fact that $\bb$ is a scalar when viewed\footnote{$\bb$ is clearly \textit{not} a spacetime scalar since its definition relies on a split of the spacetime manifold.} as a field on $S$, we obtain
\be
\Snabla_A \bb = \eta_A + \bar \eta_A ,
\label{eq_aa_etaetabar}
\ee
where the metric $\gamma_{AB}$ has been used to raise and lower the indices. 
Importantly, the sum $\eta_A + \bar \eta_A$ is also the means of conversion between a spacetime divergence and a divergence on $S$. Indeed, for any $ X \in\mathrm TS$,
\be
\nabla_a X^a = \Snabla_A X^A + (\eta_A + \bar \eta_A)X^A= e^{-\bb}\Snabla_A(e^{\bb} X^A) .
\label{eq_div}
\ee

Finally,\footnote{The remaining combination $\ell^b\nabla_a \ell_b$   identically vanishes as a consequence of the null nature of $\ell$.} we introduce the \textit{longitudinal accelerations}
\be
\bar \kappa :=\ell^a \nabla_\k \k_a
,\qquad 
\kappa :=  \k^a \nabla_{\ell} \ell_a.
\ee
as well as the \textit{normal connection} 
\be
\rom_A := {e_A}^b (k_c \nabla_b \ell^c) = - \mathring{\bar\omega}_A.
\label{eq_omega}
\ee
As we will see, this last quantity is related to the twist vector and the tangential acceleration.
The expression for these coefficients depends on the choice of null frame and chosing the null frame to be the adapted Bondi-Sachs frame will lead to simplifications.

\subsection{Einstein Tensor in Terms of the $(2+(d-2))$-Coefficients}

In this section, we use the general codimension $2$ notation, where the normal geometry is parametrized in terms of a generic basis $(\ell,\k)$. This allows us to treat $\ell$ and $\k$ on the same footing and write half of the equations without loss of information. We will specialize to the radiation frame at the very end only.  Using the Gaussi--Codazzi equations, one can show that: 
\be
-G_{\k \k} &  = (\nabla_\k-\bar\kappa) \bar\theta-(\Snabla_A+ 2\rom_A+\pa_A \bb) \bar\acc^A  +\bar\Theta : \bar\Theta,      \label{eq_G}
\\
G_{\k \ell} & = (\nabla_\k+ \bar\kappa) \theta + \Snabla_A \eta^A + \Big( \eta \cdot \eta + \bar \acc \cdot \acc + \theta\bar\theta\Big)-\tfrac12 \mathring{R}, \cr
-G_{\k  A} &  
=    \gamma_A{}^B \cL_\k \bar\eta_B +   (\nabla_{\ell}+ 2\kappa)\bar\acc_A
 - (\Snabla_B + \rom_B)(\bTheta_A{}^B - \gamma_A{}^B \bar\theta)   - \bTheta_A{}^B \eta_B,  \cr
 -\tfrac12 G_{\langle AB\rangle} & 
= [(\nabla_\k+\bar\kappa) \Theta]_{\langle AB\rangle} + \Sn_{\langle A} \eta_{B\rangle} + \eta_{\langle A}  \eta_{B\rangle} + \bar \acc_{\langle A}   \acc_{B\rangle} + \tfrac12\theta\bar\Theta_{\langle AB\rangle}+
 \tfrac12\bar\theta\Theta_{\langle AB\rangle} - \tfrac12 \mathring{R}_{\langle AB\rangle},  \nonumber
\ee
where $T_{\langle AB\rangle} := T_{(AB)}- \tfrac{\gamma_{AB}}{d-2} T$ denotes the symmetric traceless components  of a tensor $T_{AB}$ and $\mathring{R}_{AB}$ denotes the Ricci tensor of the sphere connection $\Sn$.

Another set of equations can be obtained after exchange of $\ell$ with $k$ while exchanging $(\eta,\Theta, a, \omega)$ with 
$(\bar\eta,\bTheta, \bar{a}, -\omega)$.
For instance, the symmetry $G_{\ell \k}=G_{\ell \k}$ implies the  identity
\be\label{bianc}
 (\nabla_\ell + \kappa)\bar\theta + \Snabla_A \bar\eta^A + \bar\eta \cdot\bar\eta 
 =
 (\nabla_{\k}+\bar\kappa) \theta + \Snabla_A \eta^A + \eta \cdot\eta.
\ee
Similarly, another  $\langle A B\rangle$ equation can also be obtained after exchanging $\ell$ with $k$, which gives the identity
\be
[(\nabla_\ell+ \kappa) \bar\Theta]_{\langle AB\rangle} + \Snabla_{\langle A} \bar \eta_{B\rangle} + \bar\eta_{\langle A}  \bar \eta_{B\rangle} \label{GAB=BA}.
 =
 [(\nabla_\k+\bar\kappa) \Theta]_{\langle AB\rangle} +\Snabla_{\langle A} \eta_{B\rangle} + \eta_{\langle A}  \eta_{B\rangle} 
\ee

\paragraph{Bondi-Sachs Null  Frame}
An adapted choice of null frame can simplify these equations. 
In  the Bondi frame, we select $k=K $ to be tangent to geodesics which means that $\bar{\acc}_A=0$. We also chose the additional null vector $\ell= L$ to be in the plane of $(N,K)$. This leaves us with an ambiguity in the choice of boost frame $(K,L)\to (e^{-s}K, e^{s}L)$. For the study of the radial Einstein's equation, we fix the boost frame to be such that 
\be 
K^a\pa_a =\pa_\O, \qquad L^a= N^a - \O \nu K^a. \label{Ldef}
\ee
This choice implies that $\eta_A + \rom_A=0$ and therefore $\bar\eta_A = \pa_A\bb-\eta_A$.
 This fact follows from the definition  of $N_a$  as the closed form  $\pp_a \O$, indeed: 
\be
 \rom_A +\eta_A &= e_A{}^a  K^b (\nabla_a L_b - \nabla_b L_a)
= - e_A{}^a  K^b (\nabla_a (\Omega \nu K_b) - \nabla_b ( \Omega \nu K_a))
= \Omega \nu \bar\acc_A =0.\nonumber
\ee
The evaluation of the coefficients needed to write explicitly the Einstein tensors appearing in \eqref{eq_G} in terms of the metric parameters is detailed in appendix \ref{sec:acc}. One gets
\be
\begin{dcases}
\bar\Theta_{AB}  = \tfrac12 \pp_\Omega \gamma_{AB} 
\\
e^{\bb}\Theta_{AB}  = \tfrac12 \pp_u \gamma_{AB} + \Snabla_{(A} \Upsilon_{B)} + \Omega \Phi \bar\Theta_{AB} \\
 \eta_A= -\rom_A=  \tfrac12 (\pa_A \bb - \e^{-\bb} \gamma_{AB} \pa_\O \Upsilon^B)\\
\acc_A  = \Omega(\e^{-\bb} \pp_A \Phi) \\
\bar\kappa = \pa_\O \bb\\
\kappa= \e^{-\bb} \pa_\O (\O\Phi)
\label{eq_2+2_explicit}
\end{dcases}
\ee

\subsection{Generalized news and conformal Einstein's equation\label{sec_gennews}}

From the definition \eqref{eq_gennews} of the generalized news tensor $\N_{ab}$ and the decomposition \eqref{Ldef} of $N$ onto the $(L,K)$ basis, we readily obtain
\be
\N_{ab} 
&= \Omega^{-1}\Big[ \nabla_a  L_b + \nabla_a (\Omega \nu K_b) - \nu \,g_{ab} \Big] 
\cr
&= \Omega^{-1}\Big[ \nabla_a  L_b +  \nu (L_a K_b -  \,g_{ab}) \Big] + \nabla_a(\nu K_b) +\nu^2 K_aK_b
\ee
Contracting it with $L$, $K$, and ${\gamma_A}^a$, one finds the various components of the generalized news tensor in terms of the $(2+2)$-coefficients, as well as of $\bb$ and $\Phi$.
In order to express these coefficients 
one defines 
\be\label{aspectdef}
\mu &:=   \Omega^{-1}e^{\bb} (\theta -(d-2) \nu ), \qquad \bA := \Omega^{-1}\bar\kappa=  \Omega^{-1} \pa_\O\bb \\
  \Pi_A &:=\Omega^{-1} \eta_A,\qquad  \qquad \qquad \quad 
 N_{AB} :=  \Omega^{-1}e^{\bb} \Theta_{\langle AB \rangle }.\label{aspectdef2}
\ee
Moreover, we can  evaluate the different traces of $\N$:
\bea\label{defmu}
\SN=   e^{-\bb} (\mu+\Phi\bar\theta) , \quad \hN =  e^{-\bb} (\mu+\Phi\bar\theta + [\pa_\O +\pa_\O\bb]\Phi ), \quad
\N =  e^{-\bb}  (  \mu + \Phi \bar\theta + 2 \pa_\O \Phi ),
\eea
where $\SN= \gamma_{a}{}^b \N_{b}{}^a$, $\hN= \N_a{}^b h_b{}^a$ and $\N= g^{ab} \N_{ab}$.
One key relation between the traces which will be used many times in the following is, see \eqref{omegacomp},
\be
\label{tracerel}
\hN= \SN+ A_K, \qquad A_K= \pa_\O\nu+2\pa_\O \beta \nu. 
\ee
Now using that 
$N^a\pa_a = e^{-\bb} (L+\Phi \Omega\pp_\Omega)$ we find that  
\be
\begin{array}{ccrlc}
&(KK) & \N_{KK} & = 
 - \bA &\\
&(KA) & \N_{K A} & =
 -\Pi_A&\\
&(KL)&\N_{K L}&= 
\O^{-1}(\kappa-\nu) =
e^{-\bb} \pa_\O \Phi&\\
&(AB) & \SN_{\langle AB \rangle } & =  e^{-\bb} (N_{AB} + \Phi \bar\Theta_{\langle AB \rangle })&\\
\end{array}
\label{eq_News_2+2}
\ee
the symmetry of the last term is guaranteed by the surface orthogonality condition applied to $K$. From the expressions above, it is straightforward to compute the conformal-frame  stress-energy tensor, 
$
\T_{ab} = \Big( \N g_{ab} - \N_{ab} \Big)
$ (cf.  \eqref{CEST}).
One gets 
\be
\begin{array}{ccrlc}
& (KK) & \T_{KK} & =  \bar{A} &\\
& (KA) & \T_{KA} & =  \Pi_A &\\
& (KL) & \T_{KL} & =   e^{-\bb} \left[   \mu+  \Phi \bar\theta + \pa_\O\Phi  \right] &\\
& (AB) & \rT_{\langle A B\rangle } & =  - e^{-\bb}\left(  N_{AB} + \Phi \bar\Theta_{\langle AB \rangle }\right)   &\\
\end{array}
\ee
We see that $\Pi_A$ plays the role of the conformal fluid momentum  aspect. We will see that 
\be
\E = \rN =  e^{-\bb} \left(  \mu+  \Phi \bar\theta \right)
\ee is  the conformal fluid energy  aspect while $N_{AB}$ enters the definition of the fluid's viscous  stress tensor. 
$\Phi$ is a scalar potential that shifts $(\mu,N_{AB})$ into  $(\E,\rtN_{AB}= \rN_{\langle A B \rangle})$.

The last simplification comes from imposing the Bondi gauge condition on $\sqrt{\gamma}$,
\be
\bar\theta=0,
\ee
which means that $\bar\Theta_A{}^B=\bar\Sigma_A{}^B$ is traceless.

We now have all the ingredients to evaluate Einstein's equation in the conformal frame given by $G_{ab}=(d-2) \T_{ab}$.
Here we only collect those Einstein's equation that feature in the radial evolution of the metric variables:
\be
&(KK) & 
(d-2)\bar{A}  &  =  -(\bTheta_A{}^B  \bTheta_B{}^A)      \label{eq_G2}
\\
&(KA)& 
[(d-3)  -   \O \pa_\O ]\Pi_A&  
=   
  \Snabla_B \bTheta_A{}^B   +\frac{\O}{(d-2)} \pa_A  (\bTheta_B{}^C  \bTheta_C{}^B)   \cr
&(K L) & 
\left[   (d-3) -\O\pa_\O\right] \mu   & =  e^{\bb}\left( -\tfrac12 \mathring{R} + \O \Snabla_A \Pi^A + \O^2 \Pi_A\Pi^A \right) \cr
&\langle AB\rangle
& \left[ \tfrac12 (d-4) -\O \pa_\O \right]  N_{A}{}^B   
&
=  e^{\bb}  \left(- \tfrac12 \mathring{R}_{\langle A}{}^{B\rangle} + \O \Snabla_{\langle A} \Pi^{B\rangle} + \O^2 \Pi_{\langle A}  \Pi^{B\rangle}  \right) +  \tfrac12\O \mu\bTheta_{A}{}^{B}  \nonumber
\ee
In these equations, the symbols in parenthesis on the left indicate that the corresponding equation stems from the contraction of 
the  equation with  $K^a$, $L^a$ or with the sphere's projector 
$\gamma_A{}^{a} 
$.\footnote{For the last equation we used that $\nabla_K \Theta_a{}^b = \cL_K\Theta_a{}^b + K_a (\bar\eta^c \Theta_c{}^b) + (\Theta_a{}^c\eta_c) K^b + [\Theta, \bTheta]_a{}^b$. The horizontal projection 
$e_A{}^a (\nabla_K \Theta_a{}^b)e_b{}^B$ kills the first two terms,  and that the symmetrization of the commutator vanishes 
so that the projected components satisfy $(\nabla_K \Theta)_{A}{}^{B} = \cL_K \Theta_A{}^B + [\Theta,\bTheta]_{A}{}^{B}$ and $(\nabla_K \Theta)_{\langle A}{}^{B\rangle} = \pp_\O (\Theta_{\langle A}{}^{B\rangle})$.
}
We also denoted $X_{\langle A}{}^{B \rangle} := \tfrac12 (X_A{}^B+ X^B{}_A) - \frac{1}{(d-2)} X_C{}^C \delta_A{}^B$  the symmetrized trace-free part of a codimension $2$ tensor.

It is interesting that the terms containing the expansion $\theta$ only appear inside the combination $\mu$. 
Therefore, to realize the metric expansion of the metric components, we also need the radial evolution of $\theta$. It follows from the Bianchi-identity \eqref{bianc} and reads
\be\label{bianc_bis}
\pa_\O[\e^{\bb} \theta] 
=  \mathring{\Delta}\e^{\bb} - 
2\O \Snabla_A[\e^{\bb}\Pi^A]. 
\ee
 and we used that in the Bondi gauge $\bTheta_{\langle A}{}^{B \rangle}=\bTheta_A{}^B$.
We will also need to use \eqref{GAB=BA} which relates $\Theta_{AB}$ and $\bTheta_{AB}$. In the Bondi frame, it reads 
 \be \label{N=dotC}
 (1+\O  \pa_\O) N_{A}{}^B
  & = (\cL_{e^\beta V} +  \Phi )\big[\bar\Theta_{A}{}^{B}\big] +
  \Sn_{\langle A}\Sn^{B \rangle} e^\bb \\
  & + \O\left(  \pa_\O \big(\Phi \bar\Theta_{A}{}^{B}\big)  - 2 \Snabla_{\langle A} (e^{\bb}\Pi)^{B\rangle}\right).\nonumber
\ee
These equations are the flat space analog of the  Feffermen-Graham evolution equations used in AdS \cite{Skenderis:2002wp}.
They trivialize in dimension $d=3$, where there is no codimension $2$ traceless tensor and no curvature, so $\bTheta_{AB}= \mathring{R}_{AB}=0$. In the following, we therefore assume that $d\geq 4$. In $d=4$, there is no traceless component to the codimension $2$ curvature tensor, so $\mathring{R}_{\langle AC\rangle} =0$ and the $\langle AB\rangle$ equations simplify.
In all other dimensions $d>4$, there are no further simplifications.

\subsection{Reconciling the Carrollian and codimension $2$ perspective \label{sec:reconcile}}
This final subsection summarises the relationship between the Carrollian and codimension $2$ formalism (see appendix \ref{sec:acc}).
We recall that  in \eqref{aspectdef} we have introduced the rescaled coefficients $(N_i{}^j,\Pi_i)$,  by the identities
\be
\Sigma_i{}^j = \O e^\bb  N_i{}^j,\qquad \rom_i =-\eta_i= -\O\Pi_i. 
\ee
The main point is that the codimension $2$ formalism provides a horizontal decomposition of the Carrollian tensors.
For instance, the Carrollian connection $\omega_i$ decomposes as 
\be\label{eq:AK}
\omega_i = \nu K_i + \O(  A_K K_i   - \Pi_i) , \qquad  A_K=  (\pa_\O + 2\pa_\O \beta)\nu.
\ee
Then, we express the Carrollian derivative of $V^i$ as
\be 
\hn_i V^j =\nu \delta_i{}^j +\O \hV_i{}^j,
\ee
and we can decompose the Carrollian tensors $(\bmK_i{}^j, \hV_i{}^j,  \hN_i{}^j)$ in terms of the codimension $2$ tensors as
follows
\be 
\hN_i{}^j & = e^{-\bb}\left(N_i{}^j +\Phi \bar\Sigma_i{}^j\right) +\frac{\gamma_i{}^j }{(d-2)} e^{-\bb} \left(\mu + \Phi \btheta \right) + K_i ( A_K V^j - \rJ^j) -\Pi_i V^j ,\label{hN} \\
\hV_i{}^j & = e^{-\bb}\left(N_i{}^j -\Phi \bar\Sigma_i{}^j\right) +\frac{\gamma_i{}^j }{(d-2)} e^{-\bb} \left(\mu - \Phi \btheta \right)  + K_i (A_K V^j +\rA^j) -\Pi_i V^j
\label{hV} \\
\bmK_i{}^j & =  \bar\Sigma_i{}^j +\frac{\gamma_i{}^j}{(d-2)} \btheta - K_i ( \mathring\pa^j\bb - \O\Pi^j), \label{Kt} 
\ee
where we introduce the acceleration and energy currents to be  
\be \label{Jexp}
\rA^j &:= (\mathring\pa^j + 2\mathring\pa^j\bb)\nu,\cr
\rJ^i &:= 
- (\mathring\pa^j + 2\O \Pi^j)\nu.
\ee
We have the relationships
\be 
\hN_i{}^j- \hV_i{}^j= 2\nu \, \bmK_i{}^j , \qquad  \rJ^i+\rA^i =  2\nu ( \mathring\pa^i\bb -\O\Pi^i).
\ee
 We thus introduce the \emph{asymptotic stress tensor}   
\be\label{eq:projectedCSET}
\hT_i{}^j:=\hN \delta_i{}^j- \hN_i{}^j.
\ee 
Note that the asymptotic stress tensor $\hT_i{}^j$ differs from the projection of the conformal energy-momentum tensor by a trace term, i.e. $\hT_i{}^j= h_i{}^a \T_{a}{}^b h_b{}^j - \pa_\O\nu \delta_i{}^j$ since $\N-\hN=\pa_\O\nu$.\footnote{We hope that this breach of our ``hat'' notational convention for $\T_a{}^b \to \hT_i{}^j$ does not confuse the reader.}
 As we'll see in section \ref{sec:conservationeqs}, $\hT_i{}^j$ is the subleading component of the stress tensor of the  asymptotic Carrollian fluid living on $\Sigma_\O$ \eqref{eq:CarrollStrTens}.
We can decompose it along its vertical and horizontal  components, and further decompose the latter into a traceless and trace parts. We obtain:
\be\label{eq:carrstresstens}
\hT_i{}^j &= \rtT_i{}^j + P \gamma_i{}^j  + \Pi_i V^j + K_i J^j,
\ee
where 
\be 
\rtT_i{}^j & :=- e^{-\bb}\left(N_i{}^j +\Phi \bar\Sigma_i{}^j\right)=-\rtN_i{}^j, \\
P& :=A_K +\frac{(d-3)}{(d-2)} \E,\label{P} \\
J^j& : = \E V^j  + \mathring{J}^j .\label{J}
\ee
The scalar  
\be
\mathcal{E}: =e^{-\bb} (\mu+\Phi\btheta)=\rN  =\hT_{VK}
\ee
plays the role of the energy aspect, $P$ plays the role of conformal fluid pressure, $\mathring{J}^j$ is the energy current, $\Pi_i$ the momentum  aspect, and  $\dot{\mathcal{T}}_i{}^j$ is the viscous stress-tensor of the conformal fluid. 
Finally, let us note that if $Y^i = Y^j\gamma_j{}^i$ is a  horizontal vector then we have the correspondence
\be\label{diffhnSn}
\hn_i Y^i = (\Sn_i + \rom_i +\mathring\pa_i\bb)Y^i.
\ee
This identity, combined with \eqref{div}, gives \eqref{eq_div}.

\section{Radiative expansion and  symmetries}\label{sec:radex}

We now present the analysis of the radial expansion of the Bondi-Sachs metric.
Solving the radial Einstein's equation provides  an expansion of $(\bb,\Pi_A,\mu,N_{AB})$ 
in terms of polyhomogeneous functionals of the metric around $\O=0$.

\subsection{Polyhomogeneous expansion\label{sec:polyexp}}

\paragraph{Recursive solution of the radial evolution\label{sec:recursivesol}}

The space of $n$-polyhomogenous functions, denoted $C_n^{\mathrm{poly}}$, with $n \geq 1$,
   is the space of functions $A$  of $\O$ that admit an expansion around $\O=0$ of the form
\be 
A = \sum_{p=0}^{n-1} A_{p} \O^p  + \Omega^{n} \sum_{p=0}^N \O^p \left( \sum_{q=0}^{p+1}  \ln\O^q A_{(n+p,q)} \right) +o(\O^{N+p}),
\ee 
for $N$ arbitrary.  The index $n$ labels the order at which the first logarithm  appears.\footnote{This term is $A_n \O^n + A_{(n,1)} \O^n \log\O$.} 
The space of $n$-polyhomogenous functions is stable under  logarithmic derivation, integration and products. 
If $A\in C_n^{\mathrm{poly}}$ and $ B \in C_{m}^{\mathrm{poly}}$ then 
\be
\O\pa_\O A\in C_n^{\mathrm{poly}}, \qquad \pa_\O^{-1} A :=\int_0^\O A \in C_{n+1}^{\mathrm{poly}}, 
\qquad AB \in C_{\mathrm{min}(m,n)}^{\mathrm{poly}}.
\ee
More precisely, we have that 
\be\label{eq:ppOm-inv}
\O\pa_\O (\O^a \ln^b \O) & =  a (\O^a \ln^b \O)  + b \O^a \ln \O^{b-1} \cr
\pa_\O^{-1}(\O^a \ln^b \O) &= \O^{a+1}\left(
 \sum_{c=0}^{b} C_{ab }{}^c  \ln^{b-c} \O\right), \qquad   a\neq  -1\cr
  \pa_\O^{-1}(\O^{-1} \ln^b \O) &= \frac{\ln^{b+1}\O}{b+1},
\ee
where we introduced the symbol $\pa_\O^{-1} $ to denote integration along $\O$. 
$C_{ab}{}^c$  are coefficient defined  for  $a\neq -1$ and $b\geq c$  as
\be
C_{ab}{}^c = \frac{(-1)^c}{(a+1)^{c+1}} \frac{b!}{(b-c)! }. 
\ee

Solving recursively the radial evolution equations means that the metric coefficients are $n$-polyhomogeneous functions.
Indeed, polyhomogenous functions are necessary since the operator $[(d-3) -\O\pa_\O]$   entering the LHS of the Einstein equations \eqref{eq_G2} fails to be invertible on analytic functions owing to the fact that its image excludes $\O^{d-3}$. Logarithmic anomalies develop when the RHS of Einstein's equations possess a non-zero term proportional to $\O^{d-3}$. 
Then, once a logarithmic anomaly is present, it sources new logarithmic terms. 
We now analyse the radial expansion of \eqref{eq_G2}.
Since the $\langle A B\rangle$ equation can allow radiation in odd dimensions, i.e. a non-vanishing $\bTheta_{AB \left(\frac{d-4}{2}\right)}$ only if we allow an expansion in $\sqrt{\O}$. For this reason, we restrict our analysis to $d$ even.

The resolution of \eqref{eq_G2}   implies, using the expansion just described, that  we expect ``primary'' logarithmic terms $\mu_{(d-3,1)}$,  $ \Pi_{A(d-3,1)} $   which we call 
``Coulombic anomalies'' 
and $ \bTheta_{AB \left(\frac{d-4}{2}, 1\right)}$ referred to as the  ``radiative anomaly''. 
The radiative anomaly  corresponds to a term  
$\gamma_{\left(\frac{d-2}{2}, 1\right)AB}$ in the metric expansion and, thanks to the Bianchi identity \eqref{N=dotC}, to a term $N_{\left(\frac{d-4}{2}, 1\right)AB}$ in the news expansion.
The radiative anomaly also sources, through equations the radial evolution \eqref{eq_G2}, ``secondary'' logarithmic terms 
$ \bar A_{\left(\frac{d-4}{2},1\right)}$,   $\Pi_{A \left(\frac{d-4}{2}, 1\right)}$, $\mu_{\left(\frac{d-2}{2}, 1\right)}$ and $N_{\left(\frac{d-2}{2},1\right)AB}$.

The vanishing of the radiative anomaly $ 
\bTheta_{\left(\frac{d-4}{2}, 1\right)} =0  = N_{\left(\frac{d-4}{2}, 1\right)}$ implies that the scondary logarithms vaninsh
\be
\bar{A}_{\left(\frac{d-2}2,1\right)}=\mu_{\left(\frac{d-2}2,1\right)}=
\Pi_{A \left(\frac{d-2}2,1\right)}=N_{\left(\frac{d-2}2,1\right)AB}=0.
\ee
Crucially, in $d=4$ the radiative anomaly always vanishes,
 due to the fact that in 2 dimensions the traceless part of its Ricci tensor vanishes identically, $\mathring{R}_{\langle AB\rangle }=0 $.
Similarly, in $d=4$ we also have that the source term for $\mu_{(1,1)}$ identically vanishes.\footnote{Note that in $d=4$, we have $\frac{d-2}2=d-3$, hence the secondary logarithm $\mu_{\left(\frac{d-2}{2}, 1\right)}$ which vanishes identically, is the 
same as the Coulombic logarithm $\mu_{\left(d-3, 1\right)}$. We therefore have $\mu_{(1,1)}=0$. This is a non-trivial fact that we establish in the next subsection.}

More generally, the vanishing of the radiative anomaly implies that the only source of logarithms are Coulombic: they come from $\Pi_{(d-3,1)A} $ and $\mu_{(d-3,1)}$. These terms source the radiation $N_{(d-2,1)AB}$   see \eqref{eq:Coulomban},  and therefore, assuming regular initial conditions at $u=-\infty$, $\bTheta_{(d-2,1)AB}$.

In total, this means that in dimension $d$, we have  the behaviour
\be\label{poly2}
(\Pi_A,\mu) \in  C^{\mathrm{poly}}_{d-3}, \qquad 
(N_{AB},\bTheta_{AB},\bar{A}) \in   C^{\mathrm{poly}}_{d-2}.
\ee
From (\ref{aspectdef}, \ref{aspectdef2}), we know that 
 $\gamma$ is one order up from $\bTheta$, while  $\beta$, $\O\Phi$ and $\Upsilon$ are two orders up from, respectively, $\bar A$, $\mu$ and $\Pi$, hence we conclude
\be
(\gamma_{AB}, \Upsilon^A, \O\Phi) \in  C^{\mathrm{poly}}_{d-1}, \qquad \beta \in C^{\mathrm{poly}}_{d}  \subset C^{\mathrm{poly}}_{d-1} .
\ee
 This shows that when the radiative anomaly vanishes then 
\be 
g_{ab}\in  C^{\mathrm{poly}}_{d-1}.
\ee 
This is in fact an equivalence: 
 \emph{the condition of vanishing radiative anomaly is equivalent to logarithmic simplicity.}

In practice, this means  that we have, the expansion
\begin{subequations}\label{metexp}
\begin{align}
\gamma_{AB}&= q_{AB}+ \Omega\,  \gamma_{1AB} +  \cdots + \O^{d-1} (\gamma_{(d-1)AB}+  \gamma_{((d-1),1) AB}\ln\O)+\cdots
\\
\Upsilon^A &=U^A +\Omega \Y_1^A +  \cdots + \O^{d-1} ( \Y_{(d-1)}^A+ \Y^A_{(d-1,1)} \ln \O)+\dots \\
\Phi &= F  + \Omega \Phi_1 +  \cdots + \O^{d-2}( \Phi_{(d-2)} + \Phi_{(d-2,1)} \ln \O)+\dots  \\
\bb &= \a + \O^2 \bb_2 +  \cdots + \Omega^{d}( \bb_{d} + \bb_{(d,1)} \ln \O) + \dots
\end{align}
\label{eq:expansion}%
\end{subequations}%
We have used that $\beta_1=0$ due to the Pbc. It is  also important to remember that in $d=4$ we have that $\mu_{(1,1)}=0$, hence $\Phi_{(2,1)}=0$ and $\Phi_{(3,2)}=0$; so the first logarithm is $\Phi_{(3,1)}$ in $d=4$.
See \cite{winicour1985logarithmic, chrusciel1993gravitational, Madler:2016xju, godazgar2020bms, Geiller:2022vto} for earlier results about the polyhomogeneous expansion in dimension $4$.
We see that  imposing the Penrose condition of asymptotic simplicity, i.e. the condition that the metric is differential of class $k=3$, in $d=4$, all the logarithmic terms 
$(\gamma_{(3,1) AB}, \Y^A_{(3,1)}, \Phi_{(3,1)}, \bb_{(4,1)})$  have to vanish and the metric expansion is regular.  Similarly, in $d\geq 6$, regularity of the expansion follows from demanding that the radiative \emph{and} Coulombic anomalies vanish.\footnote{These conclusions holds provided that initial conditions  at a cut, e.g. at $u=-\infty$, are also regular in $\O$ -- see next section.\label{fnt:regularinitial condition}}

\paragraph{Free data\label{sec:freedata}}

We denote by $(\a,  F, U^A)$ the value of $(\bb,  \Phi, \Upsilon^A )$ at $\Omega=0$.
These are freely chosen.
We also denote by $q_{AB}$ the metric's value at $\Omega=0$. 
In the following, the spatial indices of all {\em higher order} quantities are raised and lowered by the {\em leading order} metric $q_{AB}$ and its inverse. We will also denote the Levi-Civita derivatives associated to $q_{AB}$ by  $\D_A$ to distinguish it from $\Snabla_A$.
We also denote $\gamma_{\langle A B\rangle } $ the traceless, with respect to $q_{AB}$, component of $\gamma_{AB}$.

Here we explain the general algorithm to solve, order by order, the Einstein equations for the Bondi metric coefficients, as well as the mass, angular momentum, and radiation aspects. In doing so we will identify the free data that fully determine a spacetime metric. More details are given in $d=4$ and $d=6$ in the next section, where the logarithmic anomalies are also discussed.

Given $(\a, F,q_{AB})$, and given $\bTheta_{AB}$ up to order $k$, the order $k+1$ of $\bb$ and $ \gamma_{AB}$ are respectively  determined by the $(KK)$ Einstein equation \eqref{eq_G2}, and by the Bondi  gauge condition $\btheta=0$ (since $\tfrac12 \pa_\O\gamma_{\langle A B\rangle}$). 

Similarly, the order $k$ of $\Pi_A$ is fixed by the $(KA)$ Einstein equation \eqref{eq_G2} -- \emph{except} for $\Pi_{(d-3)A}$ which is therefore also part of the free data. 

Next, knowledge of $(\gamma_{AB}, \beta, \Pi_A)$ at order $k$ determines $\Upsilon^A$ at the order $k+1$ through the equation
\be
\label{paUps}
\pa_\O \Upsilon^A =  e^{\bb} \gamma^{AB}( \pa_B\bb - 2 \O \Pi_B).
\ee
The exception is its zero-th order $U^A$, which is (already) part of the free data.
Similarly, the $k$'s order of $\mu$ is fixed by the $(KL)$ Einstein equation \eqref{eq_G2} -- \emph{except} for $\mu_{(d-3)}$ which is therefore also part of the free data. 

Next, the quantities fixed so far determine the $(k+1)$-th order of $\theta$ via the Bianchi identity \eqref{bianc_bis} -- except for its zeroth  order which is fixed by the Penrose boundary condition as per \eqref{eq:theta=F}. The Newtonian coefficient $\Phi$ is thus determined at the same order by the identity (combine \eqref{N-vector-BS} with \eqref{aspectdef})
\be \label{thetaphi}
(d-2) \Phi  = e^\bb \theta - \O \mu.
\ee 

Next, we address how to  the order $k+1$ of $N_{AB}$ and $\bTheta_{AB}$ assuming that we know them and $(\bb,\mu,\Phi,\Upsilon^A,\Pi_A,\gamma_{AB})$ up to order $k$. First we note that equation \eqref{N=dotC} relates the quantities $N_{AB}$ and $ \mathcal{L}_V\bTheta_{AB}$ order by order.\footnote{$V$ is tangent to $\Sigma_\O$ and is determined by $(\beta, \Upsilon^A)$.} So knowledge of $N_{AB}$ at order $k$ (over the entirety of $\scri$), together with the value of $\bTheta_{AB}$ at a cut of $\Gamma_{u_C}$ of $\scri$, completely fix $\bTheta_{AB}$ at that order. Then, the only recursion relation we need to consider is the one given by the $\langle AB\rangle$ Einstein equation \eqref{eq_G2}. The recurrence then fixes $N_{AB}$ at order $k+1$ from the knowledge of $N_{AB}$ and $\bTheta_{AB}|_{u_C}$ at the order $k$ -- \emph{except} for $k+1 = \frac{d-4}{2}$.

Finally, as we are going to see in the next section, the evolution equations along $\scri$ given by the $(VN)$ and $(AN)$ components of the Einstein equations 
determine the value of $(\mu_{(d-3)},\Pi_{(d-3)A})$ in terms of their value at a cut $\Gamma_{u_C} \subset \scri$ (see equations \eqref{Tconserv} and \eqref{Eevolution}).

 Summarizing, the {\em free data}   on $\scri$ is  given by the  induced metric on $\scri$ together with the \emph{radiation aspect} $\bTheta_{\frac{d-4}{2} AB}$ 
\be
\text{free data on $\scri$:} \qquad  (\a,  F,U^A,  q_{AB},  \bTheta_{\frac{d-4}{2} AB}  ),
\ee
while the free data at a cut $\Gamma_{u_C}\subset \scri$ is given by the traceless component of the metric 
$\gamma_{\langle AB\rangle }$, or equivalently $ \bar\Theta_{AB}$, and the  
mass and angular-momentum aspects, $\mu$ and $\Pi_A$:
\be
\text{free data at a cut $\Gamma_{u_C}$:} \qquad 
(\mu_{(d-3)}, \Pi_{(d-3)A}, \bTheta_{kAB})  \,\,  \text{with} \ k\neq\tfrac{d-4}{2}.
\ee
(see similar statements in \cite{Compere:2019bua,Capone:2023roc} valid for the Bondi  gauge boundary condition).
It is customary to chose the cut $\Gamma_{-\infty} :=\scri_-$  at $u_{C}=-\infty$.

The charge aspects contained in $\bar\Theta_{AB}(u_C)$, besides the radiation data $\bar\Theta_{\tfrac{d-4}{2}AB}$, corresponds to the higher spin charges studied in \cite{Freidel:2021qpz,Freidel:2021ytz}.

\subsection{Radial expansion of  Bondi metric coefficients}\label{subsec:radexp}

In this section we discuss the radial evolution given by the Einstein equations explicitly. We start by expressing the Pbc123 and Einstein's equations at the subsequent orders in $\O$ in terms of the metric coefficients introduced in the previous section.
Details about the expansion  can be found in appendix \ref{app:radexp}.

We introduce here a notation for a subleading order of $\gamma_{AB}$ that plays a central role in $d=4$ and $d=6$:
\be\label{4dlogan}
D_A{}^B := [\bTheta_A{}^B]_1
\ee
$D_A{}^B$ is manifestly traceless in the Bondi gauge. To lower its index we use the leading order metric $q_{AB}$:
\be
D_{AB} := [\bTheta_A{}^C]_1q_{CB} =\bTheta_{1\langle AB\rangle_q} - 2 (\bTheta_0\cdot\bTheta_0)_{\langle AB\rangle_q}.
\ee
In $d=4$ the last term in the expression above vanishes\footnote{In  $d=4$, the product of two $2\times 2$ symmetric traceless matrices is pure-trace, whence  $2(\bTheta_{0}\cdot\bTheta_{0})_{A}{}^B = \delta_A{}^B (\bTheta_{0}:\bTheta_{0})$ and therefore $D_{AB} = D_{A}{}^C q_{CB}= \bTheta_{1\langle AB\rangle_q}$. \label{fnt:DAB}}  and $D_{AB} =\bTheta_{1\langle AB\rangle_q}$.

The Pbc1 \eqref{Pbc1}, which states the nullness of $\scri$, has been implemented from the get-go by having $\Phi$ multiplied by $\O$ in \eqref{eq:ds2}.
The remaining $d(d-1)/2$ Pbc2 \eqref{Pbc2}  express the finiteness at $\scri$ of $(\mu,\Pi_A,N_{AB})$ and they read:
\be
\begin{dcases}
\bb_1 = 0 & \quad (KK)
\\
\U^A_1 = q^{AB}\pp_B e^\a &\quad (KA)  
\\
\tfrac12 \pa_u q_{AB} + \nabla_{(A} U_{B)}= F q_{AB} &\quad (AB) 
\end{dcases}
\label{Pbc-components}
\ee
The LHS of the equation $(AB) $ is equal to the  leading order of the extrinsic tensor
$e^\bb \Theta_{AB}:= \frac{e^\bb}2 \gamma_{A}{}^a\gamma_{B}{}^b {\cal L}_{L} g_{ab}$ which is therefore shear-free.
The trace of this equation relates $F$ to the expansion of $\scri$:
\be
 (d-2) F =  \pa_u \ln \sqrt{q} + D_{A} U^A = e^\a\theta_0 .
\label{eq:theta=F}
\ee
Note that to insure that the foliation $\Omega$ is timelike we have to impose the condition $F\geq 0$, i.e. $\scri$ is expanding.

At the next order we use that $e^\a\theta_1=\Delta e^\a$ \eqref{bianc_bis} and 
the third  Penrose boundary condition \eqref{Pbc3} then implies 
\be\label{mupi0}
(d-3)\Pi_0^A=  D_B \bTheta_{0}^{AB}, \qquad
(d-3)\mu_0 = -\frac12 e^\a R_q.
\ee  In dimension $d=3$, the RHS of these equations vanish, meaning that  $\mu_0$ and $\Pi_0$ are the mass and angular momenta aspect not determined by the radial evolution. 

In dimension $d\geq 4$, from the Pbc3 \eqref{Pbc3}  and the relationships between $(\Phi, \mu)$ \eqref{aspectdef} and $(\Pi_A,\Upsilon_A)$ (\ref{eq_2+2_explicit},\ref{aspectdef2})  
we get that (see also appendix\ref{app:radexp})
\be
\begin{dcases}
\bb_2 = -\tfrac{1}{2(d-2)} \bTheta_{0A}{}^B\bTheta_{0B}{}^A & (KK)\\
\Phi_1 = \tfrac{1}{2(d-3)(d-2)} e^\a  R_q + \tfrac{1}{d-2} \Delta e^\a & (KL)  \\
 \Y^A_2 = - \tfrac{e^{(4-d)\a}}{(d-3)}\D_B(e^{(d-3)\a} \bTheta_{0}^{BA}) &(KA)
\end{dcases}
\label{eq:EEq-exp}
\ee
where $ R_q$ is the Ricci scalar of $q_{AB}$. 
Finally, from the expansion of \eqref{aspectdef} we deduce the relationship between the  $\Phi_2$ and $\mu_1$:
\be\label{Phi2}
- (d-2) \Phi_2 =  \mu_1+  \tfrac1{(d-3)} D_A \left(e^{\a} D_B \bTheta_{0}^{AB}\right) . 
\ee

Note that the $\langle AB\rangle$ Einstein equation \eqref{eq_G2} implies that 
\be
(d-4) N_{0  A B}
= 
  - e^{\a} {R}_{q\langle A B\rangle}.
  \label{eq:N0AB}
  \ee
 In $d=4$, this equation is trivially satisfied. While in $d\geq 6$ it means, using \eqref{N=dotC}, that 
  \be\label{eq:bTR}
 (\nabla_\ell+   F) \bTheta_{0AB}  = -\left( D_{\langle A} D_{B\rangle}e^\a+ \frac1{(d-4)} \e^{\a} {R}_{q\langle A B\rangle} \right)  =-\frac{1}{(d-4)}e^\a
 {R}_{e^{-2\a} q\langle A B\rangle},
 \ee
 where we used in the last equality the renomalization property of the curvature tensor under resscaling.
 Interestingly, the combination of 
 \eqref{eq:N0AB} and \eqref{mupi0} means that the tensor $\SN_0$ is directly related to the codimension $2$ Schouten tensor
 \be \label{eq:NS0}
 \SN_{0AB}-\nu_0 \bTheta_{0AB}
 &=   e^{-\a}N_{0AB} +e^{-b}\mu_0 \frac{q_{AB}}{(d-2)} \cr
 &= - \left( \frac{1}{(d-4)}R_{q \langle A B\rangle} + \frac{q_{AB}}{(2(d-3)(d-2)} R_q \right) \cr
  & = - \frac{1}{(d-4)}\left(R_{q AB} - \frac{q_{AB}}{2(d-3)} R_q \right) 
  = - \mathring{S}_{AB}.
 \ee 
To the next order,  the Pcb$d$ \eqref{Pbcd}, for $d>4$,  yield (see \eqref{eq:muPian})
\be 
 (d-4) \mu_1  &= \e^{\a}\left(  -\tfrac{(d-4)}{(d-3)} D_A D_B \Theta_{0}^{AB}  + \bar\Theta_0 :\, {R}_{q0}  \right)\cr
 (d-4)\Pi_{1A} &=   D_B[\bar\Theta{}^B{}_A]_1   - \tfrac{(d-4)}{2(d-2)}D_A\left( \bar\Theta_0 : \bar\Theta_0 \right).
 \label{eq:mu1-Pi1}
\ee
In $d=4$ these equations correspond to subleading vaccum equations. 
They imply that $(\mu_1,\Pi_{1A})$ are undetermined  in $d=4$ and, therefore, these quantities encode the \emph{mass} and \emph{angular momenta aspects}.
Interestingly, the RHS of the first equation vanishes in $d=4$ 
as a consequence of the fact that, in two dimensions, the Ricci curvature is pure trace.
Therefore, the $(KL)$ equation is automatically regular in $d=4$, and no logarithmic anomaly arises in the $\mu$ expansion at that order.
The RHS of the vector equation, on the other hand, is non-vanishing in $d=4 $, which implies a logarithmic anomaly
\be\label{eq:log-anom-d4}
\Pi_{(1,1)A}=  D_B D_A{}^B. 
\ee
So, we see that the divergence  of the traceless tensor $D_A{}^B $ sources the logarithmic anomaly for $\Pi_A$. 
 Since we assumed the cuts $\Gamma_u = S_{(u,\Omega=0)}$ to be spheres, demanding the $d=4$ anomaly to vanish implies that $D_A{}^B$ vanishes too \cite{Madler:2016xju}. The condition $D_A{}^B=0$ is equivalent to the demand of analyticity of $\scri$ \cite{Geroch:1977jn,Fernandez-Alvarez:2021zmp}.

The last equation to analyze is the order-1 radiative equation  (see appendix \ref{app:radm} for the detailed computations, also in $d>6$). In the polyhomogenous expansion of the trace-free $N_A{}^B$, we denote $N_{(p,q)AB} := [N_{A}{}^C]_{(p,q)}q_{CB}$, and thus obtain
\be \label{eq:N=R}
(d-6) N_{1AB} = e^b \left( -  [\mathring{R}_{\langle AD \rangle }\gamma^{DC}]_1 q_{CB} + 2 D_{\langle A} \Pi_{0B\rangle} +  e^{-b}\mu_0 \bTheta_{0 AB} \right),
\ee
In $d=4$, this equation implies (see \eqref{N_1eq}) that
\be
2N_{1AB}  = - e^b \left(\Delta   - \tfrac32  R_q \right) \bTheta_{0AB}.
\label{eq:N1-d4}
\ee
In $d=6$, it means that $N_{1 A B}$ is undetermined and therefore represents the news of radiation, and $D_A{}^B$ the radiation field. It also means that the radiative anomaly is given by \eqref{eq:rad-anomaly-d6}:\footnote{We thank F. Capone and his collaborators of \cite{Capone:2023roc} for a communication helping us correct a mistake in a previous version of our manuscript}.
\be\label{eq:rad-anomaly-d6-text}
-2 e^{-\a}N_{(1,1)AB} & =  \left(\Delta + \tfrac16 {R}_q\right)\bTheta_{0AB} - \tfrac43 D_{\langle A} (D\cdot\bTheta_0)_{B\rangle_q} \cr
& \qquad \quad + 2(W_q:\bTheta_0)_{\langle AB\rangle_q} 
- 2 (R_q\cdot \bTheta_0)_{\langle AB\rangle_q}.
\ee
where $(R_q\cdot \bTheta)_{AB} = R_{qAC}\bTheta^C{}_B$ for $R_{qAB}$ the Ricci tensor of the codimension-2 metric $q_{AB}$. Similarly,  $(W_q:\bTheta_0)_{AB}= {W}_{qACB D} \bTheta_{0}{}^{CD}$ and $W_q$ denotes the Weyl tensor of $q_{AB}$.

Finally we note that \eqref{N=dotC} means that the logarithmic radiation $N_{(1,1)}$ is sourced by $\bTheta_{(1,1)}$ with the relation\footnote{In all $d\geq 6$, the logarithmic radiation is related to the radiative anomaly by: $2N_{\left( \frac{d-4}{2},1\right)}=(\cL_{\ell}+ F)\bTheta_{( \frac{d-4}{2},1)}.$}
\be \label{eq:NbTan}
2  N_{(1,1) A}{}^B = (\cL_\ell + F)\bTheta_{(1,1)A}{}^B  ,
\ee 
while \eqref{eq:bTR} relates the time derivative of $\bTheta_0$ to the curvature. 
In section \ref{subsec:Weyl},  we provide solutions to the condition $\bTheta_{(1,1) AB}=0.$

Note that if we demand as in 
\cite{Capone:2023roc} that $b=F=U^A=0$, then  \eqref{eq:NbTan} simply becomes $2N_{(1,1)}= \pa_u \bTheta_{(1,1)}$ and one can show that 
equation \eqref{eq:rad-anomaly-d6-text} is compatible with the result of \cite{Capone:2021ouo, Capone:2023roc}.\footnote{Private communication by F. Capone: Note that in \cite{Capone:2023roc}, the anomaly  $\bTheta_{(1,1)AB}|_\text{here} = 2\gamma_{(2,1)AB}|_\text{here} = 2 h_{(2,1)AB}|_\text{there}$ and therefore $N_{(1,1)AB}|_\text{here}  = - \mathsf{H}_{(1)}|_\text{there} =\pa_uh_{(2,1)AB}|_\text{there}$. 
}

We have seen that, when the logarithmic simplicity condition $\bTheta_{(1,1)AB}=0$ is imposed, all logarithmic anomalies are Coulombic.
From \eqref{eq_G2} we can evaluate the first Coulombic logarithm to be given by 
\be \label{eq:Coulomban}
-\frac{d}2 N_{(d-2,1) A}{}^B= e^b 
D_{\langle A} \Pi_{(d-3,1)}^{B \rangle} +  \tfrac12  \mu_{(d-3,1)} \bTheta_{0A}{}^{B}
,
\ee
while \eqref{N=dotC} provides a differential equation for $\bTheta_{(d-2,1)}$
\be\label{eq:bTCoulomban}
[\cL_\ell + (d-1)F]\bTheta_{(d-2,1)\langle A}{}^{B\rangle}
&=  (d-1) N_{(d-2,1) A}{}^B
+ 
\mu_{(d-3,1)}\bTheta_{0A}{}^B
+
2
D_{\langle A}\left( e^b \Pi_{(d-3,1)}^{B \rangle}\right)
\nonumber
\\
&= 
\frac{2}{d} e^{\a} D_{\langle A} \Pi_{(d-3,1)}^{B\rangle} 
+ 2\left(D_{\langle A} e^b\right)\Pi_{(d-3,1)}^{B\rangle}
+ \frac1d \mu_{(d-3,1)} \bTheta_{0A}{}^B
\ee 
where we used that $[\pa_\O \Phi]_{(d-3,1)}= (d-2) \Phi_{(d-2,1)} = -\mu_{(d-3,1)} $  which holds thanks to \eqref{aspectdef} and the fact that $\theta$ is of higher polyhomogenous order, i.e. $\theta \in C^{\mathrm{poly}}_{d-1}$, due to \eqref{bianc_bis}.

\subsection{Simplifications in $d=4$ \label{sec:somed4simpl}}

In dimension $4$, many simplifications happen since the sphere is $2$ dimensional.
This means for instance that $\mathring{R}_{\langle AB\rangle }=0 $ and also that $(\Sigma \cdot \Sigma)_{\langle AB\rangle} =0$ for 
$\Sigma_{AB}$ a symmetric traceless tensor.
In particular this implies, as shown in \cite{Grant:2021hga}  that, once we use the  Bondi-Sachs gauge condition  \eqref{eq:bondiradframe}, we can parametrize the metric $\gamma_{AB}$ and its inverse in terms of a
traceless matrix $\cC_A{}^B$ which is symmetric  with respect to $q_{AB}$:\footnote{This means that $\cC_{A}{}^C q_{CB}= \cC_B{}^C q_{C A}$, while $\cC_A{}^A=0$.}
\be
\gamma_{AB}=\left(\rho \delta_A^C + \O \cC_A{}^C\right)  q_{CB}, \qquad 
(\gamma^{-1})^{AB}=q^{AC}\left(\rho \delta_C^B  - \O \cC_C{}^B \right),
\ee
where $ \rho:=\sqrt{1+ \frac{\O^2}2  \cC:\cC }$ with $\cC:\cC= \cC_C{}^D \cC_D{}^C$. 
 One can  check that $\gamma^{AB}\pa_\O \gamma_{AB}=0$.
The relationship between the traceless matrix $\cC_A{}^B$ and the shear tensor is 
\be\label{eq:bThetaexpansion}
q^{AB}\bTheta_{AB} = \pa_\O \rho, \qquad 
2 \bTheta_{\langle AB \rangle_q}=  (1+\O\pa_\O)(\cC_A{}^C) q_{CB}. 
\ee 
In the following, it will prove useful to use the standard  notation for the expansion of the traceless tensor $\cC_A{}^B$ 
\be\label{Ddef}
 \cC_A{}^B =: C_A{}^B + \O D_A{}^B + \cdots 
\ee
where $C_{AB}= \gamma_{1\langle AB\rangle_q}= 2\bTheta_{0AB}$ is the radiation's shear and $ D_{AB}= \gamma_{2\langle AB\rangle_q}= \bTheta_{1 \langle AB\rangle_q}$
 represent, as shown in the previous section, the source of the logarithmic anomaly.

In $d=4$, we also have the following expressions for the mass, angular-momentum  and radiation aspects 
\be \label{MPNaspects}
\rN_1 & = \E_1 = e^{-\a}\mu_1 =- 2 e^{-\a}\Phi_2 -\frac12  e^{-\a}  D_A(e^b D_B  C^{AB}) \\
\Pi_{1A} & = \frac12 \left[ \mathring{\nabla}_B \bb - \gamma_{AB} e^{-\bb}\pp_\O \Upsilon^B\right]_2,\cr 
N_{0A}{}^B &:=  e^\a\Theta_{1 \langle A}{}^{B\rangle_q} =  \frac12 (\mathcal{L}_\ell  + F) C_A{}^{B} + D_{\langle A}D^{B\rangle} e^\a.\nonumber
\ee
where $\ell = \pp_u + U^A\pp_A \SCRIeq e^\bb L$.
It will also be useful to have the expression for the acceleration (see appendix \ref{app:radexp}) 
\be 
[\kA]_0
&=  \frac12 e^{-\a}  \Delta e^{\a}  +\frac14 e^\a R_q,  \cr
[\kA]_1 &= -\E_1
-  e^{-\a} ( D_A [e^{\a} \Pi_{0}^A]  + \tfrac18  F (C:C)  ). \label{acc1}
\ee

\subsection{Other boundary conditions  and their asymptotic symmetries\label{sec:otherbc}}

The Pbc we have employed so far are the most general boundary conditions compatible with the existence of a boundary   $\scri = \pp M $ for the conformally compactified spacetime.
Most other choices made in the literature, such as: Bondi-Sachs \cite{Bondi:1960jsa,Bondi62,Sachs:1962wk}, Barnich-Troessaert \cite{Barnich:2010eb,Barnich:2016lyg }, Campiglia-Laddha \cite{Campiglia:2015yka, Compere:2018ylh, Campiglia:2020qvc}, or the most general case of BMSW \cite{Freidel:2021fxf} and Capone et al. \cite{Capone:2023roc} in higher dimensions,  impose that 
\be \label{BMSW}
\a {=} 0, \qquad F  {=} 0,\qquad U^A {=} 0.
\ee
The only exception relaxing these conditions is  \cite{Geiller:2022vto,Geiller:2024amx}. 
Note that since 
\be\label{theta2}
\e^{\bb}\theta&= F + \O\left( \Delta e^{\a} \right)  -\O^2 D_A [e^{\a} \Pi_{0}^A] + o(\O^2), 
\ee
 under these conditions the expansion of $V$ vanishes up to second order.  Moreover, due to the shear-free nature of $\scri$ \eqref{Pbc2}, we have that the BMSW boundary conditions imply the time-independence of $q_{AB}$:
 \be\label{eq:BSMW-ppu(q)}
 0\SCRIeq \mathcal{L}_V q_{AB} = \pp_u q_{AB}
 \ee
 
Given the transformations (\ref{db},\ref{dF},\ref{dV}) we see that imposing these conditions implies that 
\be
\ell[\tau]= W, \qquad \ell[W]=0, \qquad [\ell , Y]^A=0, \qquad \ell=\pa_u.
\ee
which means that $W=W(\sigma^A)$ and $Y^{A}=Y^A(\sigma^A)$ are time independent while 
$\tau= T(\sigma^A)+ uW(\sigma^A)$ where $T$ is also time independent.
The vector fields 
\be
\xi_{(T,Y,W)} = T(\sigma^A)\pa_u + Y^A(\sigma^A) \pa_A + W(\sigma^A)(\O\pa_\O + u\pa_u) 
\ee
generates the Weyl-BMS group  \cite{Freidel:2021fxf}
\be
\mathsf{BMSW}=(\mathrm{Diff}(S)\ltimes \mathbb{R}^S)\ltimes \mathbb{R}^S.
\ee
Demanding that the symmetry transformations also preserve the area forms $\delta_{\xi_{\mathsf{GBMS}}} \sqrt{q}=0$ restricts $\mathsf{BMSW}$ to the 
Generalized-BMS group \cite{Campiglia:2015yka, Compere:2018ylh, Campiglia:2020qvc} 
\be
\mathsf{GBMS}=(\mathrm{Diff}(S)\ltimes \mathbb{R}^S)
\ee
and implies that 
\be 
W_{\mathsf{GBMS}}=  \frac1{(d-2)} D_A Y^A.
\ee
The Extended-BMS group proposed by Barnich and Troessaert \cite{Barnich:2010eb,Barnich:2016lyg }
\be
\mathsf{EBMS}=(\mathrm{Conf}(S)\ltimes \mathbb{R}^S)\ltimes \mathbb{R}^S,
\ee
does not allow for general diffeomorphisms  of the sphere, but it allows for a Weyl rescaling of its metric.
It is obtained from the $\mathsf{BMSW}$ group by imposing that the symmetries preserve the sphere's conformal structure: $\delta_{\xi_{\mathsf{EBMS}}} ([\mathrm{det}(q)]^{\frac{1}{(d-2)}} q^{AB})=0$.

Denote $W_{qABCD}$ the Weyl tensor of $q_{AB}$, and $R_q$ the  Ricci scalar. Then, another natural subgroup of $\mathsf{BMSW}$ is found by demanding that the symmetry preserves the round-sphere curvature condition\footnote{In $d>4$, the round sphere conditions are $R_q = const$ and $W_{qABCD}=0$, where $W_{qABCD}$ is the Weyl tensor of $q_{AB}$. Both conditions must then be preserved by the symmetry we call Round-BMS.} $R_q = const$, i.e. $\delta_\xi R_{q}=0$. We call this group\footnote{A. Rignon-Bret has discussed this group in a seminar given at PI.} Round-BMS $\mathsf{RBMS}$. 
In $\mathsf{RBMS} \subset \mathsf{BMSW}$, one has the condition
\be
 \pa_A W_{\mathsf{RBMS}}=0.
\ee
The subgroup with $W=0$ is denoted 
\be
\mathsf{RBMS}_0 \simeq \mathrm{Diff}(S)\ltimes \mathbb{R}^S,
\ee
and $\mathsf{RBMS} = \mathsf{RMBS}_0  \ltimes \mathbb{R}$, were the $\mathbb{R}$ factor represents global rescalings.
Even though $\mathsf{RBMS}_{0}\simeq \mathsf{GBMS}$ are abstractly isomorphic as groups, the two are different subgroups of $\mathsf{BMSW}$. First, the conditions on $W$ are distinct, and second the action of $\mathrm{Diff}(S)$ on $\mathbb{R}^S$ differs in the two cases by the conformal weights of the scalar.\footnote{To be more precise, if $(Y,f)\in \mathrm{Diff}(S)\ltimes \mathbb{R}^S$, then $\delta_Y f_{(a)} = Y[f_{(a)}] + \tfrac{a}2 (\nabla_B Y^B) f_a $ with $a=0$ in $\mathsf{RBMS}_0$ and $a=-1$ in $\mathsf{GBMS}$. For a more explicit notation, one could write $\mathrm{Diff}(S)\ltimes \mathbb{R}^S_{(a)} \subset \mathsf{BMSW}$. Note that $\mathsf{RBMS}_0 \simeq \mathsf{GBMS}$ as abstract groups, because we can always identify  a density $f_a$ with a function through $f_a = \sqrt{q}^{\frac{a}{2}} f_0$.}

Finally, we also have the celebrated $\mathsf{BMS}$ group. Which for each choice of round metric $\mathring{q}_{AB}$ demands
$\delta_{\xi_{\mathsf{BMS}}} \sqrt{\mathring q}=0$ and $\delta_{\xi_{\mathsf{BMS}}} {\mathring q_{AB}}=0$. In particular, $\delta_{\xi_{BMS}} {R_{\mathring q}}=0$ as in Round-BMS.
The BMS transformations are such that 
\be 
 W_{\mathsf{BMS}}=  \frac1{(d-2)} D_A Y^A, \qquad \nabla_{\langle A} Y_{B\rangle}=0,
 \ee
 and the group is 
 \be
 \mathsf{BMS} = \mathrm{Conf}(S, \mathring{q}_{AB})\ltimes \mathbb{R}^S,
 \ee
where $\mathrm{Conf}(S,\mathring{q}_{AB})$ denotes the group generated by conformal Killing vectors of the round metric $ \mathring{q}$, which is $\mathrm{SL}(2,\mathbb{C})$ in $d=4$. Note that choices of different $\mathring{q}_{AB}$ yield isomorphic but distinct $\mathsf{BMS}_{\mathring{q}}$ subgroups of $\mathsf{GBMS}$. See \cite{SpinC}  for more on this.

\section{Evolution equations  along $\scri$, Weyl tensor and Bianchi identities}\label{sec:Weyl}

 Recall our definition \eqref{CEST} of the conformal stress tensor $\T_{ab}$: it is the conformal-frame tensor that sources the Einstein equations in that frame:
\be
G_{ab} \heq (d-2) \T_{ab}
\ee
It is a purely geometrical quantity, meaning that it is nonzero even in the absence of matter fields. 
We have seen that a projection of this tensor along $\Sigma_\O$,  corrected by a trace term, leads to the construction of the the  asymptotic stress tensor $\hT_i{}^j$ \eqref{eq:projectedCSET}.
In this section, we first relate asymptotic infinity to stretched horizons, and thus show that the asymptotic stress tensor satisfies a covariant evolution equation along that hypersurface. 
Then, we relate the radial derivative of the asymptotic stress tensor  to the components of the (physical)  Weyl tensor. 
Finally, we revisit its conservation equation in terms of  a Bianchi identity for the Weyl tensor.

\subsection{Conservation equation}\label{sec:conservationeqs}

A stretched horizon is a codimension-$1$ (sub)manifold 
equipped with a stretched Carrollian structure\footnote{To make a point on the relation between $\scri$ and a general stretched horizon, we temporarily introduce the quantity $\rho := \frac12 N^2$, instead of $\O\Sc$.  Note also that the tensor $\gamma^{ij}$ is not an independent piece of data, but we list it here for completeness. See Section \ref{sec:Carroll}.}  \eqref{eq:stretchedCarroll} $\mathscr{C}=(h_{ij}, \gamma^{ij}, V^i, K_i, \rho)$:
\be
V^iK_i=1, \qquad h_{ij}V^j=-2\rho K_i, \qquad \gamma^{ij} K_i =0
\ee
We assume that the stretched horizon is furthermore equipped with a Carrollian connection $\hn$. Recall from Section \ref{sec:Carroll-connection} that specifying this connection is equivalent to fixing the boost connection $\omega_i$ and the shear $\bTheta_{ij}$.

In \cite{Freidel:2022vjq}, the following formula was derived that controls the projection of Einstein's equation along an arbitrary stretched horizon:
\be\label{stretch-cons}
h_i{}^a G_{a N} = \hn_j ( \mathsf{N}_i{}^j  - \mathsf{N} \delta_i{}^j)  - (\bar{\mathsf{K}}_i{}^j- \bar{\mathsf{K}} \delta_i{}^j)\hn_j \rho,
\ee
where 
\be
\bar{\mathsf{K}}_{i}{}^j : =  h_i{}^a   (\nabla_a K^b) h_b{}^j
\qquad\text{and}\qquad
\mathsf{N}_i{}^j  := h_i{}^a (\nabla_a N^b) h_b{}^j.
\ee
The quantity $\bar{\mathsf{K}}_{i}{}^j$ was already introduced in Section \ref{sec:Carroll-connection}, while $\mathsf{N}_i{}^j$ is an analogue of the Weingarten map.
The projected Einstein equation \eqref{stretch-cons} can be understood as a conservation equation for the \emph{Carrollian stress tensor} 
\be
\mT_i{}^j:=\mathsf{N} \delta_i{}^j- \mathsf{N}_i{}^j .
\label{eq:CarrollStrTens}
\ee
where $\mN=\mathsf{N}_i{}^i$ is the trace of $\mathsf{N}_i{}^j$.

Now, in the conformally compactified spacetime, thanks to the Penrose boundary conditions, the projected (vacuum) Einstein's equations reduce to
\be
G_{iN} \heq (d-2) \T_{iN} = - (d-2) \hn_i \nu,
\ee
where we used \eqref{NaN} to evaluate the asymptotic stress tensor \eqref{CEST}.
Furthermore, the surfaces $\Sigma_\O$ carry a stretched Carroll structure with $\rho=\O\Sc$ (Section \ref{sec:Carroll-connection}). Using the Pbc2 \eqref{eq_gennews} $\nabla_a N^b = \O\hN_a{}^b + \Sc \delta_{a}^{b}$, we also have that $\mathsf{N}_i{}^j =  \O \hN_i{}^j +\nu \delta_i^j$ and therefore
\be \label{Asymp-Carrollian-stress}
\mT_i{}^j = (d-2) \hn_i \nu + \O \hT_i{}^j.
\ee 
We see that the asymptotic  stress tensor $\hT_i{}^j$ is the subleading component of the Carrollian stress tensor $\mT_i{}^j$.
 Combining these observations, we see that at $\scri$ the analogue of \eqref{stretch-cons} is
\be \label{stretch-cons2}
\hn_j (\hN_i{}^j-\hN \delta_i{}^j) -  (\bmK_i{}^j- \bmK \delta_i{}^j)\hn_j \nu  \heq 0.
\ee
Therefore, comparing \eqref{stretch-cons} and \eqref{stretch-cons2}, we note that if $\scri$ is non-expanding then $\nu \to 0$ at $\scri$, just like $\rho\to0$ at the null horizon.\footnote{ Note: since the statement $\nu=0$ is not conformally invariant, even at $\scri$, this construction depends on the choice of foliation $\Sigma_\O$.}
Then, in this case, the asymptotic Einstein's equations take the same form of the \emph{vacuum} stretched horizons equation of \cite{Freidel:2022vjq}, provided that we make the identification 
$(\hat\nabla_i, \mN_i{}^j,  \rho) \to (\hat\nabla_i, \hN_i{}^j, \nu)$. In particular, under this replacement, the Carrollian stress tensor $\mathsf{T}_i{}^j := \mathsf{N} \delta_i{}^j - \mathsf{N}_i{}^j$ is replaced by  $\hT_i{}^j= \hN \delta_i{}^j - \hN_i{}^j $.\footnote{Recall that $\hT_i{}^j \neq h_i{}^a \T_a{}^b h_b^j$. See the discussion below \ref{eq:projectedCSET}.}  In the correspondence, $\bmK_i{}^j$ and the Carrollian connection remain unchanged.

Given the importance of this equation for our purpose, it is interesting to give an independent and more direct derivation that follows directly from the Pbc123. As we have seen, Pbc123  implies equation \eqref{WN} for  the Weyl tensor component $W_{abcN} = \O Y_{abc}$. Furthermore,  taking the projection of these equations along the tangent indices, we get
\be 
h_i{}^ah_j{}^b (\nabla_{a} \N_{b}{}^c - \nabla_{b} \N_{a}{}^c) h_c{}^l&=  \hat{Y}_{ij}{}^l  +  h_{i}{}^l \hat{\sT}_{j } - h_{j}{}^l\hat{\sT}_{i }, 
\ee
where we denoted $\hat{Y}_{ij}{}^l:= h_i{}^a h_j{}^b Y_{ab}{}^c h_c{}^l$ and  $ \hat{\sT}_{j } := h_i{}^a {\sT}_{a N}$.
We can evaluate the terms on the LHS (see appendix \ref{app:consl}) as 
\be 
h_i{}^ah_j{}^b (\nabla_{a} \N_{b}{}^c) h_c{}^l 
&=
\hn_{i} \hN_{j}{}^l 
+ \bmK_{i}{}^l \hn_{j}\nu - \mK_{ij}   \bA{}^l,
\ee
where we denoted $ \mK_{ij} := h_i{}^a h_i^b \nabla_{a} N_b$ and introduced the dual acceleration \eqref{eq_News_2+2}
\be
\bA^i:=  -\N_K{}^a h_a{}^i =  \bA V^i + \Pi^i.
\ee
After antisymmetrizing and taking the  contraction of  these equations  along $j=l$ while using that $\hat{Y}_{ij}{}^l$ is traceless we obtain the conservation equations
\be \label{Tconserv}
\hn_{j} \hT_i{}^j
+ (\bmK_{i}{}^j - \bmK \delta_i^j) \hn_{j}\nu &= -(d-2) \hat{\sT}_{j } \heq 0.
\ee
where $\hT_i{}^j = \hN \delta_j{}^i  - \hN_{j}{}^i $ is the asymptotic stress-tensor \eqref{eq:projectedCSET}.

\paragraph{Energy Conservation} 

We can now establish the equation for the conservation of energy. It is obtained  by projecting \eqref{Tconserv} along $V$.
One first evaluates 
\be
V^i \hn_{j} \hT_i{}^j &= \hn_{j} \hT_V{}^j   -(\hn_{j}V^i) \hT_i{}^j 
= \hn_{j}(\E V^j + \rJ^j) -\nu  \hT - \O \hV_j{}^i \hT_i{}^j \cr
&= {V}[\E]  + (\Sn_i  +2 \pa_i\bb )\rJ^j - \nu[ (d-2) A_K - \E ] - (\pa_i\bb \rJ^j +\O \Pi_i \rA^i) \cr
& \hspace{1.25cm} + \O \left(e^{-2\bb} N: N  -\nu^2 \bar\Sigma: \bar\Sigma\right)
- \O \left( \E -2 \nu \btheta\right) \left( A_K-\tfrac{\E}{(d-2)}\right).
\ee
This equation is obtained by using the expressions (\ref{hN},\ref{hV},\ref{diffhnSn}), as well as the equations
\be
\hn_i V^i&= (d-1) \nu + \O  [A_K+ \E -2\nu\btheta ],\cr
 \hT&= (d-2)  [A_K+ \E ],\cr
 P&=  A_K+\E-\tfrac1{(d-2)} \E
\ee
and the identity 
\be
\pa_i\bb \rJ^i + \O \Pi_i \rA^i  
&= \bmK_V{}^j \pa_j\nu.
\ee

In the Bondi gauge, $\btheta=0$ and thus $\bTheta_{ab}=\bar\Sigma_{ab}$,  the energy conservation equation simplifies to\footnote{Recall that the combination $(d-2)\nu +\O \E =\theta$ is simply the expansion of $V$. It vanishes to second order under  the $\mathsf{BMSW}$ boundary conditions}
\be \label{Eevolution}
0&= V[\E] + (\Sn_i  +2\pa_i\bb ) \rJ^i + \big((d-2)\nu +\O \E \big) \left(  \tfrac{\E}{(d-2)} - A_K \right) + \O  \left(e^{-2\bb} N: N  -\nu^2 \bTheta: \bTheta \right).
\ee
The zeroth order of this equation is automatically satisfied owing to the Penrose boundary conditions:  using (\ref{eq:AK},\ref{Jexp}) together with the Pbc3 (in the form \eqref{eq:EEq-exp}), we derive the following expressions for the dominant contribution of $(\E,\rJ^i,A_K)$:
\be
\E_0 = -\tfrac1{2(d-3)}  R_q, \qquad \rJ_{0i}=\mathring\pa_i\nu_0, \qquad (d-2)[\kA]_0= -\E_0 +{e^{-\a} }\Delta e^\a.
\ee
And thus get that at the zeroth order in $\O$ the energy conservation equation reduces to
\be
 \ell [\E_0] -  \Delta(e^b\nu_0)   + 2 (e^b\nu_0) \E_0 =0.
\ee
where $\ell = e^\a V_0$ is the asymptotic null vector \eqref{asymptnull}. 
This equation is  \emph{identically} satisfied by the curvature $R_q = -2(d-3)\E_0$, since $e^b \nu_0$ represents the asymptotic expansion of $\ell $ : $(\cL_{\ell} q_{ac})q^{cb}=  2 (e^b \nu_0)  q_a{}^b$, see \eqref{deltaR}.

In $d=4$, the energy aspect is contained in $\E_1$, which means that the analogue of the Bondi mass loss formula corresponds to the first (and first nontrivial) order in $\O$ of the above energy conservation equation.  Note, in particular, that at this order there appears the crucial mass loss  term $N\!:\!N$.
Now, if in addition we impose the $\mathsf{BMSW}$ boundary conditions \eqref{BMSW}, equation \eqref{Eevolution} simplifies considerably: first, it allows us to introduce the \emph{Bondi mass aspect} defined as $M := - \Phi_2$; second, as per equation \eqref{theta2} the expansion of scri vanishes to second order, whence $\theta_1= [(d-2)\nu+\O\E]_1=0 $; and third equation \eqref{BMSW} immediately implies $\nu_0=0$ and thus $\rJ_{0i}=0$; moreover, from the expansion detailed in appendix \ref{app:radexp}, we also obtain  
\be \label{nuJexpansion}
 \nu_1&=\frac14  R_q, \qquad   \qquad \qquad \qquad  \quad\rJ_{1A} = - \frac{1}{4}\pa_A   R_q,\cr
 \E_1 &= 2M - \frac12 D_A D_B C^{AB}, \qquad 
 N_{AB} =\frac12 \pa_u C_{AB},
\ee
where we used that $\bTheta_0^{AB}= \tfrac12 C^{AB}$.
Finally, the BMSW boundary conditions \eqref{BMSW} imply $V=\pa_u + O(\O)$ and $\pp_u q_{AB}=0$ 
\eqref{eq:BSMW-ppu(q)}, which, together with the Pbc $\beta_1 =0$ \eqref{Pbc-components}. Putting everything together, one deduces that with BMSW boundary conditions, the first order of the equation \eqref{Eevolution} reads
\be
2 \pa_u M - D_A D_B N^{AB} - \frac14 \Delta R_q  + N:N=0,  \label{eq:masslossBMSW}
\ee
which is the Bondi mass  loss formula \cite{Compere:2018ylh, Freidel:2021qpz} (note the discrepancy in the conventions: $ \ N_{AB}|_{\mathrm{Here}} = 2 N_{AB}|_{\mathrm{There}}$).

\subsection{Stress tensor's radial evolution}
To obtain the radial evolution of the asymptotic stress tensor 
we can apply the same strategy as above and project the expression  for the Weyl components $W_{abcN}=\O Y_{abc}$ along the radial vector $K^a$. Denoting 
\be
\hat{Y}_{i}{}^l:= h_i{}^b Y_{Kb}{}^c h_c{}^l,
\ee
we obtain
\be \label{eq:nablaN=Y}
h_i{}^a K^b( \nabla_{b} \N_{a}{}^c-\nabla_{a} \N_{b}{}^c) h_c{}^l
&\heq  \hat{Y}_{i}{}^l.
\ee
The terms on the LHS can be expressed as
\begin{align}
h_i{}^a K^b(\nabla_{a} \N_{b}{}^c ) h_c{}^l&=  - (\hn_{i} +\omega_i) \bA^l 
-  
\bmK_i{}^j \hN_j{}^l 
  +  \bmK_i{}^l \pa_\O\nu \cr
  h_i{}^a K^b(\nabla_{b} \N_{a}{}^c ) h_c{}^l&=  h_i{}^a (\nabla_{K} \hN_{a}{}^c)h_c{}^l 
+ \omega_i    \hN_{K}{}^l. \label{PKN}
\end{align}
Inserting in \eqref{eq:nablaN=Y} the difference of the two expressions in \eqref{PKN} and using that 
$\pa_\O \hN_i{}^j = h_i{}^a (\nabla_{K} \hN_{a}{}^b)h_b{}^j +[\bmK, \hN]_i{}^j 
$
we finally relate the radial expansion of $\hN_i{}^j$ to the rescaled Weyl tensor:
\be \label{radhN}
\pa_\O \hN_i{}^j  +  (\hN\!\cdot \!\bmK)_i{}^j
+   \hn_{i}  \bA^j 
- \bmK_i{}^j \pa_\O \nu &\heq \hat{Y}_{i}{}^j, 
\ee
Taking the trace of this equation gives
\be
\pa_\O \hN  +  \hN_i{}^j \bmK_j{}^i
+   \hn_{i}  \bA^i 
- \bmK \pa_\O \nu &\heq 0,
\ee
which relates the radial derivatives of $\hN_i^j$---and hence the radial derivative of $\hT_i{}^j$---to the rescaled Weyl tensor.

In appendix \ref{app:radexp}, we work out the different horizontal contractions of \eqref{radhN}, which related the 
scalar, vector and tensor horizontal components of the rescaled Weyl tensor
\be\label{rY}
Y:= \hY_{VK}, \qquad \rY_{a}= \gamma_a{}^i\hY_{i K}, 
\qquad \rY_a{}^b := 
\gamma_a{}^i \hY_{i }{}^j \gamma_j^b.
\ee
to the radial derivatives of $\kA$, $\Pi_a$ and $\rN_i{}^j$, which at the appropriate order in $\O$ encode the mass, angular momentum, and radiation aspects (computations in appendix \ref{app:radexpcharge}):
\be \label{Yexpress}
Y&=   K[\kA] + V[\bA] + \nu\bA + 2\Pi\cdot \pa\beta   +\O(\kA \bA -3 \Pi\cdot\Pi) \cr
\rY_{a} &=  - \pa_\O\Pi_a +  \Sn_a \bA -  (\bTheta \cdot \Pi)_a  \cr
\rY_a{}^b & =  
\gamma_a{}^i (\pa_\O\rN_{i}{}^j) \gamma_j{}^b  + 
 ( \rN \!\cdot \!  \bTheta)_a{}^b  
 +\Sn_a \Pi^b + \bA \nu \gamma_a{}^b   
 -\bTheta_a{}^b \pa_\O\nu,
\cr
&\quad  +\O\left(\Pi_a\Pi^b+ \bA (\rN_a^b - 2\nu \bTheta_a{}^b)\right)
\ee
Taking the antisymmetric and trace components of the tensor equation, we also find that 
\be 
\rY_{[ab]} 
&= 
\Sn_{[a} \Pi_{b]} +  ( \rN \!\cdot \!  \bTheta)_{[ab]} ,
\cr
-Y 
&=  \pa_\O\E  + 
 ( \rN \! :  \!  \bTheta)  
 +\Sn_a  \Pi^a + (d-2) \bA \nu  
 -\btheta\pa_\O\nu 
\cr
& \quad +\O\left(
  \Pi\cdot \Pi + \bA  (\E-2\nu \btheta) 
\right),
\ee
where we used that $ \rY_a{}^a =- Y$, and $ \rN_{AB} = \frac1{d-2}\E \gamma_{AB} + e^{-\beta}N_{AB} + \nu \bTheta_{AB}$.

Expanding these equations in $\O$, we obtain a clear dictionary between the charge aspects and the components $\hY_i{}^j$ of the Weyl tensor. In $d=4$, it is enough to evaluate these equations at $\scri$ to obtain the covariant definition of charge aspects in terms of the Weyl tensor components.  Explicitly, in the Bondi-Sachs gauge ($\btheta=0$), using equations (\ref{eq_G2}(KK), \ref{eq:expansion}, \ref{mupi0}, \ref{eq:N0AB}, \ref{Ddef}, \ref{nuJexpansion}), we find\footnote{Recall that  in Bondi gauge $\E_1 = e^{-b}\mu_1$.} : 
\be
- Y_0 & =   \E_1  + \frac12  D_A   D_B C^{AB} + \frac12 e^{-b} N : C  \cr
& \stackrel{\text{(BMSW)}}{=} 2M + \frac12 N:C = 2 M + \frac18 \pp_u (C:C).
\cr
(\epsilon^{AB} Y_{AB})_0 
&  = \frac12 D_A D_B \tl C^{AB}+ \frac14 e^{-b}(2 N + FC):\tl C\cr
&\stackrel{\text{(BMSW)}}{=} \frac12D_AD_B\tl{C}^{AB}
+ \frac14  (\pa_u C): \tl{C}.
\ee
where the second line holds if the BMSW boundary conditions \eqref{BMSW} are imposed. We recognize here the expressions for the covariant mass and dual mass \cite{Freidel:2021qpz}, with $\tl{C}_{AB}:=\epsilon_{A}{}^C C_{CB}$.

\subsection{Bianchi evolution equations}
Now that we have expressed the relationship between the Weyl components and the components of the conformal stress tensor, we can finally use the Bianchi identity to extract all the evolution equations directly from the Weyl tensor components. This way, the evolution equations will take a very compact form.

Consider the (vacuum version of the) Bianchi identity  \eqref{Bianchi2} and project it along the normal $N$ to obtain
\be\label{Bianchi-N}
0& \heq N_c \nabla_d (\O^{3-d} W_{ab}{}^{cd}) = - \nabla_d (\O^{4-d} Y_{ab}{}^{d}) - \O^{3-d} W_{ab}{}^{cd} (\nabla_{d} N_c) = -\O^{(d-4)} \nabla_d Y_{ab}{}^{d}.
\ee
Here we used the definition \eqref{Ydef} of the projected Weyl tensor $Y_{abc} = \O^{-1}W_{abcN}$ in the second equality, and the symmetry $\nabla_{c} N_d = \nabla_d N_c$ as well as the identity $Y_{ab}{}^d N_d=0$ in the second.
We can take different projections of this identity. 
The computation is performed in appendix \ref{Bianchi-proj}. Here we report the results in terms of the projected Weyl components\footnote{Beware that $ \hY_{ij}{}^k \neq h_{jl} \hY_{i}{}^{l k}$ are different tensors. On the other hand, we have that 
$ \gamma^{jl} \hY_{il}{}^k =  \hY_{i}{}^{l k} \gamma_l{}^{j}$.
}
\be
\hY_{ij}{}^k := h_i{}^ah_j{}^b (Y_{ab}{}^c) h_c{}^k, 
\qquad
 \hY_{i}{}^{jk} := h_i{}^a (Y_{a}{}^{bc}) h_c{}^j h_c{}^k,
 \qquad 
  \hY_{i}{}^j := h_i{}^a (Y_{K a}{}^b) h_b{}^j,
\ee
namely:
\be
(\hn_k -\omega_k) \hY_{i}{}^{jk} & \heq    \hY_{N i}{}^k  \bmK_{k}{}^j  -  \gamma^{jl} (\hY\!\cdot \mK)_{l i}  \label{Bianchi-proj1}\\
(\hn_k -\omega_k) \hY_{ij}{}^k  &\heq 2 ( \hY\!\cdot \mK)_{[ji]} .\label{Bianchi-proj2}
\ee 
where we denoted $ ( \hY\!\cdot \mK)_{ij} : = \hY_i{}^k \mK_{kj}$ and   $\hY_{N i}{}^k:= h_i{}^a Y_{Na}{}^bh_b{}^k$.
Contracting \eqref{Bianchi-proj1} along $K_j$ gives the following, notable, conservation law for the tensor $\hY_i{}^j$:
\be\label{consY}
\hn_j \hY_i{}^j + \hY_{i}{}^{jk} \bmK_{kj} \heq 0. 
\ee
In $d=4$, the pull-back of these equations on $\scri$ describe the conservation of covariant energy and momenta aspects.
They reduce, after projections onto the horizontal components,  to the Bianchi conservation equation that was derived by Gerold-Held-Penrose (GHP)
in \cite{Geroch:1973am} (see \cite{Adamo:2009vu} for a nice account). The GHP Bianchi identities have recently been studied as charge conservation equations by 
 Prabhu et al. in \cite{Grant:2021sxk,Prabhu:2021cgk} and by Pranzetti-Freidel in \cite{Freidel:2021qpz}.
 Note that a higher dimensional generalization of GHP formalism was devised in \cite{Durkee:2010xq}.
 The gravitational dynamics described here provides a Carrollian analogue to this formulation adapted to the conformal frame.

\subsection{Other Weyl Tensor components} \label{subsec:Weyl}

In the previous sections, we have studied the components of the Weyl tensor $W_{abcN}$ that vanish on $\scri$.
We still have other Weyl tensor components that do not vanish on $\scri$. They are given by 
\be\label{eq:WeylAtScri}
\bar{W}_{ABCD}&:= e_A{}^a e_B{}^be_C{}^ce_D{}^d W_{abcd},\cr
\bar{W}_{ABC}&: =  e_A{}^a e_B{}^b  e_C{}^c  K^d  W_{abc d},\cr
\bar{W}_{AB  }& := e_A{}^a  e_B{}^b   K^c  K^d  W_{a c  b  d}.
\ee
where $e_A{}^a$ denotes a horizontal coordinate frame.

These components are traceless at $\scri$ with respect to the codimension-2 metric $q_{AB}$. Indeed from the metric expansion $g^{ab} =\gamma^{ab} + N^aK^b + K^a N^b - 2\O \nu K^a K^b$ and the tracelessness of $W_{abcd}$, and using that $W_{abcN}=\O Y_{abc}$, one gets that 
\be 
\gamma^{bd} W_{abcd}
=  \O (Y_{Kca} + Y_{Kac} + 2 \nu W_{aKcK}).
\ee 
Using the definition \eqref{rY} of $\rY_{AB}$ and $\rY_A$, the projection of this relations reads
\be\label{TracesW}
\gamma^{BD}\bar{W}_{ABCD} 
&=  \O (\rY_{CA}+ \rY_{AC} +2  \nu \bar{W}_{AC}),\cr
\gamma^{BC} \bar{W}_{ABC} &= - \O \rY_A,\cr
\gamma^{BD}\bar{W}_{BD} & =0.
\ee
In  $d= 
4$, the $A,B$ indices are 2-dimensional. Therefore the symmetries  and traceleness property of $\bar{W}$ implies that  $\bar{W}_{ABCD}\SCRIeq 0$ and $\bar{W}_{ABC}\SCRIeq 0$, whence we conclude that in $d=4$ the only non-zero components is $\bar{W}_{AB}$:
\be
\bar{W}_{AB}\SCRIeq - D_{AB}.
\ee
This is precisely the source of the logarithmic anomaly: $\Pi_{(1,1)A}=D_B D^B{}_A$  (\ref{eq:log-anom-d4},\ref{4dlogan}) \cite{chrusciel1993gravitational}.\footnote{This was already noted by Chrusciel et al. in \cite{chrusciel1993gravitational}. In their abstract they write: ``The occurrence of some log terms in an asymptotic expansion of the metric is related to the non-vanishing of the Weyl tensor at Scri." }  Notably, whenever the topology of $\scri$ is $S^{2}\times \mathbb{R}$,  $\Pi_{(1,1)A}$ vanishes if and only if $D_{AB}$ does \cite{Madler:2016xju}. Although generally nonvanishing in \emph{logarithmically simple} spacetimes (condition (iv$'$) in Section \ref{sec:Penrose}), this anomaly vanishes for \emph{asymptotically simple} spacetimes (condition (iv)).   Therefore, under the assumption of asymptotic simplicity we deduce, in $d=4$, the vanishing of the Weyl tensor at $\scri$, $W_{abc}{}^d\SCRIeq 0$, thus proving the anticipated assertion \eqref{Pbc4}.

In dimension  $d\geq 6$, the different Weyl tensor components are evaluated in Appendix \ref{app:Weyl} and given by
\be
\bar{W}_{ABCD}& \SCRIeq  {W}_{qABCD}, \label{eq:W=Wq}\\
\bar{W}_{ABC}&\SCRIeq  D_A \bTheta_{0BC}- D_B\bTheta_{0AC} + \frac1{(d-3)}\left((D\cdot \bTheta_0)_A q_{BC}- (D\cdot \bTheta_0)_B q_{AC}\right),\label{eq:W=DbT}\\
\bar{W}_{AB  }& = 
- (\ln\O +1)\bTheta_{(1,1)AB} - D_{AB} - (\bTheta_0\cdot\bTheta_0)_{\langle AB\rangle_q} + o(1).\label{eq:W=D}
\ee
In $d=6$, we recall that $D_{AB}=[\bTheta_{A}{}^C]_1 q_{CB} =  \bTheta_{1\langle AB\rangle_q } -  2(\bTheta_{0}\cdot\bTheta_{0})_{\langle AB\rangle_q }$ \eqref{4dlogan} is the radiation field.

Recall that the divergent logarithmic term $\bTheta_{(1,1) AB }$ vanishes identically if $d\geq8$, since the first logarithm is $\bTheta_{\left(\frac{d-4}{2},1\right)}$ and therefore the Weyl tensor is finite at $\scri$.

In $d=6$, $\bTheta_{(1,1) AB }$ is the radiative anomaly.
Since logarithmic simplicity is equivalent to the vanishing of the radiative anomaly (section \ref{sec:recursivesol}), this requirement is equivalent to asking the finiteness of the Weyl tensor at $\scri$.

Two important equations regulates the logarithm anomaly in $d=6$. The first one relates the logarithmic radiation to the radiative anomaly, (cf. that is \eqref{eq:NbTan}), namely:
\be
2N_{(1,1)}=(\cL_{\ell}+ F)\bTheta_{(1,1)}.
\ee
This means that,  assuming the regularity of the initial conditions at the initial cut\footnote{Cf. footnote \ref{fnt:regularinitial condition} .} $u=-\infty$, 
 the vanishing of the radiative anomaly implies $\bTheta_{(1,1)}=0$.

The second equation relates the asymptotic values of $\bar{W}_{ABCD}$ and $\bar{W}_{ABC}$, denoted $\bar{W}_{qABCD}$ and $\bar{W}_{qABC}$, to the logarithmic radiation $N_{(1,1)}$ given in \eqref{eq:rad-anomaly-d6-text} , that is 
\be \label{DWNtext}
2 N_{(1,1)AB}
&= -  e^b \left(D_C\bar{W}_q{}^C{}_{AB} + (\bar{W}_q: \bTheta_0)_{\langle A B \rangle_q} \right),
\ee
which is shown in appendix \ref{app:radiativeA}.
This implies that  in $d=6$ there is one more natural asymptotic condition available, which is stronger than logarithmic simplicity:  the demand that the Weyl tensor $W_{abc}{}^d$ is of Petrov type $N$ at $\scri$, i.e. that $\bar{W}_{ABCD}\SCRIeq \bar{W}_{ABC}\SCRIeq 0$. This is the boundary condition used in \cite{Capone:2023roc} near the asymptotic ends $\scri^+_{\pm}$. 
By \eqref{DWNtext} this implies that $N_{(1,1)}=0$, which implies the vanishing of $\bTheta_{(1,1)}$ under regularity of the initial conditions.

Demanding, in $d=6$ that the Weyl tensor vanishes, i.e. $W_{abcd}\SCRIeq 0$, kills radiation. Indeed, this would require that in addition to the Petrov type $N$ condition, we have $\bar{W}_{qAB}=0$  which fixes the radiation field $D_{AB}$ in terms of the boundary condition $\bTheta_{0AB}$. This is therefore stronger than the analogous condition in $d=4$, where the vanishing of the Weyl at $\scri$ is equivalent to the demand of analyticity and asymptotic simplicity.

Finally, we note one more consequence of equations \eqref{eq:W=DbT} and \eqref{DWNtext}. In $d=6$, if one asks that $\bTheta_0 = 0$,  as it is done in \cite{Hollands:2016oma,Garfinkle:2017fre}, then $\bar W_{qABC}=0$ and also $N_{(1,1)}=0$, so that $\bTheta_{(1,1)}=0$ (if one assumes regularity of the initial conditions at a cut) and the Weyl tensor is finite. This leads to a new set of asymptotic boundary conditions for the Weyl tensor that are not Petrov type $N$ ($\bar{W}_{qABCD}$ can be nonzero).
These boundary conditions however kill the possibility to have infinite dimensional BMS-like symmetry. Allowing $\bTheta_0\neq 0$ at $\scri$ allows for an infinite asymptotic symmetry group given by BMSW and allows, as shown in \cite{Kapec:2015vwa, Capone:2023roc}, for a reconciliation between asymptotic symmetries and soft theorems in $d=6$.

 We here recall a result of \cite{Capone:2023roc} that the combination of BMSW and Petrov type $N$ asymptotic boundary conditions still allows for an infinite dimensional family of non-vanishing $\bTheta_{0AB}$ parametrized by a function $T$ on the celestial sphere (and constant in $u$). Indeed, as proven in appendix \ref{app:WeylTypeN} (where a generalization to $d\geq 6$ is discussed), the vanishing of $\bar W_{qABCD}=0$ together with the choice of asymptotic radial shear
\be 
\bTheta_{0AB} \stackrel{\text{(BMSW)}}{=} - \left(D_{\langle A} D_{B \rangle_q}T + \frac{(u + T)}{2} R_{q\langle A B\rangle_q}\right)
\ee 
implies via \eqref{eq:W=DbT} that $\bar{W}_{qABC} =0$ and then via \eqref{DWNtext} that the logarithmic radiation vanishes, $N_{(1,1)}=0$. Assuming regularity of the initial conditions at a cut, we also conclude that the radiative anomaly itself vanishes, $\bTheta_{(1,1)}=0$, and therefore that this asymptotic radial shear is compatible with logarithmic simplicity and Petrov type $N$ asymptotic boundary conditions.

\section{The symplectic potential and its asymptotic renormalization}\label{sec:SP}
In the previous sections, we focused on the asymptotic equations of motion and the symmetries that diffeomorphisms preserving the Bondi-Sachs gauge induce on the asymptotic data.
It is now time to address the last ingredient necessary for the analysis of the asymptotic symmetries: the symplectic flux through $\scri$.

In this section, we renormalize the symplectic potential and show that it admits a finite limit at $\scri$. 

Let us start by recalling that the  Einstein-Hilbert symplectic potential current is given by 
\be
\tl\bftheta{}_{\mathsf{EH}}^a = \tfrac12 \sqrt{\tl g} \Big( \tl g{}^{bc} \delta \tl\Gamma^a_{bc} - \tl g^{ab} \delta \tl\Gamma^c_{bc} \big),
\label{eq:SC}
\ee
and that it satisfies the defining balance equation $
\delta \tl \bL :=  \tl \bE{}_{ab}\delta \tl g^{ab} + \pp_a {\tl\bftheta}{}_{\mathsf{EH}}^a
$ \eqref{eq:deltaL}.

\subsection{General considerations}

\paragraph{The Renormalization strategy} 
The notion of holographic renormalization of the symplectic potential was first proposed in 
\cite{Compere:2008us} in the context of AdS/CFT and then generalized and applied to renormalize the asymptotically flat symplectic structure of 4d gravity in 
\cite{Compere:2018ylh, Compere:2020lrt, Freidel:2021fxf, Chandrasekaran:2021vyu,Geiller:2024amx}. Application of this procedure in higher dimensions has been studied in \cite{Capone:2023roc}. 
The basic idea behind the  renormalization procedure is to leverage the ambiguity existing in the choice of Lagrangian $\tl \bL$ and symplectic potential $\tl \bftheta{}^a $ one can assign to a given equation of motion $\tl \bE= \tl \bE_{ab} \delta \tl g{}^{ab} +\tl \bE_\phi \delta \phi$. 
Indeed, one can shift the Lagrangian by a boundary term and the symplectic potential by a corner term according to 
\be 
 \tl \bL \to \tl \bL{}'=  \tl \bL - \pa_a \tl \bell{}^a, \qquad \tl \bftheta{}^a \to \tl\bftheta{}'^a = \tl \bftheta{}_{\mathsf{EH}}^a - \delta \tl \bell{}^a + \pa_b \tl \bvartheta{}^{ab},
\ee 
where $\tl \bvartheta{}^{ab} =- \tl \bvartheta{}^{ba}$, without affecting the equations of motion  $\tl\bE$ nor the fundamental balance equation
\be \label{balancelaw}
\delta \tl\bL{}' = \pa_a \tl\bftheta{}'^a + \tl\bE.
\ee 
Whence, the main idea behind the renormalization of the symplectic potential is to find a pair $(\tl \bell{}^a, \tl \bvartheta{}^{ab})$ such that
the new Lagrangian and symplectic potential $(\tl\bL{}', \tl\bftheta{}'^a)$ are finite when evaluated at $\scri$.

Quite remarkably, a general procedure exists that renormalizes the symplectic potential in any dimensions. This procedure mainly relies on the balance equation \eqref{balancelaw}. This procedure was first exemplified by us in the context of higher dimensional electrodynamics in \cite{Freidel:2019ohg} (see also \cite{Hopfmueller:2020yqj}). In the context of gravity a renormalization algorithm was then proposed by McNees and Zwikel \cite{McNees:2023tus}, see also Geiller and Zwikel \cite{Geiller:2024amx}. 

 The setup for the renormalization procedure is as follows.
We start by adopting radial and tangential coordinates $(\O, y^i)$ introduced earlier to decompose the symplectic potential into its flux $\tl\bftheta{}^\O$ through $\Sigma_\O$, and the transverse components $\tl\bftheta{}^i$. As a consequence of logarithmic simplicity (cf. Section \ref{sec:radex}), the various components of $\tl\bftheta{}^a$ as well as $\bL$ admit a polyhomogeneous expansion. 
To be concrete, the detailed analysis performed in the following sections  shows that 
\be \label{Poly}
\Omega^{d-2} \tl\bL \in C^{\mathrm{poly}}_{d-2},\qquad 
\O^{d-2} \tl \bftheta{}{}_{\mathsf{EH}}^\O  \in C^{\mathrm{poly}}_{d-2}, \qquad 
\O^{d-1} \tl \bftheta{}{}_{\mathsf{EH}}^i  \in C^{\mathrm{poly}}_{d-1}. 
\ee
In particular, this means that $\tl \bftheta{}^\O$ can be express when $\O\to0$ as the sum of  a divergent and a finite component \be 
\tl \bftheta^\O = \tl \bftheta{}_{\mathrm{div}}^\O + \tl \bftheta{}_{\mathrm{fin}}^\O + o(1),
\qquad
\tl \bftheta{}_{\mathrm{div}}^\O= \sum_{n=1}^{d-2} \Omega^{-n}   \bftheta{}_{\mathsf{EH}}^\O + \ln\O \bftheta{}_{\mathsf{EH}}^\O
\ee 
A similar statement holds for $\tl \bftheta{}^i $ and $ \tl\bL$.

The renormalization algorithm then proceeds as follows: One chooses the corner Lagrangian and symplectic potential to be 
\be
\tl\bell{}^\O &:= \pa_\O^{-1} \tl\bL_{\mathrm{div}} , \quad \,\,\,\quad \tl\bell{}^i=0,\\
\tl\bvartheta{}^{\O i} &:= \pa_\O^{-1} \tl\bftheta{}^i_{\mathrm{div}}, \qquad \tl\bvartheta{}^{ji} =0,
\ee
where $\pa_\O^{-1}$ is the linear operator acting on polyhomogeneous functions defined in \eqref{eq:ppOm-inv}.
The renormalized Lagrangian and transverse symplectic potential are then defined  to be  
\be 
\tl \bL_{\mathrm{R}}= \tl  \bL - \pa_\O \tl \bell{}^\O =  \tl  \bL- \tl \bL_{\mathrm{div}}, 
\qquad 
\bftheta_{\mathrm{R}}^i= \tl \bftheta{}^i - \pa_a \tl \bvartheta{}^{ai}   = \tl \bftheta{}^i - \tl \bftheta{}^i_{\mathrm{div}}.
\ee 
By construction, $(\tl \bL_{\mathrm{R}}, \bftheta_{\mathrm{R}}^i)$ are then finite. 
Then, one needs to check that the flux component of the symplectic potential
\be
\tl\bftheta{}_{\mathrm{R}}^\O = \tl \bftheta{}^\O  -\delta \tl \bell{}^\O+ \pa_i \tl \bvartheta{}^{\O i}   
\ee
is also finite. From the invariance of the balance equation, we obtain 
\be 
\pa_\O \tl\bftheta{}_{\mathrm{R}}^\O = \pa_\O \tl \bftheta{}^\O +  \pa_i \tl \bftheta{}_{\mathrm{div}}^i -\delta \tl\bL_{\mathrm{div}} =
\delta \tl\bL_R 
 -\pa_i \tl \bftheta{}_{\mathrm{R}}^i -\tl \bE.
\ee
Therefore, if we assume that we impose \emph{as boundary conditions}, i.e. as a restriction on the accessible off-shell phase space, that $\tl \bE_{\mathrm{div}}=0 $, then $\tl \bE$  is finite, and this equation implies that 
$\bftheta_{\mathrm{R}}^\O$ is also finite at $\O=0$.

One puzzle about this renormalization, emphasized in \cite{McNees:2023tus}, is that we could very well just take $\tl \bvartheta{}^{\O i} $ to be the full primitive of $\tl \bftheta{}^i$ instead of just its divergent component and similarly take $\tl\bell{}^\O$ to be the full primitive of $\tl \bL$. In this case the renormalization leads to the puzzling fact that $\tl \bftheta{}_R^i=0$ while $\pa_\O\tl \bftheta{}_R^\O \heq 0$. That is, all radial dependence has been removed in the symplectic potential on-shell. That mirrors the same (general) conclusion for the symplectic form flux, which reads $\pa_\O\delta\tl\bftheta{}^\O\heq0$ when $\bftheta{}_R^i=0$.

There are difficulties with this renormalization prescription: it is highly nonlocal to the point of yielding a vanishing Lagrangian, $\tl\bL{}'=0$.
\footnote{\label{fnt:nonlocalrenorm} 
Note that even in a standard 3+1 formulation of a field theory on $M=\Sigma \times \mathbb{R}\ni(y^i,t)$, one could formally define $(\bell^i, \bell^t) = (0, \int^t\bL)$ and $(\bvartheta^{ji},\bvartheta^{ti}) = (0,\int^t\bftheta^i)$, and thus obtain $\bL' =0$ and $\pp_t (\bftheta')^t = -\bE \heq 0$. The reason why this is not allowed is precisely that the Lagrangian counterterm would be nonlocal (in time) and therefore break the main tenet of Lagrangian mechanic and the variational principle, namely, locality. 
The underlying issue is that the variational principle from which the balance equation is derived is meaningful only for local theories, and therefore the addition of nonlocal counterterm to the Lagrangian is not allowed. For this reason we are rather looking for an (off-shell)renormalization  prescription within a space of field configurations restricted by appropriate boundary conditions that constrain the first few leading orders in the $\O$ expansion only. These boundary conditions are given by the Pbc augmented by the condition $\tl\bE_\mathrm{div}=0$.
}
The ensuing framework must involve only counterterms that depend locally on the free data allowed by the said boundary conditions, as per Section \ref{sec:freedata}. In particular, $(\tl\bell{}^a,\tl\bftheta{}^{ab})$ cannot depend on the mass and angular-momentum aspects, nor the higher spin aspects. Remarkably, as we will, see, such a framework exists and requires only the removal of the divergent parts of $\tl\bL$ and $\tl\bftheta{}^i$. 

In sum, the renormalization scheme advocated for in this paper is \emph{local }  and covariant with respect to boundary diffeomorphism, with counterterms 
$(\bell^a, \bvartheta^{ab})$ that depend (locally) only on the boundary metric data $(q_{AB}, a, U^A, F)$ and on the radiation data. This is in line with the prescription advocated for in \cite{Capone:2023roc}.

Note that there is some freedom in this construction: we could trade some of the exact terms in $\tl \bftheta{}^i_{\mathrm{div}}$ in terms of a non-vanishing boundary Lagrangian $\bell_{\mathrm{div}}^i$. We could also change the final finite answer by a finite redefinition of the symplectic potential. We will fix this ambiguity by demanding that the finite symplectic potential is written in a \emph{canonical} polarizations $\tl \bftheta_{\mathsf{can}}= P\delta Q$ where Carrollian geometry plays the role of configuration variables $Q$ while 
the charge aspects play the role of momenta $P$.

\subsection{Conformal frame expansion and renormalization}

Now that we have sketched the general strategy, we turn it into an explicit construction.
To do so,  we must express the symplectic flux and current in the conformal frame.
This rewriting will allow us to show more precisely that the {\it most divergent} components of $\tl\bftheta{}^\O$  can be absorbed in terms of total variations and divergences.

 We proceed as follows: 
(\textit{1}) First, we express the symplectic current in the conformal frame and perform a first renormalization step that turns the polarization into the canonical one; this step requires both a divergent canonical transformation, and the subtraction of a corner term with respect to the Einstein-Hilbert symplectic structure.
(\textit{2}) Second, we rewrite the symplectic flux radial evolution equation in the conformal frame 
and thus deduce the divergent symplectic flux counterterms $\delta\hat\bbeta$ and $\p_i\tl \bgamma^{\O i}$ under the assumption of logarithmic simplicity.
(\textit{3}) Third and last, we specialize to $d=4$ and, using the Pbc and radial Einstein equations, compute the counterterms introduced in the previous step explicitly.

As we will see, the first three steps of the renormalization are fully covariant and can be performed in any gauges. For the fourth step, we will specialize to the Bondi gauge.

\subsection{Step 1: The symplectic current in the conformal frame}
Introducing the rescalings 
\be 
\tl \bL= \Omega^{2-d} \bL_0,  
\qquad 
\tl \bftheta{}_{\mathsf{EH}}^a = \Omega^{1-d} \bftheta_0^a, \qquad  \tl \bE{}_{ab}\delta \tl g^{ab} = \Omega^{2-d} \bE{}_{ab}\delta g^{ab}\label{theta1a},
\ee
the (physical) Lagrangian density, symplectic potential and equations of motion can be written in terms of the conformal metric as follows (note that in our rewriting we assume that the Pbc 1 and 2, namely \eqref{Pbc1} and \eqref{Pbc2}, have been imposed):
\be\label{Ldef2}
 \bL_0 &= \sqrt{g} \left(  \tfrac12  R +(d-1) \N\right),\\
\bftheta^a_0 &=  \tfrac12 (d-2)   \sqrt{g}  \delta N^a   +   \delta (\sqrt{g} N^a )  + \Omega \bftheta^a_{\mathsf{EH}},\label{theta00}%
\\
\bE_{ab}& = \sqrt{g}( G_{ab} - (d-2)\T_{ab}) \heq0.
\ee
Here, $\bftheta_{\mathsf{EH}}^a = \tfrac12 \sqrt{g} \left( g^{bc} \delta \Gamma^a_{bc} -  g^{ab} \delta \Gamma^c_{bc} \right)$ denotes the Einstein-Hilbert symplectic potential current for the conformal metric, while $\tl \bftheta{}_{\mathsf{EH}}^a$ is the analogue quantity  for the physical metric \eqref{eq:SC}.

To establish this,  it is enough to recall how the Christoffel symbols transform under Weyl rescalings and to compute their variations considering that $\delta \O \equiv 0$ and therefore $\delta N_a \equiv 0$. Namely,
\be
\tl \Gamma^a_{bc} ={}   \Gamma^a_{bc} + \frac1\O[ N^a  g_{bc} -  N_b  \delta_{c}{}^a-N_c  \delta_{b}{}^a  ]\qquad \text{and} \qquad
 \delta \tl \Gamma^a_{bc} ={}  \delta \Gamma^a_{bc}  +  \frac1\O \delta ( N^a  g_{bc}).
\ee
\paragraph{The flux of the symplectic potential}
Next, we focus on the evaluation of the symplectic flux term $\bftheta_{0}^\O =\bftheta_0^a N_a$. Using that 
 $N^2 = N^\O=2\O\nu$, we get
\be\label{theta0}
\bftheta_{0}^\O &
= \O  \delta (d \sqrt{g} \nu  ) - (d-2) \O \nu  \delta\sqrt{g}    + \O \bftheta_{\mathsf{EH}}^\O .
\ee
In \cite{Freidel:2022vjq}, see also \cite{Parattu:2016trq}, it was shown that the symplectic potential $\bftheta_{\mathsf{EH}}^\O$ can be conveniently written in terms of  the components of the  Carrollian stress tensor $\mT_i{}^j$ \eqref{eq:CarrollStrTens} associated to the stretch Carroll structure on $\Sigma_\O$, namely\footnote{The sign convention taken here for $\mT$ is opposite to the choice made in  \cite{Freidel:2022vjq} and agrees with \cite{Chandrasekaran:2021hxc}.}
\be 
\mT_i{}^j :=\mN \delta_i{}^j  -   \mN_i{}^j,
\ee
with $\mN_i{}^j :=h_i{}^a(\nabla_a N^b)h_b{}^j$. The horizontal/vertical components of these tensors are
\be
 \mN_i{}^j &  = \rmN_i{}^j +  (\rom_i + \omega_V K_i) V^j - \O K_i \rJ^j,\cr
\mT_i{}^j  &=  \mN \gamma_i{}^j- \rmN_i{}^j   - \rom_i V^j + K_i (\rmN V^j + \O\rJ^j),
\ee
where  we used that $\omega_i = \mN_i{}^jK_j= \rom_i + \omega_V K_i$, and that $N= \rN + \omega_V$. We   denoted $\rom_i := \gamma_i{}^j \omega_j$, $\rmN_i{}^j = \gamma_i{}^a \mN_a{}^b \gamma_b{}^j$ etc the horizontal projections.
The result of \cite{Freidel:2022vjq} is then (see Appendix \ref{app:sympF} for a self-contained derivation)\footnote{Note that in our notation $\delta\gamma^{ij}$ is the \emph{variation of the inverse metric} and not the variation of the metric with raised indices (the two differ by a sign).}: 
\be 
\bftheta_{\mathsf{EH}}^\O  &=   \tfrac12  \pa_j ( \sqrt{g} \delta V^j )  -  \delta \left(  \sqrt{g} \mN \right)   \cr 
& \,\, +  \sqrt{g} \left( \tfrac12
\delta \gamma^{ij}  \left(  \rmN_{ij} - \gamma_{ij} \mN \right) +\delta K_i (\rmN V^i + \O \rJ^i)
+ \rom_i \delta V^i
+ \O   \btheta  \delta \nu
 \right). \label{thetaOFinal}
\ee

The traceless component of $\rmT_i{}^j:=\mN \gamma_i{}^j- \rmN_i{}^j $ corresponds to the  fluid viscous stress tensor
while its trace corresponds to the fluid's pressure. The combination $\rmN V^j + \O \rJ^j$ corresponds to the fluid's energy current 
and we see that its spatial component vanishes when the fluid is Carrollian at $\O=0$. 
These considerations mirror those given in Section \ref{sec:reconcile} for the  asymptotic stress tensor $\hT_i{}^j:=\hN \gamma_i{}^j- \hN_i{}^j$ \eqref{eq:carrstresstens}. Indeed $\mT$ and $\hT$ are closely related: by means of the
Penrose boundary conditions \eqref{Pbc1} and \eqref{Pbc2}, which give 
\be 
\mN_i{}^j & = \nu \delta_i{}^j + \O\hN_{i}{}^j,\cr
\mT_i{}^j & = (d-2) \Sc \delta_i{}^j + \O\hT_i{}^j,\cr
\rmN & = (d-2) \Sc + \O \E,\cr
\rmT & = (d-2)\Big[\Sc +  \O \underbrace{\left( \frac{d-3}{d-2}\E + A_K\right)}_{=:P} \Big]
\ee
From Section \ref{sec:reconcile} and Appendix \ref{sec:acc} we recall the following relations for the vector and scalar expressions entering the symplectic potential flux  \eqref{thetaOFinal}:
\be
\rom_i = -  \O\Pi_i, 
\qquad \omega_V = \nu + \O A_K.
\ee
For convenience we also recall that $\E = \rN = e^{-\beta}\mu + \bar\theta$ and $\Pi_i = - \gamma_i{}^j\hN_{jK}$ encode, at order $(d-3)$, the mass and angular momentum aspects (cf. sections \ref{sec_gennews} and \ref{sec:freedata}). 
Moreover, we have $\rJ_i =  -\hN_{V}{}^j \gamma_{ji} = - (\Sn_i +2\O \Pi_i )\Sc$ \eqref{Jexp}, and $A_K=\N_{KV}= (\pa_\O\nu + 2 \nu \pa_\O \bb)$ \eqref{omegacomp}.

Using these expressions, as well as the identity $V^i\delta K_i =\delta \beta$ with $\beta =\log (\sqrt{g/\gamma})$, we readily find the following expression for the physical symplectic potential flux through $\Sigma_\O$:
\be\label{thecan}
 \tl \bftheta{}_{\mathsf{EH}}^\O = \Omega^{2-d}\left(   \delta \bell^\O  - \pa_i \bvartheta^{\O i} \right)+ \tl\bftheta^{}\O_{\mathsf{can}}, \qquad  \tl\bftheta{}^\O_{\mathsf{can}}=
 \Omega^{3-d} \bftheta^\O_{\mathsf{can}},
 \ee
where we isolated the canonically polarized Carroll symplectic structure, which is Dirichlet in the Carrollian data $(h_{ab}, V^i, K_i,\nu)$,
  \be 
\bftheta^\O_{\mathsf{can}}   &:= \,  \sqrt{g} \left( \tfrac12
 \big(  \rN_{ij} - \gamma_{ij} \hN \big) \delta \gamma^{ij}  +\big(\rN V^i +  \rJ^i\big) \delta K_i 
+ \Pi_i \delta V^i
+    \btheta  \delta \nu
 \right), \label{thetaO1}
\ee
 as well as the local counterterms
\be\label{elllvart}
 \bell^\O =  \sqrt{g}\left( \nu - \O \hN \right), \qquad \bvartheta^{\O i} = - \frac12  \sqrt{g} \delta V^j .
 \ee
 The counterterm $ \bell^\O$ is a conformal-spacetime asymptotic analogue of the null boundary Lagrangian discussed by \cite{Parattu:2015gga, Lehner:2016vdi, Hopfmuller:2016scf, Aghapour:2018icu, Oliveri:2019gvm} in generalization of the York-Gibbons-Hawking construction \cite{York:1972sj,Gibbons:1976ue}.
 The ensuing canonically polarized symplectic structure $\bftheta^\O_{\mathsf{can}}$ thus generalizes the Dirichlet  prescription \cite{Brown:1992br} adapted to the conformal boundary \cite{Harlow:2019yfa, Chandrasekaran:2021hxc, Freidel:2022vjq}.

Quite remarkably, the   symplectic potential $\bftheta^\O_{\mathsf{can}}$ is simply the symplectic potential of a fluid associated with the projection of the   asymptotic
stress tensor \eqref{CEST} 
\be
 \hT_i{}^j  &=  \gamma_{i}{}^j \hN - \rN_{i}{}^j + \Pi_i V^j  + K_i (\rN V^j+\rJ^j).
\ee
 The analogy between $\scri$ and stretched Carroll structures is manifest in the comparison of this equation with \eqref{thetaOFinal} in the same way as it was for the conservation equations \eqref{stretch-cons} and \eqref{stretch-cons2} discussed in Section \ref{sec:conservationeqs}.

Finally, we find it interesting to assess where the first logarithm appear in the expansion of \eqref{thetaO1} in the Bondi--Sachs gauge.
Assuming logarithmic simplicity implying \eqref{poly2} and the following equations, we compute the polyhomogeneous degree of the Carrollian data:
\begin{subequations}
    \label{polyordersCarroll}
\be
K_i & = e^{\bb}\pp_i u \in C^\text{poly}_{d} ,\qquad
V^j\pp_j  = e^{-\bb}(\pp_u + \Upsilon^A\pp_A)\in C^{\mathrm{poly}}_{d-1}, \quad
 & \Sc = e^{-\bb}\Phi \in C^\mathrm{poly}_{d-2}; 
\ee
together with
\be
\gamma_{ij} & \in C^{\mathrm{poly}}_{d-1},\qquad
\Pi_i  \in C^{\mathrm{poly}}_{d-3},\qquad
\rJ_i   = -(\mathring\pa^j + 2\O\Pi^j)\Sc\in C^\mathrm{poly}_{d-2}.
\ee
While for the News tensor this means 
\begin{gather}
\dot{\N}_i{}^j  = e^{-\bb}N_i{}^j + \Sc \bTheta_i{}^j \in C^{\mathrm{poly}}_{d-2},\qquad 
\rN  = e^{-\bb} \mu \in C^{\mathrm{poly}}_{d-3}.
\end{gather}
\end{subequations}

Moreover, while both $\rN$ and $A_K = (\pp_\O + 2\pp_\O\beta)$ are in $  C^{\mathrm{poly}}_{d-3}$, we have that their sum is higher order, namely 
\be \label{eq:hatNpolylog}
\hN = \rN + A_K \in C^{\mathrm{poly}}_{d-2}.
\ee
Indeed we have that 
$\O e^{-\beta}\mu = \theta - (d-2) \nu$ and from \eqref{bianc_bis} we can establish that $\theta \in C^{\mathrm{poly}}_{d-1}$. This means that
$(A_K)_{(d-3,1)} = (\pp_\O \nu)_{(d-3,1)} = (d-2) \nu_{(d-2,1)} = - e^{-b}\mu_{(d-3,1)}=- \rN_{(d-3,1)}$. Therefore, we conclude that  $\hN_{(d-3,1)} = 0$.

Overall, in the Bondi--Sachs gauge, one thus finds $\bftheta^\O_{\mathsf{can}} \in C^{\mathrm{poly}}_{d-3}$, which means that first logarithm of $\tl \bftheta{}^\O_\mathsf{can}$ appears at a log-divergent order. 
In a logarithmically simple spacetime the logarithms are due to the Coulombic anomalies $\Pi_{(d-3,1)A}$ and $\mu_{(d-3,1)}$, from \eqref{thetaO1}, we get that the logarithmic term in $\tl \bftheta{}^\O_\mathsf{can}$ is given by 
\be
\ln\O \left[ \bftheta{}^\O_\mathsf{can} \right]_{(d-3,1)} 
& = 
\ln\O \sqrt{g}\left(%
\tfrac1{2(d-2)} \rN_{(d-3,1)}q_{ij}\delta q^{ij} + \rN_{(d-3,1)} \delta b + \Pi_{(d-3,1)A} \delta V^A %
\right)\cr
& =\ln\O \sqrt{q} \left[ \mu_{(d-3,1)}\left( - \tfrac1{d-2} \delta \ln\sqrt{q} +   \delta b \right) +  \Pi_{(d-3,1)A} e^{b}\delta(e^{-b} U^A)\right]
\ee
where we used $\rN_{(d-3,1)} = e^{-b}\mu_{(d-3,1)}$ and $\sqrt{g} = e^\beta \sqrt{\gamma}$.

In $d=4$, it so happens that $\mu_{(1,1)} = 0$ \eqref{eq:mu1-Pi1}, and thus the first logarithm is entirely due to $\Pi_{(1,1)A}=D_B D_A{}^B$ \eqref{eq:log-anom-d4}:
\be 
\ln \O  \left[\bftheta{}^\O_\mathsf{can} \right]_{(1,1)} 
\ \stackrel{(d=4)}{=} \ \ln \O \sqrt{g}   (D_B D^B{}_A) \delta(e^{-b} U^A).
\label{eq:thetalogd4}
\ee

\paragraph{First-step renormalization: canonical Lagrangian and symplectic potential} 
In order to carry through the general renormalization procedure outlined above, 
we now turn our attention to the tangential components of the symplectic potential
---which we will study in more detail in the next subsection. From (\ref{theta1a},\ref{theta00}) we get 
\be
\tl \bftheta{}^i &=  -\tfrac12 \pa_\O (  \O^{2-d} \sqrt{g}  \delta V^i )  +   \delta (\O^{1-d}  \sqrt{g} V^i )  + \O^{2-d} \bftheta_{\mathsf{can}}^i ,\cr
\bftheta_{\mathsf{can}}^i &:= \bftheta_{\mathsf{EH}}^i  + \tfrac12 \pa_\O (\sqrt{g}  \delta V^i).\label{thetai1}
\ee
This expression prompts us to define the remaining components of the counterterms  $(\bell^a, \bvartheta^{ab})$, that we partially introduced in \eqref{elllvart}, as
\be
\tl\bell{}_1^\O &:= \O^{2-d} \bell^\O, \qquad \qquad  \tl\bell{}_1^i:= \O^{1-d}  \sqrt{g} V^i ,\cr
\tl\bvartheta{}_1^{i\O } &:= \O^{2-d} \bvartheta{}^{i\O } , \qquad \quad  \mathring{\tl\bvartheta}{}_1^{ij}:=0.\label{eq:countert1}
\ee
With these counterterms, we obtain the first-step renormalized symplectic potential as
\be
\tl \bftheta{}^a_\mathsf{can}:= \tl \bftheta{}^a + \pp_b \tl\bvartheta{}^{ab}_1 - \delta \tl\bell{}_1^a
\ee
We thus decompose it into its rescaled tangential and flux components (notice the different rescalings)
\be\label{eq:tlthetacan}
\tl \bftheta{}^a_\mathsf{can} = : \O^{2-d} \bftheta_{\mathsf{can}}^i h_i{}^a + \O^{3-d} \bftheta^\O_\mathsf{can} \delta^a_\O,
\ee
where the flux component takes the canonical expression \eqref{thetaO1} and the tangential one is given in  \eqref{thetai1}.
The corresponding first-step renormalized Lagrangian, yielding the canonical symplectic flux $\tl\bftheta{}^\O_\mathsf{can}$ through $\Sigma_\O$, is given by\footnote{This expression, as others that follow from it e.g. \eqref{eq:Cevolution}, is valid in this form even in the presence of a cosmological constant $\Lambda\neq0$: in fact all terms which feature $\Lambda$ explicitly, drop from it after the first renormalization step. These terms would be the most divergent ones, up to order $\O^{-d}$.} 
\be
\tl \bL_{\mathsf{can}}&:= \tl \bL -\pa_a \tl\bell{}_1^a = \O^{2-d}  \bL_{\mathsf{can}}, \label{thetaRL1} 
\ee
The expression for $\bL_\mathsf{can}$ in the Bondi gauge, is 
\be\label{L1} 
\bL_\mathsf{can} = \sqrt{g} \left( \frac12 R + \hN + (d-1) \pp_\O \Sc + \O (\Sc \bar A +  \pp_\O \hN )\right) ,
\ee
where $\bar A$ is given in
\eqref{aspectdef} and we used \eqref{Ldef2}.
This is 
evaluated  by means of the expression $V^a = N^a - 2\O \Sc K^a$ \eqref{NVK} and the following identities (cf. (\ref{Pbc2}, \ref{determinants}, \ref{defmu}, \ref{aspectdef})):
\be
\nabla_a N^a = d\Sc + \O\N,
\quad
\nabla_a K^a = \bar\theta + \pp_\O \beta,
\quad
\N = \hN + \pp_\O \Sc,
\quad
\pp_\O\beta = \O\bar{A},
\ee
to obtain in the Bondi gauge (i.e. $\bar\theta=0$, and thus $\pp_\O \log \sqrt{g} = \pp_\O\beta$):
\be
\pp_a \tl\bell{}^a_1 = \O^{2-d} \sqrt{g}\left((d-2) \hN- \O\nu \bar{A} - \O \pa_\O (\sqrt{g} \hN) \right) .
\ee

We conclude with one last expression for $\tl\bL_\mathsf{can}$ valid if the Pbc3 \ref{Pbc3} is satisfied (cf. \eqref{eq:tildeL-EH}), namely if $\sT_{ab} := \O^{-1}(S_{ab} + \N_{ab})$ is well defined at $\scri$:
\be
\label{L2}
\tl \bL_{\mathsf{can}}  &= \pa_\O \left(\O^{3-d} \sqrt{g}   \tfrac{(d-2)}{(d-3)} \hN  \right) + \O^{3-d} \left( (d-1)\sT + \nu \bA  -\tfrac1{(d-3)} \pa_\O (\sqrt{g} \hN) \right).
\ee
 Observe that thanks to \eqref{polyordersCarroll} and \eqref{eq:hatNpolylog}, we have that $ \bL_\mathsf{can} \in C_{d-2}^{\mathrm{poly}}$, where $\tl\bL_\mathsf{can} = \O^{2-d} \bL_\mathsf{can}$.  This fact,  which is not self-evident from \eqref{L1}, is crucial for the renormalization procedure.

 \paragraph{Transverse symplectic structure}

To finalize our description of the first renormalization step, we need to give the explicit expression of $\bftheta_{\mathsf{can}}^i = \bftheta_{\mathsf{EH}}^i  + \tfrac12 \pa_\O (\sqrt{g}  \delta V^i)$ in terms of the  boundary and radiative data.
In appendix \ref{app:sympt} we prove the following 3+1 splitting of the the Einstein-Hilebrt symplectic potential:
 \be
 \bftheta_{\mathsf{EH}}^i &= \tfrac12 \sqrt{g} \left( 2 V^j  \delta \bmK_j{}^i   - V^i \delta (\bar\theta  +\bar\kappa)  
+ \gamma^{ij} \delta\omega_j
 \right) + 
\hat\bftheta{}_{\mathsf{EH}}^i \label{thetai-11},
\ee
where $\hat\bftheta{}_{\mathsf{EH}}^i $ is the codimension-$1$ symplectic potential, with $\hat\Gamma_{ij}^k$ the connection coefficients of the Carrollian connection $\hn_i$:
\be
\hat\bftheta{}_{\mathsf{EH}}^i  =  \tfrac12 \sqrt{g}(\gamma^{jk}\delta\hat\Gamma_{jk }^i- \gamma^{ji} \delta\hat\Gamma_{j k}^k).
\ee
We can further split this contribution into vertical and horizontal components with respect to the Carrollian structure $(h,\gamma, V,K)$, that is 
\be
\bftheta^i_{\mathsf{EH}}  = \bftheta^K_{\mathsf{EH}} V^i   + \mathring\bftheta{}^i_{\mathsf{EH}} ,
\ee
where $\bftheta^K_{\mathsf{EH}} := K_i \bftheta^i_{\mathsf{EH}}$ and $\mathring\bftheta{}^i_{\mathsf{EH}}  := \bftheta^j_{\mathsf{EH}} \gamma_j{}^i $.  

In appendices \ref{app:sympt}  and $\ref{app:symph}$, we provide the fully covariant expression for $\bftheta^K_{\mathsf{EH}}$  and $\mathring\bftheta{}^i_{\mathsf{EH}}$, valid in any gauge,  as well as their expansion up to first order in $\O$.
Here, we limit ourselves to the evaluation of $\hat\bftheta^K_{\mathsf{EH}}$ in Bondi gauge. 
The expansion of $\bftheta^i_{\mathsf{can}} \gamma_i{}^A$, in the Bondi gauge is performed to first order in appendix \eqref{app:symph}.

Using the Bondi gauge conditions $\bar\theta =0$ and $\delta K_i =  K_i\delta \bb$,  as well as the Pbc $\bar\kappa =\pa_\O\bb=\O\bA$ \eqref{aspectdef}, we obtain:
\be 
\hat\bftheta{}_{\mathsf{EH}}^K  = 
- \tfrac12 \sqrt{g} \left(
  \bTheta_{ij}  \delta \gamma^{ij} + \O\delta \bA\right),
 \qquad \tfrac12 \pa_\O (\sqrt{g}  \delta V^i)K_i = - \tfrac12 \O \sqrt{g}  \delta \bA .
\ee
Summing these we get the expression for $\bftheta_{\mathsf{can}}^K \in C^{\mathrm{poly}}_{d-2}$ that we were looking for
\be
\bftheta_{\mathsf{can}}^K = - \tfrac12 \sqrt{g} \left(
   \bTheta_{ij}  \delta \gamma^{ij}  + 2 \O \delta \bA\right).
\ee 
In $d=4$, the  expansion 
of this component up to the order relevant for the renormalization, 
\be
 \bftheta_{\mathsf{can}}^K = [\bftheta_{\mathsf{can}}^K]_0+ \O [\bftheta_{\mathsf{can}}^K]_1 +  \O^2 \ln \O [\bftheta_{\mathsf{can}}^K]_{(2,1)} + \O^2 [\bftheta_{\mathsf{can}}^K]_2 + o(\O),
\ee 
is given by
\be
    [\bftheta_{\mathsf{can}}^K]_0 & = 
  \tfrac14 e^\a\sqrt{q} \left( C^{AB}  \delta q_{AB} \right),\cr
[\bftheta_{\mathsf{can}}^K]_1
 &=  \tfrac12 e^\a\sqrt{q} \left(  D^{AB}  \delta q_{AB}
    \right),\cr
[\bftheta_{\mathsf{can}}^K]_{(2,1)} & 
 \stackrel{\text{BMSW}}{=} 
 \tfrac14 \pp_u^{-1}  \left( \sqrt{q} D^{\langle A} D_C D^{B\rangle C} \delta q_{AB} \right) 
\ee
where we used $\pp_u q_{AB}  \stackrel{\text{BMSW}}{=} 
0$ \eqref{eq:BSMW-ppu(q)} and that $\bTheta_{0AB}=:\frac12 C_{AB}$,  $ \bTheta_{1AB} =: D_{AB}$ \eqref{Ddef},  and $2\bA_0 = -\frac14 C:C$ \eqref{eq_G2} in $d=4$ .
Note that the last expressions comes from the Coulombic anomaly \eqref{eq:bTCoulomban}. It is important to recall that this result requires us to put the radiative anomaly to zero.

We see that,  as expected, the first component $[\bftheta^K_{\mathsf{can}}]_1$ vanishes in $d=4$ when the logarithmic anomaly $D_{AB}$ vanishes.
This fact provides a nice consistency check on our calculations: In fact, as we will see in the next section  (cf. equation \eqref{eq:Cevolution}), if $\pa_i \bftheta_{1\mathsf{can}}^i = \pp_u [\bftheta^K_{\mathsf{can}}]_1 + ...$ does not vanish, then the logarithmic term $[\bftheta^\O_{\mathsf{can}}]_{(1,1)} $  would have to be non-zero which is only possible in the presence of a nontrivial logarithmic anomaly  as per \eqref{eq:thetalogd4}.

\subsection{Step 2: Asymptotic renormalization from radial evolution}\label{subsec:renor2}

Now that the first renormalization step is achieved, we can write the fundamental balance law \eqref{balancelaw} for $\tl \bftheta{}^a_\mathsf{can} = \O^{2-d}\bftheta^i_{\mathsf{can}} h_i^a + \O^{3-d}\bftheta^\O_\mathsf{can} \delta^a_\O$ \eqref{eq:tlthetacan} as  an $\O$-evolution equation for its rescaled flux component $\bftheta^\O_{\mathsf{can}} $:
\begin{align}
   (d-3-\O\po) \bftheta^\O_{\mathsf{can}} &= \bE - \delta  \bL_{\mathsf{can}}  + \p_i \bftheta^i_{\mathsf{can}},
\label{eq:Cevolution}
\end{align}
where $\bE =  \bE_{ab}\delta g^{ab}$ are the {physical frame} equations of motion  rescaled by a factor $\O^{d-2}$, i.e. $\bE= \O^{d-2}\tl \bE$, while  $\bL_{\mathsf{can}} = \O^{d-2}\tl{\bL}_{\mathsf{can}}$ is given in \eqref{L1} or \eqref{L2}.
To renormalize, we need to impose the first $d-2$ equations of motion \eqref{Elim}, that is: $\bE^{ab}_\text{0} = 0$, which corresponds to the third Boundary condition \eqref{Pbc3} and $\bE^{ab}_{(k)}=0$ for $k=1, \dots, (d-3)$, which corresponds to the condition that  $\sT_{ab}=\O^{-1}(S_{ab}+\N_{ab})$ is well defined and its first $d-4$ orders vanish at $\scri$ \eqref{Pbcd}. 

Once these conditions are satisfied we can see (\textit{i}) from \eqref{thetaRL1}  that the leading divergence of $\tl \bL_{\mathsf{can}}$ is of order $\O^{2-d}$, and (\textit{ii}) from \eqref{L2} and \eqref{polyordersCarroll} that the first logarithm of $\tl \bL_\mathsf{can}$ is of order $\ln\O$, i.e. that $\tl \bL_\mathsf{can} =\Omega^{2-d} \bL_\mathsf{can} $ with $\bL_\mathsf{can} \in C^\text{poly}_{d-2}$
In $d=4$ this means that:
\be \nonumber
\tl \bL_{\mathsf{can}} = \O^{-2}\bL_{\mathsf{can}},
\qquad
\bL_{\mathsf{can}} = 
[\bL_{\mathsf{can}}]_0 + \O [\bL_{\mathsf{can}}]_1 + \O^2\ln\O [\bL_{\mathsf{can}}]_{(2,1)}+  \O^2[\bL_{\mathsf{can}}]_2 + o(\O^2).
\ee
Staying for concreteness in $d=4$, if one now defines the Lagrangian counterterms:
\be
\tl\bell{}^\O_2 &:= -\O^{-1} [\bL_{\mathsf{can}}]_0 + \ln\O [\bL_{\mathsf{can}}]_{1} + (\O \ln\O -\O) [\bL_{\mathsf{can}}]_{(2,1)} , \cr
\tl\bell{}^i_2 &:=0,
\ee
the fully renormalized Lagrangian $\tl\bL_R$ is given by the following expression, which by construction admits a finite limit at $\scri$:
\be
\tl \bL_R := \tl \bL_\mathsf{can} - \pp_a \tl \bell{}^a_2 
& = \O^{-2}\left(  \bL_\mathsf{can} - [\bL_{\mathsf{can}}]_0  - \O [\bL_{\mathsf{can}}]_{1} - \O^2 [\bL_{\mathsf{can}}]_{(2,1)}  \right)\cr
& \SCRIeq [ \bL_\mathsf{can}]_2.
\ee
Similarly in arbitrary dimensions we find that $\tl \bL_R \SCRIeq [ \bL_\mathsf{can}]_{d-2} $
Note that for this argument to work one needs the leading logarithmic term to be linear in $\ln\O$, which is one of the consequences of the radial equations of motion, and the leading logarithm of $\bL_\mathsf{can}$ to appear at order $(d-2,1)$ (cf. \eqref{L2} and footnote \ref{eq:hatNpolylog}).

We can do a similar analysis for $\bftheta^i_{\mathsf{can}} \in C^{\mathrm{poly}}_{d-2}$.\footnote{From \eqref{thetai-1}, one sees that the leading logarithms of $\bftheta^i_\mathsf{EH}$ come from the $\O\Pi \in C^\text{poly}_{d-2}$ \eqref{poly2} contained in $\bmK_i{}^j$ and $\omega_i$. All other terms are higher order.}
In $d=4$, it can be expanded as
\be 
\bftheta^i_{\mathsf{can}} =[\bftheta^i_{\mathsf{can}}]_0 +\O [\bftheta^i_{\mathsf{can}}]_1 + \O^2 \ln \O [\bftheta^i_{\mathsf{can}}]_{(2,1)} + \O^2 [\bftheta^i_{\mathsf{can}}]_{2} + o(\O^2).
\ee
We then define the symplectic corner counterterms 
\be
\tl\bvartheta{}_2^{\O i } & := - \O^{-1} [\bftheta_{\mathsf{can}}^i]_0 + \ln\O [\bftheta_{\mathsf{can}}^i]_1 + (\O\ln\O - \O) [\bftheta_{\mathsf{can}}^i]_{(2,1)},\qquad 
  \mathring{\tl\bvartheta}{}_2^{ij}  :=0,
 \label{eq:tlbvartheta2}
\ee
which yield the renormalized tangential symplectic potential
\be 
\tl \bftheta{}_R^i &:=  \tl \bftheta{}^i_\mathsf{can} - \delta \tl\bell{}^i_2 +\pa_\O \tl\bvartheta{}_2^{i \O } 
=[\bftheta^i_{\mathsf{can}}]_{2} + o(1).
 \label{thetaR2}
\ee
We used that $\tl \bftheta{}^i_\mathsf{can}=\O^{-2} \bftheta{}^i_\mathsf{can}$ in $d=4$.

We now turn to the flux term $\tl\bftheta{}_R^\O$ obtained from the counterterms above:
\be
\tl\bftheta{}_R^\O &:= \tl \bftheta{}^\O_\mathsf{can} -\delta \tl\bell{}_2^\O +\pa_i \tl\bvartheta{}_2^{\O i },
\ee
where $\tl \bftheta{}^\O_\mathsf{can}=\O^{-1} \bftheta{}^\O_\mathsf{can}$ in $d=4$.
We can use that the relationship \eqref{eq:Cevolution} is still satisfied after renormalization. This means that
\be
 -\pa_\O \tl \bftheta{}_R^\O = \tl \bE  - \delta  \tl \bL_R + \p_i \tl \bftheta^i_{R} = 
 [\bE]_2 %
- \delta  \left[\bL_{\mathsf{can}}\right]_2   %
+\p_i[\bftheta^i_{\mathsf{can}}]_{2}%
 + 
 o(1) .
\ee 
We conclude from this that the expansion of the renormalized symplectic potential at $\scri$ is given by
\be
\tl \bftheta{}_R^\O = [\bftheta_{\mathsf{can}}^\O]_1 - \O \left(   [\bE]_2 %
- \delta   \left[\bL_{\mathsf{can}}\right]_2 %
+ \p_i[\bftheta^i_{\mathsf{can}}]_{2} \right) + o(\O)
\ee

From this we conclude that, in $d=4$, the symplectic potential flux $\tl\bftheta{}^\O_R$ at $\scri$ is finite and simply equals the finite order of $\tl\bftheta{}^\O_\mathsf{can}$ (an explicit expression is provided in the next section):
\be \label{thetren}
\tl\bftheta{}^\O_R \SCRIeq [\bftheta^\O_\mathsf{can}]_1 .
\ee
We have thus proved the sought result, namely that in $d=4$ both the Lagrangian $\tl\bL_R$ and the symplectic potential current $\tl\bftheta{}^a_R$ have finite limits at $\scri$.

A similar conclusion applies in higher dimensions, where one finds that the renormalized symplectic flux at $\scri$ is simply given by  $\tl \bftheta{}_R^\O \SCRIeq [\bftheta_{\mathsf{can}}^\O]_{(d-3)} = [\tl\bftheta{}^\O_\mathsf{can}]_0$ and more generally, for all components (cf. \eqref{thetaR2}  and \eqref{eq:tlthetacan} ) 
\be 
\tl \bftheta{}_R^a \SCRIeq [\tl\bftheta{}_{\mathsf{can}}^a]_0.
\ee
See appendix \ref{app:ren} for a proof.

This concludes our analysis of the renormalization of the symplectic current at $\scri$ in general gauge.
Once again, we emphasize that the procedure is gauge-independent and only relies on the existence of a radial polyhomogeneous expansion  and the balance equation for the symplectic potential current.

\subsection{Explicit  renormalized symplectic potential in $d=4$}
We now specialize to $d=4$ and the Bondi-Sachs gauge, to write an explicit expression for the flux of the renormalized symplectic potential through $\scri$. We use that $\bTheta_{0AB}=\tfrac12 C_{AB}$ is the shear 
and that $\bTheta_{1AB} =: D_{AB}$ is the source of the logarithmic anomaly while $\Pi_{0A}=  \frac12 D_B  C_A{}^B$.

We start from  the explicit expression \eqref{thetaO1} and the result  \eqref{thetren}, whence we get:
\be \label{rsymp}
 \tl\bftheta^\O_{R}  &\SCRIeq \,  
\sqrt{q}e^\a
 \left(  - \tfrac12
   \rN_{0\langle A}{}^{B\rangle } \delta C_{B}{}^A
+ \Pi_{0A} \delta( e^{-\a} \Y^A_1 )
 \right)
  \cr
 &\quad +  \sqrt{q} e^\a 
 \left( -
  \tfrac12 \rN_{1\langle A}{}^{B\rangle }  \delta q_{\langle B}{}^{A\rangle}
  - \left(\tfrac12\rN_{1} -\hN_1\right) 
  \delta  \ln \sqrt{q}
  +  \rN_1 \delta \a
+ \Pi_{1A}  \delta ( e^{-b} U^A)
 \right).
\ee 
where we are using the notation
$ \delta C_A{}^B := \delta(C_{AC}q^{CB})$ and $\delta q_A{}^B := (\delta q_{AC}) q^{CB}$,\footnote{Note that since $\delta \gamma^{AB}$ is for us the variation of the inverse metric: $\delta \gamma_A{}^B :=\gamma_{AC}(\delta\gamma^{BC}) = - \delta q_A{}^B - \O( \delta C_A{}^B + [C,\delta q]_A{}^B) + O(\O^2)$ and $\delta C_A{}^A = 0$. It is also useful to observe that $[C,\delta q]_{AB}$ is skew.  } Using the equations (\ref{eq_News_2+2},\ref{Pbc-components},\ref{mupi0},\ref{Phi2},\ref{eq:N1-d4}, \ref{MPNaspects}), we can express the different coefficients entering this expression as follows\footnote{For $\rN_1$ we use the identity $D_{\langle A} \Pi_{0B\rangle}= \tfrac12 (\Delta  -  R_q)C_{AB} $  valid in $d=4$}. 
\be
e^{-\a}\Y^A_1&=  q^{AB}  D_B \a, \cr
\rN_{1\langle A B\rangle} 
&  = -D_{\langle A} \Pi_{0B\rangle} + \frac{1}{4} \left( {R}_q +  e^{-\a} \Delta e^{\a}\right) C_{AB} +e^{-\a}F  D_{AB} ,\cr
& = \frac14\left ( 2R_q + (e^{-\a}\Delta e^{\a}) - \Delta\right) C_{AB} + e^{-\a}F D_{AB}
\cr
\hN_{1} &=  -  e^{-\a} \left(  \frac12 D_A (e^{\a}  D_B C^{AB})  + \frac{F}{8} (C:C) \right) .\label{hatN1}\cr
\tfrac12\rN_{1}
&= -e^{-\a} \left(   \Phi_2 +\frac14  D_A(e^b D_B  C^{AB}) \right)\cr
\Pi_{1A} & = \frac12 \left[ \mathring{\nabla}_B \bb - \gamma_{AB} e^{-\bb}\pp_\O \Upsilon^B\right]_2,\cr
\rN_{0\langle A}{}^{B\rangle} &= e^{-\a}( N_{0A}{}^{B} + \tfrac12 F  C_{A}{}^B)  = \frac12 (\cL_\ell + 2F) C_A{}^B  + D_{\langle A} D^{B\rangle}e^b,
\ee
where due to the Pbc2 \eqref{Pbc-components} $F$ is nothing else than the expansion of $\ell = \pp_u + U^A \pp_A$, 
\be
2F = \pp_u\ln\sqrt{q} + D_A U^A.
\ee
These equations follows from the implementations of Pbc123. 
The last three expressions above encode the $d=4$ mass, angular momentum, and radiation aspects, respectively. 
The  radiation aspect $N_{0AB}$ can be written explicitly in terms of $u$-derivatives of $C_{AB}$, as shown above.

The independent variations appearing in $\tl\bftheta{}^\O_R$ are two-folds.  First, we have the radiation variations $\delta C_A{}^B$ conjugated to its own time derivative, the radiation aspect $\rN_{0 B}{}^A\sim \pa_u C_B{}^A + \cdots $.
Next, we have the variations of the Carrollian data 
$(\delta b, \delta U^A, ,\delta q_{\langle A}{}^{B\rangle}, \delta \ln\sqrt{q})$, where we have decomposed the metric variation $\delta q_A{}^B$ into traceless and trace  components.
The Carrollian data are conjugate respectively to
$(\Phi_2, \Pi_{1A},  D_{AB}, \pa_u \sqrt{q} )$ dressed by  a combination of $(b,U^A,q_{AB}, C_{AB})$ themselves.  Note that $\Phi_2$ and $\Pi_{1A}$ are the mass and angular momentum aspects respectively, and $D_{AB}$ is source of the  Coulombic anomaly in $d=4$.

 There are two points worth noting, however. First it is technically incorrect to say a quantity is conjugate to its own time derivative, \emph{when this time derivative is along the Cauchy surface itself}, here $\Sigma = \scri$. This is because the two are not independent pieces of data on $\scri$; in other words, $\pp_u C \delta C$ is not a symplectic structure presented canonically in Darboux ``coordinates" as it happens when phase space naturally takes the form of a cotangent bundle (see \cite{RielloSchiavinaAHP} for an in-depth discussion of this point). 
Second, the equation $\pa_u q_{AB}= F q_{AB}-D_{(A} U_{B)}$ (and not just its trace part considered above) is one of the Pbc2 \eqref{Pbc-components} and should therefore be imposed strongly in \eqref{rsymp} together with the leading $(d-2)=2$ orders of the equations of motion, which comprise equation \eqref{Pbc3} and $\mathscr{T}_{ab} \SCRIeq 0$ in $d=4$ (cf. the discussion under \eqref{eq:Cevolution}). However, we leave the full analysis of the ensuing pre-symplectic structure, as well as the imposition of the evolution constraints such as  \eqref{Eevolution}, to future work.

Here we conclude this section by showing that we recover the previous results of \cite{Compere:2018ylh, Freidel:2021fxf} when we impose the BMSW boundary conditions $b=F=U^A = 0$ (Section \ref{sec:otherbc}) which implies $\pa_u q_{AB}=0$.
The previous  expressions simplify considerably under  the BMSW boundary conditions.
For instance, the coefficient conjugated to $\sqrt{q} $ is
\be\label{masshere}
e^{\a}\left(\tfrac12\rN_{1}- \hN_1\right) 
&=    -\Phi_2 +\frac14  D_A(e^b D_B  C^{AB}) +  \frac{1 }{8}  F (C:C) \cr
& \stackrel{\text{(BMSW)}}{=} M + \frac14  D_A  D_B C^{AB} 
\ee
which agrees with the expression in \cite{Freidel:2021fxf}. To derive this we used that the Bondi mass is $M:=-\Phi_2$ under the BMSW boundary conditions.

Indeed, with these additional conditions the symplectic potential becomes 
\be 
\tl\bftheta^\O_{R}
\SCRIeq 
-\sqrt{q} \left( 
\frac14 \left(\pa_u C_{A}{}^B\right)\delta C_B{}^{A} 
+ \frac18\left(2R_q - \Delta\right)C_{A}{}^B \delta q_A{}^{B}
+ \left(M+\frac14 D_A D_B C^{AB}\right) \delta \ln \sqrt{q}
\right)
\ee
 We note that the variables  conjugate to $  C^{AB} $, $ q_{AB}$ and $ \sqrt{q}$ agrees with those found in 
\cite{Compere:2018ylh, Freidel:2021fxf}. 
This provides a non-trivial double check of our and their results.

It is interesting to note  that, if $F=0$ as in the BMSW boundary conditions, the source of the Coulombic anomaly $\bTheta_{1AB} = D_{AB}$ drops out of the expression for $\rN_{1AB}$ and therefore doesn't appear in the renormalized symplectic potential.
Also, if $U^A=0$, the angular momentum aspect $\Pi_{1A}$ also drops.

Finally, it is also interesting to see that for the unrestricted renormalized symplectic potential \eqref{rsymp} we have new canonical pairs:
\begin{enumerate}
\item 
The leading order redshift coefficient $b$ is conjugated to the mass aspect $\rN_1$;
\item The leading order shift coefficient $U^A$ is conjugated to the angular momentum aspect $\Pi_{1A}$; and
\item the leading order conformal structure is conjugated to the traceless tensor $\rN_{1\langle AB\rangle}$ which contains the logarithmic anomaly $ D_{AB}$. Note, however, that the logarithmic anomaly drops from $\rN_{1\langle AB\rangle}$, and hence the symplectic potential, if one fixes expansionless boundary condition for $\scri$, i.e. if $0 = \theta_0 = \Sc_0 = e^{-b}F$. 
\end{enumerate}

This concludes our analysis of the symplectic potential renormalization.

\subsection{Charge renormalization}

Having renormalized the symplectic potential, we can now straightforwardly  renormalize the symmetry charges as well \cite{Hopfmueller:2020yqj,Freidel:2021cbc}.
This is because the (corner) Noether charges associated with the gauge transformations are determined by the knowledge of the Lagrangian and symplectic structure.

Consider a diffeomorphism covariant Lagrangian, such as the Einstein Hilbert Lagrangian $\tl\bL$. The covariance can be expressed as the identity
\be
\cL_{\delta_\xi} \tl\bL = \cL_\xi \tl\bL
\ee
stating that the field-space Lie deriviative of $\tl\bL$ along the transformation encoded in the spacetime vector field $\xi$, equals the spacetime Lie derivative of $\tl\bL$ along $\xi$ itself (note that since $\tl\bL$ is a field-space 0-form, $\cL_{\delta_\xi} \tl\bL = \delta_\xi \tl\bL$). This identity holds unchanged for field-\emph{dependent} diffeomorphisms, $\delta\xi\neq0$.

Introducing the anomaly operator\footnote{In the notation of \cite{GomesRielloScipost} this anomaly operator would read $\Delta_\xi = \mathbb{L}_{\xi^\sharp} - \mathbb{i}_{(\mathbb{d}\xi)^\sharp} - \mathrm{L}_\xi$.}
\be
\Delta_\xi := \cL_{\delta_\xi} - I_{\delta\xi} - \cL_{\xi},
\ee
which generalizes the one of equation \eqref{eq:anomalyop} to higher field-space forms, we see that the diffeomorphism covariance (aka background independence) of $\tl\bL$ can be simply written as
\be
\Delta_\xi \tl\bL=0.
\ee
In general, diffeomorphism covariance of an arbitrary field-space form $\boldsymbol{\alpha}$ reads $\Delta_\xi \boldsymbol{\alpha}=0$.

Using the fundamental identity $\delta \tl\bL= \tl\bE + \rd \tl\bftheta$, and the fact that $\tl\bL$ is a top-form on spacetime, one sees that the covariance of $\tl\bL$ implies
\be\label{eq:Noether}
\rd \tl \bj_\xi = - I_\xi \tl\bE, \qquad \tl \bj_\xi:= I_\xi \tl\bftheta - \iota_\xi \tl \bL,
\ee
where $\tl \bj_\xi$ is the Noether current associated to the infinitesimal diffeomorphism $\xi^a$.
This identity has two consequences, due to Noether \cite{Noether:1918zz}: (1) $\tl\bj_\xi$ is conserved on-shell, and, due to the arbitrariness of $\xi^a$ over spacetime, (2) there exists off-shell (Bianchi) identities saying that the equations of motions are not all independent. These identities take the form $I_\xi \tl\bE = d\tl\bC_\xi$, with $\tl\bC_\xi\heq 0$ giving (upon pullback to a Cauchy surface $\Sigma$) the theory's constraints. For example, for the Einstein-Hilbert Lagrangian,  $\tl\bE = \sqrt{\tl g} \tl G_{ab}\delta \tl g^{ab}$ and $\tl\bC{}^a_\xi = \sqrt{\tl g}\tl G^a{}_b\xi^b$ thanks to the Bianchi identities $ \tl \nabla_a \tl G^{ab}\equiv 0$. 

Combining everything together, one finally deduces that
\be
\tl \bj_\xi = \tl\bC_\xi + d \tl \bq_\xi, 
\ee
meaning that the Noether charge is, on-shell, a codimension-2 (i.e. corner) integral:
\be
Q_\xi := \int_\Sigma \tl\bj_\xi \heq \int_{\pp\Sigma} \tl\bq_\xi.
\ee

This is all just a consequence of the diffeomorphism covariance of $\tl\bL$. 
We now focus on the  diffeomorphism covariance of $\tl\bftheta$, namely $\Delta_\xi \bftheta =0$. It is not hard to see that this implies (and is, in fact, equivalent to) the following would-be Hamiltonian flow equation for the (off-shell pre-)symplectic structure $\Omega_\Sigma := \int_\Sigma \delta \bftheta$:  
\be
I_\xi \Omega_\Sigma = - \delta Q_\xi + Q_{\delta\xi} + \cF_\xi - \mathcal{E}_\xi .
\ee
This would-be Hamiltonian flow equation has three potential obstructions $Q_{\delta\xi}$, $\mathcal{F}_\xi$ and $\E_\xi$, which we now define and analyze. The most trivial one is given by the last term,
\be
\mathcal{E}_\xi := \int_\Sigma \iota_\xi \tl\bE,
\ee
which vanishes on-shell, where we expect $\Omega_\Sigma$ to take the significance of a \emph{conserved} (pre)symplectic structure -- by  ``conserved'' we mean $\Sigma$-independent.
The second obstruction is due to the potential field-dependence of $\xi$, namely $Q_{\delta \xi}$. The third and last obstruction is the corner ``flux" term:
\be
\cF_\xi :=\int_{\Sigma} \rd \iota_\xi \tl\bftheta = \int_{\pa\Sigma} \iota_\xi \tl\bftheta .
\ee
This term, which is relevant only for vector fields $\xi$ which are transverse to $\partial \Sigma$, could be set to zero e.g. by imposing appropriate boundary conditions \cite{Wald:1999wa, Harlow:2019yfa}. 

At $\scri$, all these expressions are formally divergent by various powers of $\O^{-1}$. This is why we introduced a renormalized Lagrangian $\tl\bL_R = \tl \bL - \pp_a \tl\bell{}^a$ and symplectic potential current $ \tl \bftheta{}_R^a = \tl\bftheta^a - \delta \tl\bell^a + \pp_b \tl \bvartheta^{ab}$.
However, since the counterterms $\tl\bell{}^a$ and $\tl \bvartheta{}^{ab}$ depend on the conformal compactification -- which provides us with a nontrivial background structure ($\delta\O=0$, cf. Section \ref{subsec:diffeo} and equation \eqref{eq:OmegaBackground}) -- \emph{the renormalized Lagrangian and symplectic potential fail to be diffeomorphism covariant}. Their anomalous transformation is encoded in the the Lagrangian and symplectic anomalies
\be
\ba^a_\xi := -\Delta_\xi \tl\bell{}^a, \qquad \bfA_\xi^{ab} := \Delta_\xi \tl\bvartheta{}^{ab},
\ee
namely 
\be
\Delta_\xi \tl\bL_R = \pp_a \ba^a_\xi ,\qquad \Delta_\xi\tl\bftheta{}^a_R = \delta \ba^a_\xi - \ba^a_{\delta \xi} + \pp_b \bfA^{ab}_\xi.
\ee
The anomalous transformation of $\tl\bftheta_R$ can be rewritten as the following modified would-be Hamiltonian flow equation for Bondi-Sachs diffeomorphisms (see section \ref{sec:BondiSachsDiffeo}):
\be\label{eq:renormalizedflow}
I_\xi \Omega^R_\Sigma = - \delta Q^R_\xi + Q^R_{\delta\xi} + \cF^R_\xi - \mathcal{E}_\xi
\ee
where -- having eventually in mind $\Sigma = \Sigma_\O \to \scri$ -- we denoted:\footnote{We also used that $\delta \tl\bL_R = \tl\bE + \rd \tl\bftheta_R$, for unchanged equations of motion.}
\be
\Omega^R_\Sigma & := \int_{\Sigma} \delta \tl \bftheta^R\\
Q^R_\xi &:= \int_\Sigma (I_\xi \tl\bftheta_R - \iota_\xi \tl\bL_R - \ba_\xi) = Q_\xi + \int_{\pp\Sigma} (I_\xi \tl\bvartheta - \iota_\xi \tl\bell)\\
\cF^R_\xi & := \int_{\pp\Sigma} (\iota_\xi \tl\bftheta_R + \bfA_\xi).
\ee
A few remarks are in order:

(\emph{1}) Recall that the imposition of the divergent orders of the equations of motions \emph{as boundary conditions}, namely as restrictions on the available phase space, is a requirement of our renormalization algorithm (see \eqref{Elim} and Section \ref{subsec:renor2}). This observation,  together with the facts that $\xi$ is smooth at $\scri$ and its action preserves the polyhomogenous degree of the metric, readily implies that all terms in the would-be Hamiltonian flow equation \eqref{eq:renormalizedflow} have a finite limit at $\scri$. In particular, the charge $Q^R_\xi$ and the flux $\cF^R_\xi$ are finite in the asymptotic limit ``by construction''.

(\emph{2}) Even though we have the simple relation $\tl\bftheta{}_R \SCRIeq [\tl\bftheta{}_\mathsf{can}]_0$, the analogue formulas for $Q^R_\xi$ and $\cF^R_\xi$ stating that these quantities coincide with the zero-order canonical charges and fluxes, do \emph{not} hold. In fact, subleading orders of $\xi$ may combine with divergent orders in $\tl\bftheta_\mathsf{can}$ to give finite order terms that do \emph{not} enter in $Q^R_\xi$. 

(\emph{3}) The contribution $Q^R_{\delta \xi}$ is nontrivial for the Bondi-Sachs diffeomorphisms entering \eqref{eq:renormalizedflow}. In this regard, we recall that the field-\emph{in}dependent parameters that fully determine a Bondi-Sachs diffeomorphism are $(X^i,W)\in \mathfrak{X}(\scri)\times C^\infty(\scri)$, while $\xi_{(X,W)}$ is field dependent. 
More explicitly: The construction done in section \ref{sec:BondiSachsDiffeo} of the Bondi-Sachs vector field $\xi $ associated with an asymptotic pair $(X,W)$ shows, from \eqref{eq:xi.e.xpansion}, that 
 $\xi_{(X,W)} \in C^{\mathrm{poly}}_{(d-1)}$  can be constructed order by order in $\O$ from the expansion 
 \ref{metexp} of the metric coefficients. 
 
The subleading orders of $\xi_{(X,W)}$ depend on $(X,W)$ only through the parameter $\tau = X^u = e^{-\bb} X^i K_i$. This parameter is field \emph{in}dependent.\footnote{Contrary to $Y:=X-\tau \ell$ which is such that $\delta Y = - \tau\delta \ell = - \tau \delta U^A\pp_A$} We thus write
\be
\xi_{(X,W)} = \Big( X + \O\hat \xi_{\tau 1} + \O^2 \hat \xi_{\tau 2} + o(\O^2)\Big) + \Big(W +  \O w_{\tau 1} + \O^2  w_{\tau 2} + o(\O^2) \Big)  \O\pp_\O 
\ee
where (see Appendix \ref{app:BondiDiffeoExp})
\begin{subequations}
\begin{align}
&\begin{cases}
\hat \xi_{\tau 1}= - e^b (q^{AB} \pp_B\tau) \pp_A\\
w_{\tau 1} = - \frac{1}{d-2}e^b\left(2 D^A b D_A \tau + \Delta \tau \right)\O\pp_\O
\end{cases}
\end{align}
and
\begin{align}
&\begin{cases}
\hat \xi_{\tau 2}=  
(e^b \bTheta_0^{AB} D_B\tau)\pp_A \\
w_{\tau 2} = e^b\left( \frac{2}{d-2}\bTheta_0^{AB} D_A b D_B\tau + \frac{1}{d-3} D_A\bTheta^{AB}_0 D_B\tau \right)\O\pp_\O 
\end{cases}
\end{align}
\end{subequations}

(\emph{4}) To compute the renormalized charges explicitly, one needs not only the expressions for $\tl\bL_R$ and $\tl\bftheta_R$ given in the previous sections, but also the variations of the Carrollian structures entering $\tl\bL_R$ \eqref{rsymp} under the action of a Bondi-Sachs diffeomorphism. These are provided in equation \eqref{anomalies} (see also appendix \ref{app:symaction}).

(\emph{5}) Finally, since useful structural information has been obtained by applying the anomaly operator to $\tl\bL$, $\tl\bftheta$ and $\tl\bftheta_R$, we now apply it to $Q_\xi$ and $Q^R_\xi$. For example, owing to the background independence of $\tl\bftheta$ and $\tl\bL$ ($\Delta_\xi \tl\bftheta = 0 = \delta_\xi \tl\bL$), the anomaly of $Q_\xi$ encodes its equivariance:\footnote{We have defined $\Delta_\xi Q_\chi := \int_{\Sigma}\Delta_\xi \tl\bj_\chi$.}
\be
\delta \xi = \delta\chi= 0 \implies
\Delta_\xi Q_\chi = - Q_{[\xi,\chi]}.
\ee
This equivariance becomes anomalous for the renormalized charges, even after accounting for the field dependence of the Bondi-Sachs diffeomorphisms.
Indeed, using the identities\footnote{$\Delta_\xi$ is an operator of even degree, whence the bracket below are all commutators. The other brackets of interest are:
$$
[\Delta_\xi, \delta] = \cL_{\delta\xi} , \qquad [\Delta_\xi, \rd] = 0.
$$
The commutators of $\Delta_\xi$ and the Lie derivatives along $\chi$ and $\delta_\chi$ over spacetime and field space respectively, and of $\Delta_\xi$ and $\Delta_\chi$ can be computed from these.
}
\be
[\Delta_\xi, I_\chi] = I_{\delta_\chi \xi} - I_{\lbr \xi,\chi\rbr},
\qquad
[\Delta_\xi, \iota_\chi] = \iota_{\delta_\chi \xi} - \iota_{\lbr \xi,\chi\rbr},
\ee
where the bracket $\lbr\xi,\chi\rbr = [\xi,\chi]_\text{Lie} - \delta_\xi \chi + \delta_\chi \xi $ is defined in \eqref{algebroid-def}, we compute:
\be
\Delta_\xi Q_\chi - Q_{\delta_\chi \xi} = - Q_{\lbr\xi,\chi\rbr}
\ee
versus
\be
\Delta_\xi Q^R_\chi - Q^R_{\delta_\chi \xi} = - Q^R_{\lbr \xi, \chi\rbr} + \int_{\pp\Sigma} ( I_\chi \bfA_\xi  + \iota_\chi \ba_\xi). 
\ee
We thus expect the anomaly to be reflected in the charge algebra -- which is however a subtle concept since $Q^{(R)}_\xi$ is not the Hamiltonian generator of $\delta_\xi$.
(Recall that $\lbr \xi, \chi\rbr$ reflects the boundary Bondi-Sachs algebra of asymptotic symmetries $\mathfrak{B}$ as per equation \eqref{morphism}.)

We conclude our analysis of the charge renormalization here.
While it would be interesting to expand more explicitly the charges and write down the renormalized expression in terms of the expansion of the asymptotic 
stress tensor and even in terms of the Weyl tensor component, such an endeavour is beyond the scope of this article.
It would require first understanding the roles of the Lorentz transformations in the formulation of stretch horizons, which constitute the diffeomorphism anomalies \eqref{anomalies} in the conformal spacetime. It would also require a better understanding of the radial expansion near a stretched horizon before applying it to the conformal spacetime. Jai-akson has studied  these  two questions  in \cite{Jai-akson:2022gwg}, and his results will be the object of a forthcoming publication.

\section{Conclusions}

In this work, we have given a  comprehensive and exhaustive analysis of asymptotically flat spacetimes in terms of the Penrose conformal frame. We have performed this study   within the most general boundary  conditions compatible with the definition of conformal infinity, and for a polyhomogenous expansion of the metric in the conformal factor. 
We have extracted from this formulation a key object, the asymptotic stress tensor, which can be extracted from a projection of the Conformal stress tensor sourcing the Einstein's equations and be seen as subleading component of the Carrollian fluid stress tensor associated with the stretched infinity. We also have shown that its radial expansion contains information about the charge aspects. We have shown that after the first renormalization step, the asymptotic boundary of the conformal spacetime can be understood as a stretched horizon equipped with a stretched Carrollian structure.  We have also represented the charge aspects, their  conservation laws and their radial expansion as Bianchi identities for the asymptotic Weyl tensor.
These structures have allowed us to provide a covariant renormalization of the symplectic potential. We presented an explicit renormalization in dimension $4$ and in  a minimal Bondi gauge which generalizes the previous renormalizations of \cite{Compere:2018ylh, Freidel:2021fxf} and agrees with them when restricted.  (Our Bondi gauge fixes bulk diffeomorphism freedom completely, while leaving untouched the full group of boundary diffeomorphisms as well as the first nontrivial order of the radial diffeomorphisms, which yield boundary Weyl transformations).
These results have allowed us to draw a deep connection between the geometry of asymptotic infinity and the geometry of finite distance stretched horizons \cite{Freidel:2022vjq, Freidel-Jai-24 }. See also the recent work \cite{Ashtekar:2024mme} for a discussion of horizons and null infinity.

Several questions remains to be developed: First, as mentioned above, a more explicit renormalization of the charge aspect would be welcomed. Such analysis requires a deeper understanding of stretched horizons symmetry charges associated with finite hypersurfaces.

Second, the charge renormalization that we have performed allows to remove divergences when the limit $\O\to 0$ is taken.
It does not guarantee, however, that the charges have a finite limit when $u\to \pm \infty$. More work is needed to ensure this fundamental property. Presumably along the line of the interesting work \cite{Peraza:2023ivy}. 

 In the context of renormalized symplectic potential and asymptotic charges for in the Bondi gauge in $d=4$, we also need to highlight the work \cite{Geiller:2024amx} by Geiller and Zwikel which builds on their \cite{Geiller:2022vto}, and appeared when this manuscript was being completed.

Our analysis has been done for asymptotically flat spacetimes. However, most of our formalism could be applied to the challenging study of asymptotically de-Sitter or AdS spacetimes with leaky boundary conditions \cite{Poole:2018koa, Compere:2020lrt, Ruzziconi:2020cjt, Geiller:2022vto, Senovilla:2022pym, Erfani:2022lnr} Since these spaces also admit a conformal completion. 

Finally, it would be interesting to connect the work done here with the analysis of symmetry done purely at spacelike infinity
\cite{Henneaux:2019yax, AH:2020rfq,Chakraborty:2021sbc, Fuentealba:2022xsz} and understand whether the extension described here 
can survive the matching of spacelike with null infinity as they do in \cite{Compere:2023qoa} for BMS.

\section*{Acknowledgements}
LF would like to thank his collaborators Luca Ciambelli, Elliot Gesteau, Puttarak Jai-akson,  Jerzy Kowalski-Glikman, Rob Leigh, Faroogh Moosavian, Djordje Minic, Javier Peraza, Andrea Puhm, Daniele Pranzetti,  Ana-Maria Raclariu and Romain Ruzziconi and  mentees 
Nicolas Cresto, Robin Oberfrank, Eirini Telali for their patience and their teachings. 
LF would like to dedicate this work to Joy Rosenthal; a beautiful and gentle soul who has left us too early and suddenly. 
AR would like to thank F. Hopfm\"uller for contributions to this work in its early stages, and R. Myers for his support in bringing this work to fruition.
We would also like to specially thank F. Capone and P. Mitra for pointing out a correction to the first version of this paper and for detailed explanations of their works.
Research at Perimeter Institute is supported in part by the Government of Canada through the Department of Innovation, Science and Economic Development Canada and by the Province of Ontario through the Ministry of Colleges and Universities. This work was supported by the Simons Collaboration on Celestial Holography.

\appendix

\section{Schouten and Weyl tensors, Bianchi identities}
\subsection{Schouten tensor}
The schouten tensor $S_{ab}$ is such that 
\bea 
R_{abcd} &=& W_{abcd} + g_{ac} S_{bd} + g_{bd}S_{ac}-g_{bc} S_{ad} -g_{ad}S_{bc},\cr
R_{ac} &=&  g_{ac} S + (d-2) S_{ac},\cr
R &=&  2(d-1)  S. 
\eea
In other words we have  
\be
G_{ab} = (d-2)(S_{ab} - g_{ab} S), \qquad 
S_{ab}= \frac1{(d-2)} \left( R_{ab} - \frac{g_{ab}}{2(d-1)} R\right). 
\ee
Under a transformation $\tl {g}_{ab} \to {g}_{ab}= \Omega^2 \tl {g}_{ab}$ we have 
$\tl {S}_{ab} \to S_{ab}$ where
\bea
{S}_{ab}&=& \tl {S}_{ab}- {\nabla}_a \Upsilon_b - \Upsilon_a\Upsilon_b + \frac{g_{ab}}{2} \Upsilon^2, \cr
&=&  \tl {S}_{ab} - \frac1{\Omega} {\nabla}_a N_b  +\frac{g_{ab}}{2 \Omega^2} N^2, \cr
S& =& \Omega^{-2} (\tl {S} - \frac1{\Omega} \nabla_c N^c  +\frac{d}{2 \Omega^2} N^2),\cr
G_{ab} &=& \tl {G}_{ab} - (d-2)
\left( \frac1{\Omega} (\nabla_a N_b-g_{ab} \nabla_c N^c) + \frac1{\Omega^2}\frac{(d-1) }{2 } g_{ab} N^2 \right),
\eea
where  we temporarily introduced $\Upsilon_a = \frac{\nabla_a \Omega}{\Omega}$.
\be
\tilde{\nabla}_a \Upsilon_b =  {\nabla}_a \Upsilon_b - (2 \Upsilon_a \Upsilon_b - \Upsilon^2 g_{ab}).
\ee

\subsection{Bianchi identities}\label{app:Bianchi}
Given $\tl g_{ab} =\O^{-2} g_{ab}$ we have 
\be
\tl \Gamma_{ab}{}^c = \Gamma_{ab}^c + C_{ab}^c, \qquad 
C_{ab}^c = - (N_a \delta_b^c+N_b\delta_a^c - N^c g_{ab}) \O^{-1}. 
\ee 
The  Bianchi identity for the Weyl tensor is 
\be 
\nabla_d W_{abc}{}^d = 2 (3-d) \nabla_{[a} S_{b]c}.
\ee 
From this we can evaluate that 
\be 
\tl \nabla_d W_{abc}{}^d 
&=  
\nabla_d W_{abc}{}^d - C_{da}^e W_{ebc}{}^d -C_{db}^e W_{aec}{}^d-C_{dc}^e W_{abe}{}^d + C_{de}^d W_{abc}{}^e\cr
&= \nabla_d W_{abc}{}^d + \O^{-1}  (N_d \delta_a^e+N_a\delta_d^e - N^e g_{da}) W_{ebc}{}^d
+ \O^{-1}  (N_d \delta_b^e+N_b\delta_d^e - N^e g_{db}) W_{aec}{}^d\cr
& \quad +\O^{-1}  (N_d \delta_c^e+N_c\delta_d^e - N^e g_{dc}) W_{abe}{}^d  - \O^{-1}  (d N_e )  W_{abc}{}^e \cr
&= \nabla_d W_{abc}{}^d + 3\O^{-1}  W_{abc N}
+0
- \O^{-1} (W_{Nbca} + W_{aNcb}+ W_{abNc}) - \O^{-1} 
(d N_e )  W_{abc}{}^e \cr
&= \nabla_d W_{abc}{}^d +  (3-d) Y_{abc}\cr
&= (3-d) ( 2 \nabla_{[a} S_{b]c} + Y_{abc}).
\ee
We used the Bianchi identity $ W_{Nbca} + W_{aNcb}+ W_{abNc}=0$ and we defined 
$Y_{abc} :=  W_{abcN} \O^{-1}$ .
We then evaluate 
\bea
\tl \nabla_{[a} \tl S_{b] c} &=&  \nabla_{[a} \tl S_{b] c}   - C_{c [a }^d \tl S_{b] d} \\
&=&\nabla_{[a} \tl S_{b] c}   +\O^{-1} (  N_{[a}   \tl S_{b] c}  -  g_{c[a}  \tl S_{b] N}   )
\eea
Since $\tl W_{abc}{}^{d}= W_{abc}{}^{d}$ this means that we have two different evaluations of the Weyl tensor  divergence
\bea
\tl \nabla_d W_{abc}{}^d &=&   (3-d) ( 2 \nabla_{[a} S_{b]c} + Y_{abc} ) \\
\tl \nabla_d W_{abc}{}^d &=&  (3-d) (  2 \nabla_{[a} \tl S_{b]c} + \O^{-1} 2(  N_{[a}   \tl S_{b] c}  -  g_{c[a}  \tl S_{b] N}   )
)
\eea
Taking the difference and using that  $\tl S_{ab}=S_{ab} +\N_{ab}$ we get that 
\be 
 W_{abcN}  = \O Y_{abc} = 2\O \nabla_{[a} \N_{b]c}+ 2(  N_{[a}   \tl S_{b] c}  -  g_{c[a}  \tl S_{b] N}   ).
\ee
This establishes \eqref{WN}.
Imposing Pbc3 \eqref{Pbc3},  stating $\tilde S_{ab} \SCRIeq 0$, means that  $Y_{abc}:= \O^{-1} W_{abcN}$  is regular at $\scri$ and satisfies
\bea
 \nabla_d W_{abc}{}^d  + (3-d) Y_{abc} &=& (3-d)\left(Y_{abc}  +2 \nabla_{[a} S_{b]c}\right) \\
 &=& 2(3-d) \left( 2  N_{[a}   \sT_{b] c} -  g_{c[a}  \sT_{b] N} + \O \nabla_{[a} \sT_{b]c} \right)  .
\eea
If we further assume that $W_{abcd}\SCRIeq 0$ we can define $D_{abcd}:= \O^{-1} W_{abcd}$ and the Bianchi identity now reads
\be 
 \O \nabla_d D_{abc}{}^d  + (4-d) Y_{abc} &= 2(3-d) \left( 2  N_{[a}   \sT_{b] c} -  g_{c[a}  \sT_{b] N} + \O \nabla_{[a} \sT_{b]c} \right)  
\ee
The conservation law  for $\tl T^{\mathsf{mat}}_{ab} =\O^{d-3} \Tm_{ab}$ 
is
\be 
\tl \nabla_a \tl T^{\mathsf{mat}}{}_b{}^a & =  \nabla_a (\O^{d-1} T^{\mathsf{mat}}{}_b{}^a) + \O^{d-1}[ C_{ac}^a T^{\mathsf{mat}}{}_b{}^c - C_{ab}^c T^{\mathsf{mat}}{}_{c}{}^a], \\
&= \O^{d-2} ( \O \nabla_aT^{\mathsf{mat}}{}_b{}^a  -   T^{\mathsf{mat}}{}_{bN} +  
N_b T^{\mathsf{mat}} ).
\ee 

\subsection{Weyl Tensor} \label{app:Weyl}

The Weyl tensor is given by the Riemann tensor minus the Kulkarni-Nomizu product of the metric with the Schouten tensor
\be 
W_{abcd}= R_{abcd} - S_{ac} g_{bd}-  S_{bd}g_{ac}+S_{ad}g_{bc}+S_{bc} g_{ad}.
\ee 
The different components of the Weyl tensor defined in  section \ref{subsec:Weyl}, are given in a general null  frame by 
\be
\bar{W}_{ABCD} & =
R_{ABCD} +S_{AD}g_{BC}+S_{BC} {\gamma}_{AD}- S_{AC} {\gamma}_{BD}-  S_{BD}{\gamma}_{AC}\cr
\bar{W}_{ABC}&= R_{A B C K }  +  S_{AK} \gamma_{BC}-S_{B K } \gamma_{AC} , 
\cr
\bar{W}_{AB  }& := R_{A K B K } - S_{KK} \gamma_{AB} = R_{\langle A |K| B \rangle  K }, 
\ee
where we use the shorthand notation $V_A:=e_A{}^a V_a$.
We now evaluates these components in terms of the $(d-2)+2$ coefficients \ref{sec:2+2EEq}. 
Recall that our notation is such that 
$\mathring{\nabla}_A$ is the derivative operator compatible with $\gamma_{AB}$ and therefore we denote 
$\mathring{R}_{ABCD}$ the curvature tensor of $\mathring{\nabla}$ and similarly for $\mathring{S}_{AB}$ which is the Schouten tensor of $\mathring{\nabla}$.

\paragraph*{Evaluation of $\bar{W}_{ABCD}$.}

Denoting the Kulkarni-Nomizu product  by  $\circ$,  with
\be 
(\bTheta \circ\Theta)_{ABCD}:=
\bTheta_{AC}\Theta_{BD} + \Theta_{AC}\bTheta_{BD} - \bTheta_{BC}\Theta_{AD} - \Theta_{BC}\bTheta_{AD},
\ee 
one has 
\be
R_{ABCD} &= \mathring{R}_{ABCD}- (\bTheta \circ\Theta)_{ABCD} 
\cr
&=\mathring{W}_{ABCD} 
- (\bTheta \circ\Theta)_{ABCD} + ( \mathring{S}\circ \gamma)_{ABCD},
\ee 
where $\mathring{S}_{AB}$ is the codimension-$2$ Schouten tensor (of the metric $\gamma_{AB}$), namely
\be 
\mathring{S}_{AB} = \frac1{d-4}\left( \mathring{R}_{AB} - \frac{1}{2(d-3)} \mathring{R} \gamma_{AB}\right) = \frac1{d-4} \mathring{R}_{\langle AB\rangle } + \frac{1}{2(d-2)(d-3)} \mathring{R} \gamma_{AB}.
\ee 
From this expression,
\be
\bar{W}_{ABCD} &= {R}_{ABCD}
- ( S\circ \gamma)_{ABCD}
\cr
&= \mathring{W}_{ABCD}
- (\bTheta \circ\Theta)_{ABCD} + ( (\mathring{S}-S)\circ \gamma)_{ABCD}.
\ee
Now, recall that $N_a = L_a + \O \nu K_b$, as well as the definition of $\mathcal{N}_{ab}$ \eqref{eq_gennews}, and that $\Theta_{AB}$ and $\bTheta_{AB}$ are the deformation tensors of $L$ and $K$ respectively. Combining these pieces of information, we obtain the relationship
\be
\Theta_{AB} = \nu \gamma_{AB} + \O(\SN_{AB} - \nu \bTheta_{AB}).
\ee
Recall as well the fact that the Einstein equations for the physical frame Ricci $\tl{R}_{ab} =0$ when expressed in terms of the Schouten tensor read $\tl{S}_{ab}=0$. Owing to \eqref{tlS} their projection on the sphere thus reads
\be
S_{AB} + \SN_{AB} = 0.
\ee
(Note that we impose the leading order(s) of these equation as the Pbc3(d) boundary conditions, (\ref{Pbc3},\ref{Pbcd}).)

Thus, reinserting these equations in the above expression for the projected Weyl tensor, we obtain
\be
\bar{W}_{ABCD} =  \mathring{W}_{ABCD} + ( (\mathring{S} + \cL)\circ \gamma)_{ABCD} 
 - \O( \bTheta\circ \cL )_{ABCD},
\ee
where  we denoted (cf. the definition (\ref{defmu}, \ref{eq_News_2+2} of the news tensor)
\be
\mathcal{L}_{AB}:= \SN_{AB}-\nu \bTheta_{AB}= e^{-\beta} \left( \tfrac{1}{(d-2)} \mu \gamma_{AB} + N_{AB}\right).
\ee
Evaluating at $\scri$ in $d\geq6$ by means of \eqref{mupi0} and \eqref{eq:N0AB}, one finds $\cL_{AB} \SCRIeq - \mathring{S}_{AB}$ \eqref{eq:NS0}, and thus:
\be
\bar{W}_{ABCD} \SCRIeq \mathring{W}_{ABCD}.
\ee
 In $d=4$, $\bar{W}_{ABCD}$ is traceless at $\scri$ and therefore for symmetry reasons it vanishes identically there. Note that the above equation therefore remains valid also in $d=4$.

\paragraph*{Evaluation of $\bar{W}_{ABC}$.}

 Using that $\omega_A=-\O\Pi_A$ (\ref{eq_2+2_explicit},\ref{aspectdef2}), and $\cN_{KA} = - \Pi_A$ \eqref{eq_News_2+2}, one gets on shell of the Einstein equations ($0 \EOMeq \tl{S}_{ab} = S_{ab} + \cN_{ab}$) that 
\be 
R_{ABCK} &=(\Sn_A-\O\Pi_A)\bTheta_{BC}-
(\Sn_B-\O\Pi_B)\bTheta_{AC} \cr
S_{AK} &= \Pi_A,
\ee 
whence
\be 
\bar{W}_{ABC}&= (\Sn_A-\O\Pi_A)\bTheta_{BC}-
(\Sn_B-\O\Pi_B)\bTheta_{AC}  +  \Pi_A \gamma_{BC}-\Pi_B \gamma_{AC}.
\ee 
Given that $(d-3)\Pi_{0A}= D_B \bTheta_0^B{}_A$ \eqref{mupi0} we obtain the desired result
\be 
\bar{W}_{ABC} \SCRIeq  D_A \bTheta_{0BC}- D_B\bTheta_{0AC} + \frac1{(d-3)}\left((D\cdot \bTheta_0)_A q_{BC}- (D\cdot \bTheta_0)_B q_{AC}\right).
\ee
THis expression can be seen to vanish  identically in $d=4$.

\paragraph*{Evaluation of $\bar{W}_{AB}$.}

Using that in the Bondi-Sachs gauge $\bar{a}_A=0$, i.e. the  sphere components of the acceleration of $K$  vanish, we get that 
\be 
R_{AKBK}&= -\gamma_A{}^a\gamma_B{}^b (\cL_K- \O \bA) \bTheta_{ab} + (\bTheta \!\cdot\!\bTheta)_{AB}  \cr
& = -(\cL_K \bTheta_A{}^C) \gamma_{CB} - (\bTheta\cdot\bTheta)_{AB} + \O \bA\ \bTheta_{AB} .
\ee 
 On-shell $S_{KK} \EOMeq - \cN_{KK}$ which is, in Bondi gauge,  equal to $\bA$ \eqref{eq_News_2+2} . Thus, using the $(KK)$ component of \eqref{eq_G2}, i.e. $(d-2)\bA = - \bTheta:\bTheta$, we have
\be
\bar{W}_{AB} = R_{AKB K} - \bA \gamma_{AB} = - (\pa_\O \bTheta_A{}^C) \gamma_{CB}  - (\bTheta\cdot\bTheta)_{\langle AB\rangle } + \O \bA\ \bTheta_{AB}
\ee
which is (as expected) traceless. 
At leading order this expression gives (recall \eqref{4dlogan})
\be
\bar{W}_{AB} = - (\ln\O +1)\bTheta_{(1,1)AB} - D_{AB} - (\bTheta_0\cdot\bTheta_0)_{\langle AB\rangle_q} + o(1).
\ee
As observed elsewhere the product of two traceless $2\times2$ matrices is pure-trace hence the third term in the expression above vanishes in $d=4$.

\paragraph*{Consistency checks}

Note that we can evaluate the trace 
\be 
\bar{W}_{AB}{}^B
&= (d-3)\Pi_A -\Sn_B \bTheta_{A}{}^B +
\O\Pi_B\bTheta_{A}{}^B \cr
&=   \O \left( \pa_\O \Pi_A + \tfrac{1}{(d-2)} \pa_A  (\bTheta:  \bTheta) +  \bTheta_A{}^B\Pi_B
\right)
\ee 
where we used the $(KA)$ equation of \eqref{eq_G2}. We know (see eq.\eqref{TracesW}) that this trace is equal to $-\O \mathring{Y}_A$.  We can therefore deduce from this an expression for $\mathring{Y}_A$ which agrees with \eqref{Yexpress} since $\bar{A}= -\tfrac{1}{(d-2)} (\bTheta:  \bTheta)$. This provides a nice consistency check on our expressions. 

 Similarly, from the tracelessness of the spacetime Weyl, one sees that
\be
\bar{W}_{ADB}{}^D = 2\O (\rY_{(AB)} + \nu \bar W_{AB}).
\ee
Taking the trace of $\bar{W}_{ABCD}$, we obtain
\be \label{eq:trW}
\bar{W}_{ADB}{}^D &= 
(d-4)\mathcal{L}_{AB} + \mathcal{L}{\gamma_{AB}} +\mathring{R}_{AB}
- \O \left( \cL  \bTheta_{AB} - 2 (\bTheta \!\cdot \cL )_{(AB)}\right)
\ee
Replacing the definition of $\cL_{AB}$ in terms of $\mu$ and $N_{AB}$, one finds that 
\be
(d-4)\mathcal{L}_{AB} + \mathcal{L}\gamma_{AB} +\mathring{R}_{AB} &= \left(2(d-3) e^{-\beta}\mu +\mathring{R} \right)\frac{\gamma_{AB}}{(d-2)} + \left((d-4)e^{-\beta} N_{AB} +\mathring{R}_{\langle AB \rangle}\right).
\ee
 Combing with the $(KL)$ and $(AB)$ components of the radial Einstein's equations \eqref{eq_G2} we find: 
\be
(d-4)\mathcal{L}_A{}^B + \mathcal{L}\gamma_A{}^B+ \mathring{R}_A{}^B & = 2 \O \left( e^{-\beta}\pp_\O (e^\beta \mathcal{L}_A{}^B) + \Snabla_{(A} \Pi^{B)} + \O \Pi_{A} \Pi^{B} +\frac12 e^{-\beta}\mu \bTheta_A{}^B\right).
\ee
 Thus, recalling $\pp_\O\beta = \O\bA $ and $\cL = \E = \rN = e^{-\bb}\mu$, we finally deduce
\be
\bar{W}_{AD}{}^{BD} = 2\O \left(  \pp_\O \mathcal{L}_A{}^B + ( \bTheta\cdot\cL)_{(A}{}^{B)} + \Snabla_{(A} \Pi^{B)} + \O\left(\bA \mathcal{L}_A{}^B  +  \Pi_{A} \Pi^{B} \right)  \right) .
\ee

We can now check the equality $\bar{W}_{ADB}{}^D = 2\O (\rY_{(AB)}+\nu\bar{W}_{AB})$, by using $\rY_a{}^b$ as independently computed in \eqref{Yexpress}. 
Indeed:
\be
\rY_{(AB)} + \nu \bar{W}_{AB} 
& =  (\pa_\O\rN_{A}{}^C) \gamma_{CB} + 
 ( \rN \!\cdot \!  \bTheta)_{(AB)}  
 +\Sn_{(A} \Pi_{B)} + \bA \nu \gamma_{AB}
 -\bTheta_{AB} \pa_\O\nu,
\cr
&\quad  +\O\left(\Pi_A\Pi_B+ \bA (\rN_{AB} - 2\nu \bTheta_{AB}) 
\right) 
- \nu (\pa_\O \bTheta_A{}^C) \gamma_{CB} \cr
& \quad -\nu (\bTheta\cdot\bTheta)_{\langle AB\rangle } + \O \nu \bA\ \bTheta_{AB}\cr
& = \left( (\pa_\O\rN_{A}{}^C) \gamma_{CB}  +\bTheta_{AB} \pa_\O\nu - \nu (\pa_\O \bTheta_A{}^C) \gamma_{CB} \right) \cr
&\quad + \left( (\rN\cdot\bTheta)_{(AB)} +\bA \nu \gamma_{AB} - \nu (\bTheta\cdot\bTheta)_{\langle AB\rangle} \right) \cr
&\quad +\Sn_{(A} \Pi_{B)} + \O\left( \Pi_A \Pi_B + \bA (\rN_{AB} - \nu \bTheta_{AB})\right)\cr
& = \bar{W}_{ADB}{}^D,
\ee
where in the last step we recognized that $\bA = - \frac{1}{d-2}\bTheta:\bTheta$ \eqref{eq_G2} to deduce
\be
\bA \nu \gamma_{AB} - \nu (\bTheta\cdot\bTheta)_{\langle AB\rangle} = - \nu (\bTheta\cdot\bTheta)_{AB}.
\ee

\section{Acceleration and twists}\label{sec:acc}

We first evaluate the acceleration and connection coefficients appearing in \eqref{eq_2+2_explicit}.
We use that the  expressions of $K$ and $L $ are given by 
 \be
K_a\rd x^a = e^\bb \rd u,\qquad  L_a\rd x^a = (N_a-\Omega \nu K_a)\rd x^a = \rd \O - \O \Phi \rd u,  \qquad \Phi= \nu e^\bb.
\ee
Taking the differential of these equality we get that 
\be
\nabla_a K_b -\nabla_b K_a &= (\nabla_a\bb) K_b- (\nabla_b\bb) K_a,\cr
 \nabla_a L_b -\nabla_b L_a &= -e^{-\bb}(\nabla_a(\O \Phi) K_b- \nabla_b(\O \Phi) K_a).
\ee
Therefore after contraction we get the accelerations and twists
\be 
\nabla_K K_b &= K[\bb] K_b, \cr
\nabla_L L_b &=  e^{-\bb}( \nabla_b(\O\Phi) - L[\O\Phi] K_b) =  e^{-\bb}(\Snabla_b[\O \Phi] + K[\O \Phi] L_b), \cr 
\nabla_L K_b &= (L^a \nabla_b K_a) - (\Snabla_b\bb + L_b K[\bb] ), \cr
\nabla_K L_b &= (K^a \nabla_b L_a) - e^{-\bb} K[\O \Phi] K_b.
\ee 
After projection this gives 
\be
\bar{\kappa}= \pa_\O\bb, \qquad \bar{\acc}_A=0, \qquad \bar\eta_A =  {\omega}_A+\mathring\pa_A \bb,
\cr
\kappa = \e^{-\bb} \pa_\O (\O\Phi) , \qquad {\acc}_A= \O e^{-\bb} \Snabla_A \Phi  , \qquad \eta_A =  -{\omega}_A.
\ee
We also have that 
\be
\bar\eta_A -\eta_A= e^{-\bb} q_{AB}(\pa_\O \Upsilon^B) 
\ee
This follows from the fact that 
the vector fields are $K^a\pa_a =\pa_\O$ and $L^a\pa_a = e^{-\bb} ( \pa_u + \Y^A\pa_A +  \Phi \O \pa_\O )$ and therefore 
\be
[K,L] &= \nabla_K L-\nabla_LK = -\bar\kappa L + \kappa K + (\bar\eta-\eta)^a \mathring\pa_a,\cr
&= [\pa_\O, e^{-\bb}(\pa_u+\Upsilon^A\pa_A + \O\Phi \pa_\O)]\cr
&= - (\pa_\O \bb) L + e^{-\bb} \pa_\O(\O\Phi) K + e^{-\bb} (\pa_\O \Upsilon^A) \pa_A. 
\ee
form which we get 
\be
\bar\kappa=   (\pa_\O \bb), \qquad \kappa = e^{-\bb} \pa_\O(\O\Phi), \qquad (\bar\eta-\eta)^A =e^{-\bb} (\pa_\O \Upsilon^A).
\ee
\subsection{Accelerations}\label{acc2} 
In this section we evaluate some of the acceleration coefficients $\nabla_V V^a$ and also $\bmK_{V}^a$
We define 
\be
-\O \rJ_i &:= \gamma_i{}^a\nabla_V N_a,\cr
\omega_V&:= K^b\nabla_V N_b.
\ee
which means that 
\be
\mN_{V}{}^b=\nu V^b +\O \hN_V{}^b &= (\nabla_V N^a) h_a{}^b  := \omega_V V^b - \O\rJ^b.
\ee
We find that
\be
\omega_V&= K^b\nabla_N N_b - 2\O\nu K^b\nabla_K N_b\cr
&=  N^a \nabla_K N_a + 2\O \nu N_b\nabla_K K^b \cr
& = (K +2\bar\kappa)[\O\nu]=\nu + \O (K +2\bar\kappa)[\nu].
\ee
This means that we have 
\be
\omega_V &: = \nu +\O A_K, \quad \mathrm{with}\quad   \kA =(\pa_\O +2\pa_\O\beta)[\nu].
\ee
And therefore 
\be\N_{V}{}^i = \kA V^i - \rJ^i. 
\ee
We also have that $\omega_V$ is related to the longitudinal acceleration of $V$
\be
K_i \hn_V V^i = K_a \hn_V N^a  - 2K_a \hn_V (\O\nu K^a) =  K^a \hn_V N_a = \omega_V.
\ee
Now we also establish that 
\be
- \rJ_i &=  \gamma_i{}^j (\nabla_j - 2 \omega_j) \nu,\cr
\bmK_{V}{}^i &= \nabla_V K^a h_a{}^i = - \bar\eta^i = -(\mathring\pa^i\bb -\O \Pi^i), \cr
\hn_V V^i & = \omega_VV^i + \gamma^{ij}(\pa_j \nu + 2\nu  \pa_j\bb).
\ee
Using that $ V^b = N^b + 2\O \nu K^b$,
one first look at 
\be 
 \nabla_V N_a 
&= V^b \nabla_a N_b = N^b\nabla_a N_b - 2\O \nu K^b \nabla_a N_b \cr
&=\nabla_a(\O \nu)  -2\O \nu (- \bar\kappa N_a +\omega_a )
\cr
&= N_a (\pa_\O   \nu + \pa_\O \beta)[\O \nu] + \O( \hn_a -2\omega_a)\nu).
\ee
Therefore 
\be 
\nabla_V N^a h_a{}^b=  V^b \underbrace{(K + 2\bar\kappa)[\O\nu]}_{=\omega_V}  +  \O  \underbrace{\gamma^{ba}(\hn_a -2\omega_a)\nu}_{=-\rJ^b}.
\ee
We also have 
\be 
\bmK_{V}{}^a= \nabla_V K^a h_a{}^b &=\nabla_L K^a h_a{}^b  - \O\nu   \nabla_K K^a h_a{}^b = \nabla_L K^a \gamma_a{}^b = - \bar\eta^b .
\ee
Finally, we have that 
\be
\hn_V V^a &= \nabla_V N^b h_b{}^a - 2\O\nu K_{V}{}^a \cr
&= \omega_V V^b  + \gamma^{ab}(\hn_b -2\omega_b)\nu + 2\O \nu \bar\eta^a \cr
&= \omega_V V^b + \gamma^{ab}(\hn_b +2(\bar\eta-\omega_b)_b)\nu\cr
&=\omega_V V^b + \gamma^{ab}(\hn_b +2(\bar\eta+\eta_b)_b)\nu\cr
&= \omega_V V^b + \gamma^{ab}(\hn_b \nu + 2\nu \hn_b\bb).
\ee
We conclude using that $\gamma_a{}^b \omega_b =-\eta_b$ and $\Pi_a +\bar\eta_a =\eta_a+\bar\eta_a$.

\subsection{News coefficients}\label{sec:News}
We evaluate the news tensor coefficients $\N_{aK}$. Using that $\nabla_a N_b =\nabla_b N_a$ and that $N^2 =2 \O \nu$ we get 
\be
\nabla_N N_a &=   N^b (\nabla_a N_b )= \nabla_a ( \O \nu)  =   N_a \nu +  \O \nabla_a \nu,\\
 K^a \nabla_N N_a &= \pa_\O (\O\nu) =\kappa - \O\nu  \bar\kappa=\nu + \O\pa_\O \nu,\\
\nabla_K N_a &=   (K^b \nabla_a N_b )= \omega_V K_a - \bar\kappa N_a + \rom_a.
\ee
One uses that $ \nabla_a K_b -\nabla_b K_a = \nabla_a \beta K_b - \nabla_b \beta K_a$ to establish that 
\be
 N^b = V^b + 2\O \nu K^b. 
\ee
to get 
\be
\nabla_K N_b = \nu K_b +\O \N_{bK}&=K^c \nabla_b N_c = K_c \nabla_b N^c \cr
&= K_c \nabla_b V^c   + 2 K_c \nabla_b (\O \nu K^c) \cr
&= N_b (K_c \nabla_K V^c ) + h_b{}^d (K_c \nabla_d V^c)\cr
&= - N_b K[\bb] + \omega_b.
\ee
Therefore 
\be
\O \N_{a K} =  - N_b K[\bb]  + \omega_a -\nu K_a.
\ee
Decomposing this we get 
\be
 \N_{KK}&= - \O^{-1}\pa_\O \bb,\cr
  \N_{KV}&= \O^{-1}(\omega_V- \nu) = \O^{-1} (\kappa - \Phi -V[\bb])\cr
  \N_{KA}&=  \O^{-1} \Pi_A.
\ee
We can also evaluate  $\N_{KV}$ 
\be
\N_{KV} &=   (\N_{KN} -2\O \nu \N_{KK})\cr
&= (\pa_\O \nu  +2 \nu \pa_\O \bb)\cr
&= e^{-\bb} (\pa_\O \Phi  + \Phi \pa_\O \bb)
\ee
And this means that we have 
\be
\omega_V &= (1+ \O \pa_\O) \nu +  2 \nu \O\pa_\O \nu = e^{-2\bb}(1+ \O \pa_\O)[e^\bb \Phi].\cr
\kappa &= (1+ \O \pa_\O) \Phi  +  V[\bb] + \Phi \O \pa_\O \bb \cr
&=  e^{\bb}\left(( 1+ \O \pa_\O)\nu + N[\bb]  \right)
\ee

\subsection{Generalized news}\label{app:news-news}
We evaluate the relationship between the extrinsic curvature  $\mK_{ij}:= h_i{}^a h_j{}^b\nabla_a N_b$ 
and the projected news \eqref{Nproj}
\be
\mK_{ij}=\nu h_{ij} + \O \K_{ij}&:= h_i{}^a h_j{}^b\nabla_a N_b = h_i{}^a \nabla_a N^c g_{cd} h_j{}^b
\cr
&= h_i{}^a \nabla_a N^d h_d{}^c g_{cd} h_j{}^b  + h_i{}^a \nabla_a N^d N_d K^c g_{cd} h_j{}^b \cr
&= h_i{}^a \nabla_a N^d h_d{}^c g_{cd} h_j{}^b  + h_i{}^a \nabla_a N^d N_d K^c g_{cd} h_j{}^b \cr
&= \nu h_{ij}+  \hN_i{}^d h_{dj} + \O \hn_i \nu K_j.
\ee
Therefore 
\be
 \K_{ij} =  \hN_{(i}{}^k h_{j)k} +  \hn_{(i} \nu K_{j)}.
\ee 
While on the other hand we have 
\be
\hN_i{}^j &= h_i{}^a (\nabla_a N^b) h_b{}^j = 
h_i{}^a (\nabla_a V^b) h_b{}^j  + h_i{}^a (\nabla_a (2\rho K_b)  g^{bc} h_c{}^j \cr
&= \hn_i V^j  + 2 \rho  (\hn_i  K_l )  \gamma^{lj}  \cr
&= \hn_i V^j  +  2 \rho  \bmK_i{}^j . 
\ee

We also have that 
\be
\N_{ab} &= \rN_{ab} - \Pi_a N_b - N_a\Pi_b - \rJ_a K_b -K_a \rJ_b \cr
 & +
\kA (K_a N_b + K_bN_a) + K_aK_b [V[\rho]+ 2\nu \kA] - N_aN_b \bA,\cr
\hN_a{}^b&=  \rN_{a}{}^b 
- \Pi_a V^b  -K_a \rJ^b  +
 \kA K_a V^b  \cr
\rN_{a}{}^b &= \rtN_{a}{}^b +\frac{\E}{(d-2)} \gamma_a{}^b.
\ee

\subsection{Covariant derivatives} \label{subsec:covder}

Here, we summarize the relations involving the spacetime covariant derivative $\nabla_a$, the rigged covariant derivative $D_a$, and the horizontal covariant derivative $\Sn_A$.  We use the results of \cite{Freidel:2022vjq}. First, let us provide the general form of the spacetime covariant derivative of the tangential vector $V$, the transverse vector $K$, and their combination $N = V + 2\rho K$ that will become handy in the computations. We denote $\rho = N^2/2=\O\nu$ and $\varphi_a=\mathring\pa_a\bb$.
\begin{align}
\nabla_a N^b = \ & \mN_a{}^b   +\hn_a\rho K^b
+ N_a\left( K[\rho]  K^b + \rom^b  -  \bar\kappa V^b\right), \label{nabla-N}\\
\nabla_a K^b = \ & \bmK_a{}^b  -\omega_a K^b  +N_a  \bar{\kappa}  K^b,  \label{nabla-k}\\
\nabla_a V^b = \ & \hn_a V^b -(\hn_a\rho - 2\omega_a \rho)K^b
- N_a\left( (K[\rho] + 2\rho \bar\kappa) K^b - \rom^b +  \bar{\kappa} V^b\right). \label{nabla-v}
\end{align}
where $\hn_a V^b:= h_a{}^c (\nabla_c V^d)  h_d{}^b $ and similarly we have 
$\bmK_a{}^b  := h_a{}^c (\nabla_c K^d) h_d{}^b $,  and $\mN_a{}^b = h_a{}^c (\nabla_a N^b)h_d{}^b$.
Using that 
\be\label{ome=}
\omega_a =\rom_a + \omega_V K_a, \qquad \omega_V= \pa_\O\rho + 2\rho \bar\kappa= [\kappa+\rho \bar\kappa].
\ee
we get that the projections of (\ref{nabla-v},\ref{nabla-k},\ref{nabla-N}) are given by 
\begin{align}
\mN_a{}^b & = \rmN_a{}^b  + (\rom_a + \omega_V K_a) V^b - K_a \sJ^b,\\
\bmK_a{}^b &  = \bTheta_a{}^b - K_a (\rom^b + \varphi^b), \label{D-k} \\
\hn_a V^b &  =  \rmV_a{}^b + (\rom_a + \omega_V K_a) V^b + K_a\sA^b, \label{D-ell}
\end{align}
where we recalled the acceleration and currents are 
\be 
\omega_V &:= (K+ 2\pa_\O \bb)\rho,\label{KV}\\
\sA^a &:= (\Sn^a + 2\varphi^a)\rho,\label{A}\\
\sJ^a &:= 2\rho (\rom^b + \varphi^b)-\sA^a = -(\Sn^a - 2\rom^a)\rho .\label{J=} \\
\varphi_a &=\mathring{\pa}_a \beta.
\ee 
and we denoted 
\be
\mathsf{V}_a{}^b = (\rmN_a{}^b-2\rho \bTheta_a{}^b).
\ee
Taking the trace gives $\mN := \rmN + \kappa$. Denoting $\rtmN_a{}^b$ the traceless component of $\rmN_a{}^b= \rtmN_a{}^b + \frac{\gamma_a{}^b}{(d-2)} \rmN $  we get 
\be
\mT_a{}^b &=- \rtmN_a{}^b + \gamma_a{}^b \left(\omega_V +\tfrac{(d-3)}{(d-2)}\rmN\right)    - \rom_a  V^b + K_a (\sJ^b  + \rmN V^b)
\ee
We see that $T_V{}^b = \sJ^b  + \rmN V^b$ is the full energy current.
We can also write the full expression for $\nabla_a N_b$ and $\bmK_{ab}$ as 
\be
 \nabla_a N_b &= \rmN_{ab} + \rom_a N_b + N_a\rom_b - \sJ_a K_b -K_a \sJ_b \cr
 & +
 \omega_V (K_a N_b + K_bN_a) + K_aK_b [V[\rho]+ 2\rho \omega_V] - N_aN_b \bar\kappa, \label{nablaN}
 \\
  \nabla_a K_b &= \bTheta_{ab} -K_a (\rom_b+\varphi_b) -K_b \rom_a  +N_aN_b \bar\kappa ,
  \label{nablaK}
  \ee
where the symmetry of $ \nabla_a N_b$ is ensured by the identities (\ref{KV}, \ref{J=}).
Since $g_{ab} = \gamma_{ab} + K_aN_b +N_aK_b - 2\rho K_aK_b$ and using that $\rho=\O\nu$, and 
$\omega_V =\nu + \O \kA$, $\bar\kappa = \O \bA$, while $\rmN_{ab}= \nu \gamma_{ab} +\O \rN_{ab}$, $\sJ^a=\O\rJ^a$  and $\rom_a=-\O\Pi_a$ we get that 
\be
\N_{ab} &= \rN_{ab} - \Pi_a N_b - N_a\Pi_b - \rJ_a K_b -K_a \rJ_b \cr
 & +
 \kA (K_a N_b + K_bN_a) + K_aK_b [V[\rho]+ 2 \nu \kA] - N_aN_b \bA.\cr
 \hN_{a}{}^b &= \rN_{a}{}^b - \Pi_a V^b   -K_a \rJ^b +
 \kA K_a V^b. 
 \ee
As we have seen, there are three layers of covariant derivatives, $\nabla_a, \hn_a$ and $\Sn_a$. To connect them, we first look at the spacetime covariant derivative of the horizontal basis $e_A$ along another horizontal basis. One can verify that it is given by
\begin{align}
\nabla_{e_A} e_B{}^c = \mathring\Gamma^C_{AB} e_C{}^c - \bTheta_{AB} V{}^c - \rmN_{AB} K{}^c. \label{del-e}
\end{align}
Using the decomposition of the spacetime metric  and the Leibniz rule, we express the spacetime divergence of the horizontal basis as
\begin{equation}
\begin{aligned}
\hn_a e_A{}^a & = \left( K_a V^b + q^{BC} e_{Ba} e_C{}^b    \right) \nabla_b e_A{}^a  = \mathring\Gamma^B_{BA}  + \rom_A +\varphi_A,\cr 
\nabla_a e_A{}^a & =  (N_a K^b +h_a{}^b)\nabla_b e_A{}^a = \hn_a e_A{}^a - \rom_A 
 = \mathring\Gamma^B_{BA}  + \varphi_A . 
\end{aligned}
\end{equation}
 Following from these results, the covariant derivative of a generic horizontal vector fields $X^a := X^A e_A{}^a$ projected onto the horizontal subspace is 
\begin{align}
e^B{}_a\nabla_{e_A} X^a & = e_A[ X^B] + X^C e^B{}_b \nabla_{e_A} e_C{}^b = \Sn_A X^B. 
\end{align}
Furthermore, the spacetime and Carrollian divergence of the horizontal vector are 
\begin{equation}
\begin{aligned}
\nabla_a \left( X^A e_A{}^a \right) &= e_A[ X^A] + X^A \nabla_a e_A{}^a = \left(\Sn_A +\varphi_A \right)X^A.\cr
\hn_a X^a = \hn_a (X^A e_A{}^a) &= \Sn_A X^A + (\rom_A + \varphi_A) X^A.
\end{aligned}
\end{equation} 

In addition, let us also look at the rigged covariant derivative of the horizontal basis. We can show by recalling that $h_a{}^b = \gamma_a{}^b + K_a V^b$ and $\gamma_a{}^b = e^A{}_a e_A{}^b$ the following relation
\begin{equation}
\label{D-e_A}
\begin{aligned}
\hn_b e_A{}^c &=h_b{}^d h_c{}^a \nabla_d e_A{}^c \\
&= (\gamma_b{}^d + K_bV^d) \nabla_d e_A{}^d (\gamma_d{}^c + K_d V^c) \\
&= \Sn_b e_A{}^c + K_b( \hn_V e_A{}^d)\gamma_d{}^c + (\gamma_b{}^d\hn_d e_A{}^d K_d) V^c + K_b( \hn_V e_A{}^d) K_d  V^c\cr
&=e_b{}^B  \mathring\Gamma^C_{BA} e_C{}^a   + \rmV_A{}^B e_B{}^c K_b - 
e_b{}^B \bTheta_{BA} V^c   + (\rom_A + \varphi_A) K_b V^c\cr
&=   \mathring\Gamma^c_{bA}    + \rmV_A{}^c K_b - 
\bTheta_{bA} V^c   + (\rom_A + \varphi_A) K_b V^c
\end{aligned} 
\end{equation}
We used that $[V, e_A]= \varphi_A V$ hence $\hn_V e_A= \hn_{e_A}V + \varphi_A V$. 

For completeness, let us also compute the Carrollian covariant derivative of the co-frame ${e}^A$. Using that 
$\hn_a (V^b e_b{}^A) = V^b \hn_a e_b{}^A +  (\hn_a V^b)e_b{}^A =0$ and $\hn_a (e_B{}^b e_b{}^A{}) = e_B{}^b \hn_a e_b{}^A + e_b{}^A \hn_a e_B{}^b =0$ we hence write 
\begin{align}
\hn_a e_b{}^A &=  K_b V^c \hn_a e_c{}^A+ \gamma_b{}^c \hn_a e_c{}^A   \cr
&= - ( \hn_a V^c)e_c{}^A K_b - (e_b{}^B \hn_a e_B{}^c) e_c{}^B \\
& = - (\mathsf{V}_a{}^A + K_a \sA^A)K_b - \mathring\Gamma^A_{ab}-\mathsf{V}_b{}^A K_a.
\end{align}
Following from $\gamma_c{}^b = e_c{}^A e_A{}^b$, we can then show that the Carrollian covariant derivative of the codimension $2$ Carrollian projector is 
\begin{align}
\hn_a \gamma_c{}^b &= e^A{}_c \hn_ae_A{}^b + e_A{}^b \hn_a e^A{}_c \cr
&= \left( \mathring\Gamma^b_{ac} +(- \bTheta_{ac}+(\Pi_a + \rom_c)K_a)V^b + K_a \mathsf{V}_c{}^b \right) - \left( (\mathsf{V}_a{}^b + K_a \sA^b)K_c + \mathring\Gamma^b_{ac}+\theta_c{}^b K_a\right) \cr
& = -( \bTheta_{ac}- K_a (\rom_a + \varphi_c))V^b - (\mathsf{V}_a{}^b + K_a \sA^b)K_c. 
\end{align}
This result can also be obtain by simply using that $\gamma_c{}^b = h_c{}^b - K_c V^b$ and that $\hn_a h_c{}^b =0$.

\section{Conformal stress energy: conservation and radial expansion}

\subsection{Conservation Law}\label{app:consl}
In this section we work out the consequences of the equation \eqref{WN}
\be
 \nabla_{[a} \N_{b]}{}^c =  Y_{ab}{}^c +2 ( \delta^c_{[a}\sT_{b] N}-N_{[a} \sT_{b]}{}^c ), 
\ee
where $ Y_{abc} := \O^{-1} W_{abcN} $.  The goal is to project it along the tangent or transverse indices.
To warmup for the establishment of the conservation laws \eqref{stretch-cons2} we  define 
\be 
\bmK_i{}^j &:= h_i{}^a (\nabla_a K^b ) h_b{}^j , \cr 
\mK_{ij} &:= h_i{}^a h_j{}^b \nabla_a N_b =\nu h_{ij} +\O \hK_{ij},\cr
\mN_i{}^j&: = h_i{}^a (\nabla_a N^b ) h_b{}^j =\nu \delta_i{}^j +\O \hN_i{}^j.
\ee 
and given an arbitrary vector $Z^a$ we first evaluate
\be
h_i{}^a (\nabla_a Z^b) h_b{}^j &
= \hn_i \hat{Z}^j + Z_N \bmK_i{}^j, \cr
h_i{}^ah_j{}^b (\nabla_a Z_b) &= \hn_i \hat{Z}_j + \mK_{ij} Z_K, \cr
h_i{}^a K^b (\nabla_a Z_b)  & = (\hn_i +\omega_i) Z_K - \bmK_i{}^j \hat{Z}_j, \cr
h_i{}^a  (\nabla_a Z^b) N_b & = (\hn_i -\omega_i) Z_N - \mK_{ij} \hat{Z}^j.
\ee
where we denote $\hat{Z}_i = h_i{}^a Z_a$, $\hat{Z}^i = Z^a h_a{}^i $, $Z_K = Z_a K^a$ and $Z_N = Z^a N_a$.
The rules applies by linearity to tensors since we have  $ h_i{}^ah_j^b(\nabla_a ( N_b Z_{KN} K^c) h_c{}^l =0$ which means that the evaluation vanishes if there is more than one normal $N_a$ or $K^b$ involved.
Using this we can evaluate
\be
h_i{}^ah_j{}^b (\nabla_{a} \N_{b}{}^c) h_c{}^l &= \hn_{i} \hN_{j}{}^l 
+ \bmK_{i}{}^l \hn_{j}\nu - \mK_{ij}   \bA^l \cr
h_i{}^a K^b(\nabla_{a} \N_{b}{}^c ) h_c{}^l&=  -(\hn_{i}+\omega_i)  \bA^l
- \bmK_i{}^j \hN_j{}^l 
+ \bmK_i{}^l \pa_\O\nu,\cr
h_i{}^a K^b(\nabla_{b} \N_{a}{}^c ) h_c{}^l&=  h_i{}^a (\nabla_{K} \hN_{a}{}^c)h_c{}^l 
- \omega_i   \bA^l \label{PPK}
\ee
where we denote $-\bA^l= \N_{K}{}^c h_c{}^l$ and we used $\N_{KN}= \hn_\O \nu$.
Taking the skew symmetric combination  of the first combination gives
\be 
\hn_{i} \hN_{j}{}^l -\hn_{j} \hN_{i}{}^l 
+ \bmK_{i}{}^l \hn_{j }\nu - \bmK_{j}{}^l \hn_{i }\nu &=  \hat{Y}_{ij}{}^l  +  h_{i}{}^l \hat{\sT}_{j } - h_{j}{}^l\hat{\sT}_{i },
\ee
where we denote $\hat{Y}_{ij}{}^l:= h_i{}^a h_j{}^b Y_{ab}{}^c h_c{}^l$ and  $ \hat{\sT}_{j } := h_i{}^a {\sT}_{a N}$.
Taking the contraction $j=l$ and using that $\hat{Y}_{ij}{}^j=0$  we get the conservation law we where looking for
\be 
\hn_{i} \hN_{j}{}^i -\hn_{j} \hN
+ (\bmK\delta_i^j - \bmK_{j}{}^i) \hn_{i }\nu &=    (d-1) \hat{\sT}_{j } - \hat{\sT}_{j } \heq 0.
\ee
We can now consider the antisymetrisation of the last two terms of \eqref{PPK}. It gives 
\be\label{radhN0}
 h_i{}^a (\nabla_{K} \hN_{a}{}^c)h_c{}^l 
+   \hn_{i}  \bA^l
+ \bmK_i{}^j \hN_j{}^l 
- \bmK_i{}^l \pa_\O\nu &\heq \hat{Y}_{i}{}^l 
\ee
where we defined 
\be 
\hat{Y}_{i}{}^l:= h_i{}^b Y_{Kb}{}^c h_c{}^l, \qquad \bA^i:= - \N_{K}{}^a h_a{}^i = \bA V^i +\Pi^i.
\ee 
One can use that 
\be
\pa_\O \hN_i{}^l = h_i{}^a(\cL_K \hN_{a}{}^c) h_c{}^l = h_i{}^a( \nabla_K \hN_a{}^c) h_c{}^l + \bmK_i{}^j \hN_j{}^l - \hN_i{}^j \bmK_j{}^l.
\ee
Therefore, we finally get that 
\be \label{eq:hYonshell}
\pa_\O \hN_i{}^l  +  \hN_i{}^j \bmK_j{}^l
 +  \hn_{i}  \bA^l 
- \bmK_i{}^l \pa_\O \nu &\heq \hat{Y}_{i}{}^l, 
\ee
which relates the radial expansion of $\hN$ to the Weyl tensor. 
Taking the contraction $i=l$ of this equation gives
\be
\pa_\O \hN  +  \hN_i{}^j \bmK_j{}^i
+   \hn_{i}  \bA^i 
- \bmK \pa_\O\nu &\heq 0.
\ee
We can then use that $\bar{A} = \O^{-1} \bar\kappa$
\be
 \N_{K}{}^a h_a{}^i = -\bA^i= -(\bar{A} V^i +\Pi^i).
 \ee
From which we get the radial evolution for the asymptotic stress tensor
\be
  \pa_\O \hT_i{}^j  +  (\hN \!\cdot \!\bmK)_i{}^j - \delta_i{}^j (\hN \! : \!\bmK)
+   \hn_{i}\bA^j - \delta_i^j(\hn_k\bA^k)
+(\bmK\delta_i^j- \bmK_i{}^j) \pa_\O \nu &\heq \hat{Y}_{i}{}^j.
\ee

\subsection{Radial expansion of the charges}\label{app:radexpcharge}

In this section we perform the codimension $2$ expansion of \eqref{radhN0}.
We use that (\ref{hN},\ref{hV},\ref{Kt})
\be
 \hN_{i}{}^j&= \rN_{i}{}^j    -K_i \rJ^j + (\kA K_i- \Pi_i )V^j. \cr
 \bmK_j{}^i &= \bTheta_i{}^j  - K_i(
\mathring\pa^j \beta - \O \Pi^j) \cr
\hV_i{}^j & = \hN_i{}^j - 2\nu \bmK_i{}^j\cr
\rJ^i + \rA^i & = 2\nu (\mathring{\pa}^i\beta -\O\Pi^i)
\ee
and
\be
\hn_iV^j & = \nu\delta_i^j +\O\hV_i{}^j,\cr
\nabla_K V^i &=- \O\bA^i   = - \O(\bA V^i + \Pi^i)\cr
\nabla_K K_j & = \O \bA K_j \cr
\gamma_a{}^i(\nabla_i K_j) V^j &= \O \Pi_a
\cr
 \hN_{i K} & = A_K K_i -\Pi_i,\cr 
\hN_V{}^j & = A_K V^j -\rJ^j.
\ee
We can now expand the different components entering 
\eqref{radhN0}:
\be
\nabla_{K} \hN_{i}{}^j 
&=  \nabla_{K}\rN_{i}{}^j  -(\nabla_K\Pi_i) V^j + K_i \left( K[\kA] V^j - \kA \nabla_K \rJ^j\right) \cr
&+ \O \left( \Pi_i\Pi^j  -K_i ( \bA  \rJ^j + \kA \Pi^j ) + \bA \Pi_i V^j \right), 
  \cr
  \hn_i \bA^j &= 
  (\hn_i \bA) V^j +\hn_i \Pi^j + \bA \nu \delta_i^j             \cr
&  + \O \bA  (\rV_i{}^j + K_i  \mathring{A}^j -\Pi_i V^j + K_i \kA V^j  ), \cr
(\bmK\!\cdot \!\hN)_i{}^j 
&= (\bTheta \!\cdot \! \rN)_i{}^j  - (\bTheta\cdot \Pi)_i V^j - K_i ( \mathring\pa \bb \!\cdot \! \rN)^j + K_i ( \mathring\pa\bb \!\cdot \! \Pi) V^j \cr
& + \O \left(  K_i ( \Pi \!\cdot \! \rN)^j - K_i ( \Pi \!\cdot \! \Pi) V^j \right), \cr
- \bmK_i{}^l \pa_\O\nu&= 
\left(-\bTheta_i{}^j  + K_i \mathring\pa^j\bb    
- \O K_i \Pi^j \right)\pa_\O\nu.
\ee
We also compute: 
\be
 \nabla_K \Pi_i 
 & =  \gamma_i{}^j \nabla_K \Pi_j - K_i (\Pi_j \nabla_K V^j)\cr
& = \gamma_i{}^j \nabla_K \Pi_j  + \O  (\Pi\cdot \Pi) K_i \cr
 (\nabla_{K} \hN_{i}{}^j ) K_j
&  = K[\kA] K_i -  \nabla_K \Pi_i + \O \bA \Pi_i \cr
& = (K[\kA]-  \O \Pi\cdot \Pi) K_i - \gamma_i{}^j \nabla_K \Pi_j + \O \bA \Pi_i \cr
   K_i \nabla_V\bA^i &  =  \Pi\cdot \pp \beta + \nu \bA +\O(\bA \kA - \Pi\cdot\Pi)
\ee

Defining 
\be
Y:=\hat{Y}_{VK}, \qquad \rY_{a}= \gamma_a{}^i\hat{Y}_{i K}, 
\qquad \rY_a{}^b = 
\gamma_a{}^i\hat{Y}_{i }{}^j \gamma_j^b,
\ee
 using \eqref{eq:hYonshell}, we get that on shell:
\be
\rY_a{}^b 
&  = 
\gamma_a{}^i (\pa_\O\rN_{i}{}^j) \gamma_j{}^b  + 
 ( \rN \!\cdot \!  \bTheta)_a{}^b  
 +\Sn_a \Pi^b + \bA \nu \gamma_a{}^b   
 -\bTheta_a{}^b \pa_\O\nu,
\cr
&\quad  +\O\left(\Pi_a\Pi^b+ \bA (\rN_a^b - 2\nu \bTheta_a{}^b) 
\right), \cr
\rY
& = \pp_\O \E   + 
 (\bTheta \!: \! \rN) 
 +\Sn\cdot \Pi + \bA \nu (d-2)   
 -\btheta \pa_\O\nu
 +\O\left(
\Pi \!\cdot \! \Pi  + \bA  (\E-2\nu \btheta)
\right),\cr
\rY_{a}
&  = \gamma_a{}^i( \nabla_K \hN_i{}^l + \bmK_i{}^j \hN_j{}^l)K_l + \gamma_a^i(\hn_i \bA^j )K_j \cr
& = -\gamma_a{}^i \nabla_K \Pi_i + \O\bA \Pi_a - \bTheta_a{}^j\Pi_j +  \Sn_a \bA - \gamma_a{}^i \nabla_i K_j \bA^j \cr
& = -\gamma_a{}^i (\nabla_K \Pi_i  + \bmK_i{}^j\Pi_j)  + \Snabla_a \bA - (\bTheta\cdot\Pi)_a
\cr
&  = -\gamma_a{}^i \cL_K \Pi_i  + \Snabla_a \bA - (\bTheta\cdot\Pi)_a 
\cr
&  = -\pp_\O \Pi_a  + \Snabla_a \bA - (\bTheta\cdot\Pi)_a 
\cr
Y
&  = V^i ( \nabla_K \hN_i{}^l + \bmK_i{}^j \hN_j{}^l)K_l + (\nabla_V \bA^l) K_l\cr
&  = K[\kA] - V^i \nabla_K \Pi_i - (\mathring\pa^i \beta - \O\Pi^i)(\kA K_i - \Pi_i) + V[\bA] + \bA (\nabla_V V^i) K_i  + (\nabla_V\Pi^i) K_i \cr
&=  K[\kA] + V[\bA] - \O \Pi\cdot \Pi + (\mathring\pa^i \beta - \O\Pi^i) \Pi_i + \bA (\nu +\O A_K) +  \Pi_i(
\mathring\pa^j \beta - \O \Pi^j)\cr 
&=  K[\kA] + V[\bA] + \nu\bA + 2\Pi\cdot \pa\beta   +\O(\kA \bA -3 \Pi\cdot\Pi) 
\ee
where we used $\rN = \E$.

\subsection{Projections of $\nabla_c Y_{ab}{}^c$} \label{Bianchi-proj}
We start from the identity \eqref{Bianchi-N}
\be
0& \heq  h_i{}^a (\nabla_c Y_{a}{}^{bc})  h_b{}^j\cr
&=    h_i{}^a \nabla_c (\hY_{a}{}^{bc} + N_a \hY_{K}{}^{bc} + K^b \hY_{a N}{}^c + N_a K^b Y_{KN}{}^c )  h_b{}^j  \cr
&= (\hn_k -\omega_k) \hY_{i}{}^{jk} +  h_i{}^a (\nabla_c N_a)  \hY_{K}{}^{jc}   + \hY_{i N}{}^c  \nabla_c K^b h_b{}^j \cr
&= (\hn_k -\omega_k) \hY_{i}{}^{jk} +    \hY_{K}{}^{jk}  \mK_{ki} + \hY_{i N}{}^k  \bmK_{k}{}^j \cr
&= (\hn_k -\omega_k) \hY_{i}{}^{jk} +   \gamma^{jl} \hY_{l}{}^{k} \mK_{ki}  + \hY_{i N}{}^k  \bmK_{k}{}^j \cr
&= (\hn_k -\omega_k) \hY_{i}{}^{jk} +   \gamma^{jl} (\hY\!\cdot \mK)_{l i}  - \hY_{N i}{}^k  \bmK_{k}{}^j.\label{Bianchi-proj-A1}
\ee
where we use the notation $ \hY_{K}{}^{bc} : = Y_{K}{}^{de} h_d{}^bh_c{}^e$ and $\hY_{a N}{}^c = h_a{}^d Y_{dN}{}^e h_c{}^e$. 
And we have used that 
\be
 \hY_{K}{}^{ij } = \gamma^{ik} \hY_k{}^j. 
\ee
Contracting \eqref{Bianchi-proj-A1} along $K_j$  gives 
\be
0& \heq  [(\hn_k -\omega_k) \hY_{i}{}^{jk}] K_j \cr
&= \hn_k \hY_{i K }{}^{k} - \hY_{i}{}^{jk} (\hn_k +\omega_k) K_j  \cr
&= - \hn_k \hY_i{}^k - \hY_{i}{}^{jk} \bmK_{kj} \cr
&=  - \hn_k \hY_i{}^k - \hY_{i}{}^{jk} \bTheta_{kj}  + \hY_{i}{}^{jk} K_k (\rom_j+\varphi_j).\cr
&= - \hn_k \hY_i{}^k - \hY_{i}{}^{jk} \bTheta_{kj}  + \hY_{i j K}(\rom^j+\varphi^j),
\label{Bianchi-proj-A2}
 \ee
  where we used \eqref{nablaK}. 
 The Bianchi identity implies that 
 \be
 \hY_{ijK}= \hY_{ji}-\hY_{ij} 
 \ee
 where $\hY_{ij}:= \hY_i{}^k h_{kj}$.

We now consider 
\be
0& \heq  h_i{}^a  h_j{}^b(\nabla_c Y_{ab}{}^{c})  \cr
&=    h_i{}^a  h_j{}^b \nabla_c (\hY_{ab}{}^{c}  + N_a \hY_{K b}{}^{c} + N_b \hY_{a K}{}^c + N_a N_b Y_{KK}{}^c )   \cr
&= (\hn_k -\omega_k) \hY_{ij}{}^k +   h_i{}^a    (\nabla_c N_a)  \hY_{K j}{}^{c}   + \hY_{i K}{}^c  h_j{}^b  \nabla_c N_b  \cr
&= (\hn_k -\omega_k) \hY_{ij}{}^k +    \hY_{Kj}{}^{k}  \mK_{ki} + \hY_{i K }{}^k  \mK_{kj}\cr
&= (\hn_k -\omega_k) \hY_{ij}{}^k 
- 2 ( \hY\!\cdot \mK)_{[ij]}.
\ee

\section{Symmetries}

\subsection{Generalized brackets}\label{app:brackets}

We first establish \eqref{algebroid-def}.
 Denoting $\delta_1 \equiv \delta_{\xi_1}$, $\LL_1 \equiv \LL_{\xi_1}$ and $\delta_1 2 \equiv \delta_{\xi_1}(\xi_2)$, one finds
\begin{align*}
[\delta_{\xi_1} , \delta_{\xi_2}]  \tl g_{ab} 
&=  \delta_{\xi_1}  (\delta_{\xi_2} \tl g_{ab}) - (1\leftrightarrow 2)
 = \delta_{\xi_1}  (\LL_{\xi_2}  \tl  g_{ab}) - (1\leftrightarrow 2)\\
 &=  \LL_{\delta_{\xi_1}\xi_2}   \tl g_{ab}  +  \LL_{\xi_2}   (\delta_{\xi_1} \tl g_{ab}) - (1\leftrightarrow 2) \\
&= \delta_{\delta_{\xi_1}\xi_2} \tl g_{ab} - \delta_{\delta_{\xi_2}\xi_1} \tl g_{ab} +[ \LL_{\xi_2} , \LL_{\xi_1}  ] \tl g_{ab}   \\
 &= \left( 
\delta_{\delta_{\xi_1}\xi_2 }  - \delta_{\delta_{\xi_2}\xi_1 }   - \delta_{[\xi_1,\xi_2]_{\mathrm{Lie}}} \right)( \tl g_{ab})  = -\delta_{\lbr \xi_1,  \xi_2\rbr} \tl g_{ab} .
\end{align*}
We then prove Jacobi identity.
\be
\lbr \lbr  \xi_1,\xi_2\rbr, \xi_3 \rbr &= [\lbr  \xi_1,\xi_2\rbr,\xi_3]_\text{Lie}  - \delta_{\lbr  \xi_1,\xi_2\rbr}\xi_3 +\delta_{\xi_3} \lbr  \xi_1,\xi_2\rbr\cr
&= [[  \xi_1,\xi_2]_\text{Lie},\xi_3]_\text{Lie} - [\delta_{\xi_1} \xi_2,\xi_3]_\text{Lie} + [\delta_{\xi_2} \xi_1,\xi_3]_\text{Lie}   - \delta_{\lbr  \xi_1,\xi_2\rbr}\xi_3 +\delta_{\xi_3} \lbr  \xi_1,\xi_2\rbr\cr
&= [[  \xi_1,\xi_2]_\text{Lie},\xi_3]_\text{Lie} - [\delta_{\xi_1} \xi_2,\xi_3]_\text{Lie} + [\delta_{\xi_2} \xi_1,\xi_3]_\text{Lie}   - \delta_{\lbr  \xi_1,\xi_2\rbr}\xi_3 +\delta_{\xi_3} [  \xi_1,\xi_2]_\text{Lie} \cr
&- \delta_{\xi_3}\delta_{\xi_1} \xi_2 + \delta_{\xi_3}\delta_{\xi_2} \xi_1. 
\ee
After summing over cyclic permutation of  triples  we get 
\be & \lbr \lbr  \xi_1,\xi_2\rbr, \xi_3 \rbr +\text{cycl.}\cr
&= [[  \xi_1,\xi_2]_\text{Lie},\xi_3]_\text{Lie} +\text{cycl.}\cr 
& - [\delta_{\xi_3} \xi_1,\xi_2]_\text{Lie} + [\delta_{\xi_3} \xi_2,\xi_1]_\text{Lie}  + \delta_{\xi_3} [  \xi_1,\xi_2]_\text{Lie} +\text{cycl.} \cr
&  - \delta_{\lbr  \xi_1,\xi_2\rbr}\xi_3  - \delta_{\xi_1}\delta_{\xi_2} \xi_3 + \delta_{\xi_1}\delta_{\xi_3} \xi_2 +\text{cycl.} \cr
&=0
\ee

 \subsection{Symmetry transformations }\label{app:symaction}

 The condition on the transformation reads
 \be
 \pa_\O \hxi^i + e^\bb \gamma^{ij} \pa_j \xi{}^u =0. 
 \ee
 The symmetry transformations are such that 
 \be
\delta_\xi N_a =0, \qquad  \delta_\xi K^b =0.  
 \ee
 From this we can evaluate 
 $\delta_\xi K_a$, $\delta_\xi V^a$, $\delta_\xi \nu$ $\delta_\xi h_{ab}$ and $\delta_\xi \gamma^{ab}$ which are the building block 
 of the symplectic structure. We have that $ (h_{ij}, \gamma^{ij},K_i, V^i, \nu)$ are respectively of conformal weight $(+2,-2,+1,-1, -1)$
 We find that  for each field $O$ of conformal dimension $s$ we have that the transformation under field space diffeomorphism is 
 \be
 \L_\xi O= \cL_\hxi O + \W_\xi (\O\pa_\O - s)O +   \Delta_\xi O 
 \ee
 where $ \Delta_\xi $ is the anomaly operator. This operator depends on one form $H_\xi$, one  horizontal vectors $G_\xi$ and a scalar
 $B_\xi$ 
 \be 
 H_{\xi i } := \hat\pa_i \W_\xi, \qquad  G_{\xi}^i=-\pa_\O\hxi^i  = e^\bb \gamma^{ij}\hat\pa_j\xi^u, \qquad B_{\xi} := \pa_\O \W_\xi, \qquad 
 \ee
 \be
  \Delta_\xi K_i&= + \O B_\xi K_i - G_{\xi i}\cr
\Delta_\xi V^i &= - \O( B_\xi V^i  + H_{\xi }^i - 2 \nu G_{\xi }^i)\cr
 \Delta_\xi h_{ij} &= \O( K_i H_{\xi j}+ K_j H_{\xi i})  \cr
 \Delta_\xi \gamma^{ij} &=  V^i  G_\xi^j  + V^j G_\xi^i\cr
  \Delta_\xi \nu &=   -( 2  \nu B_\xi +   H_{\xi V} ) 
 \ee
 where $G_{\xi i}= h_{ij} G_\xi^j$ and $H_\xi^i =\gamma^{ij} H_{j\xi}$.
 The parameters $(B_\xi, G_\xi^i, H_{i \xi})$ parametrized bulk Lorentz transformations viewed from the boundary. They preserve the defining relations of the stretched Carrollian geometry
 \be\label{Carrolrelapp}
 V_i K^i=1, \qquad \gamma^{ij} K_j=0,\qquad h_{ij} V^j = -2\O\nu K_i.
 \ee
 Indeed we have 
 \be
 \Delta_\xi ( K_i V^i) &= (\Delta_\xi  K_i) V^i+  K_i (\Delta_\xi V^i) \cr
 &= (\O B_\xi K_i - G_{\xi i})V^i - \O K_i  ( B_\xi V^i  + H_{\xi }^i - 2\nu G_{\xi }^i) \cr
 &= - G_{\xi i} V^i - \O K_i H_{\xi }^i=0\cr
 \Delta_\xi (\gamma^{ij} K_j) 
 &= \Delta_\xi\gamma^{ij} K_j +  \gamma^{ij}\Delta_\xi K_j \cr
 &= (V^i  G_\xi^j  + V^j G_\xi^i) K_j + \gamma^{ij} (\O B_\xi K_j - G_{\xi j}) \cr
 &= G_\xi^i - \gamma^{ij}G_{\xi j} =0. \cr
 \Delta_\xi (h_{ij} V^j)  &=  (\Delta_\xi  h_{ij} )V^j  +   h_{ij}  \Delta_\xi V^j   \cr
 &=  \O (K_i H_{\xi j}+ K_j H_{\xi i})V^j  - \O   h_{ij}  ( B_\xi V^i  + H_{\xi }^i - 2\nu G_{\xi }^i) \cr
  &=  \O (V^j H_{\xi j} + 2\O \nu  B_\xi )K_i + \O (H_{\xi i}- h_{ij} H_\xi^j)+ 2\O \nu G_{\xi i}  \cr
  &= \O  (2 H_{\xi V} + 4 \O \nu  B_\xi) K_i + 2\O\nu (G_{\xi i} -B_\xi K_i )  \cr
  &= - 2 \O (\Delta_\xi\nu)  K_i - 2\O\nu (\Delta_\xi K_i).\cr
 &= \Delta_\xi (2\O\nu K_i).
 \ee
 We used that $G_\xi^j K_j=0= H_\xi^i K_i$ and $(H_{\xi i}- h_{ij} H_\xi^j) = H_{\xi V} K_i$, together with the Carrollian relations
 \eqref{Carrolrelapp}.
 
 One finds that 
 \be 
 \delta_\xi K_a = (\delta_\xi \bb) K_a, \qquad \delta_\xi \bb = \xi[\bb] + e^{\bb} V[\xi^u] +(\O\pa_\O -1) \W_\xi.
 \ee 
 We also evaluate 
 \be
 \cL_\xi K_a &=  (\xi[\bb] + e^{\bb} V[\xi^u] ) K_a + e^\bb \mathring\pa_a\xi^u \cr
 \Delta_\xi K_a&= \left((\O\pa_\O -1) \W_\xi\right) K_a - e^\bb \mathring\pa_a\xi^u.
 \ee
 where $\Delta_\xi = \delta_\xi -\cL_\xi$ is the anomaly.
 For $V$ we have 
 \be
 \cL_\xi V^i&= \W_\xi \O\pa_\O V^i +[\hxi,V]^i ,\cr
\Delta_\xi V^i &=  V^i (1 -\O\pa_\O ) \W_\xi 
 + 2\O \nu \gamma^{ij} e^\bb \pa_j \hxi^u - \O \gamma^{ij} \pa_j \W_\xi 
\ee

\paragraph*{Proofs}
 \be 
 \delta_\hxi K_a&= \delta_{\hxi} g_{a\O} = \hxi^j \pa_j g_{a\O}  +(\pa_a \hxi^j) g_{j \O} + (\pa_\O \hxi^j) g_{ja}\cr
 &= \hxi^j \pa_j K_a   +(\pa_a \hxi^j) K_j    - e^\bb \gamma^{ij} (\pa_j \hxi{}^u)  g_{ja} \cr
 \delta_\hxi K_\O &=    (\pa_\O \hxi^j) K_j    - e^\bb \gamma^{ij} (\pa_j \hxi{}^u)  g_{j\O}
 =    e^\bb (\pa_\O \hxi^u)=0  \cr
 \delta_\hxi K_A &= (\pa_A \hxi^u) e^\bb    - e^\bb \gamma^{ij} (\pa_j \hxi{}^u)  g_{jA}=0 \cr
 \delta_\hxi K_u &=   \hxi^j \pa_j K_u   +(\pa_u \hxi^u) e^\bb    - e^\bb \gamma^{ij} (\pa_j \hxi{}^u)  g_{ju}\cr
 &= \hxi^j \pa_j e^\bb    +(\pa_u \hxi^u) e^\bb    + e^\bb \Y^A (\pa_A \hxi{}^u)  g_{ju}\cr
 &= \hxi[e^\bb] + e^{2\bb} V[\xi^u]\cr
 &= e^{\bb} ( \hxi[\bb] + e^{\bb} V[\xi^u])
 \ee
 Looking at the Weyl action on gets 
 \be 
 \delta_w K_a&= \delta_{w} g_{a\O} =  \xi^\O \pa_\O g_{a\O}  +(\pa_a\xi^\O )  g_{\O \O} + (\pa_\O \hxi^\O) g_{\O a}- 2 w g_{a\O} \cr
 &= w(\O \pa_\O \bb)  K_a + (\pa_\O w\O) K_a - 2 w K_a\cr
  &= w(\O \pa_\O \bb)  K_a + (-1 + \O\pa_\O) w  K_a \cr
  \cL_\xi K_a &= \iota_\xi \rd \alpha \wedge K+ \rd \iota_\xi K\cr
  &= \xi[\bb] K_a - \pa_a\bb e^\bb \xi^u + \pa_a (e^\bb \xi_u)\cr
  &=  \xi[\bb] K_a + e^\bb  \pa_a \xi_u \cr
  &= \xi[\bb] K_a + e^\bb  (N_a K[\xi_u]+ K_a V[\xi^u]+ \mathring\pa_a \xi^u) \cr
 \ee

We first evaluate that 
\bea
\delta_\xi N^a &=& \delta_\xi g^{a\Omega}= [\xi, N]^a- g^{ab}\pa_b\xi^\Omega  
+ 2 \W_\xi N^a\cr
&=&  [\xi, N]^a- g^{a\O}\pa_\O(\O\W_\xi)   - g^{aj}\pa_j(\O\W_\xi)  
+ 2 \W_\xi N^a\cr
&=&[\xi, N]^a + N^a (1 -\O\pa_\O ) \W_\xi 
- \O g^{aj} \pa_j \W_\xi.
\eea
where $[\xi,N]^a = \xi^b \pa_b N^a -N^b\pa_b\xi^a$ is the $d$ dimensional Lie bracket and 
$[\xi,V]^j = \xi^i \pa_i V^j -V^i\pa_i\xi^j$ is the codimension $1$ one.
\be
\delta_\xi V^a &= \delta_\xi N^b  h_b{}^a=
(\xi^c \pa_c N^b  - N^c\pa_c \xi^b) h_b{}^a + V^a (1 -\O\pa_\O ) \W_\xi 
- \O h_b{}^a g^{bc}h_c{}^j \pa_j \W_\xi \cr
&= (\xi^c \pa_c V^b  - V^c\pa_c \xi^b) h_b{}^a 
+(\xi^c \pa_c (2\O \nu K ^b)  - 2\O \nu K ^c\pa_c \xi^b) h_b{}^a+ V^a (1 -\O\pa_\O ) \W_\xi 
- \O \gamma^{aj} \pa_j \W_\xi \cr
\delta_\xi V^i &= \W_\xi \O\pa_\O V^i +[\hxi,V]^i - 2\O \nu \pa_\O\hxi^i + V^i (1 -\O\pa_\O ) \W_\xi 
- \O \gamma^{ij} \pa_j \W_\xi 
\ee

In the following given a vector $\xi^a\pa_a=\xi^i\pa_i + \Omega \W_\xi \pa_\Omega$ where $\xi^i\pa_i=\xi^u\pa_u + \xi^A\pa_A$.
We start with the variation of $N^\Omega$ which gives 
\bea
 \delta_\xi N^\Omega &=&  \hxi^j \pa_j N^\O +  \W_\xi \O\pa_\O N^\O -N^\O \pa_\O (\O \W_\xi)    -V^j \pa_j \O \W_\xi 
\cr
& & +  N^\O (1 -\O\pa_\O ) \W_\xi  - \O V^j \pa_j \W_\xi \cr
&=& \hxi^j \pa_j N^\O +  \W_\xi \O\pa_\O N^\O -  2\O N^\O \pa_\O \W_\xi    - 2  \O V^j \pa_j \W_\xi \cr
&=& 2\O ( \hxi^j \pa_j \nu +  \W_\xi (\O\pa_\O +1)  \nu -  2 \nu  \O \pa_\O \W_\xi    - V^j \pa_j  \W_\xi ).
 \eea
 Therefore we have that 
 \be 
 \delta_\xi \nu &=   \hxi[\nu]  + \W_\xi (\O\pa_\O +1)\nu   \underbrace{-(2 \nu( \O \pa_\O) \W_\xi   + V[\W_\xi ])}_{\Delta_\xi \nu}. 
 \ee
  Then we use that $V^i =N^i$ and focus on 
 \be
 \delta_\xi V^i &=\xi^b \pa_b V^i -N^b\pa_b\xi^i + V^i (1 -\O\pa_\O ) \W_\xi 
- \O g^{ij} \pa_j \W_\xi \cr
&=\xi^b \pa_b V^i -V^j\pa_j\xi^i -N^\O \pa_\O\xi^i  + V^i (1 -\O\pa_\O ) \W_\xi 
- \O g^{ij} \pa_j \W_\xi \cr
&=\xi^b \pa_b V^i -V^j\pa_j\xi^i + 2\nu \O ( e^\bb \gamma^{ij} \pa_j \tau)   + V^i (1 -\O\pa_\O ) \W_\xi 
- \O \gamma^{ij} \pa_j \W_\xi \cr
&=\xi^b \pa_b V^i -V^j\pa_j\xi^i  + V^i (1 -\O\pa_\O ) \W_\xi 
 +  \O  \gamma^{ij}  (2\nu  e^\bb\pa_j\tau  - \pa_j \W_\xi ),   \cr
  \cL_\xi V^i &=\xi^b \pa_b V^i -V^j\pa_j\xi^i  - \O V[\W_\xi] K^i.
 \ee
 \be
 \delta_\xi h_{ij} &= \delta_\xi( h_i{}^a h_j{}^b g_{ab}) =  h_i{}^a h_j{}^b (\cL_\xi g_{ab} -2 \W_\xi g_{ab})\cr
 &= \xi^c \pa_c g_{ij} + \pa_i \xi^c g_{cj} + \pa_j \xi^c g_{ci} \cr
 &=  \W_\xi \O \pa_\O g_{ij}  + \hxi^k \pa_k g_{ij} + \pa_i \hxi^k g_{kj} + \pa_j \hxi^k g_{ki}
 +  \pa_i \xi^\O g_{\O j} + \pa_j \xi^\O g_{\O i}\cr
 &= \W_\xi [\O \pa_\O -2] h_{ij} + \cL_\xi h_{ij} +  \O(\pa_i \W_\xi K_j +\pa_j\W_\xi K_i ),
 \ee
 \be
 \delta_\xi h^{ij} &= \delta_\xi  g^{ab} h_a{}^i h_b{}^j  =  (\cL_\xi g^{ab} +2 \W_\xi g^{ab}) h_a{}^i h_b{}^j\cr
 &= \xi^c \pa_c g^{ij} - g^{ic} \pa_c \xi^j  - g^{jc} \pa_c \xi^i  + 2 \W_\xi h^{ij}\cr
  &= \xi^c \pa_c h^{ij} - h^{ik} \pa_c \xi^j  - h^{jk} \pa_c \xi^i  + 2 \W_\xi h^{ij}
  - g^{i\O} \pa_\O \xi^j  - g^{j\O} \pa_\O \xi^i \cr
  &= \hxi^k \pa_k h^{ij} - h^{ik} \pa_c \hxi^j  - h^{jk} \pa_c \hxi^i  +  \W_\xi ( \O\pa_\O +2) h^{ij}
  +  V^i  e^\bb \mathring\pa^j \xi^u  + V^j e^\bb \mathring\pa^j \xi^u.
 \ee
 \be
 \delta_\xi N^2 &= \delta_\xi  g^{ab} N_a N_b =  (\cL_\xi g^{ab} +2 \W_\xi g^{ab}) N_a N_b\cr
 &= \xi^c \pa_c g^{\O \O } - 2 g^{\O c} \pa_c \xi^\O     + 2 \W_\xi  g^{\O\O} \cr
  &= \hxi^i \pa_i g^{\O \O } - 2 g^{\O i} \pa_i \xi^\O   +  \xi^\O \pa_\O g^{\O \O } - 2 g^{\O \O} \pa_\O \xi^\O     + 2 \W_\xi  g^{\O\O} \cr
    &= 2\O ( \hxi^i \pa_i \nu  -  V^i \pa_i \W_\xi)  + 2\O( \W_\xi  \pa_\O (\O \nu)  - 2 \nu  \pa_\O (\O \W_\xi)     + 2  w_\xi \nu)  \cr
    &= 2\O ( \hxi^i \pa_i \nu  -  V^i \pa_i \W_\xi)  + 2\O(  \W_\xi (1+ \O\pa_\O)\nu  - 2 \nu  \O \pa_\O \W_\xi     ). 
 \ee

\section{Radial expansion}\label{app:radexp}
We detail here the evaluation of the first orders in the expansion of the Bondi-Sachs metric coefficients.
We make use of the radial evolution equations
\be 
(d-2)\pa_\O \bb &= - \O (\bTheta:\bTheta)\cr 
[(d-3)  -   \O \pa_\O ]\Pi_A&  
=   
  \Snabla_B \bTheta_A{}^B   +\frac{\O}{(d-2)} \pa_A  (\bTheta_B{}^C  \bTheta_C{}^B)   \cr
\left[   (d-3) -\O\pa_\O\right] \mu   & =  e^{\bb}\left( -\tfrac12 \mathring{R} + \O \Snabla_A \Pi^A + \O^2 \Pi_A\Pi^A \right) \cr
 \left[ \tfrac12 (d-4) -\O \pa_\O \right]  N_{A}{}^B   & 
=  e^{\bb}  \left(- \tfrac12 \mathring{R}_{\langle AC\rangle} + \O \Snabla_{\langle A} \Pi_{C\rangle} + \O^2 \Pi_{\langle A}  \Pi_{C\rangle}  \right)\gamma^{CB} +  \tfrac12\O \mu\bTheta_{A}{}^{B} 
\ee
\be
e^\bb \rN_{\langle AB \rangle } & =   (N_{AB} + \Phi \bar\Theta_{\langle AB \rangle }), \cr 
 \rN &= e^{-\bb}\mu =  -\tfrac1{2(d-3)} {R}_q + \O  e^{-\a}\mu_1  + o(\O)
,\cr
\kA&=  \pa_\O \nu + 2\nu \pa_\O \bb = e^{-\bb} (\pa_\O\Phi + \Phi \pa_\O \bb),\cr
\hN&= \rN+\kA.
\ee
From \eqref{aspectdef} and \eqref{bianc_bis}
\be
(d-2) \bb &= (d-2) \a - \tfrac12 \O^2  (\bTheta_0:\bTheta_0) +o(\O^2) \cr
\pa_\O[\e^{\bb}\theta] 
& =  \mathring{\Delta}\e^{\bb} -  2\O \Snabla_A[\e^{\bb}\Pi^A]\cr
\e^{\bb}\theta&= F + \O\left( \Delta e^{\a} \right)  -\O^2 D_A [e^{\a} \Pi_{0}^A] + o(\O^2) \cr
e^\a \theta_0=F &= \tfrac12 q^{AB}\pa_u q_{AB} + D_A U^A,\cr
(d-2) \Phi &=    \e^{\bb}\theta -\O \mu \cr
&=   F + \O \left(  \Delta e^{\a}  -\mu_0\right)  -\O^2 ( D_A [e^{\a} \Pi_{0}^A] +\mu_1)  +  o(\O^2)  \cr
(d-2) \nu 
&=   e^{-\a}F + \O e^{-\a}\left(  \Delta e^{\a}  -\mu_0\right)  -\O^2 e^{-\a} \left( D_A [e^{\a} \Pi_{0}^A] +\mu_1- \frac{F}{2}(\bTheta_0:\bTheta_0) \right)  +  o(\O^2)  \cr
(d-2)e^{-\a} \pa_\O \Phi
&=  e^{-\a}\left(  \Delta e^{\a}  -\mu_0\right)  
- 2\O e^{-\a}  ( D_A [e^{\a} \Pi_{0}^A] + \mu_1 )  +  o(\O^2)  \cr
(d-2)\kA
&=  e^{-\a}\left(  \Delta e^{\a}  -\mu_0\right)  
- 2\O e^{-\a} ( D_A [e^{\a} \Pi_{0}^A] + \mu_1 + \tfrac12  F (\bTheta_0:\bTheta_0)  )  +  o(\O^2)  \cr
(d-2)\hN  &= e^{-\a}\left(  \Delta e^{\a} +  (d-3) \mu_0\right)  \cr
&+ \O e^{-\a} \left( - 2 D_A [e^{\a} \Pi_{0}^A] + (d-4) \mu_1 - F (\bTheta_0:\bTheta_0)\right)  +  o(\O^2)
\ee
For the vector we have $\eta_A =\O \Pi_A$ and 
\be
e^{-\bb} \pa_\O \Y^A &=  \gamma^{AB} (\pa_B\bb - 2\O \Pi_B)\cr
 \Y^A &=  U^A + \O e^{\a}q^{AB} \pa_B\a - \O^2 e^{\a}\left(  \bTheta_{0B}{}^C \pa_C \a +\Pi_{0B} \right) 
\ee
\subsection{Order $0$}
 The vector equation at order $0$  is 
 \be
 (d-3) \Pi_{0A}=   D_B \bTheta_{0A}{}^B\label{pi0}
 \ee
 This means that \a
 \be
     (d-3)  q_{AB}  \Y_2^B  =  -[ e^a D_B \bTheta_{0A}{}^B +   (d-3)\bTheta_{0A}{}^B \pa_B e^a ].
 \ee
 Similarly 
 \be\label{mu0}
(d-3)\mu_0   & =  -e^{\a}\tfrac12 {R}_q
 \ee
and 
 \be
(d-2) \Phi_1  
 &= e^\a\theta_1- \mu_0\cr
 &= {\Delta}\e^\a + \tfrac{e^{\a}}{2(d-3)}  {R}_q.
 \ee
 Note that 
 \be
 R(e^\a q) = e^{-\a} R(q) -2 e^{-\a} \Delta \a 
 \ee
 
 \subsection{Order $1$}
 At order 1 for the vector equation $(KA)$ gives 
 \be
 (d-4)\Pi_{1A} =   \left[\Sn_B {\bar\Theta^B}_A \right]_1 + \tfrac{1}{d-2}D_A\left( \bar\Theta_0 : \bar\Theta_0 \right)
 \ee
 In order to evaluate the momenta aspect we have to expand to first order the derivative $\Snabla{}_B {\bar\Theta^B}_A $. We use that 
\begin{subequations}
\begin{align} 
[\Snabla{}_A v^B]_1 &= D_A v^B_{1} +  \left(D_A\bar{\Theta}^B_{0 C}+D_C \bar{\Theta}^B_{0 A}-D^B \bar{\Theta}_{0 AC} \right) v_{0}^C,\\
[\Snabla{}_A v_B]_1 &= D_A v_{1B}  -  \left(D_A \bar{\Theta}^C_{0 B} +D_B \bar{\Theta}^C_{0 A} -D^C \bar{\Theta}_{0 AB} \right) v_{0C},
\end{align}
\end{subequations}
to conclude that  for a symmetric tensor $T_{AB}=T_{(AB)}$, and when the shear is traceless, we have  
\be
 [\Snabla{}_B T^B{}_A]_1 = D_B T_1^B{}_A
  - (D_A  \bar\Theta_0^C{}_B)   T_0^{B}{}_C.
\ee
This means that 
\be
\left[\Snabla{}_B {\bar\Theta^B}_A \right]_1 = D_B[\bar\Theta{}^B{}_A]_1   -  \tfrac12 D_A (\bar\Theta_0 :\bar\Theta_0), 
\ee
which gives 
\be
 (d-4)\Pi_{1A} =   D_B[\bar\Theta{}^B{}_A]_1   - \tfrac{(d-4)}{2(d-2)}D_A\left( \bar\Theta_0 : \bar\Theta_0 \right)
 \ee
 
 Next we study the  scalar equation $(KL)$ at order $1$.
 \be
 (d-4) \mu_1 = \e^{\a}\left(   D_A \Pi^A_0 -\tfrac12 [\mathring{R}]_1   \right).
 \ee
 Recalling the general formula
\be\label{deltaR}
\delta\mathring{R} =  \Snabla{}^B\Snabla{}_A\Big(  (\delta \gamma)^A{}_B - \delta^A{}_{B}   (\delta \gamma)^C{}_C \Big)- (\delta \gamma)^A{}_B \mathring{R}^B{}_A,
\ee
where we denote $ (\delta \gamma)^A{}_B = \gamma^{AC}\delta \gamma_{CB}$.
The fact  that $\gamma_0^{AC} \gamma_{1CB}= 2 \bar\Theta_0^A{}_{B} $, and that $\btheta_0=0$ implies that
\be
\tfrac12 \mathring{R}_1
& =
D_AD_B\bar\Theta_{0}^{AB}
-\bar\Theta_0 :\, {R}_{q0}.
\ee
Using also that $
D_A \Pi_{0}^A = \frac1{(d-3)} D_A D_B \Theta_{0}^{AB} $ 
therefore implies that 
\be 
 (d-4) \mu_1 &= \e^{\a}\left(   D_A \Pi^A_0 -\tfrac12 [\mathring{R}]_1   \right)\cr
 &= \e^{\a}\left(  -\tfrac{(d-4)}{(d-3)} D_A D_B \Theta_{0}^{AB}  + \bar\Theta_0 :\, {R}_{q0}  \right).
\ee
This equation  emphasizes the fact that both its left and right hand sides vanish identically in $d=4$, as a consequence of the fact that in two dimensions the Ricci curvature is pure trace, i.e. that $\stackrel{S}{R}_{AB} \;\stackrel{d=4}{=}\; K q_{AB}$ where $K$ is the Gaussian curvature.
Therefore, the $(KL)$ equation is automatically regular in $d=4$.

 \begin{equation}
\begin{array}{cllc} \label{eq:muPian}
& (KL) & (d-4) \mu_1 = \e^{\a}\left(   D_A \Pi^A_0 -\tfrac12 [\mathring{R}]_1   \right)&\\
& (KA) & (d-4)\Pi_{1A} =   \left[\Sn_B {\bar\Theta^B}_A \right]_1 
 + \tfrac{1}{d-2}D_A\left( \bar\Theta_0 : \bar\Theta_0 \right)&\\
\end{array}
\end{equation}

\subsection{Radiative modes}\label{app:radm}

Finally we have the radiative terms.
Note that the tensor we are expanding in powers of $\O$ is $N_A{}^B$, with one index up and one index down. Indices are then lowered by $q_{AB}$, i.e. $N_{(k)AB}:= N_{(k)A}{}^C q_{CB}$. 

The zeroth order of the $\langle AB\rangle$ component of \eqref{eq_G2} gives:
\be
(d-4) N_{0AB} = - e^b\mathring{R}_{\langle AB\rangle}.
\ee
This is trivially satisfied in $d=4$, where the codimension-2 hypersurfaces are 2-dimensional and $\mathring{R}_{AB}$ is pure trace.  In other words, in $d=4$, the would-be radiative anomaly ($- \mathring{R}_{\langle AB\rangle}$) automatically vanishes and there is no need to introduce a ``radiative logarithm''.

The next order gives 
\be\label{eq:radanomaly_d6}
(d-6) N_{1AB} = e^b \left( -  [\mathring{R}_{\langle AD \rangle} \gamma^{DC}]_1 q_{CB} + 2 D_{\langle A} \Pi_{0B\rangle} +  e^{-b}\mu_0 \bTheta_{AB} \right),
\ee
where, using that $\gamma_{1AB} = 2 \bTheta_{0AB}$, one finds
\be
[\mathring{R}_{\langle AD \rangle} \gamma^{DC}]_1 q_{CB} = 
\left[\mathring{R}_{AD } \gamma^{DC} -\frac{\delta_A^C}{(d-2)} \mathring{R}_{BD } \gamma^{DB}\right]_1 q_{CB}= \mathring{R}_{1\langle AB\rangle} - 2 ({R}_q\cdot \bTheta_0)_{\langle AB\rangle}.
\ee
To compute $R_{1\langle AB\rangle}$ we proceed as follows. First recall that 
\be
\delta \mathring{R}_{ab} := 
\frac12 (\Sn_c \Sn_a \delta \gamma_{bd} + \Sn_c \Sn_b \delta \gamma_{ad}-\Sn_c \Sn_d \delta \gamma_{ab}
- \Sn_a \Sn_b \delta \gamma_{cd}) \gamma^{cd}.
\ee
For a variation $\delta \gamma_{AB} = 2 \bTheta_{AB}$ that corresponds to a radial derivative one gets 
\be
\pa_\O \mathring{R}_{AB}&= 
\Sn_C \Sn_A \bTheta_{B}{}^C +\Sn_C \Sn_B \bTheta_{A}{}^C - \mathring\Delta \bTheta_{AB}
-\Sn_A \Sn_B \btheta \cr
&=  \Sn_A (\Sn \cdot \bTheta)_{B} + \Sn_B( \Sn \cdot \bTheta)_A - \mathring\Delta \bTheta_{AB}
-\Sn_A \Sn_B \btheta \cr
& - 2 \mathring{R}_{ACB}{}^D \bTheta_{D}{}^C  + \mathring{R}_{AD} \bTheta_{B}{}^D + \mathring{R}_{BD} \bTheta_{A}{}^D
\ee
where we used that 
in the Bondi-Sachs gauge $\btheta=0$. This implies that
\be
\pa_\O\left(\mathring{R}_{\langle A}{}^{B \rangle} \right)
 &=    2 D_{\langle A} (D\cdot \bTheta)^{B\rangle} - \Delta \bTheta_{A}{}^{B} - 2 (\mathring{R}: \bTheta)_{\langle A  }{}^{B\rangle }.
\ee
where 
$(\mathring{R}: \bTheta)_{AB} := 
R_{ACBD}\bTheta^{CD}$.\footnote{Note that 
$ \pa_\O\left(\mathring{R}_{\langle A}{}^{B \rangle} \right)\neq \left( \pa_\O\mathring{R}\right)_{\langle A}{}^{B \rangle}  $}
This means that 
\be
[\mathring{R}_{\langle A}{}^{C \rangle}]_1 q_{CB} 
 &=    2 D_{\langle A} (D\cdot \bTheta_0)_{B\rangle} - \Delta \bTheta_{0AB} - 2 R_{q\langle A}{}^C{}_{B\rangle}{}^D \bTheta_{0CD} .
\ee
In $d=4$, the codimension-2 Riemann is $R_{qACBD} = \frac12 R_q (q_{AB} q_{CD} - q_{AD}q_{CB})$, whence recalling \eqref{pi0} and \eqref{mu0} we find:
\be\label{N_1eq}
2 N_{1AB} 
& = e^b\left( -\Delta \bTheta_{0AB}  + \tfrac32 R_q \bTheta_{0AB}  \right)
\ee

In $d>4$, using the decomposition of the Riemann into the Weyl and Schouten tensors, $R_{q AC}{}^{BD} = W_{qAC}{}^{BD} + 4 S_{q[A}{}^{[B}\delta_{C]}{}^{D]}$, we first find
\be
[\mathring{R}_{\langle A}{}^{C \rangle}]_1 q_{CB} 
 & = 2 D_{\langle A} (D\cdot \bTheta_0)_{B\rangle_q} - \Delta \bTheta_{0AB} - 2 (W_q: \bTheta_0)_{\langle AB\rangle_q} + 4(S_q \cdot \bTheta_0)_{\langle AB\rangle_q}  
\ee
and then, recalling the relationship between the Ricci and Schouten tensor (for the codimension-2 metric),
\be
S_{qAB} = \frac{1}{d-4} \left(R_{qAB} - \frac{1}{2(d-3)}R_q q_{AB}\right),
\ee
we conclude
\be
[\mathring{R}_{\langle A}{}^{C \rangle}]_1 q_{CB} 
 & = 2 D_{\langle A} (D\cdot \bTheta_0)_{B\rangle_q} - \Delta \bTheta_{0AB}\cr
 & \quad - 2 (W_q: \bTheta_0)_{\langle AB\rangle_q} + \tfrac{4}{d-4}(R_q \cdot \bTheta)_{\langle AB\rangle_q} - \tfrac{2}{(d-4)(d-3)}R_q \bTheta_{0AB} 
\ee

Thus, in $d>6$, and recalling \eqref{pi0} and \eqref{mu0}, one finds
\be
(d-6) N_{1AB} & = e^b \left( \Delta \bTheta_{0AB}  - \tfrac{2(d-4)}{d-3} D_{\langle A} (D\cdot\bTheta_0)_{B\rangle_q} + 2(W_q:\bTheta_0)_{\langle AB\rangle_q}
\right.\cr & \quad\quad\quad  \left.
- \tfrac{4}{d-4} (R_q\cdot \bTheta_0)_{\langle AB\rangle_q} -  \tfrac{(d-8)}{2(d-3)(d-4)} {R}_q\bTheta_{AB} \right).
\ee
Finally, in $d=6$, we see that the rhs of \eqref{eq:radanomaly_d6} gives instead the radiative anomaly, corresponding to the following logarithmic term in the expansion of $N_A{}^B$:
\be\label{eq:rad-anomaly-d6}
- 2 N_{(1,1)AB} & = e^b \big( \left(\Delta +\tfrac16 R_q\right)\bTheta_{0AB} - \tfrac43 D_{\langle A} (D\cdot\bTheta_0)_{B\rangle_q}\cr
& + 2(W_q:\bTheta_0)_{\langle AB\rangle_q} 
- 2 (R_q\cdot \bTheta_0)_{\langle AB\rangle_q}  \big).
\ee

\subsubsection{Radiative anomaly}\label{app:radiativeA}

One recall that at $\scri$ we have (see \eqref{eq:W=DbT})
\be
\bar{W}_{q CAB}&=  D_C\bTheta_{0AB} + \tfrac1{d-3}(D\cdot\bTheta_0)_C q_{AB} - (A\leftrightarrow C)
\ee
Therefore 
\be
D_C\bar{W}_q{}^C{}_{AB}&=  \Delta \bTheta_{0AB}  - D_C D_A \bTheta_{0}{}^C{}_{B}
-  \tfrac1{d-3}D_B (D\cdot \bTheta_0)_A
+ \tfrac1{d-3}(D\!\cdot D\!\cdot\! \bTheta_0) q_{AB} 
\cr
&= \Delta \bTheta_{0AB}  -  D_A (D \cdot  \bTheta_{0})_{B}
-  \tfrac1{d-3}D_B (D\cdot \bTheta_0)_A
+ \tfrac1{d-3}(D\!\cdot D\!\cdot\! \bTheta_0) q_{AB}\cr
&\quad + (\text{Riem}_q: \bTheta_0)_{AB} - (\text{Ric}_q \cdot \bTheta_0)_{AB}.
\ee
Since $\bar{W}_{ABC}$ is traceless we have that $D_C\bar{W}_{q}{}^C{}_{AB} = D_C\bar{W}_{q}{}^C{}_{\langle AB\rangle_q}$, hence
\be
D_C\bar{W}_q^C{}_{AB}&=
\Delta \bTheta_{0AB}  
- \tfrac{(d-2)}{(d-3)} D_{\langle A} (D \cdot  \bTheta_{0})_{B\rangle_q}
\cr
&\quad + (W_q: \bTheta_0)_{\langle A B \rangle_q} - \tfrac{(d-2)}{(d-4)}(R_q \cdot \bTheta_0)_{\langle A B \rangle_q} + \tfrac{1}{(d-4)(d-3)} R_q \bTheta_{0AB}.
\ee
When $d=6$, using \eqref{eq:rad-anomaly-d6-text} we find that  
\be \label{DWN}
D_C\bar{W}^C{}_{AB}&=
(\Delta+\tfrac{1}{6} R_q) \bTheta_{0AB}  
- \tfrac{4}{3} D_{\langle A} (D \cdot  \bTheta_{0})_{B\rangle_q}
+ (W_q: \bTheta_0)_{\langle A B \rangle_q} - 2(R_q \cdot \bTheta_0)_{\langle A B \rangle_q} \cr
&= - 2 e^{-b} N_{(1,1)AB} - (W_q: \bTheta_0)_{\langle A B \rangle_q}.
\ee

\subsection{Radial shear and asymptotic Weyl of Petrov type $N$ in $d\geq 6$}\label{app:WeylTypeN}

Let $d\geq6 $ and consier a conformally flat codimension-2 metric $q$:
\be
\bar{W}_{qABCD} =0.
\ee
We are going to look for a family of solutions such that 
\be
\bar{W}_{qABC} =0,
\qquad
\bTheta_{(1,1)AB}=0.
\ee

In $d\geq 8$, $\bTheta_{(1,1)AB}=0$ identically and it is enough to show that $\bar{W}_{qABC} =0$.
In $d=6$ we reach the same conclusion thanks to equation \eqref{DWNtext} (i.e. equation \eqref{DWN}) and the condition $\bar{W}_{qABCD} =0$.

Note that the codimension-2 Bianchi identity
\be
D_D \bar{W}_{qABC}{}^{D} = D_A S_{qBC} - D_B S_{qAC}
\ee
can be rewritten as
\be
(d-4) D_D \bar{W}_{qABCD} = D_A R_{qBC} - \frac{1}{d-3} (D\cdot R_{q})_A q_{BC} - (A\leftrightarrow B).
\ee
Now, denote $R_{q\langle{AB}\rangle} =: (\text{Ric}^\text{TF}_q)_{AB}$, and observe that the contracted Bianchi identity $2 (D\cdot R_q)_A=D_A R_q $ gives
\be
2(D\cdot\text{Ric}_q^\text{TF})_A = \frac{d-4}{d-2} D_A R_q, 
\ee
whence
\be\label{eq:BianchiWRTF}
(d-4) D_D \bar{W}_{qABC}{}^{D} = D_A \text{Ric}^\text{TF}_{qBC} + \frac{1}{d-3} (D\cdot \text{Ric}^\text{TF}_{q})_A q_{BC} - (A\leftrightarrow B).
\ee
Thus we see that setting $\bTheta_{0AB} = f R_{q\langle AB\rangle}$ with $D_A f = 0$, equation \eqref{eq:W=DbT} gives $\bar{W}_{qABC} = f D_D \bar{W}_{qABC}{}^D$ which thus vanishes if $\bar{W}_{qABCD}$ does. Compare this with \eqref{eq:bTR}, namely:
\be \label{timeder}
(\cL_{\ell} + F) \bTheta_{0AB} 
= - D_{\langle A} D_{B\rangle} e^\a - \frac{1}{d-4}e^\a R_{q\langle AB\rangle} \stackrel{\text{(BMSW)}}{\implies} \pp_u \bTheta_{0AB} = - \frac{1}{d-4} R_{q\langle AB\rangle}.
\ee
Therefore, if the BMSW boundary condition are satisfied ($\a =F=U^A =0$) as in \cite{Capone:2023roc}, we see that setting
\be
\bTheta_{0AB} \stackrel{\text{(BMSW)}}{=} \frac{u}{d-4} R_{q\langle AB\rangle}
\ee
 implies via \eqref{eq:BianchiWRTF} and \eqref{eq:W=DbT} that $\bar{W}_{qABC}=0$ if $\bar{W}_{qABCD}=0$, and is compatible with \eqref{timeder} (recall that BMSW boundary conditions imply that $\pp_u q_{AB} = 0$ \eqref{eq:BSMW-ppu(q)}.

Now, set $\check q_{AB} = e^{-2\phi} q_{AB}$ for $\phi$ a function on  $\scri$. The previous computation gives
\be
\check D_D {\bar{W}}_{\check qABC}{}^D = \check D_A \text{Ric}^\text{TF}_{\check qBC} + \frac{1}{d-3} (\check D\cdot \text{Ric}^\text{TF}_{\check q})_A \check q_{BC} - (A\leftrightarrow B).
\ee
Using the Weyl transformation properties of the covariant derivative, we find:
\be
\check D_D \check{\bar{W}}_{qABC}{}^D = e^\phi D_A (e^{-\phi}\text{Ric}^\text{TF}_{\check qBC}) + \frac{1}{d-3} e^\phi( D\cdot e^{-\phi}\text{Ric}^\text{TF}_{\check q})_A  q_{BC} - (A\leftrightarrow B)
\ee
where now all contractions are taken with respect to $q_{AB}$ and its inverse.

Finally, we note that 
\be
e^{-\phi}\text{Ric}^\text{TF}_{\check qAB} = e^{-\phi}\text{Ric}^\text{TF}_{qAB} + (d-4) D_{\langle A} D_{B\rangle} e^{-\phi}.
\ee
Denoting $T \equiv e^{-\phi}$, and using the conformal invariance of the Weyl, ${\bar{W}}_{\check qABC}{}^D = \bar{W}_{qABC}{}^D$,  we conclude that 
\be
(d-4)T \check D_D {\bar{W}}_{qABC}{}^D & = D_A \left( \frac{T}{d-4} R_{q\langle BC\rangle_q}  + D_{\langle B}D_{C\rangle}T\right) \cr & \quad + \frac1{d-3}  D^D\cdot\left( \frac{T}{d-4} R_{q\langle DA\rangle} + D_{\langle D}D_{A\rangle} T\right)\cr & \quad   - (A \leftrightarrow B)
\ee
This shows that setting
\be
\bTheta_{0AB}  = \frac{T}{d-4}  R_{q\langle AB\rangle_q} + D_{\langle A}D_{B\rangle}T
\ee
one has $\bar{W}_{qABCD} = 0 \implies \bar{W}_{qABC}=0$ \eqref{eq:W=DbT}, as well.

Combining the previous two results, one concludes that with BMSW boundary condition, setting
\be
\bTheta_{0AB}  \stackrel{\text{(BMSW)}}{=} \frac{T+u}{d-4}  R_{q\langle AB\rangle_q} + D_{\langle A}D_{B\rangle}T
\ee
one has $\bar{W}_{qABCD} = 0 \implies \bar{W}_{qABC}=0$ \eqref{eq:W=DbT}.
 Since BMSW boundary conditions imply $\pp_uq_{AB}=0$, compatibility with the evolution equation \eqref{timeder} finally gives us that 
\be
\pp_u T = 0.
\ee

\subsection{Bondi diffeomorphisms}\label{app:BondiDiffeoExp}

In this appendix we address the radial expansion of Bondi-Sachs diffeomorphisms. Recall that the field-\emph{in}dependent parameters that fully determine a Bondi-Sachs diffeomorphism are $(X^i,W)\in \mathfrak{X}(\scri)\times C^\infty(\scri)$, while the spacetime Bondi-Sachs diffeomorphism itself, denoted $\xi_{(X,W)}$, is field dependent (see section \ref{sec:BondiSachsDiffeo}). This is because a diffeomorphism preserving the Bondi-Sachs gauge is completely fixed by its ``boundary value" $(X,W)$ by means of the radial differential equation \eqref{evol} whose solution is \eqref{eq:xi.e.xpansion}.

Recall that a Bondi-Sachs diffeomorphism takes the form ($\hat \xi^i := \xi^a h_a{}^i$)
\be
    \xi_{(X,W)} & = \hat\xi_{(X,W)} + w_{(X,W)} \O\pp_\O,
\ee
where ($Y^A := X^i \gamma_i{}^A$)
\be
\begin{cases}
\hat \xi \SCRIeq X =: \tau \ell + Y\\
w \SCRIeq W
\end{cases}
\ee
and (see  \eqref{eq:xi.e.xpansion})
\begin{align}    
\hat \xi_{(X,W)} & = \tau\pp_u + (Y^A + \tau U^A - G^{AB}\pp_B\tau) \pp_A \\
w_{(X,W)} & = W +  \frac1{d-2}\left( (U^A - \Upsilon^A)\pp_A \tau - D_A ( G^{AB}\pp_B\tau)\right)
\end{align}
for
\be
G^{AB} := \int_0^\O e^\beta \gamma^{AB} & = \int_0^\O \left(e^b( q^{AB} - 2\O' \bTheta^{AB}_0) + O(\O'{}^2)\right)\\& = \O e^b q^{AB} - \O^2 e^b\bTheta^{AB}_0 + O(\O^3).
\ee
Therefore we see that the subleading orders of $\xi_{(X,W)}$ depend on $(X,W)$ only through the parameter $\tau = X^u = e^{-\bb} X^i K_i$. This parameter is field \emph{in}dependent (contrary to $Y$, which is such that $\delta Y = \delta (X-\tau\ell) = - \tau\delta \ell = - \tau \delta U^A\pp_A$). We thus write
\be
\xi_{(X,W)} = \Big( X + \O\hat \xi_{\tau 1} + \O^2 \hat \xi_{\tau 2} + O(\O^3)\Big) + \Big(W +  \O w_{\tau 1} + \O^2  w_{\tau 2} + O(\O^3) \Big)  \O\pp_\O 
\ee

Using the expressions above and the expansion to second order of $\Upsilon^A$ (equations \ref{Pbc-components}, \ref{eq:EEq-exp}), we obtain
\be
\hat \xi_{\tau 1} & = - e^b (q^{AB} \pp_B\tau) \pp_A \\
w_{\tau 1} & = \frac1{d-2}( - \Upsilon^A_1 D_A \tau - D_A(e^b D^A\tau) ) \notag\\
& = \frac1{d-2}( - D^A e^b D_A \tau - D_A(e^b D^A\tau) )\notag\\
& =- \frac{e^b}{d-2}\left(2 D^A b D_A \tau + \Delta \tau \right)
\ee
and
\be
\hat \xi_{\tau 2} & = (e^b \bTheta_0^{AB} D_B\tau)\pp_A\\
w_{\tau 2} & = \frac1{d-2}\left( - \Upsilon^A_2\pp_A \tau + D_A(e^b\bTheta^{AB}_0\pp_B\tau) \right)\notag\\
& = \frac1{d-2}\left(\frac{e^{(4-d)b}}{d-3} D_B(e^{(d-3)b}\bTheta_0^{BA})\pp_A \tau + D_A(e^b\bTheta^{AB}_0\pp_B\tau) \right)\notag\\
& = \frac{e^b}{d-2}\left( \bTheta_0^{AB} D_A b D_B\tau + \frac1{d-3} D_A\bTheta^{AB}_0 \pp_B\tau + \bTheta_0^{AB} D_A b D_B + D_A\bTheta^{AB}_0 \pp_B\tau   \right)\notag\\
& = \frac{e^b}{d-2}\left( 2\bTheta_0^{AB} D_A b D_B\tau + \frac{d-2}{d-3} D_A\bTheta^{AB}_0 \pp_B\tau \right)\notag\\
& = e^b\left( \frac{2}{d-2}\bTheta_0^{AB} D_A b D_B\tau + \frac{1}{d-3} D_A\bTheta^{AB}_0 D_B\tau \right).
\ee

\section{Symplectic potentials}\label{app:symp}
Here we first describe the metric and connection coefficients, then construct the symplectic flux and then the tangential components of the symplectic potential.

\subsection{Metric and connection}\label{app:metGN}
The inverse metric in Bondi coordinates  is given by
\be
g^{ab}= \left(
\begin{matrix}
g^{\O \O} & g^{\O u} & g^{\O A} \\
 & g^{uu} & g^{uA}\\
 & & g^{AB}
\end{matrix}\right)=\left(
\begin{matrix}
\O e^{-\bb}\Phi  & e^{-\bb} & e^{-\bb}\Y^A \\
 &  0 & 0\\
 & &  \gamma^{AB}
\end{matrix}\right)
\ee
The Carrollian metric and its dual are
\be
h_{ij}
=
\left(
\begin{matrix}
 &-2\O e^\bb\Phi + \Y_A \Y^A & -\Y_A\\
 & -\Y_A &  \gamma_{AB}
\end{matrix}\right), \quad
\gamma^{ij} = 
\left(
\begin{matrix}
 &  0 & 0\\
 & 0 &  \gamma^{AB}
\end{matrix}\right)
\ee
while the codimension $2$ projector is 
\be
\gamma_i{}^j =  \left(
\begin{matrix}
 & \gamma_{u}{}^u & \gamma_{u}{}^B\\
 & \gamma_{A}{}^u & \gamma_{A}{}^B
\end{matrix}\right)=\left(
\begin{matrix}
 &  0 & -\Y^B\\
 &  0 &  \delta_{A}{}^B
\end{matrix}\right)
\ee
The connection coefficients are 
\be
\Gamma_{\O\O}^\O &= \bar\kappa, \cr
\Gamma_{\O\O}^i &= 0,\cr
\Gamma_{\O i}^\O &= -\omega_i, \cr
\Gamma_{i \O }^j &=  \bmK_i{}^j, \cr
 \Gamma_{ij}^\O  & = - \mK_{ij},\cr
  \Gamma_{ij}^k  & = \hat\Gamma_{ij}^k,
  \cr
  \Gamma_{ij}^i &= \omega_j +\hat\pa_j \ln \sqrt{g}.
\ee
where $\hat\Gamma_{ij}^k =h_i{}^ah_j{}^b \Gamma_{ab}^c h_c{}^k$ denotes the Carrolian connection coefficients and 
and we have that 
\be
\hat\Gamma_{ij}^l K_l = (\omega_i +\hat\pa_i \bb) K_j - \bmK_{ij} =  (\omega_i +\hat\pa_i \bb) K_j  +(\omega_j + \hat\pa_j\bb) K_i - \bTheta_{ij}
\ee
We recall that 
\be
\bmK_{i}{}^j &:= h_i{}^a (\nabla_a K^b) h_b{}^j   = \bTheta_{i}{}^j - k_i \bar\eta^j,\cr
\mN_{i}{}^j&:= h_i{}^a  (\nabla_a N^b)h_b{}^j =\hn_i V^j + 2\O\nu \bmK_i{}^j, \cr
\mK_{i j}  &:= h_i{}^a h_j{}^b (\nabla_a N_b)    = \mN_{i}{}^k h_{kj} +\O  \hn_i \nu  K_j.
\ee
\be
 \mK_{i V} 
&= \O (\hn_i-2 \omega_i)\nu, 
\cr
\mK_{V V} &= \O [(\hn_V-2 \omega_V)\nu], \cr
 \gamma_i{}^j \mK_{V j} &= \gamma_i{}^j (\Sn_j - 2\rom_j)\nu,
 \cr
\mN_{V }{}^i \gamma_i{}^j &= \mK_{V i} \gamma^{ij} := -\O \rJ^j,   \cr
\mN_{i K} 
& =\omega_i,
\cr
\gamma_i{}^j\mN_{j K} 
& :=\rom_i, \cr 
\omega_V :=\mN_{V K}
&= \nu + \O[ K[\nu] - 2\nu \bar\kappa ] = \kappa - \O \nu \bar\kappa.
\ee
\be
\mN_i{}^j &= \rmN_i{}^j - K_i \rJ^j  + \rom_i V^j + K_i V^j  \omega_{V},\cr
\mN &= \rmN  +  \omega_{V}. \cr
\mT_i{}^j &= \underbrace{\mN \gamma_i{}^j-\rmN_i{}^j }_{\rmT_i{}^j}  - \rom_i V^j + K_i (\rJ^j   +    \rmN V^j).
\ee

\subsection{Symplectic Flux } \label{app:sympF}

In this section as in the rest of the paper we use the convention  that early latin index $a,b$ are $d$-dimensional, mid level latin endex $i,j,k$ are 
$d-1$ dimensional and capital Latin index $A,B,C$ are $d-2$ dimensional.
If a vector such as $V^i$ is tangent to $\Sigma_\O$ we have that $V^i=V^ah_a{}^i$ and we can either view it intrinsically as a vector  $V^i$ or extrinsically as a embedded vector $V^a = V^b h_b{}^a$. Similarly a vector $Y^A$ is tangent to $\Sigma_\O$ we have that $V^A=V^a q_a{}^A$ and we can either view it intrinsically as a vector  $V^A$ or extrinsically as a embedded vector $V^a = V^b q_b{}^a$.
The Bondi coordinates are such that $h_\O{}^i=0=h_i{}^\O$, however we do not have the same property for the codimension $2$ projector since 
$\gamma_i{}^u=0$ but $\gamma_u{}^A =\U^A \neq 0$. 
For the codimension two tensors such as $\gamma^{ij} := g^{ab}h_a{}^ih_b^j $ we have
\be
h^{ij} := g^{ab}h_a{}^ih_b^j  = \gamma^{ab}h_a{}^ih_b^j, 
\qquad
 h_{ij} = h_i{}^a h_j{}^b g_{ab}=   h_i{}^a h_j{}^b \gamma_{ab}-2 \rho  k_i k_j
\ee
which are equal to $\gamma^{ij}= \gamma^{AB} e_A{}^i e_B{}^j $.
The frame $e_A{}^j$ denotes the coordinate two frame. Its dual $e^A=\rd y^i e_i{}^A$ is such that $\hat\rd e^A=0$ (See \cite{Freidel:2022bai} for conventions on the choice of Carrollian frame).
Repated indices are summed over. For instance we have that $\Gamma_{a b}^{a} = \Gamma_{\O b}^{\O}+ \Gamma_{j b}^{j}$ to express the decomposition of a 4d sum into a 3d one.

 Finally, note that in our notation $\delta\gamma^{ij}$ is the \emph{variation of the inverse metric} and not the variation of the metric with raised indices (the two differ by a sign):
\be
\delta\gamma^{ij} : = \delta(\gamma^{-1}).
\ee

To evaluate the symplectic flux one starts with
\be 
\bftheta^\O &= \tfrac12 \sqrt{g} \left( g^{ab} \delta \Gamma_{ab}^\O - g^{a\O}\delta\Gamma_{b a}^b \right)\cr
&= \tfrac12 \sqrt{g} \left( g^{a\O} \delta \Gamma_{a\O}^\O - g^{a\O}\delta\Gamma_{\O a}^\O 
+g^{a j} \delta \Gamma_{a j }^\O - g^{a\O}\delta\Gamma_{j a}^j  \right)\cr
&= \tfrac12 \sqrt{g} \left( g^{a j} \delta \Gamma_{a j }^\O - g^{a\O}\delta\Gamma_{j a}^j  \right)\cr
&= \tfrac12 \sqrt{g} \left( g^{\O j} \delta \Gamma_{\O j }^\O - g^{\O\O}\delta\Gamma_{j \O}^j 
+g^{i j} \delta \Gamma_{i j }^\O - g^{i \O}\delta\Gamma_{j i}^j \right)\cr
&= \tfrac12 \sqrt{g} \left( V^i (\delta \Gamma_{\O i }^\O-  \delta\Gamma_{j i}^j )  - g^{\O\O}\delta\Gamma_{j \O }^j 
+\gamma^{i j} \delta \Gamma_{i j }^\O \right) \label{thetaOin}
\ee
And then one uses that 
$
\Gamma_{\O i }^\O
=-\omega_i ,
$
to establish that 
\be 
  \delta [ \nabla_j V^j ]  &=  \nabla_j  \delta V^j  +  V^i  (\delta \Gamma_{j i}^j -  \delta \Gamma_{\O i }^\O )  - 2  V^i \delta \omega_i,
\ee 
which allows us to evaluate the first term in \eqref{thetaOin}.
To evaluate the last two terms one uses that 
\be 
 \Gamma_{i \O }^j =  \bmK_i{}^j 
 ,   \qquad
 \Gamma_{ij}^\O = - \mK_{ij}.
\ee
So this gives  the first implementation of the symplectic potential
\be 
\bftheta^\O &= - \frac12 \sqrt{g} \left( \delta [ \nabla_j V^j ]  -  \nabla_j  \delta V^j   + 2\O \nu \delta\btheta  +2  V^i \delta \omega_i
+ \gamma^{i j} \delta \mK_{ij}  \right)
\ee
This expression is not yet in the form we need for the renormalization. We need to precisely isolate the total variation and total derivative terms and present the symplectic potential in the geometric polarization where the configuration variables are the Carrollian geometric variables $(\bb,\nu, V^i,\gamma^{AB})$. This is achieved as follow
\be
\bftheta^\O 
&=   \tfrac12  \pa_j ( \sqrt{g} \delta V^j )  - \tfrac12 \delta \left(  \sqrt{g} \nabla_j V^j \right)   
+ \tfrac12 (\delta \sqrt{g}) \nabla_j V^j    
- \tfrac12 \sqrt{g} \left( 2\O \nu \delta\btheta   +2  V^i \delta \omega_i + \gamma^{i j} \delta \mK_{ij}  \right)
\cr
&=   \tfrac12  \pa_j ( \sqrt{g} \delta V^j )  - 
\tfrac12 \delta \left(  \sqrt{g} [(\nabla_j +2\omega_j)V^j + \gamma^{ij}\mK_{ij} +2\O \nu \btheta   ] \right) \cr 
& \,\,+ \tfrac12 (\delta \sqrt{g}) \nabla_j V^j + \delta (\sqrt{g}  V^j) \omega_j+ \O \delta( \sqrt{g} \btheta) \nu +  \tfrac12 \delta (\sqrt{g} \gamma^{ij}) \mK_{ij} 
 \cr
&=   \tfrac12  \pa_j ( \sqrt{g} \delta V^j )  - \tfrac12 \delta \left(  \sqrt{g} [(\nabla_j +2\omega_j)V^j+  \gamma^{ij}\mK_{ij}  +2\O \nu \btheta ] \right)
\cr
&\,\,   + 
\tfrac12 \sqrt{g} \delta \bb  [(\nabla_j +2\omega_j)V^j  + \gamma^{ij}\mK_{ij}+ 2\O \nu \btheta ] 
 \cr 
& \,\, +  \tfrac12 \sqrt{g} 
\delta \gamma^{ij}  \left(\mK_{ij} - \tfrac12 \gamma_{ij} [(\nabla_j +2\omega_j)V^j+ \gamma^{ij}\mK_{ij} +2\O \nu \btheta ]\right)
\cr
&\,\, + \sqrt{g} \left( \omega_i \delta V^i
+ \O   \btheta   \delta \nu
 \right). \label{thetaOFullF}
\ee
where we used that $\sqrt{g}=e^\bb \sqrt{\gamma}$ hence
\be 
\delta \sqrt{g} = \sqrt{g} \left[ \delta \bb - \tfrac12 \delta \gamma^{AB}  \gamma_{AB} \right].
\ee

We introduce a new vector  and tensor 
\be
\mathring\iota_i :=  -\gamma_i{}^a \mK_{aV}, \qquad \mathring\mK_{ij} := \gamma_i{}^a\gamma_j^{b}\mK_{ab}.
\ee
This enters in the evaluation of 
\be
\delta \gamma^{ij}  \mK_{ij} &=  
\delta \gamma^{ij}  (\gamma_i{}^a\gamma_j^{b} \mK_{ij}
+ \gamma_i{}^a \mK_{aV} K_j + K_i \gamma_j{}^b \mK_{Vb} + K_i K_j \mK_{VV})\cr
& = \delta \gamma^{ij}   \mathring\mK_{ij} + 2  \delta K_i  \mathring\iota^i.
\ee
The symplectic flux  expression can be further simplified by using that $V^a=N^a-2\O\nu K^a$ which implies that 
\be
\nabla_a V^a &= \nabla_a N^a -\nabla_a[ 2\O \nu K^a] \cr
&=[\gamma^{ba} + K^b V^a +  N^b K^a ] \nabla_a N_b -2\nu - 2\O [ K [\nu] + (\bar\kappa +\bar\theta) \nu]\cr
&= \gamma^{ab} \nabla_a N_b +  \left[ K_b \nabla_V N^b +  N^b \nabla_K N_b- 2\nu - 2\O [ K [\nu] + (\bar\kappa +\bar\theta) \nu] \right],\cr
&= \gamma^{ab} \rmK_{ab} +  \left[ V^b \nabla_K N_b +  N^b \nabla_K N_b- 2\nu - 2\O [ K [\nu] + (\bar\kappa +\bar\theta) \nu] \right],
\ee
Then we evaluate 
\be 
N^b \nabla_K N_b &=\nu +\O K[\nu],
\qquad 
\omega_V =  \mN_{VK} 
= \nu +\O ( K[\nu] +2\bar\kappa \nu ),
\ee
where we denote $\omega_V=\omega_a V^a$.
Therefore we conclude that 
\be
\tfrac12 [\nabla_a V^a + \gamma^{ab} \nabla_a N_b]+  \O \bar\theta \nu&=\gamma^{ab} \nabla_a N_b .
\ee
Finally we can use that 
\be
\gamma^{ab}\rmK_{ab} + \omega_V  = (\gamma^{ab} + V^a K^b ) \nabla_a N_b
=(\gamma_b{}^{a} +  K_b V^a ) \nabla_a N^b = h_a{}^b \hmN_a{}^b = \hmN,  
\ee
and the symplectic flux is  given by 
\be 
\bftheta^\O  &=   \tfrac12  \pa_j ( \sqrt{g} \delta V^j )  -  \delta \left(  \sqrt{g} \hmN \right)  + 
\sqrt{g} \delta \bb  \hmN  \cr 
& \,\, +  \sqrt{g} \left( \tfrac12
\delta \gamma^{AB}  \left(  \rmK_{AB} - \gamma_{AB} \hmN \right) + \delta K_i \mathring\iota^i
+ \omega_i \delta V^i
+ \O   \btheta  \delta \nu
 \right). \label{thetaOFullFinal}
\ee
One last simplification comes from using that $\delta \alpha =  \delta K_i V^i =-\delta V^i K_i$. We can also write that 
$\omega_i = \omega_V K_i + \rom_i $ which means that 
\be
\delta \bb  \hmN  +  \omega_i \delta V^i  =  \delta K_i (  \hmN   - \omega_V  ) V^i    + \rom_i \delta V^i = \delta K_i (  \rmN   V^i )   + \rom_i \delta V^i
\ee
Therefore, we finally get 
\be 
\bftheta^\O  &=   \tfrac12  \pa_j ( \sqrt{g} \delta V^j )  -  \delta \left(  \sqrt{g} \hmN \right)   \cr 
& \,\, +  \sqrt{g} \left( \tfrac12
\delta \gamma^{AB}  \left(  \rmK_{AB} - \gamma_{AB} \hmN \right) +\delta K_i (\rmN V^i+\mathring\iota^i)
+ \rom_i \delta V^i
+ \O   \btheta  \delta \nu
 \right). \label{thetaOFullFinal2}
\ee

\subsection{Stretched Horizon expression}
The expression derived by Freidel and Jai-akson in \cite{Freidel:2022vjq}  for the symplectic potential reads (formula (124) there), after the change of sign convention for $T_a{}^b$
\be 
\Theta^\O &= -\sqrt{g} \left[ T_a{}^b \left( V^a (h_b{}^c \delta K_c) - K_b (\delta V^c)  \gamma_c{}^a + \frac12 \gamma^{ac} \gamma_b{}^d \delta \gamma_{cd} \right) +\btheta \delta \rho \right]
\ee
We now use the decomposing of the stress tensor as 
\be
\hmN_i{}^j &= \rmN_i{}^j + \omega_i V^j - K_i \mathring\iota^j \cr
\hmN &= \rmN  + \omega_V \cr
\mT_i{}^j  &=   -\rmN_i{}^j  + (\rmN  + \omega_V)\gamma_i{}^j  - \gamma_i{}^j\omega_i V^j + K_i (\rmN V^j +\mathring\iota^j) \cr
  \mT_i{}^j  &=  \rmT_i{}^j - \rom_i V^j + K_i \iota^j, \qquad \iota^j =  \rmN V^j +\mathring\iota^j
\ee
to get the simpler expression 
\be
\Theta^\O &= -\sqrt{g} \left[ \frac12  \rmT^{ij} \delta \gamma_{ij} + \rom_i \delta V^i 
+  \iota^i  \delta K_i  +\btheta \delta \rho \right]
\ee
Which is up to an overall sign the same as \eqref{thetaOFullFinal2}

\subsection{Transverse symplectic potential} \label{app:sympt}
In this subsection we evaluate the transverse symplectic potential
\be 
\bftheta^i_{\mathsf{EH}} &= \tfrac12 \sqrt{g} \left( g^{ab} \delta \Gamma_{ab}^i - g^{ai}\delta\Gamma_{ab}^b \right)\cr
&= \tfrac12 \sqrt{g} \left( g^{\O\O } \delta \Gamma_{\O\O}^i + 2 g^{\O j  } \delta \Gamma_{\O j}^i + 
g^{jk}\delta\Gamma_{jk}^i - g^{\O i} (\delta\Gamma_{\O \O}^\O +\delta\Gamma_{\O j}^j)  - g^{ji} (\delta\Gamma_{j \O}^\O +\delta\Gamma_{j k}^k) \right)\cr
&= \tfrac12 \sqrt{g} \left( 2 V^j  \delta \bmK_j{}^i   - V^i \delta (\bar\theta  +\bar\kappa)
- \gamma^{ji} \delta\Gamma_{j \O}^\O
+ 
\gamma^{jk}\delta\hat\Gamma_{jk }^i- \gamma^{ji} \delta\hat\Gamma_{j k}^k  \right)\cr
&= \tfrac12 \sqrt{g} \left( 2 V^j  \delta \bmK_j{}^i   - V^i \delta (\bar\theta  +\bar\kappa)  
+ \gamma^{ij} \delta\omega_j
 \right) + 
\hat\bftheta{}_{\mathsf{EH}}^i \label{thetai-1}
\ee
where $\hat\bftheta{}_{\mathsf{EH}}^i $ represents the codimension $1$ symplectic potential 
\be
\hat\bftheta{}_{\mathsf{EH}}^i  =  \tfrac12 \sqrt{g}(\gamma^{jk}\delta\hat\Gamma_{jk }^i- \gamma^{ji} \delta\hat\Gamma_{j k}^k).
\ee
Now from the definition
\be
\hn_i K_j =-\omega_i K_j -K_i(\omega_j+\hat\pa_j\bb) + \bTheta_{ij} 
\ee
we can evaluate its projection along $K$:
\be
 K_l \hat\bftheta{}_{\mathsf{EH}}^l &=  \tfrac12 \sqrt{g}\left( \gamma^{ij} (\delta \Gamma_{ij}^l) K_l= \gamma^{ij} \hn_i \delta K_j - \gamma^{ij} \delta (\hn_i K_j)\right) \cr
&= \tfrac12 \sqrt{g}\left(\hn_i ( \gamma^{ij} \delta K_j) -  (\hn_i  \gamma^{ij}) \delta K_j   + \gamma^{ij} \delta (\omega_i K_j +K_i(\omega_j+\hat\pa_j\bb) - \bTheta_{ij} )\right)\cr
&= \tfrac12 \sqrt{g}\left(\hn_i ( \gamma^{ij} \delta K_j) + ( \bmK_i{}^i V^j +\bmK_i{}^j V^j) \delta K_j   
+ (2\rom^i +\mathring\pa^i\bb)  \delta K_i  -\gamma^{ij}\delta \bTheta_{ij} \right)\cr
&= \tfrac12 \sqrt{g}\left( (\hn_i-\omega_i) ( \gamma^{ij} \delta K_j) +  \btheta  V^i\delta K_i   
+  2\rom^i  \delta K_i  -\gamma^{ij}\delta \bTheta_{ij} \right).
\ee
In the last equality we use that $\bmK_{V}{}^i = -(\rom^i +\mathring\pa^i\bb)$.
We can now combine with \eqref{thetai-1} to get 
\be 
K_i \bftheta^i_{\mathsf{EH}} &=
\tfrac12 \sqrt{g} \left( 2 V^j  (\delta \bmK_j{}^i) K_i  -  \delta (\bar\theta  +\bar\kappa)  
 \right) + 
K_i \hat\bftheta{}_{\mathsf{EH}}^i \cr
&=  
\tfrac12 \sqrt{g} \left( 2(2\rom^i +\mathring\pa^i\bb)  \delta K_i  -  \delta (\bar\theta  +\bar\kappa)  
+\nabla_i ( \gamma^{ij} \delta K_j) +  \btheta  V^i\delta K_i   
  -\gamma^{ij}\delta \bTheta_{ij}  \right) 
\cr
&=   \tfrac12 \pa_i (\sqrt{g} \gamma^{ij} \delta K_j) 
- \tfrac12 \sqrt{g} \left( 2(2\rom^i +\mathring\pa^i\bb)  \delta K_i 
+\btheta  V^i\delta K_i   
 +  \delta \bar\kappa  
 +   \bTheta_{ij}  \delta \gamma^{ij} \right)
\ee
where we used that $\nabla_i V^i=\hn_i-\omega_i V^i$, $\bmK_i{}^j K_j =0$ and $\gamma^{ij} K_j=0$. 
In the Bondi coordinate system since $K_i= e^\bb \delta_i^u$ we restrict to variation for $K_i$ which are such that $\delta K_i =\bb K_i$.
Imposing such a restriction on the  variation  implies that 
\be
 \bftheta^K_{\mathsf{EH}} = 
- \tfrac12 \sqrt{g} \left( 
\btheta  \delta  \bb
 +  \delta \bar\kappa  
 +   \bTheta_{ij}  \delta \gamma^{ij} \right).
\ee
where we denote  $\bftheta^K_{\mathsf{EH}}=K_i\bftheta^i_{\mathsf{EH}}$. 
To compute $\bftheta_{\mathsf{can}}^K$ we need to evaluate  the expression
\be
 \tfrac12 \pa_\O (\sqrt{g}  \delta V^i) K_i 
 &=  \tfrac12 \sqrt{g}  \left(\pa_\O \bb  \delta V^i K_i +  (\delta \pa_\O V^i) K_i\right)\cr
 &= \tfrac12 \sqrt{g} \left(- \pa_\O \bb   \delta \bb  +     \delta (-\pa_\O\bb  V^i + (\pa_\O V^j) \gamma_j{}^i) K_i\right)\cr
 &= \tfrac12 \sqrt{g} \left(- \pa_\O \bb   \delta \bb  - \pa_\O \delta \bb    + \pa_\O\bb  \delta \bb -((\pa_\O V^j) \gamma_j{}^i)\delta K_i  \right)\cr
 &= -\tfrac12 \sqrt{g} \left(  \pa_\O \delta \bb     \right) = -\tfrac12 \sqrt{g} \left(   \delta \bar\kappa     \right),
\ee
where we use the Bondi gauge condition $\pa_\O \sqrt{g}  = \pa_\O \bb \sqrt{g} $ and $\gamma_i{}^j\delta K_j=0$.
This gives us 
\be 
\bftheta^K_{1} = - \tfrac12 \sqrt{g} \left( 
\btheta  \delta \alpha   
 +  2 \delta \bar\kappa  
 +   \bTheta_{ij}  \delta \gamma^{ij} \right)
\ee 
We have that $\bar\kappa =\pa_\O\bb \heq   -\frac{\O}{(d-2)} (\bTheta_A{}^B  \bTheta_B{}^A)$ so we can now perform the expansion 
of this component of the symplectic potential in the Bondi gauge where $\btheta=0$. We have 
\be
 \bftheta^K_{1} = [\bftheta^K_{1}]_0+ \O[\bftheta^K_{1}]_1 + O(\O)
\ee
where
\be 
 [\bftheta^K_{1}]_0 &= - \tfrac12 e^\a\sqrt{q} \left(     \bTheta_{0ij}  \delta q^{ij} \right)  = 
  \tfrac12 e^\a\sqrt{q} \left( \bTheta_{0}^{ij}  \delta q_{ij} \right)\cr
 [\bftheta^K_{1}]_1&= 
-  \tfrac12 e^\a\sqrt{q} \left(  
2  \delta \bar\kappa_1  
 +   \bTheta_{0ij}  \delta \gamma_1^{ij}+    \bTheta_{1ij}  \delta q^{ij} \right) \cr
 &= -  \tfrac12 e^\a\sqrt{q} \left(  
- \frac{2}{(d-2)}  \delta  (\bTheta_0 : \bTheta_0)  
 +  2 \bTheta_{0ij}  \delta \bTheta_0^{ij}+    \bTheta_{1ij}  \delta q^{ij} \right) \cr
 &=  \tfrac12 e^\a\sqrt{q} \left(  \bTheta_{1}^{ij}  \delta q_{ij}
 + \frac{(d-4)}{(d-2)}  \delta  (\bTheta_0 : \bTheta_0)  
    \right)
\ee

\subsection{Horizontal  component} \label{app:symph}
The projection $\gamma_i{}^j$ components are $ \gamma_i{}^u=0$ $\gamma_u{}^A =\Y^A$ and $\gamma_A{}^B =\delta_A{}^B$.
\be
\hat\bftheta{}_{\mathsf{EH}}^i \gamma_i{}^A &=  \tfrac12 \sqrt{g}(\gamma^{jk}(\delta\hat\Gamma_{jk }^i)\gamma_i{}^A  -\gamma^{Aj} \delta\hat\Gamma_{j k}^k).
\ee
To evaluate this 
one uses that 
\be
\gamma_B{}^j \gamma_{C}{}^k\hat\Gamma_{jk }^i = \mathring\Gamma_{BC }^i \gamma_D{}^i - \bTheta_{BC } V^i,
\ee
where $ \mathring\Gamma_{ij}^k =\gamma_i{}^a\gamma_j{}^a \hat\Gamma_{ab}^c \gamma_c^k$ are the connection coefficients of the codimension $2$ connection $\Sn_i$.
Now in the Bondi-Sachs frame where $ \gamma_i{}^j\delta K_j=0$ we have that
\be 
\gamma^{jk} \delta(\hat\Gamma_{jk }^i)\gamma_i{}^A &=  \gamma^{BC}  \delta(\mathring\Gamma_{BC }^i  - \bTheta_{BC } V^i)\gamma_i{}^A \cr
&= \gamma^{BC}  \delta(\mathring\Gamma_{BC }^i)\gamma_i{}^A   - \btheta \delta V^i \gamma_i{}^A
\ee
Then we focus on 
\be 
\gamma^{Aj} \delta\hat\Gamma_{j k}^k  &= 
 \gamma^{Aj} \delta(\omega_j +\hat\pa_j \bb)  + \gamma^{AB} \delta \mathring \Gamma_{B C}^C 
\ee
Combining the two together, and using agian that $ \gamma_i{}^j\delta K_j=0$, gives then
\be
\hat\bftheta{}_{\mathsf{EH}}^i \gamma_i{}^A &=    - \tfrac12 \sqrt{g} \left(\btheta \delta V^i \gamma_i{}^A + \gamma^{Aj} \delta(\rom_j +\mathring\pa_j \bb)\right)  + e^\bb \mathring\bftheta{}_{\mathsf{EH}}^A
\ee
where $\mathring\bftheta{}_{\mathsf{EH}}^A$ is the Einstein-Hilbert symplectic potential for the codimension $2$ connection compatible with $\gamma$:
\be
\mathring\bftheta{}_{\mathsf{EH}}^A :=  
\tfrac12 \sqrt{\gamma}(\gamma^{BC}(\delta\mathring \Gamma_{BC }^A)  -\gamma^{B A} \delta\mathring\Gamma_{BC}^C)
\ee
One can also evaluate 
\be
 \tfrac12 \pa_\O (\sqrt{g} \delta V^i ) \gamma_i{}^A &=  \pa_\O (\sqrt{g} \delta V^i  \gamma_i{}^A )- (\sqrt{g} \delta V^i )  \pa_\O\gamma_i{}^A \cr
 &= \tfrac12  \pa_\O (\sqrt{\gamma} \delta \Y^A  )- \sqrt{g} \delta V^i \bmK_i{}^A\cr
 &= \tfrac12  \pa_\O (\sqrt{\gamma} \delta \Y^A  )+ \sqrt{g} (\delta V^i K_i) (\rom^A+ \mathring\pa^A \bb) - \sqrt{g} \delta V^i \bTheta_i{}^A \cr
 &= \tfrac12  \sqrt{\gamma}  ( \pa_\O \delta \Y^A  -  \delta \Y^B\bTheta_B{}^A - \delta e^\bb (\rom^A+ \mathring\pa^A \bb) ).
\ee
Adding the two we get in the Bondi gauge where $\btheta=0$ that
\be
\hat\bftheta{}_{1}^i \gamma_i{}^A &=     \tfrac12 \sqrt{\gamma}  \left( \delta(\pa_\O \Y^A)  -  \delta \Y^B\bTheta_B{}^A - \gamma^{AB} \delta(e^\bb(\mathring\pa_B \bb +\O \Pi_B))\right)  + e^\bb \mathring\bftheta{}_{\mathsf{EH}}^A \cr
&=   \tfrac12 \sqrt{\gamma}  \left( \delta \gamma^{AB} e^\bb (\mathring\pa_B\bb- 2 \O  \Pi_B )  -  \delta \Y^B\bTheta_B{}^A 
- 3\O \gamma^{AB} \delta(e^\bb  \Pi^A )  \right)+ e^\bb \mathring\bftheta{}_{\mathsf{EH}}^A,
\ee
where we  used that $\pa_\O \Y^A= e^\bb (\mathring\pa^A\bb- 2 \O  \Pi^A )$.
\be \label{thetaTexp}
[\hat\bftheta{}_{1}^i \gamma_i{}^A]_0 &=     \tfrac12 \sqrt{q}  \left( e^\a \delta \gamma^{AB}  (\mathring\pa_B\a )  -  \delta U^B\bTheta_{0B}{}^A 
 \right)  + e^\a [\mathring\bftheta{}_{\mathsf{EH}}^A]_0 \\
 [\hat\bftheta{}_{1}^i \gamma_i{}^A]_1 &= \tfrac12  \sqrt{\gamma}  \left( 2\delta \bTheta_0^{AB} e^\a (\mathring\pa_B\a ) 
 -2 \delta q^{AB}  \Pi_{0B}  -  \delta U^B\bTheta_{1B}{}^A-\delta(q^{AB}\pp_B e^\a) \bTheta_{0B}{}^A 
- 3q^{AB} \delta(e^\a  \Pi_0^A )  \right)+ e^\a [\mathring\bftheta{}_{\mathsf{EH}}^A]_1 \nonumber
\ee
and we can use that $(d-3)\Pi_{0A}=  D_B \bTheta_{0A}{}^B$.

\section{General renormalization}\label{app:ren}
As we mentioned earlier, the above procedure can be generalized to the case where the metric requires a more general  expansion (polyhomogenous) that not only involves powers of $\O$ but also powers of its logarithm  $\O$ -- provided they appear multiplied by a $\O^{d-2}$ prefactor.
This fact can be summarized in the following lemma, whose proof can be found in  \cite{Freidel:2019ohg}[App. A].

\paragraph*{Lemma}
Consider a quantity $\tl X := \O^{-n} X$ which satisfies the following radial evolution equation
\be
(n- \O\po)X = Z
\label{eq:radialXY}
\ee
defined in terms of a polyhomogenous source $Z(\O)$ of the form $Z(\O) = \sum_{k=0}^n \O^k Z_k + \O^{n+1} P(\O)$, with $P(\O) = \sum_{p,q\geq 0} h_{p,q}\O^p ( \ln\O)^q$. Notice that in this parametrization, the $Z_k$'s constitute the divergent part of $\tl Z = \O^{-n} Z$ which -- for our future convenience -- we denote as
\be
\lceil Z\rceil_n =  \sum_{k=0}^n \O^k Z_k.
\label{eq:ceil}
\ee
(Although the term $k=n$  not divergent, it still plays a special role, on which we comment below.)

The most general solution to this equation is given by
\be
X = C_n(Z) + \O^n\Big( \mathcal X - \frac1{n!} \int_0^\O P(\O')\d \O'\Big),
\ee
where $\mathcal X$ is a {\it free} integration constant and  the counterterm $C_n$ is given by:
\be
C_n(Z) := \sum_{k=0}^{n-1} \frac{(n -1 - k)!}{n!} \O^k \pp_\O^k Z - \O^n \ln \O\, Z_n.
\ee
In other words, being a first order differential equation, $(n-\O\po)X=Z$ admits a unique solution once an arbitrary initial condition $\mathcal X$  is provided. The important feature of the operator $(n-\O\po)$ is that this ``initial condition'' appears at order $\O^n$ in the power-expansion of $X$.

The counterterm $C_n(Z)$ contributes to the divergent part of $\tl X = \O^{-n}X$, so that the renormalized quantity
\be
\tl X_R := \O^{-n}( X - C_n(Z)) 
\ee
 has a finite limit at $\O\to0$:
\be
\tl X_R \xrightarrow{\O\to0} \cal X.
\ee
Note that this procedure defines a renormalized $\tl X_R$ at all values of $\O$, not only in the limit $\O\to 0$. 

Finally, a comment on the term $Z_n$. The fact that $Z$ admits a Taylor expansion with no logarithmic terms does {\it not} mean that $X$ will not develop any logarithmic term. Indeed, if $Z = \O^n Z_n$,  one finds that $C_n(Z)=-\O^n\ln \O\, Z_n$  and $X = \O^n \mathcal X - \O^n \ln \O\, Z_n$. In this case we refer to $Z_n$ as the anomaly. It is precisely in this sense that the vector anomaly $D_B\bTheta_1^{BA}$  makes its appearance in the radial evolution equation for  $\Y^A$.

\paragraph*{The renormalized symplectic flux}
Combining this general result on equations of the form \eqref{eq:radialXY}, together with the radial evolution equation for $\bftheta^\O_{\mathsf{can}}$ \eqref{eq:Cevolution}, we find that -- on-shell of the equations of motion -- one can define a renormalized symplectic flux as
\be
\tl\bftheta{}^\O_R :=\O^{3-d}\left(  \tl\bftheta{}_{\mathsf{can}}^\O - C_{d-3}( \lceil \bZ \rceil_{d-3} ) \right)
\ee
where,
\be
\bZ =  \bE -  \delta \bL_{\mathsf{can}}  + \p_i \bftheta^i_\mathsf{can} 
\ee
and $\lceil \bZ \rceil_{n}$ is as in \eqref{eq:ceil}. 

The renormalized flux $\tl\bftheta{}^\O_R$ has the following finite limit  on $\scri$:
\be
 \lim_{\O\to 0} \tl\bftheta{}^\O_R = \left.\frac{\pp^{d-3} \tl\bftheta{}^\O_\mathsf{can}}{(d-3)!}\right|_{\O=0} =  [\tl\bftheta{}^\O_\mathsf{can}]_0.
\label{eq:finiteflux}
\ee
Therefore, the punchline, couldn't have been simpler: the renormalized symplectic flux through $\scri$, $[\tl\bftheta{}^\O_\mathsf{can}]_0$, coincides with the zero-th order of $\tl\bftheta{}^\O_R$ in the $\O$ expansion.

\bibliographystyle{bib-style2}
\bibliography{Conformal}

\providecommand{\href}[2]{#2}\begingroup\raggedright\begin{thebibliography}{100}

\bibitem{Bondi:1960jsa}
H.~Bondi, \emph{{Gravitational Waves in General Relativity}}, \href{http://dx.doi.org/10.1038/186535a0}{\emph{Nature} {\bfseries 186} (1960) 535--535}.

\bibitem{Bondi62}
H.~Bondi, M.~van~der Burg and A.~Metzner, \emph{Gravitational waves in general relativity. 7. waves from axisymmetric isolated systems}, {\emph{Proc.Roy.Soc.Lond.} {\bfseries A269} (1962) 21--52}.

\bibitem{Sachs:1961zz}
R.~K. Sachs, \emph{{Gravitational waves in general relativity. 6. The outgoing radiation condition}}, \href{http://dx.doi.org/10.1098/rspa.1961.0202}{\emph{Proc. Roy. Soc. Lond. A} {\bfseries 264} (1961) 309--338}.

\bibitem{Sachs:1962wk}
R.~K. Sachs, \emph{{Gravitational waves in general relativity. 8. Waves in asymptotically flat space-times}}, \href{http://dx.doi.org/10.1098/rspa.1962.0206}{\emph{Proc. Roy. Soc. Lond.} {\bfseries A270} (1962) 103--126}.

\bibitem{Penrose:1962ij}
R.~Penrose, \emph{{Asymptotic properties of fields and space-times}}, \href{http://dx.doi.org/10.1103/PhysRevLett.10.66}{\emph{Phys. Rev. Lett.} {\bfseries 10} (1963) 66--68}.

\bibitem{Penrose:1965am}
R.~Penrose, \emph{{Zero rest mass fields including gravitation: Asymptotic behavior}}, \href{http://dx.doi.org/10.1098/rspa.1965.0058}{\emph{Proc. Roy. Soc. Lond.} {\bfseries A284} (1965) 159}.

\bibitem{PenroseRindler2}
R.~Penrose and W.~Rindler, \emph{{Spinors And Space-Time. Vol. 2: Spinor And Twistor Methods In Space-Time Geometry}}.
\newblock CUP, 1986.

\bibitem{Geroch:1977jn}
R.~Geroch, \emph{Asymptotic structure of space-time},  in \emph{Asymptotic Structure of Space-Time} (F.~P. Esposito and L.~Witten, eds.), (Boston, MA), Springer US, 1977.
\newblock \href{http://dx.doi.org/10.1007/978-1-4684-2343-3_1}{DOI}.

\bibitem{friedrich2002conformal}
H.~Friedrich, \emph{Conformal einstein evolution},  in \emph{The conformal structure of space-time: Geometry, analysis, numerics}, pp.~1--50.
\newblock Springer, 2002.

\bibitem{LevyLeblond1965}
J.-M. L\'evy-Leblond, \emph{Une nouvelle limite non-relativiste du groupe de {Poincar\'e}}, {\emph{Annales de l'institut Henri Poincar\'e. Section A, Physique Th\'eorique} {\bfseries 3} (1965) 1--12}.

\bibitem{Duval:2014uva}
C.~Duval, G.~W. Gibbons and P.~A. Horvathy, \emph{{Conformal Carroll groups and BMS symmetry}}, \href{http://dx.doi.org/10.1088/0264-9381/31/9/092001}{\emph{Class. Quant. Grav.} {\bfseries 31} (2014) 092001}, [\href{https://arxiv.org/abs/1402.5894}{{\ttfamily 1402.5894}}].

\bibitem{Adamo:2009vu}
T.~M. Adamo, C.~N. Kozameh and E.~T. Newman, \emph{Null geodesic congruences, asymptotically flat space-times and their physical interpretation}, \href{http://dx.doi.org/10.12942/lrr-2009-6}{\emph{Living Rev. Rel.} {\bfseries 12} (2009) 6}, [\href{https://arxiv.org/abs/0906.2155}{{\ttfamily 0906.2155}}].

\bibitem{Madler:2016xju}
T.~M\"adler and J.~Winicour, \emph{{Bondi-Sachs Formalism}}, \href{http://dx.doi.org/10.4249/scholarpedia.33528}{\emph{Scholarpedia} {\bfseries 11} (2016) 33528}, [\href{https://arxiv.org/abs/1609.01731}{{\ttfamily 1609.01731}}].

\bibitem{NP62}
E.~Newman and R.~Penrose, \emph{An approach to gravitational radiation by a method of spin coefficients}, {\emph{J.Math.Phys.} {\bfseries 3} (1962) 566--578}.

\bibitem{friedrich1983cauchy}
H.~Friedrich, \emph{Cauchy problems for the conformal vacuum field equations in general relativity}, {\emph{Communications in Mathematical Physics} {\bfseries 91} (1983) 445--472}.

\bibitem{winicour1985logarithmic}
J.~Winicour, \emph{Logarithmic asymptotic flatness}, {\emph{Foundations of Physics} {\bfseries 15} (1985) 605--616}.

\bibitem{andersson1993hyperboloidal}
L.~Andersson and P.~T. Chru{\'s}ciel, \emph{Hyperboloidal cauchy data for vacuum einstein equations and obstructions to smoothness of null infinity}, {\emph{Physical review letters} {\bfseries 70} (1993) 2829}.

\bibitem{chrusciel1993gravitational}
P.~T. Chrusciel, M.~A. MacCallum and D.~B. Singleton, \emph{Gravitational waves in general relativity: Xiv. bondi expansions and the``polyhomogeneity''of$\backslash$scri},  [\href{https://arxiv.org/abs/9305021}{{\ttfamily 9305021}}].

\bibitem{Friedrich:2017cjg}
H.~Friedrich, \emph{{Peeling or not peeling\textemdash{}is that the question?}}, \href{http://dx.doi.org/10.1088/1361-6382/aaafdb}{\emph{Class. Quant. Grav.} {\bfseries 35} (2018) 083001}, [\href{https://arxiv.org/abs/1709.07709}{{\ttfamily 1709.07709}}].

\bibitem{Tafel:2021bxa}
J.~Tafel, \emph{{The Einstein metrics with smooth scri}}, \href{http://dx.doi.org/10.1007/s10714-022-02986-5}{\emph{Gen. Rel. Grav.} {\bfseries 54} (2022) 103}, [\href{https://arxiv.org/abs/2110.07688}{{\ttfamily 2110.07688}}].

\bibitem{kehrberger2021case}
L.~M.~A. Kehrberger, \emph{The case against smooth null infinity i: Heuristics and counter-examples},  vol.~23, pp.~829--921, 2022.
\newblock \href{http://dx.doi.org/10.1007/s00023-021-01108-2}{DOI}.

\bibitem{Ashtekar:1978zz}
A.~Ashtekar and R.~O. Hansen, \emph{{A unified treatment of null and spatial infinity in general relativity. I - Universal structure, asymptotic symmetries, and conserved quantities at spatial infinity}}, \href{http://dx.doi.org/10.1063/1.523863}{\emph{J. Math. Phys.} {\bfseries 19} (1978) 1542--1566}.

\bibitem{Ashtekar:1981sf}
A.~Ashtekar, \emph{{Asymptotic Quantization of the Gravitational Field}}, \href{http://dx.doi.org/10.1103/PhysRevLett.46.573}{\emph{Phys. Rev. Lett.} {\bfseries 46} (1981) 573--576}.

\bibitem{Ashtekar:1981bq}
A.~Ashtekar and M.~Streubel, \emph{{Symplectic Geometry of Radiative Modes and Conserved Quantities at Null Infinity}}, \href{http://dx.doi.org/10.1098/rspa.1981.0109}{\emph{Proc. Roy. Soc. Lond. A} {\bfseries 376} (1981) 585--607}.

\bibitem{Ashtekar:1990gc}
A.~Ashtekar, L.~Bombelli and O.~Reula, \emph{The covariant phase space of asymptotically flat gravitational fields},  in \emph{Mechanics, Analysis and Geometry: 200 Years After Lagrange} (M.~Francaviglia, ed.), North-Holland Delta Series, pp.~417 -- 450.
\newblock Elsevier, Amsterdam, 1991.
\newblock \href{http://dx.doi.org/10.1016/B978-0-444-88958-4.50021-5}{DOI}.

\bibitem{Wald:1999wa}
R.~M. Wald and A.~Zoupas, \emph{{A General definition of 'conserved quantities' in general relativity and other theories of gravity}}, \href{http://dx.doi.org/10.1103/PhysRevD.61.084027}{\emph{Phys. Rev. D} {\bfseries 61} (2000) 084027}, [\href{https://arxiv.org/abs/gr-qc/9911095}{{\ttfamily gr-qc/9911095}}].

\bibitem{Ciambelli:2018xat}
L.~Ciambelli, C.~Marteau, A.~C. Petkou, P.~M. Petropoulos and K.~Siampos, \emph{{Covariant Galilean versus Carrollian hydrodynamics from relativistic fluids}}, \href{http://dx.doi.org/10.1088/1361-6382/aacf1a}{\emph{Class. Quant. Grav.} {\bfseries 35} (2018) 165001}, [\href{https://arxiv.org/abs/1802.05286}{{\ttfamily 1802.05286}}].

\bibitem{Chandrasekaran:2021hxc}
V.~Chandrasekaran, E.~E. Flanagan, I.~Shehzad and A.~J. Speranza, \emph{{Brown-York charges at null boundaries}}, \href{http://dx.doi.org/10.1007/JHEP01(2022)029}{\emph{JHEP} {\bfseries 01} (2022) 029}, [\href{https://arxiv.org/abs/2109.11567}{{\ttfamily 2109.11567}}].

\bibitem{Freidel:2022vjq}
L.~Freidel and P.~Jai-akson, \emph{{Carrollian hydrodynamics and symplectic structure on stretched horizons}},  [\href{https://arxiv.org/abs/2211.06415}{{\ttfamily 2211.06415}}].

\bibitem{Mars:2022gsa}
M.~Mars and G.~S\'anchez-P\'erez, \emph{{Double null data and the characteristic problem in general relativity}}, \href{http://dx.doi.org/10.1088/1751-8121/acb098}{\emph{J. Phys. A} {\bfseries 56} (2023) 035203}, [\href{https://arxiv.org/abs/2205.15267}{{\ttfamily 2205.15267}}].

\bibitem{Freidel:2023bnj}
L.~Freidel, M.~Geiller and W.~Wieland, \emph{Corner Symmetry and Quantum Geometry}, pp.~1--36.
\newblock Springer Nature Singapore, Singapore, 2023.
\newblock \href{https://arxiv.org/abs/2302.12799}{{\ttfamily 2302.12799}}.
\newblock 10.1007/978-981-19-3079-9\_107-1.

\bibitem{Strominger:2013jfa}
A.~Strominger, \emph{{On BMS Invariance of Gravitational Scattering}}, \href{http://dx.doi.org/10.1007/JHEP07(2014)152}{\emph{JHEP} {\bfseries 07} (2014) 152}, [\href{https://arxiv.org/abs/1312.2229}{{\ttfamily 1312.2229}}].

\bibitem{Strominger:2014pwa}
A.~Strominger and A.~Zhiboedov, \emph{{Gravitational Memory, BMS Supertranslations and Soft Theorems}}, \href{http://dx.doi.org/10.1007/JHEP01(2016)086}{\emph{JHEP} {\bfseries 01} (2016) 086}, [\href{https://arxiv.org/abs/1411.5745}{{\ttfamily 1411.5745}}].

\bibitem{Weinberg:1965nx}
S.~Weinberg, \emph{{Infrared photons and gravitons}}, \href{http://dx.doi.org/10.1103/PhysRev.140.B516}{\emph{Phys. Rev.} {\bfseries 140} (1965) B516--B524}.

\bibitem{Pasterski:2015tva}
S.~Pasterski, A.~Strominger and A.~Zhiboedov, \emph{{New Gravitational Memories}}, \href{http://dx.doi.org/10.1007/JHEP12(2016)053}{\emph{JHEP} {\bfseries 12} (2016) 053}, [\href{https://arxiv.org/abs/1502.06120}{{\ttfamily 1502.06120}}].

\bibitem{Pasterski:2016qvg}
S.~Pasterski, S.-H. Shao and A.~Strominger, \emph{{Flat Space Amplitudes and Conformal Symmetry of the Celestial Sphere}}, \href{http://dx.doi.org/10.1103/PhysRevD.96.065026}{\emph{Phys. Rev. D} {\bfseries 96} (2017) 065026}, [\href{https://arxiv.org/abs/1701.00049}{{\ttfamily 1701.00049}}].

\bibitem{Kapec:2016jld}
D.~Kapec, P.~Mitra, A.-M. Raclariu and A.~Strominger, \emph{{2D Stress Tensor for 4D Gravity}}, \href{http://dx.doi.org/10.1103/PhysRevLett.119.121601}{\emph{Phys. Rev. Lett.} {\bfseries 119} (2017) 121601}, [\href{https://arxiv.org/abs/1609.00282}{{\ttfamily 1609.00282}}].

\bibitem{Donnay:2020guq}
L.~Donnay, S.~Pasterski and A.~Puhm, \emph{{Asymptotic Symmetries and Celestial CFT}}, \href{http://dx.doi.org/10.1007/JHEP09(2020)176}{\emph{JHEP} {\bfseries 09} (2020) 176}, [\href{https://arxiv.org/abs/2005.08990}{{\ttfamily 2005.08990}}].

\bibitem{Cachazo:2014fwa}
F.~Cachazo and A.~Strominger, \emph{{Evidence for a New Soft Graviton Theorem}},  [\href{https://arxiv.org/abs/1404.4091}{{\ttfamily 1404.4091}}].

\bibitem{Guevara:2021abz}
A.~Guevara, E.~Himwich, M.~Pate and A.~Strominger, \emph{{Holographic Symmetry Algebras for Gauge Theory and Gravity}},  [\href{https://arxiv.org/abs/2103.03961}{{\ttfamily 2103.03961}}].

\bibitem{Strominger:2021lvk}
A.~Strominger, \emph{{w(1+infinity) and the Celestial Sphere}},  [\href{https://arxiv.org/abs/2105.14346}{{\ttfamily 2105.14346}}].

\bibitem{Freidel:2021ytz}
L.~Freidel, D.~Pranzetti and A.-M. Raclariu, \emph{Higher spin dynamics in gravity and ${w}_{1+\ensuremath{\infty} }$ celestial symmetries}, \href{http://dx.doi.org/10.1103/PhysRevD.106.086013}{\emph{Phys. Rev. D} {\bfseries 106} (Oct, 2022) 086013}, [\href{https://arxiv.org/abs/2112.15573}{{\ttfamily 2112.15573}}].

\bibitem{Strominger:2017zoo}
A.~Strominger, \emph{{Lectures on the Infrared Structure of Gravity and Gauge Theory}},  [\href{https://arxiv.org/abs/1703.05448}{{\ttfamily 1703.05448}}].

\bibitem{Raclariu:2021zjz}
A.-M. Raclariu, \emph{{Lectures on Celestial Holography}},  [\href{https://arxiv.org/abs/2107.02075}{{\ttfamily 2107.02075}}].

\bibitem{Pasterski:2021rjz}
S.~Pasterski, \emph{Lectures on celestial amplitudes}, \href{http://dx.doi.org/10.1140/epjc/s10052-021-09846-7}{\emph{The European Physical Journal C} {\bfseries 81} (2021) 1062}, [\href{https://arxiv.org/abs/2108.04801}{{\ttfamily 2108.04801}}].

\bibitem{Donnay:2023mrd}
L.~Donnay, \emph{{Celestial holography: An asymptotic symmetry perspective}},  [\href{https://arxiv.org/abs/2310.12922}{{\ttfamily 2310.12922}}].

\bibitem{Barnich:2010eb}
G.~Barnich and C.~Troessaert, \emph{{Aspects of the BMS/CFT correspondence}}, \href{http://dx.doi.org/10.1007/JHEP05(2010)062}{\emph{JHEP} {\bfseries 05} (2010) 062}, [\href{https://arxiv.org/abs/1001.1541}{{\ttfamily 1001.1541}}].

\bibitem{Barnich:2016lyg}
G.~Barnich and C.~Troessaert, \emph{{Finite BMS transformations}}, \href{http://dx.doi.org/10.1007/JHEP03(2016)167}{\emph{JHEP} {\bfseries 03} (2016) 167}, [\href{https://arxiv.org/abs/1601.04090}{{\ttfamily 1601.04090}}].

\bibitem{Campiglia:2015yka}
M.~Campiglia and A.~Laddha, \emph{{New symmetries for the Gravitational S-matrix}}, \href{http://dx.doi.org/10.1007/JHEP04(2015)076}{\emph{JHEP} {\bfseries 04} (2015) 076}, [\href{https://arxiv.org/abs/1502.02318}{{\ttfamily 1502.02318}}].

\bibitem{Campiglia:2020qvc}
M.~Campiglia and J.~Peraza, \emph{{Generalized BMS charge algebra}}, \href{http://dx.doi.org/10.1103/PhysRevD.101.104039}{\emph{Phys. Rev. D} {\bfseries 101} (2020) 104039}, [\href{https://arxiv.org/abs/2002.06691}{{\ttfamily 2002.06691}}].

\bibitem{Compere:2008us}
G.~Compere and D.~Marolf, \emph{{Setting the boundary free in AdS/CFT}}, \href{http://dx.doi.org/10.1088/0264-9381/25/19/195014}{\emph{Class. Quant. Grav.} {\bfseries 25} (2008) 195014}, [\href{https://arxiv.org/abs/0805.1902}{{\ttfamily 0805.1902}}].

\bibitem{Flanagan:2015pxa}
E.~E. Flanagan and D.~A. Nichols, \emph{{Conserved charges of the extended Bondi-Metzner-Sachs algebra}}, \href{http://dx.doi.org/10.1103/PhysRevD.95.044002}{\emph{Phys. Rev. D} {\bfseries 95} (2017) 044002}, [\href{https://arxiv.org/abs/1510.03386}{{\ttfamily 1510.03386}}].

\bibitem{Compere:2018ylh}
G.~Comp\`{e}re, A.~Fiorucci and R.~Ruzziconi, \emph{{Superboost transitions, refraction memory and super-Lorentz charge algebra}}, \href{http://dx.doi.org/10.1007/JHEP11(2018)200}{\emph{JHEP} {\bfseries 11} (2018) 200}, [\href{https://arxiv.org/abs/1810.00377}{{\ttfamily 1810.00377}}].

\bibitem{Compere:2020lrt}
G.~Comp\`ere, A.~Fiorucci and R.~Ruzziconi, \emph{{The $\Lambda$-BMS$_4$ charge algebra}}, \href{http://dx.doi.org/10.1007/JHEP10(2020)205}{\emph{JHEP} {\bfseries 10} (2020) 205}, [\href{https://arxiv.org/abs/2004.10769}{{\ttfamily 2004.10769}}].

\bibitem{Freidel:2021fxf}
L.~Freidel, R.~Oliveri, D.~Pranzetti and S.~Speziale, \emph{{The Weyl BMS group and Einstein\textquoteright{}s equations}}, \href{http://dx.doi.org/10.1007/JHEP07(2021)170}{\emph{JHEP} {\bfseries 07} (2021) 170}, [\href{https://arxiv.org/abs/2104.05793}{{\ttfamily 2104.05793}}].

\bibitem{Chandrasekaran:2021vyu}
V.~Chandrasekaran, E.~E. Flanagan, I.~Shehzad and A.~J. Speranza, \emph{{A general framework for gravitational charges and holographic renormalization}}, \href{http://dx.doi.org/10.1142/S0217751X22501056}{\emph{Int. J. Mod. Phys. A} {\bfseries 37} (2022) 2250105}, [\href{https://arxiv.org/abs/2111.11974}{{\ttfamily 2111.11974}}].

\bibitem{Ciambelli:2018wre}
L.~Ciambelli, C.~Marteau, A.~C. Petkou, P.~M. Petropoulos and K.~Siampos, \emph{{Flat holography and Carrollian fluids}}, \href{http://dx.doi.org/10.1007/JHEP07(2018)165}{\emph{JHEP} {\bfseries 07} (2018) 165}, [\href{https://arxiv.org/abs/1802.06809}{{\ttfamily 1802.06809}}].

\bibitem{Ciambelli:2019lap}
L.~Ciambelli, R.~G. Leigh, C.~Marteau and P.~M. Petropoulos, \emph{{Carroll Structures, Null Geometry and Conformal Isometries}}, \href{http://dx.doi.org/10.1103/PhysRevD.100.046010}{\emph{Phys. Rev. D} {\bfseries 100} (2019) 046010}, [\href{https://arxiv.org/abs/1905.02221}{{\ttfamily 1905.02221}}].

\bibitem{Ciambelli:2018ojf}
L.~Ciambelli and C.~Marteau, \emph{{Carrollian conservation laws and Ricci-flat gravity}}, \href{http://dx.doi.org/10.1088/1361-6382/ab0d37}{\emph{Class. Quant. Grav.} {\bfseries 36} (2019) 085004}, [\href{https://arxiv.org/abs/1810.11037}{{\ttfamily 1810.11037}}].

\bibitem{Donnay:2022aba}
L.~Donnay, A.~Fiorucci, Y.~Herfray and R.~Ruzziconi, \emph{{Carrollian Perspective on Celestial Holography}}, \href{http://dx.doi.org/10.1103/PhysRevLett.129.071602}{\emph{Phys. Rev. Lett.} {\bfseries 129} (2022) 071602}, [\href{https://arxiv.org/abs/2202.04702}{{\ttfamily 2202.04702}}].

\bibitem{Donnay:2022wvx}
L.~Donnay, A.~Fiorucci, Y.~Herfray and R.~Ruzziconi, \emph{{Bridging Carrollian and celestial holography}}, \href{http://dx.doi.org/10.1103/PhysRevD.107.126027}{\emph{Phys. Rev. D} {\bfseries 107} (2023) 126027}, [\href{https://arxiv.org/abs/2212.12553}{{\ttfamily 2212.12553}}].

\bibitem{Bagchi:2022emh}
A.~Bagchi, S.~Banerjee, R.~Basu and S.~Dutta, \emph{{Scattering Amplitudes: Celestial and Carrollian}}, \href{http://dx.doi.org/10.1103/PhysRevLett.128.241601}{\emph{Phys. Rev. Lett.} {\bfseries 128} (2022) 241601}, [\href{https://arxiv.org/abs/2202.08438}{{\ttfamily 2202.08438}}].

\bibitem{Gupta1966}
N.~D.~S. Gupta, \emph{On an analogue of the galilei group}, \href{http://dx.doi.org/10.1007/bf02740871}{\emph{Il Nuovo Cimento A Series 10} {\bfseries 44} (1966) 512--517}.

\bibitem{Henneaux1979a}
M.~Henneaux, \emph{{Geometry of zero signature spacetime}}, {\emph{Bull.Soc.Math.Belg. 31 47-63} (1979) }.

\bibitem{Penna:2018gfx}
R.~F. Penna, \emph{{Near-horizon Carroll symmetry and black hole Love numbers}},  [\href{https://arxiv.org/abs/1812.05643}{{\ttfamily 1812.05643}}].

\bibitem{Donnay:2019jiz}
L.~Donnay and C.~Marteau, \emph{{Carrollian Physics at the Black Hole Horizon}}, \href{http://dx.doi.org/10.1088/1361-6382/ab2fd5}{\emph{Class. Quant. Grav.} {\bfseries 36} (2019) 165002}, [\href{https://arxiv.org/abs/1903.09654}{{\ttfamily 1903.09654}}].

\bibitem{Parattu:2015gga}
K.~Parattu, S.~Chakraborty, B.~R. Majhi and T.~Padmanabhan, \emph{{A Boundary Term for the Gravitational Action with Null Boundaries}}, \href{http://dx.doi.org/10.1007/s10714-016-2093-7}{\emph{Gen. Rel. Grav.} {\bfseries 48} (2016) 94}, [\href{https://arxiv.org/abs/1501.01053}{{\ttfamily 1501.01053}}].

\bibitem{Donnay:2015abr}
L.~Donnay, G.~Giribet, H.~A. Gonzalez and M.~Pino, \emph{{Supertranslations and Superrotations at the Black Hole Horizon}}, \href{http://dx.doi.org/10.1103/PhysRevLett.116.091101}{\emph{Phys. Rev. Lett.} {\bfseries 116} (2016) 091101}, [\href{https://arxiv.org/abs/1511.08687}{{\ttfamily 1511.08687}}].

\bibitem{Donnay:2016ejv}
L.~Donnay, G.~Giribet, H.~A. Gonz\'alez and M.~Pino, \emph{{Extended Symmetries at the Black Hole Horizon}}, \href{http://dx.doi.org/10.1007/JHEP09(2016)100}{\emph{JHEP} {\bfseries 09} (2016) 100}, [\href{https://arxiv.org/abs/1607.05703}{{\ttfamily 1607.05703}}].

\bibitem{Parattu:2016trq}
K.~Parattu, S.~Chakraborty and T.~Padmanabhan, \emph{{Variational Principle for Gravity with Null and Non-null boundaries: A Unified Boundary Counter-term}}, \href{http://dx.doi.org/10.1140/epjc/s10052-016-3979-y}{\emph{Eur. Phys. J. C} {\bfseries 76} (2016) 129}, [\href{https://arxiv.org/abs/1602.07546}{{\ttfamily 1602.07546}}].

\bibitem{Lehner:2016vdi}
L.~Lehner, R.~C. Myers, E.~Poisson and R.~D. Sorkin, \emph{{Gravitational action with null boundaries}}, \href{http://dx.doi.org/10.1103/PhysRevD.94.084046}{\emph{Phys. Rev. D} {\bfseries 94} (2016) 084046}, [\href{https://arxiv.org/abs/1609.00207}{{\ttfamily 1609.00207}}].

\bibitem{Hopfmuller:2016scf}
F.~Hopfmuller and L.~Freidel, \emph{{Gravity Degrees of Freedom on a Null Surface}}, \href{http://dx.doi.org/10.1103/PhysRevD.95.104006}{\emph{Phys. Rev.} {\bfseries D95} (2017) 104006}, [\href{https://arxiv.org/abs/1611.03096}{{\ttfamily 1611.03096}}].

\bibitem{Wieland:2017cmf}
W.~Wieland, \emph{{Fock representation of gravitational boundary modes and the discreteness of the area spectrum}}, \href{http://dx.doi.org/10.1007/s00023-017-0598-6}{\emph{Annales Henri Poincare} {\bfseries 18} (2017) 3695--3717}, [\href{https://arxiv.org/abs/1706.00479}{{\ttfamily 1706.00479}}].

\bibitem{Wieland:2017zkf}
W.~Wieland, \emph{{New boundary variables for classical and quantum gravity on a null surface}}, \href{http://dx.doi.org/10.1088/1361-6382/aa8d06}{\emph{Class. Quant. Grav.} {\bfseries 34} (2017) 215008}, [\href{https://arxiv.org/abs/1704.07391}{{\ttfamily 1704.07391}}].

\bibitem{Hopfmuller:2018fni}
F.~Hopfm\"uller and L.~Freidel, \emph{{Null Conservation Laws for Gravity}}, \href{http://dx.doi.org/10.1103/PhysRevD.97.124029}{\emph{Phys. Rev. D} {\bfseries 97} (2018) 124029}, [\href{https://arxiv.org/abs/1802.06135}{{\ttfamily 1802.06135}}].

\bibitem{Chandrasekaran:2018aop}
V.~Chandrasekaran, E.~E. Flanagan and K.~Prabhu, \emph{{Symmetries and charges of general relativity at null boundaries}}, \href{http://dx.doi.org/10.1007/JHEP11(2018)125}{\emph{JHEP} {\bfseries 11} (2018) 125}, [\href{https://arxiv.org/abs/1807.11499}{{\ttfamily 1807.11499}}].

\bibitem{Oliveri:2019gvm}
R.~Oliveri and S.~Speziale, \emph{{Boundary effects in General Relativity with tetrad variables}}, \href{http://dx.doi.org/10.1007/s10714-020-02733-8}{\emph{Gen. Rel. Grav.} {\bfseries 52} (2020) 83}, [\href{https://arxiv.org/abs/1912.01016}{{\ttfamily 1912.01016}}].

\bibitem{Adami:2020ugu}
H.~Adami, M.~M. Sheikh-Jabbari, V.~Taghiloo, H.~Yavartanoo and C.~Zwikel, \emph{{Symmetries at null boundaries: two and three dimensional gravity cases}}, \href{http://dx.doi.org/10.1007/JHEP10(2020)107}{\emph{JHEP} {\bfseries 10} (2020) 107}, [\href{https://arxiv.org/abs/2007.12759}{{\ttfamily 2007.12759}}].

\bibitem{Adami:2021nnf}
H.~Adami, D.~Grumiller, M.~M. Sheikh-Jabbari, V.~Taghiloo, H.~Yavartanoo and C.~Zwikel, \emph{{Null boundary phase space: slicings, news \& memory}}, \href{http://dx.doi.org/10.1007/JHEP11(2021)155}{\emph{JHEP} {\bfseries 11} (2021) 155}, [\href{https://arxiv.org/abs/2110.04218}{{\ttfamily 2110.04218}}].

\bibitem{Ashtekar:2021kqj}
A.~Ashtekar, N.~Khera, M.~Kolanowski and J.~Lewandowski, \emph{{Charges and fluxes on (perturbed) non-expanding horizons}}, \href{http://dx.doi.org/10.1007/JHEP02(2022)066}{\emph{JHEP} {\bfseries 02} (2022) 066}, [\href{https://arxiv.org/abs/2112.05608}{{\ttfamily 2112.05608}}].

\bibitem{Chandrasekaran:2020wwn}
V.~Chandrasekaran and A.~J. Speranza, \emph{{Anomalies in gravitational charge algebras of null boundaries and black hole entropy}}, \href{http://dx.doi.org/10.1007/JHEP01(2021)137}{\emph{JHEP} {\bfseries 01} (2021) 137}, [\href{https://arxiv.org/abs/2009.10739}{{\ttfamily 2009.10739}}].

\bibitem{Odak:2022ndm}
G.~Odak, A.~Rignon-Bret and S.~Speziale, \emph{{Wald-Zoupas prescription with soft anomalies}}, \href{http://dx.doi.org/10.1103/PhysRevD.107.084028}{\emph{Phys. Rev. D} {\bfseries 107} (2023) 084028}, [\href{https://arxiv.org/abs/2212.07947}{{\ttfamily 2212.07947}}].

\bibitem{Sheikh-Jabbari:2022mqi}
M.~M. Sheikh-Jabbari, \emph{{On symplectic form for null boundary phase space}}, \href{http://dx.doi.org/10.1007/s10714-022-02997-2}{\emph{Gen. Rel. Grav.} {\bfseries 54} (2022) 140}, [\href{https://arxiv.org/abs/2209.05043}{{\ttfamily 2209.05043}}].

\bibitem{Ciambelli:2023mir}
L.~Ciambelli, L.~Freidel and R.~G. Leigh, \emph{{Null Raychaudhuri: Canonical Structure and the Dressing Time}},  [\href{https://arxiv.org/abs/2309.03932}{{\ttfamily 2309.03932}}].

\bibitem{Chandrasekaran:2023vzb}
V.~Chandrasekaran and {\'E}.~{\'E}. Flanagan, \emph{Horizon phase spaces in general relativity}, \href{http://dx.doi.org/10.1007/JHEP07(2024)017}{\emph{Journal of High Energy Physics} {\bfseries 2024} (2024) 17}, [\href{https://arxiv.org/abs/2309.03871}{{\ttfamily 2309.03871}}].

\bibitem{Odak:2023pga}
G.~Odak, A.~Rignon-Bret and S.~Speziale, \emph{General gravitational charges on null hypersurfaces}, \href{http://dx.doi.org/10.1007/JHEP12(2023)038}{\emph{Journal of High Energy Physics} {\bfseries 2023} (2023) 38}, [\href{https://arxiv.org/abs/2309.03854}{{\ttfamily 2309.03854}}].

\bibitem{Brown:1992br}
J.~D. Brown and J.~W. York, Jr., \emph{{Quasilocal energy and conserved charges derived from the gravitational action}}, \href{http://dx.doi.org/10.1103/PhysRevD.47.1407}{\emph{Phys. Rev. D} {\bfseries 47} (1993) 1407--1419}, [\href{https://arxiv.org/abs/gr-qc/9209012}{{\ttfamily gr-qc/9209012}}].

\bibitem{Jai-akson:2022gwg}
P.~Jai-akson, \emph{{Edge Modes and Carrollian Hydrodynamics on Stretched Horizons}}.
\newblock PhD thesis, U. Waterloo, https://uwspace.uwaterloo.ca/handle/10012/18824, 10, 2022.

\bibitem{Geiller:2022vto}
M.~Geiller and C.~Zwikel, \emph{{The partial Bondi gauge: Further enlarging the asymptotic structure of gravity}}, \href{http://dx.doi.org/10.21468/SciPostPhys.13.5.108}{\emph{SciPost Phys.} {\bfseries 13} (2022) 108}, [\href{https://arxiv.org/abs/2205.11401}{{\ttfamily 2205.11401}}].

\bibitem{Freidel:2019ohg}
L.~Freidel, F.~Hopfm\"uller and A.~Riello, \emph{{Asymptotic Renormalization in Flat Space: Symplectic Potential and Charges of Electromagnetism}}, \href{http://dx.doi.org/10.1007/JHEP10(2019)126}{\emph{JHEP} {\bfseries 10} (2019) 126}, [\href{https://arxiv.org/abs/1904.04384}{{\ttfamily 1904.04384}}].

\bibitem{Hopfmueller:2020yqj}
F.~Hopfmueller, \emph{{Canonical Structure and Conservation Laws of General Relativity on Null Surfaces and at Null Infinity}}.
\newblock PhD thesis, U. Waterloo, https://uwspace.uwaterloo.ca/handle/10012/16633, 2020.

\bibitem{McNees:2023tus}
R.~McNees and C.~Zwikel, \emph{{Finite charges from the bulk action}}, \href{http://dx.doi.org/10.1007/JHEP08(2023)154}{\emph{JHEP} {\bfseries 08} (2023) 154}, [\href{https://arxiv.org/abs/2306.16451}{{\ttfamily 2306.16451}}].

\bibitem{Geiller:2024amx}
M.~Geiller and C.~Zwikel, \emph{{The partial Bondi gauge: Gauge fixings and asymptotic charges}}, \href{http://dx.doi.org/10.21468/SciPostPhys.16.3.076}{\emph{SciPost Phys.} {\bfseries 16} (2024) 076}, [\href{https://arxiv.org/abs/2401.09540}{{\ttfamily 2401.09540}}].

\bibitem{Hollands:2003xp}
S.~Hollands and A.~Ishibashi, \emph{{Asymptotic flatness at null infinity in higher dimensional gravity}},  in \emph{{7th Hungarian Relativity Workshop (RW 2003)}}, pp.~51--61, 11, 2003.
\newblock \href{https://arxiv.org/abs/hep-th/0311178}{{\ttfamily hep-th/0311178}}.

\bibitem{Godazgar:2012zq}
M.~Godazgar and H.~S. Reall, \emph{{Peeling of the Weyl tensor and gravitational radiation in higher dimensions}}, \href{http://dx.doi.org/10.1103/PhysRevD.85.084021}{\emph{Phys. Rev. D} {\bfseries 85} (2012) 084021}, [\href{https://arxiv.org/abs/1201.4373}{{\ttfamily 1201.4373}}].

\bibitem{Hollands:2016oma}
S.~Hollands, A.~Ishibashi and R.~M. Wald, \emph{{BMS Supertranslations and Memory in Four and Higher Dimensions}}, \href{http://dx.doi.org/10.1088/1361-6382/aa777a}{\emph{Class. Quant. Grav.} {\bfseries 34} (2017) 155005}, [\href{https://arxiv.org/abs/1612.03290}{{\ttfamily 1612.03290}}].

\bibitem{Kapec:2017gsg}
D.~Kapec and P.~Mitra, \emph{{A $d$-Dimensional Stress Tensor for Mink$_{d+2}$ Gravity}}, \href{http://dx.doi.org/10.1007/JHEP05(2018)186}{\emph{JHEP} {\bfseries 05} (2018) 186}, [\href{https://arxiv.org/abs/1711.04371}{{\ttfamily 1711.04371}}].

\bibitem{Pate:2017fgt}
M.~Pate, A.-M. Raclariu and A.~Strominger, \emph{{Gravitational Memory in Higher Dimensions}}, \href{http://dx.doi.org/10.1007/JHEP06(2018)138}{\emph{JHEP} {\bfseries 06} (2018) 138}, [\href{https://arxiv.org/abs/1712.01204}{{\ttfamily 1712.01204}}].

\bibitem{Cameron:2021fhd}
P.~Cameron and P.~T. Chru\'sciel, \emph{{Asymptotic flatness in higher dimensions}}, \href{http://dx.doi.org/10.1063/5.0083728}{\emph{J. Math. Phys.} {\bfseries 63} (2022) 032501}, [\href{https://arxiv.org/abs/2112.13140}{{\ttfamily 2112.13140}}].

\bibitem{Capone:2021ouo}
F.~Capone, \emph{{General null asymptotics and superrotation-compatible configuration spaces in $d\ge4$}}, \href{http://dx.doi.org/10.1007/JHEP10(2021)158}{\emph{JHEP} {\bfseries 10} (2021) 158}, [\href{https://arxiv.org/abs/2108.01203}{{\ttfamily 2108.01203}}].

\bibitem{Fuentealba:2022yqt}
O.~Fuentealba, M.~Henneaux, J.~Matulich and C.~Troessaert, \emph{{Asymptotic structure of the gravitational field in five spacetime dimensions: Hamiltonian analysis}}, \href{http://dx.doi.org/10.1007/JHEP07(2022)149}{\emph{JHEP} {\bfseries 07} (2022) 149}, [\href{https://arxiv.org/abs/2206.04972}{{\ttfamily 2206.04972}}].

\bibitem{Capone:2023roc}
F.~Capone, P.~Mitra, A.~Poole and B.~Tomova, \emph{Phase space renormalization and finite bms charges in six dimensions}, \href{http://dx.doi.org/10.1007/JHEP11(2023)034}{\emph{Journal of High Energy Physics} {\bfseries 2023} (2023) 34}, [\href{https://arxiv.org/abs/2304.09330}{{\ttfamily 2304.09330}}].

\bibitem{Colferai:2020rte}
D.~Colferai and S.~Lionetti, \emph{{Asymptotic symmetries and the subleading soft graviton theorem in higher dimensions}}, \href{http://dx.doi.org/10.1103/PhysRevD.104.064010}{\emph{Phys. Rev. D} {\bfseries 104} (2021) 064010}, [\href{https://arxiv.org/abs/2005.03439}{{\ttfamily 2005.03439}}].

\bibitem{Hollands:2004ac}
S.~Hollands and R.~M. Wald, \emph{{Conformal null infinity does not exist for radiating solutions in odd spacetime dimensions}}, \href{http://dx.doi.org/10.1088/0264-9381/21/22/008}{\emph{Class. Quant. Grav.} {\bfseries 21} (2004) 5139--5146}, [\href{https://arxiv.org/abs/gr-qc/0407014}{{\ttfamily gr-qc/0407014}}].

\bibitem{Tanabe:2011es}
K.~Tanabe, S.~Kinoshita and T.~Shiromizu, \emph{{Asymptotic flatness at null infinity in arbitrary dimensions}}, \href{http://dx.doi.org/10.1103/PhysRevD.84.044055}{\emph{Phys. Rev. D} {\bfseries 84} (2011) 044055}, [\href{https://arxiv.org/abs/1104.0303}{{\ttfamily 1104.0303}}].

\bibitem{Garfinkle:2017fre}
D.~Garfinkle, S.~Hollands, A.~Ishibashi, A.~Tolish and R.~M. Wald, \emph{{The Memory Effect for Particle Scattering in Even Spacetime Dimensions}}, \href{http://dx.doi.org/10.1088/1361-6382/aa777b}{\emph{Class. Quant. Grav.} {\bfseries 34} (2017) 145015}, [\href{https://arxiv.org/abs/1702.00095}{{\ttfamily 1702.00095}}].

\bibitem{Kapec:2015vwa}
D.~Kapec, V.~Lysov, S.~Pasterski and A.~Strominger, \emph{{Higher-dimensional supertranslations and Weinberg\textquoteright{}s soft graviton theorem}}, \href{http://dx.doi.org/10.4310/AMSA.2017.v2.n1.a2}{\emph{Ann. Math. Sci. Appl.} {\bfseries 02} (2017) 69--94}, [\href{https://arxiv.org/abs/1502.07644}{{\ttfamily 1502.07644}}].

\bibitem{Durkee:2010xq}
M.~Durkee, V.~Pravda, A.~Pravdova and H.~S. Reall, \emph{{Generalization of the Geroch-Held-Penrose formalism to higher dimensions}}, \href{http://dx.doi.org/10.1088/0264-9381/27/21/215010}{\emph{Class. Quant. Grav.} {\bfseries 27} (2010) 215010}, [\href{https://arxiv.org/abs/1002.4826}{{\ttfamily 1002.4826}}].

\bibitem{Grant:2021sxk}
A.~M. Grant, K.~Prabhu and I.~Shehzad, \emph{{The Wald\textendash{}Zoupas prescription for asymptotic charges at null infinity in general relativity}}, \href{http://dx.doi.org/10.1088/1361-6382/ac571a}{\emph{Class. Quant. Grav.} {\bfseries 39} (2022) 085002}, [\href{https://arxiv.org/abs/2105.05919}{{\ttfamily 2105.05919}}].

\bibitem{Prabhu:2021cgk}
K.~Prabhu and I.~Shehzad, \emph{{Conservation of asymptotic charges from past to future null infinity: Lorentz charges in general relativity}}, \href{http://dx.doi.org/10.1007/JHEP08(2022)029}{\emph{JHEP} {\bfseries 08} (2022) 029}, [\href{https://arxiv.org/abs/2110.04900}{{\ttfamily 2110.04900}}].

\bibitem{Mohamed:2023jwv}
M.~M.~A. Mohamed, K.~Prabhu and J.~A.~V. Kroon, \emph{{BMS-supertranslation charges at the critical sets of null infinity}}, \href{http://dx.doi.org/10.1063/5.0187927}{\emph{J. Math. Phys.} {\bfseries 65} (2024) 032501}, [\href{https://arxiv.org/abs/2311.07294}{{\ttfamily 2311.07294}}].

\bibitem{frauendiener2004conformal}
J.~Frauendiener, \emph{Conformal infinity}, {\emph{Living Reviews in Relativity} {\bfseries 7} (2004) 1}.

\bibitem{Penrose:1980yx}
R.~Penrose, \emph{{Null Hypersurface Initial Data For Classical Fields Of Arbitrary Spin And For General Relativity}}, \href{http://dx.doi.org/10.1007/BF00756234}{\emph{Aerospace Research Laboratories Report 63-56 (1963). Reprinted in: Gen. Rel. Grav.} {\bfseries 12} (1980) 225--264}.

\bibitem{Fernandez-Alvarez:2021zmp}
F.~Fern\'andez-\'Alvarez and J.~M.~M. Senovilla, \emph{{Asymptotic structure with vanishing cosmological constant}}, \href{http://dx.doi.org/10.1088/1361-6382/ac387e}{\emph{Class. Quant. Grav.} {\bfseries 39} (2022) 165011}, [\href{https://arxiv.org/abs/2105.09166}{{\ttfamily 2105.09166}}].

\bibitem{Bieri:2020pee}
L.~Bieri, \emph{{New structures in gravitational radiation}}, \href{http://dx.doi.org/10.4310/ATMP.2022.v26.n3.a1}{\emph{Adv. Theor. Math. Phys.} {\bfseries 26} (2022) 531--594}, [\href{https://arxiv.org/abs/2010.07418}{{\ttfamily 2010.07418}}].

\bibitem{Geroch:1973am}
R.~P. Geroch, A.~Held and R.~Penrose, \emph{{A space-time calculus based on pairs of null directions}}, \href{http://dx.doi.org/10.1063/1.1666410}{\emph{J. Math. Phys.} {\bfseries 14} (1973) 874--881}.

\bibitem{BongaPrabhu2020}
B.~Bonga, A.~M. Grant and K.~Prabhu, \emph{Angular momentum at null infinity in einstein-maxwell theory}, \href{http://dx.doi.org/10.1103/PhysRevD.101.044013}{\emph{Phys. Rev. D} {\bfseries 101} (Feb, 2020) 044013}.

\bibitem{Blanchet:1986dk}
L.~Blanchet, \emph{{Radiative gravitational fields in general relativity. 2. Asymptotic behaviour at future null infinity}}, \href{http://dx.doi.org/10.1098/rspa.1987.0022}{\emph{Proc. Roy. Soc. Lond. A} {\bfseries 409} (1987) 383--399}.

\bibitem{friedrich2004smoothness}
H.~Friedrich, \emph{Smoothness at null infinity and the structure of initial data},  in \emph{The Einstein Equations and the Large Scale Behavior of Gravitational Fields: 50 Years of the Cauchy Problem in General Relativity}, pp.~121--203.
\newblock Springer, 2004.

\bibitem{acena2011conformal}
A.~E. Acena and J.~A.~V. Kroon, \emph{Conformal extensions for stationary spacetimes}, {\emph{Classical and quantum gravity} {\bfseries 28} (2011) 225023}.

\bibitem{curry2018introduction}
S.~N. Curry and A.~R. Gover, \emph{An introduction to conformal geometry and tractor calculus, with a view to applications in general relativity}, {\emph{Asymptotic analysis in general relativity} {\bfseries 443} (2018) 86}.

\bibitem{Fernandez-Alvarez:2021yog}
F.~Fern\'andez-\'Alvarez and J.~M.~M. Senovilla, \emph{{Asymptotic structure with a positive cosmological constant}}, \href{http://dx.doi.org/10.1088/1361-6382/ac395b}{\emph{Class. Quant. Grav.} {\bfseries 39} (2022) 165012}, [\href{https://arxiv.org/abs/2105.09167}{{\ttfamily 2105.09167}}].

\bibitem{Herfray:2021xyp}
Y.~Herfray, \emph{Tractor geometry of asymptotically flat spacetimes}, \href{http://dx.doi.org/10.1007/s00023-022-01174-0}{\emph{Annales Henri Poincar{\'e}} {\bfseries 23} (2022) 3265--3310}, [\href{https://arxiv.org/abs/2103.10405}{{\ttfamily 2103.10405}}].

\bibitem{Barnich:2009se}
G.~Barnich and C.~Troessaert, \emph{{Symmetries of asymptotically flat 4 dimensional spacetimes at null infinity revisited}}, \href{http://dx.doi.org/10.1103/PhysRevLett.105.111103}{\emph{Phys. Rev. Lett.} {\bfseries 105} (2010) 111103}, [\href{https://arxiv.org/abs/0909.2617}{{\ttfamily 0909.2617}}].

\bibitem{Barnich:2011mi}
G.~Barnich and C.~Troessaert, \emph{{BMS charge algebra}}, \href{http://dx.doi.org/10.1007/JHEP12(2011)105}{\emph{JHEP} {\bfseries 12} (2011) 105}, [\href{https://arxiv.org/abs/1106.0213}{{\ttfamily 1106.0213}}].

\bibitem{Paetz:2013nga}
T.-T. Paetz, \emph{{Conformally covariant systems of wave equations and their equivalence to Einstein's field equations}}, \href{http://dx.doi.org/10.1007/s00023-014-0359-8}{\emph{Annales Henri Poincare} {\bfseries 16} (2015) 2059--2129}, [\href{https://arxiv.org/abs/1306.6204}{{\ttfamily 1306.6204}}].

\bibitem{newman1962behavior}
E.~T. Newman and T.~W. Unti, \emph{Behavior of asymptotically flat empty spaces}, {\emph{Journal of Mathematical Physics} {\bfseries 3} (1962) 891--901}.

\bibitem{Barnich:2011ty}
G.~Barnich and P.-H. Lambert, \emph{{A Note on the Newman-Unti group and the BMS charge algebra in terms of Newman-Penrose coefficients}}, \href{http://dx.doi.org/10.1155/2012/197385}{\emph{Adv. Math. Phys.} {\bfseries 2012} (2012) 197385}, [\href{https://arxiv.org/abs/1102.0589}{{\ttfamily 1102.0589}}].

\bibitem{Mars:1993mj}
M.~Mars and J.~M.~M. Senovilla, \emph{{Geometry of general hypersurfaces in space-time: Junction conditions}}, \href{http://dx.doi.org/10.1088/0264-9381/10/9/026}{\emph{Class. Quant. Grav.} {\bfseries 10} (1993) 1865--1897}, [\href{https://arxiv.org/abs/gr-qc/0201054}{{\ttfamily gr-qc/0201054}}].

\bibitem{Freidel:2022bai}
L.~Freidel and P.~Jai-akson, \emph{{Carrollian hydrodynamics from symmetries}}, \href{http://dx.doi.org/10.1088/1361-6382/acb194}{\emph{Class. Quant. Grav.} {\bfseries 40} (2023) 055009}, [\href{https://arxiv.org/abs/2209.03328}{{\ttfamily 2209.03328}}].

\bibitem{Mars:2023hty}
M.~Mars and G.~S\'anchez-P\'erez, \emph{{Covariant definition of double null data and geometric uniqueness of the characteristic initial value problem}}, \href{http://dx.doi.org/10.1088/1751-8121/acd312}{\emph{J. Phys. A} {\bfseries 56} (2023) 255203}, [\href{https://arxiv.org/abs/2301.02722}{{\ttfamily 2301.02722}}].

\bibitem{Manzano:2023oub}
M.~Manzano and M.~Mars, \emph{{The Constraint Tensor: General Definition and Properties}},  [\href{https://arxiv.org/abs/2309.14813}{{\ttfamily 2309.14813}}].

\bibitem{Duval:2014uoa}
C.~Duval, G.~W. Gibbons, P.~A. Horvathy and P.~M. Zhang, \emph{{Carroll versus Newton and Galilei: two dual non-Einsteinian concepts of time}}, \href{http://dx.doi.org/10.1088/0264-9381/31/8/085016}{\emph{Class. Quant. Grav.} {\bfseries 31} (2014) 085016}, [\href{https://arxiv.org/abs/1402.0657}{{\ttfamily 1402.0657}}].

\bibitem{Adami:2020amw}
H.~Adami, D.~Grumiller, S.~Sadeghian, M.~M. Sheikh-Jabbari and C.~Zwikel, \emph{{T-Witts from the horizon}}, \href{http://dx.doi.org/10.1007/JHEP04(2020)128}{\emph{JHEP} {\bfseries 04} (2020) 128}, [\href{https://arxiv.org/abs/2002.08346}{{\ttfamily 2002.08346}}].

\bibitem{Adami:2021kvx}
H.~Adami, M.~M. Sheikh-Jabbari, V.~Taghiloo and H.~Yavartanoo, \emph{{Null surface thermodynamics}}, \href{http://dx.doi.org/10.1103/PhysRevD.105.066004}{\emph{Phys. Rev. D} {\bfseries 105} (2022) 066004}, [\href{https://arxiv.org/abs/2110.04224}{{\ttfamily 2110.04224}}].

\bibitem{Petkou:2022bmz}
A.~C. Petkou, P.~M. Petropoulos, D.~R. Betancour and K.~Siampos, \emph{{Relativistic fluids, hydrodynamic frames and their Galilean versus Carrollian avatars}}, \href{http://dx.doi.org/10.1007/JHEP09(2022)162}{\emph{JHEP} {\bfseries 09} (2022) 162}, [\href{https://arxiv.org/abs/2205.09142}{{\ttfamily 2205.09142}}].

\bibitem{Herfray:2020rvq}
Y.~Herfray, \emph{{Asymptotic shear and the intrinsic conformal geometry of null-infinity}}, \href{http://dx.doi.org/10.1063/5.0003616}{\emph{J. Math. Phys.} {\bfseries 61} (2020) 072502}, [\href{https://arxiv.org/abs/2001.01281}{{\ttfamily 2001.01281}}].

\bibitem{LOJAFERNANDES2002119}
R.~{Loja Fernandes}, \emph{Lie algebroids, holonomy and characteristic classes}, \href{http://dx.doi.org/https://doi.org/10.1006/aima.2001.2070}{\emph{Advances in Mathematics} {\bfseries 170} (2002) 119--179}.

\bibitem{BarnichAlgebroid}
G.~Barnich, \emph{{A Note on Gauge Systems from the Point of View of Lie Algebroids}}, \href{http://dx.doi.org/10.1063/1.3527427}{\emph{AIP Conference Proceedings} {\bfseries 1307} (11, 2010) 7--18}, [\href{https://arxiv.org/abs/https://pubs.aip.org/aip/acp/article-pdf/1307/1/7/11783091/7\_1\_online.pdf}{{\ttfamily https://pubs.aip.org/aip/acp/article-pdf/1307/1/7/11783091/7\_1\_online.pdf}}].

\bibitem{GomesRielloScipost}
H.~Gomes and A.~Riello, \emph{{The quasilocal degrees of freedom of Yang-Mills theory}}, \href{http://dx.doi.org/10.21468/SciPostPhys.10.6.130}{\emph{SciPost Phys.} {\bfseries 10} (2021) 130}.

\bibitem{RielloSchiavinaATMP}
A.~Riello and M.~Schiavina, \emph{Hamiltonian gauge theory with corners: constraint reduction and flux superselection},  [\href{https://arxiv.org/abs/2207.00568}{{\ttfamily 2207.00568}}].

\bibitem{bergmannkomar1972coordinate}
P.~G. Bergmann and A.~Komar, \emph{The coordinate group symmetries of general relativity}, {\emph{International Journal of Theoretical Physics} {\bfseries 5} (1972) 15--28}.

\bibitem{Salisbury1983}
D.~C. Salisbury and K.~Sundermeyer, \emph{Realization in phase space of general coordinate transformations}, \href{http://dx.doi.org/10.1103/PhysRevD.27.740}{\emph{Phys. Rev. D} {\bfseries 27} (Feb, 1983) 740--756}.

\bibitem{Freidel:2021cbc}
L.~Freidel, R.~Oliveri, D.~Pranzetti and S.~Speziale, \emph{{Extended corner symmetry, charge bracket and Einstein\textquoteright{}s equations}}, \href{http://dx.doi.org/10.1007/JHEP09(2021)083}{\emph{JHEP} {\bfseries 09} (2021) 083}, [\href{https://arxiv.org/abs/2104.12881}{{\ttfamily 2104.12881}}].

\bibitem{Hansen:2021fxi}
D.~Hansen, N.~A. Obers, G.~Oling and B.~T. S\o{}gaard, \emph{{Carroll Expansion of General Relativity}}, \href{http://dx.doi.org/10.21468/SciPostPhys.13.3.055}{\emph{SciPost Phys.} {\bfseries 13} (2022) 055}, [\href{https://arxiv.org/abs/2112.12684}{{\ttfamily 2112.12684}}].

\bibitem{Skenderis:2002wp}
K.~Skenderis, \emph{{Lecture notes on holographic renormalization}}, \href{http://dx.doi.org/10.1088/0264-9381/19/22/306}{\emph{Class. Quant. Grav.} {\bfseries 19} (2002) 5849--5876}, [\href{https://arxiv.org/abs/hep-th/0209067}{{\ttfamily hep-th/0209067}}].

\bibitem{godazgar2020bms}
M.~Godazgar and G.~Macaulay, \emph{Bms charges in polyhomogeneous spacetimes}, {\emph{Physical Review D} {\bfseries 102} (2020) 064036}.

\bibitem{Compere:2019bua}
G.~Comp\`ere, A.~Fiorucci and R.~Ruzziconi, \emph{{The $\Lambda$-BMS$_4$ group of dS$_4$ and new boundary conditions for AdS$_4$}}, \href{http://dx.doi.org/10.1088/1361-6382/ab3d4b}{\emph{Class. Quant. Grav.} {\bfseries 36} (2019) 195017}, [\href{https://arxiv.org/abs/1905.00971}{{\ttfamily 1905.00971}}].

\bibitem{Freidel:2021qpz}
L.~Freidel and D.~Pranzetti, \emph{{Gravity from symmetry: duality and impulsive waves}}, \href{http://dx.doi.org/10.1007/JHEP04(2022)125}{\emph{JHEP} {\bfseries 04} (2022) 125}, [\href{https://arxiv.org/abs/2109.06342}{{\ttfamily 2109.06342}}].

\bibitem{Grant:2021hga}
A.~M. Grant and D.~A. Nichols, \emph{{Persistent gravitational wave observables: Curve deviation in asymptotically flat spacetimes}}, \href{http://dx.doi.org/10.1103/PhysRevD.105.024056}{\emph{Phys. Rev. D} {\bfseries 105} (2022) 024056}, [\href{https://arxiv.org/abs/2109.03832}{{\ttfamily 2109.03832}}].

\bibitem{SpinC}
L.~Freidel, S.~F. Moosavian and D.~Pranzetti, \emph{{On the definition of the spin charge in asymptotically-flat spacetimes}},  [\href{https://arxiv.org/abs/2403.19547}{{\ttfamily 2403.19547}}].

\bibitem{Aghapour:2018icu}
S.~Aghapour, G.~Jafari and M.~Golshani, \emph{{On variational principle and canonical structure of gravitational theory in double-foliation formalism}}, \href{http://dx.doi.org/10.1088/1361-6382/aaef9e}{\emph{Class. Quant. Grav.} {\bfseries 36} (2019) 015012}, [\href{https://arxiv.org/abs/1808.07352}{{\ttfamily 1808.07352}}].

\bibitem{York:1972sj}
J.~W. York, Jr., \emph{{Role of conformal three geometry in the dynamics of gravitation}}, \href{http://dx.doi.org/10.1103/PhysRevLett.28.1082}{\emph{Phys. Rev. Lett.} {\bfseries 28} (1972) 1082--1085}.

\bibitem{Gibbons:1976ue}
G.~W. Gibbons and S.~W. Hawking, \emph{{Action Integrals and Partition Functions in Quantum Gravity}}, \href{http://dx.doi.org/10.1103/PhysRevD.15.2752}{\emph{Phys. Rev. D} {\bfseries 15} (1977) 2752--2756}.

\bibitem{Harlow:2019yfa}
D.~Harlow and J.-Q. Wu, \emph{{Covariant phase space with boundaries}}, \href{http://dx.doi.org/10.1007/JHEP10(2020)146}{\emph{JHEP} {\bfseries 10} (2020) 146}, [\href{https://arxiv.org/abs/1906.08616}{{\ttfamily 1906.08616}}].

\bibitem{RielloSchiavinaAHP}
A.~Riello and M.~Schiavina, \emph{Null hamiltonian yang--mills theory: Soft symmetries and memory as superselection}, \href{http://dx.doi.org/10.1007/s00023-024-01443-0}{\emph{Annales Henri Poincar{\'e}} (2024) }, [\href{https://arxiv.org/abs/2303.03531}{{\ttfamily 2303.03531}}].

\bibitem{Noether:1918zz}
E.~Noether, \emph{{Invariant Variation Problems}}, \href{http://dx.doi.org/10.1080/00411457108231446}{\emph{Gott. Nachr.} {\bfseries 1918} (1918) 235--257}, [\href{https://arxiv.org/abs/physics/0503066}{{\ttfamily physics/0503066}}].

\bibitem{Freidel-Jai-24}
L.~Freidel and P.~Jai-akson, ``The symplectic geometry of stretched horizons.'' To appear.

\bibitem{Ashtekar:2024mme}
A.~Ashtekar and S.~Speziale, \emph{{Horizons and null infinity: A fugue in four voices}}, \href{http://dx.doi.org/10.1103/PhysRevD.109.L061501}{\emph{Phys. Rev. D} {\bfseries 109} (2024) L061501}, [\href{https://arxiv.org/abs/2401.15618}{{\ttfamily 2401.15618}}].

\bibitem{Peraza:2023ivy}
J.~Peraza, \emph{{Renormalized electric and magnetic charges for O(r$^{n}$) large gauge symmetries}}, \href{http://dx.doi.org/10.1007/JHEP01(2024)175}{\emph{JHEP} {\bfseries 01} (2024) 175}, [\href{https://arxiv.org/abs/2301.05671}{{\ttfamily 2301.05671}}].

\bibitem{Poole:2018koa}
A.~Poole, K.~Skenderis and M.~Taylor, \emph{{(A)dS$\mathbf{_4}$ in Bondi gauge}}, \href{http://dx.doi.org/10.1088/1361-6382/ab117c}{\emph{Class. Quant. Grav.} {\bfseries 36} (2019) 095005}, [\href{https://arxiv.org/abs/1812.05369}{{\ttfamily 1812.05369}}].

\bibitem{Ruzziconi:2020cjt}
R.~Ruzziconi, \emph{{On the Various Extensions of the BMS Group}}.
\newblock PhD thesis, Universit{\'e} libre de Bruxelles, 9, 2020.
\newblock \href{https://arxiv.org/abs/2009.01926}{{\ttfamily 2009.01926}}.

\bibitem{Senovilla:2022pym}
J.~M.~M. Senovilla, \emph{{Gravitational Radiation at Infinity with Non-Negative Cosmological Constant}}, \href{http://dx.doi.org/10.3390/universe8090478}{\emph{Universe} {\bfseries 8} (2022) 478}, [\href{https://arxiv.org/abs/2208.05436}{{\ttfamily 2208.05436}}].

\bibitem{Erfani:2022lnr}
D.-S. Erfani, \emph{{Bondi news in de Sitter space-time}},  [\href{https://arxiv.org/abs/2204.05960}{{\ttfamily 2204.05960}}].

\bibitem{Henneaux:2019yax}
M.~Henneaux and C.~Troessaert, \emph{{The asymptotic structure of gravity at spatial infinity in four spacetime dimensions}}, \href{http://dx.doi.org/10.1134/S0081543820030104}{\emph{Proc. Steklov Inst. Math.} {\bfseries 309} (2020) 127}, [\href{https://arxiv.org/abs/1904.04495}{{\ttfamily 1904.04495}}].

\bibitem{AH:2020rfq}
A.~A.~H., A.~Khairnar and A.~Kundu, \emph{{Generalized BMS algebra at timelike infinity}}, \href{http://dx.doi.org/10.1103/PhysRevD.103.104030}{\emph{Phys. Rev. D} {\bfseries 103} (2021) 104030}, [\href{https://arxiv.org/abs/2005.05209}{{\ttfamily 2005.05209}}].

\bibitem{Chakraborty:2021sbc}
S.~Chakraborty, D.~Ghosh, S.~J. Hoque, A.~Khairnar and A.~Virmani, \emph{{Supertranslations at timelike infinity}}, \href{http://dx.doi.org/10.1007/JHEP02(2022)022}{\emph{JHEP} {\bfseries 02} (2022) 022}, [\href{https://arxiv.org/abs/2111.08907}{{\ttfamily 2111.08907}}].

\bibitem{Fuentealba:2022xsz}
O.~Fuentealba, M.~Henneaux and C.~Troessaert, \emph{{Logarithmic supertranslations and supertranslation-invariant Lorentz charges}}, \href{http://dx.doi.org/10.1007/JHEP02(2023)248}{\emph{JHEP} {\bfseries 02} (2023) 248}, [\href{https://arxiv.org/abs/2211.10941}{{\ttfamily 2211.10941}}].

\bibitem{Compere:2023qoa}
G.~Comp\`ere, S.~E. Gralla and H.~Wei, \emph{{An asymptotic framework for gravitational scattering}}, \href{http://dx.doi.org/10.1088/1361-6382/acf5c1}{\emph{Class. Quant. Grav.} {\bfseries 40} (2023) 205018}, [\href{https://arxiv.org/abs/2303.17124}{{\ttfamily 2303.17124}}].

\end{thebibliography}\endgroup

\end{document}